\documentclass[12pt,a4paper,openright]{report}
\usepackage{amsmath}
\usepackage{subcaption}
\usepackage{amsmath}
\usepackage{mathtools}
\usepackage{amsfonts}
\usepackage{amssymb}
\usepackage{fontenc}
\usepackage{makecell}
\usepackage[T1]{fontenc}
\usepackage{pifont}
\usepackage[utf8]{inputenc}
\usepackage[french]{babel}
\usepackage{lipsum}
\usepackage{bbold}
\usepackage{pdfpages}
\usepackage{setspace}
\usepackage{float}
\usepackage{url}
\usepackage{wasysym}
\usepackage{tcolorbox}
\usepackage{pst-node}
\usepackage{fancybox}
\usepackage{lmodern}
\usepackage{graphicx,wrapfig}
\usepackage{titlesec}
\usepackage{sidecap}
\usepackage[tikz]{bclogo}
\usepackage{varwidth}
\usepackage[Conny]{fncychap} 
\usepackage{epigraph}
\usepackage[vlined,algoruled]{algorithm2e}
\usepackage{lipsum}
\usepackage{graphicx}
\usepackage{caption}
\usepackage{listings}
\usepackage{lipsum}
\usepackage{hyperref}
\hypersetup{
	colorlinks=true,
	linkcolor=blue,
	filecolor=magenta,      
	urlcolor=cyan,
	pdftitle={Overleaf Example},
	pdfpagemode=FullScreen,
	citecolor=blue
}
\usepackage[toc]{appendix}
\usepackage[titles]{tocloft}
\usepackage{amsmath}
\usepackage{tcolorbox}
\usepackage{braket}
\usetikzlibrary{calc}
\usepackage{minitoc}
\usepackage{braket}
\usepackage[twoside,left=1.7cm,right=1.7cm,top=1.5cm,bottom=1.5cm]{geometry}
\mtcsetrules{*}{off} 
\mtcsettitle{minitoc}{\vrule height0ptwidth0pt depth15pt\relax Sommaire}
\usepackage{silence} 
\usepackage{setspace} 
\usepackage{fancyhdr}
\usepackage{enumitem}
\usepackage{inputenc}
\usepackage[T1]{fontenc}
\makeatletter
\newcommand*{\rom}[1]{\expandafter\@slowromancap\romannumeral #1@}
\makeatother
\usepackage{fancyhdr} 

\newcounter{subsubsubsection}[subsubsection]

\makeatletter
\renewcommand\paragraph{\@startsection{paragraph}{5}{\z@}%
	{3.25ex \@plus1ex \@minus.2ex}%
	{-1em}%
	{\normalfont\normalsize\bfseries}}
\renewcommand\subparagraph{\@startsection{subparagraph}{6}{\parindent}%
	{3.25ex \@plus1ex \@minus .2ex}%
	{-1em}%
	{\normalfont\normalsize\bfseries}}
\def\toclevel@subsubsubsection{4}
\def\toclevel@paragraph{5}
\def\toclevel@paragraph{6}
\def\l@subsubsubsection{\@dottedtocline{4}{7em}{4em}}
\def\l@paragraph{\@dottedtocline{5}{10em}{5em}}
\def\l@subparagraph{\@dottedtocline{6}{14em}{6em}}
\makeatother

\newtheorem{Mth}{Theorem}[section]
\newtheorem{MD}{Definition}[section]
\newtheorem{LM}{Lemma}[subsection]

\newtheorem{ME}{Exemple}[section]
\newtheorem{MPreuve}{Proof}[section]
\newtheorem{MC}{Corollaire}[section]

\newcounter{example}[section]

\usepackage{amsmath}
\makeatletter
\renewcommand{\section}{\@startsection {section}{1}{\z@}%
	{-4.5ex \@plus -1ex \@minus -.2ex}%
	{2.3ex \@plus.2ex}%
	{\normalfont\normalsize\sffamily\bfseries}}
\makeatother
\makeatletter
\renewcommand{\subsection}{\@startsection {subsection}{1}{\z@}%
	{-4.5ex \@plus -1ex \@minus -.2ex}%
	{2.3ex \@plus.2ex}%
	{\normalfont\normalsize\sffamily\bfseries}}
\makeatother

\newcolumntype{R}[1]{>{\raggedleft\arraybackslash }b{#1}}
\newcolumntype{L}[1]{>{\raggedright\arraybackslash }b{#1}}
\newcolumntype{C}[1]{>{\centering\arraybackslash }b{#1}}
\usepackage{enumitem}
\newlist{abbrv}{itemize}{1}
\setlist[abbrv,1]{label=,labelwidth=1in,align=parleft,itemsep=0.1\baselineskip,leftmargin=!}
\usepackage{longtable}

\newcommand{\chaptertoc}[1]{\chapter*{#1}
	\addcontentsline{toc}{chapter}{#1}
	\markboth{\slshape\MakeUppercase{#1}}{\slshape\MakeUppercase{#1}}}
\providecommand{\openone}{\leavevmode\hbox{\small1\kern0pt\normalsize1}}
\makeatletter
\patchcmd{\@makechapterhead}{\vspace*{50\p@}}{\vspace*{25\p@}}{}{}
\patchcmd{\@makeschapterhead}{\vspace*{50\p@}}{\vspace*{25\p@}}{}{}
\makeatother

\usepackage{lettrine}
\usepackage{blindtext} 
\usepackage{epigraph}
\usepackage{xintexpr}

\title{Epigraph example}
\author{Overleaf}

\usepackage{transparent}
\usepackage{eso-pic}
\usepackage{tikz}
\usetikzlibrary{calc}

\AddToShipoutPicture*{
	\unitlength=0.5cm
	\put(2,1){
		\parbox[b][\paperheight]{\paperwidth}{%
			\vfill
			\centering
			{\transparent{0.5}\includegraphics[width=1.5\textwidth]{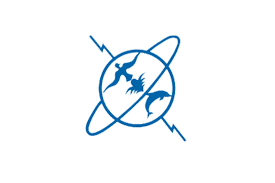}}
			
		}
	}
}
\usepackage[object=vectorian]{pgfornament} 
\usepackage{lipsum,tikz}

\newcounter{boxlblcounter}  
\newenvironment{boxlabel}
{\begin{list}
		{\arabic{boxlblcounter}}
		{\usecounter{boxlblcounter}
			\setlength{\labelwidth}{3em}
			\setlength{\labelsep}{0em}
			\setlength{\itemsep}{2pt}
			\setlength{\leftmargin}{1.5cm}
			\setlength{\rightmargin}{2cm}
			\setlength{\itemindent}{0em} 
			
		}
	}
	{\end{list}}
\dominitoc
\begin{document}
	\begin{titlepage}
		\begin{figure}
			\begin{center}
			\vspace{-1cm}	\includegraphics[width=0.6\linewidth]{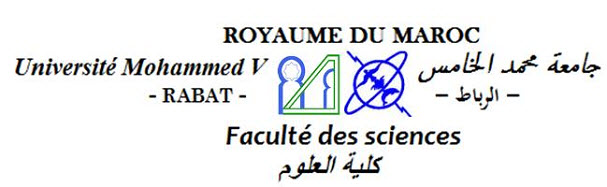}
			\end{center}		
		\end{figure}
		\vspace{-3cm} 	
		\begin{center}
			\begin{minipage}{18cm}
				\vspace*{-2.5cm}
				\begin{center}
					{\textcolor{blue}{\fontfamily{pnc}{\selectfont
								{ \textit{CENTRE D’ETUDES DOCTORALES - SCIENCES ET TECHNOLOGIES}}
								\vskip .2cm
								\noindent\hrule height 2pt\vskip 0.2ex\nobreak 
								{\textcolor{green}{
										\noindent\hrule height 2pt \vskip 0.2ex}}
					}}}
				\end{center}
				\begin{flushright}
					\textbf{N° d’ordre:} 3663
				\end{flushright}
			\end{minipage}
			
		\end{center}
		\begin{center}
			{\Huge \textbf{THÈSE}}\\
			\vspace*{0.2cm}
			En vue de l’obtention du : {\large \textbf{\textit{DOCTORAT}}}
		\end{center}
	\vspace{0.2cm}
		\begin{minipage}{18cm}
			
			{\textcolor{black}{\fontfamily{pnc}{\selectfont
	{\textbf{Structure de Recherche} : {{Physique des Hautes Énergies- Modélisation et Simulation}}\\
	\textbf{Discipline} :{ {Physique / Physics}}\\
	\textbf{Spécialité} : {{Physique Quantique-Statistique / Quantum Physics-Statistical }}}}
					\vspace{0.2cm}
					{\textcolor{blue}{
							\noindent\hrule height 2pt\vskip 0.2ex\nobreak} 
						{\textcolor{green}{
								\noindent\hrule height 2pt \vskip 0.2ex}}
			}}}
		\end{minipage}
		
		\begin{minipage}{18cm}
			\vspace{0.15cm}
			\begin{center}
				\textbf{Présentée et soutenue le : 23/07/2022 par :}
				\normalsize  \\ \vspace{0.15cm}
				\large \textbf{{\Large \underline{Lahcen BAKMOU}}}
			\end{center}
		\end{minipage}
		\vspace{0.28cm}
		\begin{center}
			\begin{minipage}{17cm}
				\begin{center}
					{\LARGE \textbf{Advantages of quantum mechanics in \\
					\vspace{0.2cm}the estimation theory}}
				\end{center}
			\end{minipage}
		\end{center}
	\vspace{0.13cm}
		\begin{center}
			\large \textbf{JURY}
		\end{center}
	\vspace{-1cm}
		\begin{center}
			\begin{tabular}{l l l c}
    \hspace{-1cm}	{\textbf{Mohamed BENNAI}}& {PES} &{Université Hassan \rom{2}, Faculté } &  {Président/Rapporteur}\\
	\vspace*{0.12cm}
		\hspace{-1cm}	{\textbf{}}& {} &{des Sciences, Ben M'Sik, Casablanca.} &  {}\\
	\vspace*{0.12cm}
	\hspace{-1cm}	{\textbf{Rachid AHL LAAMARA}}& {PH} &{Université Mohammed \rom{5}, Faculté} &\hspace{0.3cm} {Rapporteur/Examinateur}\\	
	\vspace*{0.12cm}
	\hspace{-1cm}	{\textbf{}}& {} &{des sciences, Rabat.} &  {}\\
	\vspace*{0.12cm}
	\hspace{-1cm}	{\textbf{Elmehdi SALMANI }}& {PH} &{Université Mohammed \rom{5}, Faculté} &  {Rapporteur/Examinateur}\\
	\vspace*{0.12cm}
	\hspace{-1cm}	{\textbf{}}& {} &{ des sciences, Rabat} &  {}\\
	\vspace*{0.12cm}
	\hspace{-1cm}	{\textbf{Morad EL BAZ}}& {PES} &{Université Mohammed \rom{5}, Faculté } &\hspace{0.4cm} {Examinateur}\\
	\vspace*{0.12cm}
	\hspace{-1cm}	{\textbf{}}& {} &{des sciences, Rabat.} &  {}\\
	\vspace*{0.12cm}
	\hspace{-1cm}	{\textbf{Mohammed EL FALAKI}}& {PH} &{Université Chouaib Doukkali, Faculté } & \hspace{0.4cm} {Examinateur}\\
	\vspace*{0.12cm}
	\hspace{-1cm}	{\textbf{}}& {} &{ des Sciences, El Jadida.} &  {}\\
	\vspace*{0.12cm}
	\hspace{-1cm}	{\textbf{Mohammed DAOUD}}& {PES} &{Université Ibn Tofail, Faculté } &  {Directeur de thèse}\\
			\hspace{-1cm}	{\textbf{}}& {} &{ des Sciences, Kénitra.} &  {}\\
			\end{tabular}
		\end{center}
\vspace{0.3cm}
		\begin{center}
			\begin{minipage}{15cm}
				\begin{center}
					{\textcolor{black}{\fontfamily{pnc}{\selectfont
								{ Année Universitaire : {2021/2022}}
								\vspace{0.2cm}
								\noindent\hrule height 1.79pt\vskip 0.2ex\nobreak
					}}}
				\end{center}
			\end{minipage}
		
			{\large \XBox} Faculté des Sciences, avenue Ibn Battouta, BP. 1014 RP, Rabat –Maroc\\
			\phone \hspace*{0.2cm}00212(0) 37 77 18 76, \hspace*{0.1cm} {\large \bell}Fax:\hspace*{0.1cm} 00212(0) 37 77 42 61 ; http://www.fsr.um5.ac.ma
		\end{center}
	\end{titlepage}
\newgeometry{left=1.8cm,right=1.8cm,top=2.5cm,bottom=2.5cm}
	\pagenumbering{roman} 
	\chaptertoc{Acknowledgment}
	\vspace{-2cm}
\textit{During the preparation of this thesis, many things changed in my life, both personally and scientifically. Before diving into the content of this thesis, I would like to express my gratitude and thanks to those who have been present in this period of my life and have contributed directly or indirectly to this final work.}

\textit{First, I would like to thank \textbf{Mr. El Hassan SAIDI} Professor at the Faculty of Science of Rabat for giving me the opportunity to enroll in a Master's degree in mathematical physics, which in turn allowed me to enroll in the Ph.D. program at the Laboratory of High Energy Physics, Modelisation, and Simulation (LPHE-MS). I appreciate their advice and encouragement throughout this work. My sincere thanks for their availability as well as the efforts they have always made for the success of several activities of the LPHE-MS.}
	
\textit{In addition to the director of LPHE-MS laboratory, I would like to warmly thank my supervisor \textbf{Mr. Mohammed DAOUD} Professor at faculty of science of Kenitra for his magnificent guidance throughout my graduate studies. Without him, this thesis would never have been possible. I appreciate his continuous support and guidance with formidable motivation and support in my studies for my Master’s and Ph.D.}

\textit{My regards also go to \textbf{Mr. Mohamed BENNAI}, professor at the Faculty of Science Ben M’Sik of Casablanca, who accepted the presidency of the jury also agreeing to be a reporter of my modest work and for his bearing the hardships of traveling from Casablanca to Rabat to be with the jury members. I also acknowledge him for his help, discussion, encouragement, and motivation.}

\textit{Let me also thank the rapporteurs of the thesis: I will start with \textbf{Mr. Rachid AHL LAAMARA}, professor at the Faculty of Sciences of Rabat. I would like to thank him warmly for welcoming me to the LPHE-MS and for accepting the invitation to be the rapporteur and examiner of my thesis. His continuous encouragement gave me the strength and courage to complete this thesis as well as for his teaching of the group theory course and Lie algebra course.}

\textit{My acknowledgment also goes to \textbf{Mr.  Elmehdi SALMANI}, Professor at the Faculty of Sciences of Rabat, for accepting to be the reporter of my thesis.  I thank him for his interest and responsibility as an examiner of this work. Yours sincerely.}

\textit{My sincere gratitude also goes to \textbf{Mr. Morad EL BAZ}, professor at the faculty of sciences of Rabat for accepting to be with the jury members of this thesis who accepted as an examiner. I thank him for his interest and responsibility as an examiner. Please accept, Sir, the expression of my respect and deep gratitude.}

\textit{My sincere gratitude also goes to \textbf{ Mr. Mohammed EL FALAKI}, Professor at at the Faculty of Sciences of El Jadida, for accepting to be with the jury members of this thesis. I thank him for his interest and responsibility as an examiner as well as for having endured the difficulties of the trip from El Jadida to Rabat to be with the jury members of this thesis.}

\textit{I would also like to thank all members, past and present, of the LPHE-MS, with whom I have had the pleasure to interact, for the many valuable discussions and shared moments.}
	
\textit{Most importantly, none of this could have happened without my family. My father, mother, brothers, and sisters, I sincerely thank them for their encouragement, motivation, and support which gave me the strength and courage to complete this thesis.}

\newgeometry{left=2cm,right=2cm,lines=60,top=0.3in}
\setstretch{1.4}
\chaptertoc{Abstract}
	\vspace{-3cm}	\lettrine[lines=2]Q{uantum} estimation theory is a reformulation of random statistical theory with the modern language of quantum mechanics. Since the mathematical language of quantum mechanics is operator theory, then the probability distribution functions of conventional statistics are replaced by the density operator appearing in its quantum counterpart. Thus, the density operator plays a role similar to that of probability distribution functions in classical probability theory and statistics. However, the use of the probability distribution functions in classical theories is founded on premises that seem intuitively clear enough. Whereas in quantum theory, the situation with operators is different due to its non-commutativity nature. By exploiting this difference, quantum estimation theory aims to attain ultra-measurement precision that would otherwise be impossible with classical resources. In this thesis, we reviewed all the fundamental principles of classical estimation theory. Next, we extend our analysis to quantum estimation theory. Due to the non-commutativity of quantum mechanics, we prove the different families of QFIs and the corresponding QCRBs. We compared these bounds and discussed their accessibility in the single-parameter and multiparameter estimation cases. We also introduce HCRB as the most informative alternative bound suitable for multiparameter estimation protocols. Since the quantum state of light is the most accessible in practice, we studied the quantum estimation theory with the formalism of these types of quantum states. We formulate, with complete generality, the quantum estimation theory for Gaussian states in terms of their first and second moments.  Furthermore, we address the motivation behind using Gaussian quantum resources and their advantages in reaching the standard quantum limits under realistic noise. In this context, we propose and analyze a measurement scheme that aims to exploit quantum Gaussian entangled states to estimate the displacement parameters under a noisy Gaussian environment. 
	
\newgeometry{left=2cm,right=2cm,lines=60,top=0.3in}
\chaptertoc{Résumé en français}
\vspace{-3cm}
	\lettrine[lines=2]L{a théorie} de l'estimation quantique est une reformulation de la théorie statistique aléatoire avec le langage moderne de la mécanique quantique. Puisque le langage mathématique de la mécanique quantique est basé sur la théorie des opérateurs, la fonction de densité de probabilité des statistiques conventionnelles est remplacée par l'opérateur de densité apparaissant dans sa contrepartie quantique. Ainsi, l'opérateur de densité joue un rôle similaire à celui de la fonction de densité de probabilité dans la théorie classique des probabilités et des statistiques. Cependant, l'utilisation des fonctions de distribution de probabilité dans les théories classiques est fondée sur des prémisses qui semblent intuitivement assez claires. Alors qu'en théorie quantique, la situation des opérateurs est différente en raison de leur nature non-commutative. En exploitant cette différence, la théorie de l'estimation quantique vise à atteindre une ultra-précision de mesure qui serait autrement impossible avec les ressources classiques. Dans cette thèse, nous avons passé en revue tous les principes fondamentaux de la théorie de l'estimation classique. Ensuite, nous étendons notre analyse à la théorie de l'estimation quantique. En raison de la non-commutativité de la mécanique quantique, nous prouvons les différentes familles de QFIs et les QCRBs correspondants. Nous avons comparé ces bornes et discuté de leur accessibilité dans les cas d'estimation à un et plusieurs paramètres. Nous présentons également le HCRB comme la limite alternative la plus informative adaptée aux protocoles d'estimation multiparamètres. L'état quantique de la lumière étant le plus accessible en pratique, nous avons étudié la théorie de l'estimation quantique avec le formalisme de ces types d'états quantiques. Nous formulons, avec une généralité complète, la théorie de l'estimation quantique pour les états gaussiens en termes de leurs premiers et seconds moments.  En outre, nous abordons la motivation derrière l'utilisation des ressources quantiques gaussiennes et leurs avantages pour atteindre les limites quantiques standard sous un bruit réaliste. Dans ce contexte, nous proposons et analysons un schéma de mesure qui vise à exploiter les états quantiques gaussiens intriqués pour estimer les paramètres de déplacement dans un environnement gaussien bruyant.\\
\underline{\bf Mots clés:} Théorie de l'estimation classique, Théorie de l'estimation quantique, Etat Gaussien, Canaux gaussiens bruyants, limite quantique standard, intrication.\\
 \newgeometry{left=2cm,right=2cm,top=2cm,bottom=2cm}
\chaptertoc{List of publications}
\section*{{\large Publications included in the content of this thesis}}
\begin*{}{}
\makeatletter
\providecommand \@ifxundefined [1]{%
	\@ifx{#1\undefined}
}%
\providecommand \@ifnum [1]{%
	\ifnum #1\expandafter \@firstoftwo
	\else \expandafter \@secondoftwo
	\fi
}%
\providecommand \@ifx [1]{%
	\ifx #1\expandafter \@firstoftwo
	\else \expandafter \@secondoftwo
	\fi
}%
\providecommand \natexlab [1]{#1}%
\providecommand \enquote  [1]{``#1''}%
\providecommand \bibnamefont  [1]{#1}%
\providecommand \bibfnamefont [1]{#1}%
\providecommand \citenamefont [1]{#1}%
\providecommand \href@noop [0]{\@secondoftwo}%
\providecommand \href [0]{\begingroup \@sanitize@url \@href}%
\providecommand \@href[1]{\@@startlink{#1}\@@href}%
\providecommand \@@href[1]{\endgroup#1\@@endlink}%
\providecommand \@sanitize@url [0]{\catcode `\\12\catcode `\$12\catcode
	`\&12\catcode `\#12\catcode `\^12\catcode `\_12\catcode `\%12\relax}%
\providecommand \@@startlink[1]{}%
\providecommand \@@endlink[0]{}%
\providecommand \url  [0]{\begingroup\@sanitize@url \@url }%
\providecommand \@url [1]{\endgroup\@href {#1}{\urlprefix }}%
\providecommand \urlprefix  [0]{URL }%
\providecommand \Eprint [0]{\href }%
\providecommand \doibase [0]{http://dx.doi.org/}%
\providecommand \selectlanguage [0]{\@gobble}%
\providecommand \bibinfo  [0]{\@secondoftwo}%
\providecommand \bibfield  [0]{\@secondoftwo}%
\providecommand \translation [1]{[#1]}%
\providecommand \BibitemOpen [0]{}%
\providecommand \bibitemStop [0]{}%
\providecommand \bibitemNoStop [0]{.\EOS\space}%
\providecommand \EOS [0]{\spacefactor3000\relax}%
\providecommand \BibitemShut  [1]{\csname bibitem#1\endcsname}%
\let\auto@bib@innerbib\@empty
\begin{boxlabel}
	\item \BibitemOpen
	\bibfield  {author} {\bibinfo {author}  { \bibnamefont{Bakmou Lahcen, }\
			\bibfnamefont {Daoud Mohammed,}}\ }\bibfield  {title} {\enquote {\bibinfo {title} {Multiparameter quantum estimation theory in quantum Gaussian states},}\ }\href {\doibase 10.1088/1751-8121/aba770} {\bibfield  {journal} {\bibinfo  {journal} {Journal of Physics A: Mathematical and Theoretical}\ }\textbf {\bibinfo {volume} {53}},\ \bibinfo {pages} {385301} (\bibinfo {year} {2020})}\BibitemShut
	{NoStop}%
	\item \bibfield  {author} {\bibinfo {author} \hspace{0.2cm} { \bibnamefont{Bakmou Lahcen, }\
			\bibfnamefont {Daoud Mohammed,}}\ }\bibfield  {title} {\enquote {\bibinfo {title} {Ultimate precision of joint parameter estimation under noisy Gaussian environment},}\ }\href {\doibase 10.1016/j.physleta.2022.127947} {\bibfield  {journal} {\bibinfo  {journal} {Physics Letters A}\ }\textbf {\bibinfo {volume} {}},\ \bibinfo {pages} {127947} (\bibinfo {year} {2022})}\BibitemShut
	{NoStop}%
\end{boxlabel}
\end*{}
\section*{{\large Publications beyond the scope of this thesis}}
\begin*{}{}
\makeatletter
\providecommand \@ifxundefined [1]{%
	\@ifx{#1\undefined}
}%
\providecommand \@ifnum [1]{%
	\ifnum #1\expandafter \@firstoftwo
	\else \expandafter \@secondoftwo
	\fi
}%
\providecommand \@ifx [1]{%
	\ifx #1\expandafter \@firstoftwo
	\else \expandafter \@secondoftwo
	\fi
}%
\providecommand \natexlab [1]{#1}%
\providecommand \enquote  [1]{``#1''}%
\providecommand \bibnamefont  [1]{#1}%
\providecommand \bibfnamefont [1]{#1}%
\providecommand \citenamefont [1]{#1}%
\providecommand \href@noop [0]{\@secondoftwo}%
\providecommand \href [0]{\begingroup \@sanitize@url \@href}%
\providecommand \@href[1]{\@@startlink{#1}\@@href}%
\providecommand \@@href[1]{\endgroup#1\@@endlink}%
\providecommand \@sanitize@url [0]{\catcode `\\12\catcode `\$12\catcode
	`\&12\catcode `\#12\catcode `\^12\catcode `\_12\catcode `\%12\relax}%
\providecommand \@@startlink[1]{}%
\providecommand \@@endlink[0]{}%
\providecommand \url  [0]{\begingroup\@sanitize@url \@url }%
\providecommand \@url [1]{\endgroup\@href {#1}{\urlprefix }}%
\providecommand \urlprefix  [0]{URL }%
\providecommand \Eprint [0]{\href }%
\providecommand \doibase [0]{http://dx.doi.org/}%
\providecommand \selectlanguage [0]{\@gobble}%
\providecommand \bibinfo  [0]{\@secondoftwo}%
\providecommand \bibfield  [0]{\@secondoftwo}%
\providecommand \translation [1]{[#1]}%
\providecommand \BibitemOpen [0]{}%
\providecommand \bibitemStop [0]{}%
\providecommand \bibitemNoStop [0]{.\EOS\space}%
\providecommand \EOS [0]{\spacefactor3000\relax}%
\providecommand \BibitemShut  [1]{\csname bibitem#1\endcsname}%
\let\auto@bib@innerbib\@empty
\begin{boxlabel}
	\item\BibitemOpen
	\bibfield  {author} {\bibinfo {author}  { \bibnamefont{Bakmou Lahcen, }\
			\bibfnamefont {Daoud Mohammed, Slaoui Abdallah, Ahl Laamara, Rachid}}\ }\bibfield  {title} {\enquote {\bibinfo {title} {Quantum Fisher information matrix in Heisenberg XY model},}\ }\href {\doibase 10.1007/s11128-019-2282-x} {\bibfield  {journal} {\bibinfo  {journal} {Quantum Information Processing}\ }\textbf {\bibinfo {volume} {18}},\ \bibinfo {pages} {1--20} (\bibinfo {year} {2019})}\BibitemShut
	{NoStop}%
	\item \bibfield  {author} {\bibinfo {author} \hspace{0.2cm} { \bibnamefont{Slaoui Abdallah, }\
			\bibfnamefont {Bakmou Lahcen, Daoud Mohammed, Ahl Laamara, Rachid}}\ }\bibfield  {title} {\enquote {\bibinfo {title} {A comparative study of local quantum Fisher information and local quantum uncertainty in Heisenberg XY model},}\ }\href {\doibase 10.1016/j.physleta.2022.127947} {\bibfield  {journal} {\bibinfo  {journal} {Physics Letters A}\ }\textbf {\bibinfo {volume} {383}},\ \bibinfo {pages} {2241--2247} (\bibinfo {year} {2019})}\BibitemShut
	{NoStop}%
	\item \bibfield  {author} {\bibinfo {author} \hspace{0.2cm} { \bibnamefont{Y. Lahlou}\
			\bibfnamefont {L. Bakmou, B. Maroufi, M. Daoud,}}\ }\bibfield  {title} {\enquote {\bibinfo {title} {Quantifying quantum correlations in noisy Gaussian channels},}\ }\href {\doibase 10.1007/s11128-022-03597-4} {\bibfield  {journal} {\bibinfo  {journal} {Quantum Information Processing}\ }\textbf {\bibinfo {volume} {21}},\ \bibinfo {pages} {1--15} (\bibinfo {year} {2022})}\BibitemShut
	{NoStop}%
\end{boxlabel}
\end*{}
\vspace{0.7cm}
\begin*{}{}
\makeatletter
\providecommand \@ifxundefined [1]{%
	\@ifx{#1\undefined}
}%
\providecommand \@ifnum [1]{%
	\ifnum #1\expandafter \@firstoftwo
	\else \expandafter \@secondoftwo
	\fi
}%
\providecommand \@ifx [1]{%
	\ifx #1\expandafter \@firstoftwo
	\else \expandafter \@secondoftwo
	\fi
}%
\providecommand \natexlab [1]{#1}%
\providecommand \enquote  [1]{``#1''}%
\providecommand \bibnamefont  [1]{#1}%
\providecommand \bibfnamefont [1]{#1}%
\providecommand \citenamefont [1]{#1}%
\providecommand \href@noop [0]{\@secondoftwo}%
\providecommand \href [0]{\begingroup \@sanitize@url \@href}%
\providecommand \@href[1]{\@@startlink{#1}\@@href}%
\providecommand \@@href[1]{\endgroup#1\@@endlink}%
\providecommand \@sanitize@url [0]{\catcode `\\12\catcode `\$12\catcode
	`\&12\catcode `\#12\catcode `\^12\catcode `\_12\catcode `\%12\relax}%
\providecommand \@@startlink[1]{}%
\providecommand \@@endlink[0]{}%
\providecommand \url  [0]{\begingroup\@sanitize@url \@url }%
\providecommand \@url [1]{\endgroup\@href {#1}{\urlprefix }}%
\providecommand \urlprefix  [0]{URL }%
\providecommand \Eprint [0]{\href }%
\providecommand \doibase [0]{http://dx.doi.org/}%
\providecommand \selectlanguage [0]{\@gobble}%
\providecommand \bibinfo  [0]{\@secondoftwo}%
\providecommand \bibfield  [0]{\@secondoftwo}%
\providecommand \translation [1]{[#1]}%
\providecommand \BibitemOpen [0]{}%
\providecommand \bibitemStop [0]{}%
\providecommand \bibitemNoStop [0]{.\EOS\space}%
\providecommand \EOS [0]{\spacefactor3000\relax}%
\providecommand \BibitemShut  [1]{\csname bibitem#1\endcsname}%
\let\auto@bib@innerbib\@empty
\end*{}


	
	 \newgeometry{left=1.7cm,right=1.7cm,top=2.5cm,bottom=2.5cm}
	\renewcommand\contentsname{Contents}
	\tableofcontents
	\addstarredchapter{Contents} 
	\adjustmtc
	\addstarredchapter{List of Abbreviations}
	\adjustmtc
	 \newgeometry{left=1.7cm,right=1.7cm,top=2.5cm,bottom=2.5cm}
	\chaptertoc{{List of Abbreviations}}
	\renewcommand*{\arraystretch}{1.37}
	\begin{longtable}{@{}l @{\hspace{5mm}} l }
		PDF   \hspace{3cm}              &Probability density function\\
		PMF   \hspace{3cm}              &Probability mass function\\
		CCRB   \hspace{3cm}             &Classical Cramér-Rao bound\\
		CFI    \hspace{3cm}             &Classical Fisher information\\
		CFIM   \hspace{3cm}             &Classical Fisher information matrix\\
		MSE    \hspace{3cm}             &Mean square error \\
		MLE    \hspace{3cm}             &Maximum likelihood estimator\\
		iid    \hspace{3cm}             &independent and identical distribution\\
		CRLB    \hspace{3cm}            &Cramér-Rao lower bound\\
		POVM    \hspace{3cm}            &Positive operator-valued measures\\
		PVM     \hspace{3cm}            &Projection valued measures\\
		QFI     \hspace{3cm}            &Quantum Fisher information\\
		QCRB    \hspace{3cm}            &Quantum Cramér-Rao bound\\
		LD      \hspace{3cm}            &logarithmic derivative\\
		SLD    \hspace{3cm}             &Symmetric logarithmic derivative\\
		RLD     \hspace{3cm}            &Right logarithmic derivative\\
		QFIM     \hspace{3cm}           &Quantum Fisher information matrix\\
		LIGO      \hspace{3cm}          &Laser Interferometer Gravitational-Wave Observatory\\
		HCRB   \hspace{3cm}             & Holevo Cramér-Rao bound \\
		WMSE   \hspace{3cm}             &Weight mean square error\\
		SQL    \hspace{3cm}             &Standard quantum limit\\
		HL      \hspace{3cm}            &Heisenberg limit\\
		CV      \hspace{3cm}            &Continuous variables\\
		LO      \hspace{3cm}            &Local oscillator\\
		OPA     \hspace{3cm}            &Optical parametric amplifier\\
		BS      \hspace{3cm}            &Beam splitter\\
		HDB     \hspace{3cm}            &Homodyne detection bound\\
		EPR     \hspace{3cm}            &Einstein-Podolsky-Rosen\\
		TMSV    \hspace{3cm}            &Two-mode squeezed vacuum\\
		TMDV    \hspace{3cm}            &Two-mode displacement vacuum\\
		TMST    \hspace{3cm}            &Two-mode displacement thermal\\
		TMDT    \hspace{3cm}            &Two-mode displacement thermal\\
		LOCC    \hspace{3cm}            & Local operations and classical communications\\
		ALD    \hspace{3cm}             & anti-symmetric logarithmic derivative\\
	\end{longtable}

	\chaptertoc{{List of Symbols}}
			\vspace{-2cm}
	\renewcommand*{\arraystretch}{1.37}
	\begin{longtable}{@{}l @{\hspace{5mm}} l }
	Bold-face characters&\hspace{3cm} denote vectors or matrices.\\	
	Mild-face characters& \hspace{3cm}	denote scalar.\\
	$\theta_{N}^{est}\left(X\right)$ & \hspace{3cm} denote estimator\\	
	$\theta_{N}^{est}\left(x\right)$ & \hspace{3cm} denote the outcome of estimator\\
	$\theta_{N}^{MLE}\left(X\right)$ & \hspace{3cm} denote the maximum likelihood estimator\\
	$\theta_{N}^{MLE}\left(x\right)$ & \hspace{3cm} denote the outcome of maximum likelihood estimator\\
	$*$ & \hspace{3cm} denote complex conjugate of scalar\\
	$\bar{(..)}$ & \hspace{3cm} denote complex conjugate of operator\\
	$\sim$& \hspace{3cm}  denotes is distributed according to\\
	$(...)^{\top}$& \hspace{3cm} denote transpose\\
	$\hat{ (...)}$ &\hspace{3cm} denote operator\\
	$\mathtt{Var}_{\theta}$&\hspace{3cm} denote the variance of an estimator\\
	$\mathtt{E}_{\theta}$&\hspace{3cm} denote the expectation value of an estimator\\
	$\mathtt{Cov}_{\boldsymbol{\theta}}$&\hspace{3cm} denote the covariance of the estimator vector\\
	$\left[\boldsymbol{A}\right]_{ij}$&\hspace{3cm} denote the $ij^{\text{th}}$ elements of matrix $\boldsymbol{A}$\\
		$\left[\boldsymbol{b}\right]_i$& \hspace{3cm} denote $i^{\text{th}}$ element of vector $\boldsymbol{b}$\\
	$\operatorname{Tr \left[...\right]}$ & \hspace{3cm} denote the trace of a matrix\\
	$(...)^\dagger$ & \hspace{3cm} denote adjoint\\
	$(...)^{-1}$& \hspace{3cm} denote inverses\\
	$(...)^{+}$& \hspace{3cm} denote generalized inverses\\
	$\mathfrak{Re}$& \hspace{3cm} denote the real part\\
	$\mathfrak{Im}$& \hspace{3cm} denote the imaginary part\\
	$ {\left\langle ... \right\rangle}$& \hspace{3cm} denote the average value of an operator\\
	$\left[.,.\right], \quad \left\{.,.\right\}$ &\hspace{3cm} denote respectively commutator, anti-commutator\\
	$\mathtt{vec}\left[...\right]$&\hspace{3cm} denote the vec-operator\\
    $\bigotimes, \quad \bigoplus$& \hspace{3cm} denote respectively tensor product, direct sum\\
	$\operatorname{TrAbs}\left[...\right]$& \hspace{3cm} denotes the absolute sum of the eigenvalues of an operator\\
    $\left\|...\right\|_1$& \hspace{3cm} denote the trace norm of a operator\\
 $\left\|.\right\|_{\infty}$& \hspace{3cm} denoted the largest eigenvalue of a operator\\
	
	\end{longtable}
	\thispagestyle{empty}
	\newpage\pagenumbering{arabic} 
	\renewcommand{\thechapter}{\Roman{chapter}}
	\setcounter{chapter}{-1}
		 \newgeometry{left=1.7cm,right=1.8cm,top=2.55cm,bottom=2.55cm}
	\chapter{General Introduction}
\epigraph{\textit{\textbf{{ "You can, for example, never predict what any one man will do, but you can say with precision what an average number will do. Individuals vary, but the percentages remain constant. That's what the statistician says."}}}}{\textit{Sherlock Holmes}}
In the beginning, the estimation theory appeared at the heart of many electronic signal processing systems designed to extract information. These included radars \cite{skolnik1980introduction}, sonar \cite{knight1981digital}, control noise in communication \cite{proakis2001digital}, and image analysis \cite{banham1997digital}. All these applications share the common problem of needing to estimate the values of a set of parameters characterizing them. For example, in the case of the radar system, we wish to determine the rank of an aircraft using airport surveillance radar. In the sonar application, we are also interested in the position of ships in maritime navigation. Another application is noise control, in which we are interested in damping fluctuation. In all these applications and others, we are faced with the problem of inference the values of unknown parameters based on the data set analysis \cite{cox2006principles, kay1993fundamentals}. Mathematically, we have the N-point data set that depends on the unknown parameters, denoted $\boldsymbol{\theta}$, that we will estimate. We hope to determine the values of $\boldsymbol{\theta}$ based on the data set. This is the problem of estimation theory which is the subject of this thesis.
	
	The first step in finding the best estimate values of parameters Q is to model, mathematically, the data sets. As that last is inherently random, we described it by its probability distribution function (PDF). Indeed, the PDF is parameterized by unknown parameters. Thus, we have the class of PDFs that belongs to the usual probability distribution laws. From this PDF, one can try to guess the values of $\boldsymbol{\theta}$ from a realization of the random variables $\boldsymbol{X}$ by calculating the estimator $\boldsymbol{\theta}^{est}$. The most common performance criterion in estimation theory is the mean square error (MSE), which is reduced to the variance in the cases that satisfy the unbiased condition. Under this condition, the best ultimate precision is bounded by the reciprocal of the Fisher information - a measure of how much information the measurable data contains about the unknown parameters \cite{fisher1925theory}. In the framework of estimation theory, this bound is called the Cramér-Rao bound \cite{cramer2016mathematical}, which allows setting a lower bound on the variance of all the possible estimators.  In this direction, the estimators that attain the Cramér-Rao lower bound (CRLB) of the MSE or the variance are named efficient estimators and provide the most accurate estimate from the data sample. This fact means that the fundamental goal of estimation theory is always to find an estimate that saturates the CRB and then improves the accuracy. This goal is motivated by the increased need for more accurate and sensitive detectors.
	
	As ultra-precise measurement schemes require the finest possible revolution technology in the detection, they are then convinced to be limited by the fundamental building block describing the physical nature at the microscope levels. More precisely, improving detection precision requires exploiting the resources of quantum mechanics, which deals with physical phenomena at the nanoscopic scale. In this context, the resources of quantum theory have succeeded in reaching a limit of accuracy that is impossible to deduce using its classical counterpart. In the terminology of quantum mechanics, the estimation theory is known as quantum estimation theory or quantum metrology \cite{paris2009quantum, giovannetti2011advances}. It was formally adopted using the correspondence rule postulated by Niels Bohr in the so-called Copenhagen interpretation of quantum mechanics \cite{bohr1928quantum, heisenberg1958copenhagen, faye2002copenhagen}. This rule also called the correspondence principle, stipulates that a new scientific theory should be able to explain the phenomenon under consideration as long as the earlier theory is valid. For example, Einstein's special relativity satisfies the correspondence principle; because it is reduced to classical mechanics in the limit of velocities small compared to the speed of light. As well as, the theory of general relativity is reduced to Newtonian gravity in the limit of weak gravitational fields. Also, statistical mechanics reproduces thermodynamics when the number of particles is large. Therefore, according to this principle, quantum estimation theory must be reproduce its classical counterpart within certain limits. In estimation terminology, this limit is called the optimal limit, where there is a coincidence between the classical and quantum metrology \cite{paris2009quantum}.
	
	Quantum metrology was initially proposed by the pioneering works of Helstrom \cite{helstrom1969quantum} and Holevo \cite{holevo2011probabilistic} in the 1970s. It has then expanded considerably at both theoretical and experimental levels. In particular, recent progress in quantum estimation theory has been stimulated by the quantum technology revolution, which aims at developing brand-new technologies exploiting quantum phenomena. It has been applied to design high-precision measurements at the quantum frontier by providing theoretical tools for various estimation purposes. These applications include standard frequency estimation \cite{boss2017quantum}, Unruh-Hawking effect estimation \cite{aspachs2010optimal}, magnetic field detection \cite{zhang2014fitting}, the optimal estimation of phases \cite{vidrighin2014joint}, temperature \cite{monras2011measurement}. In the quantum version of the estimation theory, the unknown parameters $\boldsymbol{\theta }$ are encoded in the quantum state that describe the quantum statistical model. Thus, the PDFs that appear in classical statistics are replaced by the quantum density operators that arise in the quantum counterparts. In quantum theory, nanoscale objects generally do not commute with each other, as well as the quantum states can be in a linear superposition of other states. Both results predict that quantum metrology has a more mathematically involved theory than classical metrology, and many notions in the classical theory can find their analogs in the quantum theory. In this context, the CCRB and CFI are part of these concepts and are the principal keys of quantum metrology.
	
	As mentioned earlier, the fundamental keys of estimation theory are the CCRB and CFI, which give a lower bound for the MSE or variance of an unbiased estimator. In the quantum case, due to the non-commutativity of quantum mechanics, several quantum Cramér-Rao bounds (QCRB)s are known, so several families of quantum Fisher information (QFI)s. In particular, the two best studied in the literature are the symmetric logarithmic derivative quantum Fisher information (SLD-QFI) and right logarithmic Fisher information (RLD-QFI), each of which has a corresponding QCRB \cite{suzuki2018classification, katariya2021geometric}. In this sense, early works in quantum metrology are devoted to a single estimation parameter \cite{helstrom1969quantum, boixo2007generalized, abur2004power}. In this case, the SLD-QFI has been extensively studied, as it provides a tight bound of precision and always attains the lower bound. However, the quantum statistical model for the multi-parameter case is more challenging, as it requires dealing with the trade-off between the MSEs of the respective parameters simultaneously. This means, in fact, that we cannot always simultaneously achieve the optimal estimate of the parameters. This difficulty is due to the incompatibility that arises from the inherent quantum nature of the underlying physical system \cite{carmeli2019quantum, sidhu2021tight, albarelli2020perspective}. Attempting to decipher the optimization problem or at least limit it, Yuan \& Lax \cite{yuen1973multiple} derived the bound on the estimation performance by introducing the RLD-QCRB in the expectation parameter estimation theory for quantum Gaussian models. But wait! this success is not generalized to all quantum statistical models. Soon after, A. S. Holevo introduced and studied another Cramér-Rao bound (HCRB) that, in a sense, unifies SLD and RLD-QCRBs \cite{holevo2011probabilistic, holevo1977commutation}. In addition, the HCRB  is the most fundamental scalar lower bound imposed by quantum mechanics on the weighted mean square error (WMSE). The HCRB represents the best precision attainable with a collective measurements on an asymptotically large number of identical copies of a quantum state \cite{albarelli2019evaluating}. 
	
	Despite its success, the HCRB has been used more as a mathematical object in asymptotic quantum statistics than applied to concrete metrological estimation problems. There are several reasons for this. First, the HCRB is hard to evaluate since it is defined via a minimization procedure and is usually not written in an explicit form. Second, implementing collective measurements is generally a difficult task. However, applications of HCRB in metrological problems do exist, although they are few. J. Suzuki found closed-form results for parameter estimation with qubits \cite{suzuki2016explicit} and explored the connections between different types of metrological bounds in the specific case of the two-parameter estimation theory. For pure states, the HCRB has been achieved by single-copy measurements\cite{bradshaw2017tight}. Also, it has been used as a tool to define the precision of state estimation for finite-dimensional quantum systems \cite{yang2019attaining}. Without any doubt, the HCRB is the most relevant point in multi-parameter estimation theory. Recently, an increasing number of true multi-parameter protocols have been explored. For instance, thermometry \cite{correa2015individual}, microscopy \cite{hess2006ultra}, super-resolution quantum imaging \cite{tsang2016quantum}, as well as gravitational-wave detection by the use of large interferometers such as VIRGO \cite{acernese2014advanced} and LIGO \cite{abbott2009ligo}. All these applications require using the procedure of multi-parameter quantum metrology, and therefore the need for generally attainable bounds is urgent. More recently, Albarelli et al. have numerically investigated the HCRB for finite-dimensional systems \cite{albarelli2019evaluating}. They recast the evaluation of HCRB as a semi-definite program, which is an optimization problem that can be implemented efficiently. But generally, to date, no general analytical expression for HCRB is known, and then the theory of multi-parameter quantum estimation is not yet completed. This will be an incentive to emerge the new works in this field.

	However, the above attempt did not treat the performance bounds of quantum estimation with infinite-dimensional quantum systems, which has a fundamental importance in many areas of science and new quantum technology. Applications include; distance measurements with laser range finders or radar \cite{kruapech2010laser}, measurement of the shape and composition of objects in microscopy and spectroscopy \cite{boss2017quantum, schmitt2017submillihertz}, and angular velocities with laser gyroscopes \cite{ciminelli2010photonic}. In all these applications, the estimation parameters are encoded in the state of light and then tries to extract its values by suitably detecting the light. The state of light is described using a bosonic mode of the electromagnetic field, which is, in turn,  described as a quantum harmonic oscillator with quadratic field operators that satisfy the Heisenberg uncertainty \cite{heisenberg1949physical, robertson1929uncertainty}. Since quadratic field operators have a continuous eigenvalue spectrum,  the bosonic mode is a continuous-variable system, which cannot be represented on a finite-dimensional Hilbert space. In this case, the quantum statistical model has an operator density acting in the Hilbert space of infinity-dimensional. Due to the infinite dimensionality of these types of quantum statistical models,  evaluating different precision bounds becomes more challenging from a mathematical point of view. Most remarkably, these difficulties are alleviated by the Gaussian state formalism \cite{weedbrook2012gaussian, ferraro2005gaussian, olivares2012quantum}. 
	
	Like all disciplines of quantum information theory, quantum metrology also seeks to take advantage of the simplicity and power of the Gaussian state formalism to bring us closer to understanding more precise estimation with continuous variables systems. In addition to quantum optics, where they have proven an applicable description of the quantum state available in laboratories: laser fields manipulated with passive and active linear optical elements \cite{schuller2010plasmonics, jia2014monolithic}, Gaussian states also appear naturally in the description of optomechanical and nanomechanical oscillators \cite{tian2010optical, nunnenkamp2011single} as well as gases of cold atom and ionic traps \cite{harter2014cold}. Moreover, Gaussian states are easily accessible from a theoretical point of view since they are fully described by a finite number of degrees of freedom. All these great successes are the motivation that led to the development of many works devoted to quantum metrology with the Gaussian state formalism. The first partial results of these works include those by A. Monras, who succeeded in unifying, firstly, the SLD-QFI with the Gaussian formalization using phase space analysis \cite{monras2013phase}. Soon after,  R. Nichols et al \cite{nichols2018multiparameter}, and D. Šafránek \cite{vsafranek2018estimation} succeeded in generalizing these results to the multiparameter case, i.e. finding the SLD-QFIM. Bakmou et al \cite{bakmou2020multiparameter} succeeded in expressing both SLD and RLD-QFIMs in terms of the first and second moments of the Gaussian state. All these successes have many practical applications, including the ones examined by M. G. Genoni et al. on the optimal estimation of joint parameters in phase space \cite{genoni2013optimal}, one by M. Aspachs et al. on optical phase estimation \cite{aspachs2009phase}, and the one by R. Demkowicz-Dobrzański et al. on the quantum limits in optical interferometry \cite{demkowicz2015quantum}. 
	
	In many of these settings, the estimation problem can be modeled as the estimation of unknown parameters encoded on a probe field initialized in a Gaussian state. That last has been preserved its Gaussian characteristic during certain transformations; even unitary ones such as that associated with the symplectic forms in the phase space \cite{weedbrook2012gaussian, ferraro2005gaussian, olivares2012quantum}, or non-unitary ones such as those describing the noisy dynamics Markovian \cite{bellman1957markovian} and non-Markovian \cite{diosi1998non} produced by the inevitable interaction with the environment. Both unitary and noisy transformations of these kinds of quantum states are known as Gaussian channels. In these quantum channels, the task of quantum metrology focuses on making ultra-precision measurements exploiting quantum resources in the initial preparation state. Among these resources, we find entanglement \cite{horodecki2009quantum} and superposition \cite{friedman2000quantum}. According to references \cite{genoni2013optimal, giovannetti2011advances, hyllus2012fisher, przysikezna2015quantum}, the entanglement is resource use to improve the precision of channel parameter estimates. For example,  it has been used for estimating squeezed parameter \cite{gao2014bounds},  joint estimation of displacement parameters \cite{genoni2013optimal}, and phase estimation \cite{dorner2009optimal}. However, all these applications are devoted to estimating the unknown parameters imposed by unitary transformations. Concerning open quantum systems and non-unitary channels, quantum metrology with quantum entanglement has been applied extensively for finite-dimensional systems. For example, it has been exploited to optimally estimate the noise parameter of depolarizing \cite{fujiwara2001quantum} or amplitude-damping channels \cite{ji2008parameter}. Unfortunately, quantum metrology generally for infinite-dimensional systems remains limited, except for the estimation parameters of a loss bosonic channel \cite{monras2007optimal} and optimal phase estimation in a Gaussian environment \cite{oh2019optimal}. Moreover, these mentioned protocols are devoted to single parameter estimation. On the other hand, to the best of our knowledge, very little is known about the use of multi-parameter cases in noisy Gaussian channels. This is one of the issues that we will address in this thesis (see also \cite{bakmou2022ultimate}).
	
	This thesis aims at exploiting the advantage of quantum mechanical resources to achieve ultra-precision measurement that is not possible using classical statistics. In order to accomplish this purpose, we will structure our thesis as follows:
	
	In the first chapter \ref{Ch. 1}, we will review all fundamental concepts of classical estimation theory, such as; the probability distribution of a random variable, classical statistical model, and estimator notion. Then, in the next, we introduce the estimation problems and proven the CCR and CFI in both single and multiparameter cases. We end this chapter by discussing the maximum likelihood principle as an appropriate statistical technique for discovering an efficient estimator. This chapter is presented in a textbook-like method with many examples, which could help readers, step by step, to understand the classical estimation theory.
	
	In the second chapter \ref{Ch. 2}, we will reformulate the estimation theory using the terms of quantum mechanics. And then introduce the quantum statistical model. Using Born's rule as a link to connect classical and quantum statistics, and due to the non-commutativity of quantum mechanics, we prove the different families of QFIs that are SLD and RLD-QFIs. With this in place, we will establish the various SLD and RLD-QCRBs in single and multiparameter estimations problems. Since both SLD and  RLD-QCRBs are not always saturable due to the incompatibility between the estimated parameters, we introduce and discuss the HCRB, which is most appropriate in such cases. We end this chapter by discussing the classification of the quantum statistical model based on the relationship between the different QCRBs and HCRB.
	
	Chapter \ref{Ch. 3} of this thesis is devoted to integrating the quantum metrology and continuous-variable system, specifically quantum Gaussian states, into the unified framework. The most efficient and appropriate way to realize that is to use a phase-space analysis. In this direction, we will review the basic concepts of Gaussian states and their operations. Then, we will provide the analytical expression of the central quantities in multiparameter quantum estimation theory, namely the SLD and RLD-QFIMs. We will investigate the estimation of the optical phase as an example to illustrate how to use these results. This example will clarify the different precision limits available in quantum metrology, which are standard quantum limit \cite{giovannetti2004quantum} and Heisenberg limit \cite{giovannetti2006quantum}. This chapter is presented with many technical details, given in the appendix as a supplementary chapter \ref{Ch. 5}. These details could help readers to understand quantum metrology with phase space analysis.
	
	In chapter \ref{Ch. 4}, we provide a brief review of Gaussian non-unitary channels. We will start with single-mode radiation after extending to the multi-mode radiation case. Then, we present some relevant measurements of continuous-variable systems; homodyne and heterodyne detections. Next, we propose and analyze a measurement scheme of the multiparameter protocol under environmental fluctuation. This scheme aims to exploit entanglement as a resource of quantum mechanics to improve the estimation precision of displacement parameters under the noisy Gaussian channel. To realize this scheme with the quantum states available in labs, we limit our analysis to a general two-mode squeezed displaced thermal state. We conclude this chapter by discussing the role of entanglement in enhancing precision, even with the inevitable existence of environmental fluctuations.
	\newgeometry{left=1.7cm,right=1.8cm,top=2.55cm,bottom=2.55cm}
	\chapter{Classical estimation theory}\label{Ch. 1}
	
	\section{Introduction}
	In the models of estimation problems, the classical estimation theory aims to infer the values of each set of unknown parameters from a data set of empirical observation values. Ordinarily, in the many statistical inference problems, we do not have any information on the probability distribution of the observation values, which contrasts with the case of the models of estimation problems. In that latter and according to the usual probability laws \cite{von1981probability, loeve2017probability}, the probability distribution is well-known in a mathematical statement. Thus, what is not known is the values of the sets of parameters on which the probability distribution depends. In the direction of extracting the best-estimates values of the unknown parameters, the estimation theory generalizes its problem in the following question: "How should using the data observation to infer the best-estimates values of the unknown parameters?" To answer this question, it should define, firstly, what is meant by best-estimates values. The natural and reasonable definition should be that the best-estimates values of the parameters are around nearest to the actual values of the estimates parameters. In addition to this and by consideration,  it is necessary to note that the estimated values of the parameters are the function of the empirical observation values of a specific sample random variable. This last was following unavoidably a particular usual law of the probability distribution function. Since the observation of the data is the inherent realization of a hazard, we should then use statistical methods to obtain the best-estimated values. In this context, there are different statistical methods that we can use to measure the ability of specific distribution, hence measuring the concentration around the actual values of the estimated parameters. In general, each of these methods focuses on various characteristics of the sample random variable and gives distinct criteria for the best estimate. The most relevant and extended statistical methods are ones based on the minimization of MSE or minimized the variance of unbiased estimator \cite{kay1993fundamentals, casella2021statistical, chatfield1995model}. This method is that we will use during this thesis.
	
	This chapter covers the basic concepts of classical estimation theory and introduces many fundamentals that are later necessary. We will review some basic concepts such as; probability distribution of a random variable, statistical estimation model, estimator notion. Next, we will present the most relevant keys of this theory, which are known as CCR bound or CCR inequality and classical Fisher information. In the sets of the proposed estimators, the CCR inequality plays a crucial role to fixes a lower bound on the MSE or the variance. The lower bound implied, in fact, the saturation of CCR inequality. This saturation is not, in general, possible except for some estimation models that satisfy an attainment condition. We will end this chapter by detailing the maximum likelihood principle as a method of determining appropriate statistical techniques for finding an efficient estimator, which is an estimator that satisfies the attainment condition and eventually extracts the best-estimate values of unknown parameters. 
	\section{Parameter estimation problem and probability distributions}
	An experimenter uses information from a sample $X$ to make inferences about unknown parameters. This sample can be associated with a random variable so that the performances of $N$ experiments can be modeled by $X_1, X_2, ..., X_N$, and the realizations (results observed in the experiment) are modeled by $x_1, x_2,...,x_N$. It will be necessary to distinguish during this thesis: we have manipulated the random variables $X_1, X_2, ..., X_N$ whose realizations we do not know. During the measurement task in the experiment, we observe $N$ results $x_1, x_2, ..., x_N$, which are respectively realizations of $X_1, X_2, ..., X_N$, and we then coil to extract information from these data and estimate the unknown parameters. Generally, upper case letters designate random variables and lower case letters correspond to observations and thus to real values.
	
	One of the main purposes of a statistician is to draw conclusions about a certain population by conducting experiments. In general, to study a random experiment, it is necessary to identify all possible outcomes or, in statistical terminology, the probability space.
	\subsection{Probability Space}
	In probability theory, a probability space or probability triple $\left(\mathbf{\Omega}, \mathcal{A}, p\right)$ is a mathematical construct that provides a formal model of a random experiment, consists of three main elements:\\
	$\bullet$ The sample space denoted $\mathbf{\Omega}$ is the set of all outcomes that can be obtained in a random experiment.\\
	$\bullet$ Event space, which is a set of events $\mathcal{A}$ so that for each event there is a result in the sample space.\\
	$\bullet$ A probability distribution function $p\left(\mathcal{A}_i\right)$, which assigns a probability to each event in the event space $\mathcal{A}_i \in \mathcal{A}$.
	
	Now, we consider in a probability space $\left( {\boldsymbol{\Omega} \left( \boldsymbol{X} \right), \mathcal{A}\left( \boldsymbol{x} \right), p\left( \boldsymbol{x} \right)} \right)$, a random experiment whose results are described by the realization of random variable $\boldsymbol{X}$ and $p\left(\boldsymbol{x}\right)$ the probability distribution corresponding to $\boldsymbol{X}$. In fact, there are various ways to approach the problem of constructing a probability distribution $p\left(\boldsymbol{x}\right)$.  If the functional form of $p\left(\boldsymbol{x}\right)$ is already known or can be guessed with close and reasonable accuracy, then the parametric approach is quite suitable. In this context, suppose that $\boldsymbol{X}$ is the random variable subject to a law that belongs to a family of usual laws, i.e. the law of $\boldsymbol{X}$ belongs to a family of parametric laws of statistical description. More precisely, consider that the probability distribution function $p\left(\boldsymbol{x}\right)$ is parameterized by a number of parameters $\boldsymbol{\theta}$ which implies $p\left(\boldsymbol{x}\right) = p\left(\boldsymbol{x}; \boldsymbol{\theta}\right)$, where $\boldsymbol{\theta} = \left(\theta_1, \theta_2, ..., \theta_m\right) \in \Theta$ with $\Theta$ is an open subset of $\mathbb{R}^m$ called a parameter space. In this way, any ignorance of $p\left(\boldsymbol{x}\right)$ reduces to ignorance of the actual parameters $\boldsymbol{\theta}$, which is a considerable simplification of the problem.
	\subsection{Classical statistical model}
	A classical statistical model $\mathcal{S}$ is a family of probability densities on $\boldsymbol{X}$, parameterized by $m$ reals parameters $\boldsymbol{\theta}  \in \Theta  \subset {\mathbb{R}^m}$ and denoted by
	\begin{equation}
		\mathcal{S} = \left\{ {\forall \boldsymbol{x} \in \boldsymbol{X},\quad p\left(\boldsymbol{x};\boldsymbol{\theta}\right):\to \left( {\Omega \left( \boldsymbol{X} \right),\mathcal{A}\left( \boldsymbol{x} \right),p\left(\boldsymbol{x}\right)} \right) \in \Theta } \right\},
	\end{equation}
	where the parameterization map $\left( {\Omega \left( \boldsymbol{X} \right), \mathcal{A}\left( \boldsymbol{x} \right), p\left(\boldsymbol{x}; \boldsymbol{\theta}\right)} \right) \to \Theta$ is injective,  so that the map $p\left(\boldsymbol{x}, \boldsymbol{\theta}\right)$ can be derivable $k^{th}$ times with respect to the parameters, i.e. all possible derivatives $\partial_{1}^{k_{1}} \cdots \partial_{m}^{k_{m}} p\left(\boldsymbol{x}; \boldsymbol{\theta}\right)$ exist (with ${\partial _i} = {\partial _{{\theta _i}}}$). Besides, note that whatever $\boldsymbol{x}$ belongs to $\boldsymbol{X}$, the map $p\left(\boldsymbol{x};\boldsymbol{\theta}\right)$ is always positive and normalized such that
	\begin{equation}
		\forall \boldsymbol{x} \in \boldsymbol{X} \quad \text { and } \quad \forall \boldsymbol{\theta} \in \Theta ; \quad p\left(\boldsymbol{x}; \boldsymbol{\theta}\right)>0, \quad \sum_{\boldsymbol{x}} p\left(\boldsymbol{x}; \boldsymbol{\theta}\right)=1 \quad \text { if } \boldsymbol{X} \text { is discrete },
	\end{equation}
	\begin{equation}
		\forall \boldsymbol{x} \in \boldsymbol{X} \quad \text { and } \quad \forall \boldsymbol{\theta} \in \Theta ; \quad p\left(\boldsymbol{x}; \boldsymbol{\theta}\right)>0, \quad \int_{\boldsymbol{x}} p\left(\boldsymbol{x}; \boldsymbol{\theta}\right)d\boldsymbol{x}=1 \quad \text{ if }\boldsymbol{X} \text { is continuous }.
	\end{equation}
	The parameterization map $p\left(\boldsymbol{x}; \boldsymbol{\theta}\right)$ is called the probability mass function (P.M.F). In case  of $\boldsymbol{X}$ is continuous, the map $p\left(\boldsymbol{x}; \boldsymbol{\theta}\right)$ knowing as the probability density function (P.D.F). Throughout this thesis, we denote the both (P.D.F) and (P.M.F) by probability distribution function (PDF). In the following, we present some examples of classical statistical models and their associated parameterized PDF.
	\begin{ME}\label{EX. 1}{(\textbf{Bernoulli distribution} $\boldsymbol{X} \sim B\left(N, \theta\right)$)}\\
		A coin is tossed $N$ times successfully where the probability of getting the tails is $\theta$  (and the probability of getting heads is $1-\theta$). Note that this experiment is identical to the one used in epidemiology when we study a certain characteristic (disease) of individuals in a population, such that a proportion $\theta$ of individuals have the disease characteristic and $1-\theta$ not having it. Then, the problem is that we do not know the value of the parameter $\theta$, but by performing successively $N$ experiments, we will be able to estimate the value of $\theta$? In this context, we call $X_i$ the random variable at the $i^{th}$ toss, such that $x_i = 1$ if tails (disease) comes out, and $x_i = 0$ otherwise. Hence, we write
		\begin{equation}
			\mathcal{S}=\left\{ {x = {{\{ 1,0\} }^N},\quad p\left( {{x_i};\theta } \right) = {\theta ^{{x_i}}}{(1 - \theta )^{1 - {x_i}}}, \quad \text{where} \quad \theta  \in \Theta  = \left[ {0;1} \right],\quad m = 1,} \right\},
		\end{equation}
		with $x_i$ is the number of the head obtained in $N$ tosses. This classical statistics model is known as a Bernoulli distribution of parameter $\theta$ and denoted $\boldsymbol{X}  \sim B\left(N, \theta\right)$.
	\end{ME}
	\begin{ME}\label{EX. 2}(\textbf{Normal distribution} $\boldsymbol{X}\sim N \left(\mu, \sigma^2\right)$)\\
		The normal distribution is among the most widely used probability distributions for modeling natural experiments resulting from multiple random events. It is also called Gaussian law. More formally, the normal distribution is a complete continuous probability distribution that depends on two unknown parameters, namely the mathematical expectation that is a real number denoted $\mu$, and the standard deviation $\sigma$ or the variance denoted $\sigma^2$. The PDF of the normal distribution of expectation $\mu$ and the variance $\sigma^2$ is given by
		\begin{equation}
			p\left(\boldsymbol{x} ; \boldsymbol{\theta}\right)=\frac{1}{\sqrt{2 \pi \theta_{2}}} \exp \left(-\frac{\left(x-\theta_{1}\right)^{2}}{2 \theta_{2}}\right)=\frac{1}{\sqrt{2 \pi} \sigma} \exp \left(-\frac{(x-\mu)^{2}}{2 \sigma^{2}}\right). \label{Eq. 1.5}
		\end{equation}
		Therefore, the statistical model of this probability distribution is
		{\small \begin{equation}
				\mathcal{S} = \left\{ {\boldsymbol{x} = \mathbb{R},\quad p(\boldsymbol{x};\boldsymbol{\theta}) = \frac{1}{{\sqrt {2\pi {\sigma ^2}} }}\exp \left( { - \frac{{{{\left( {x- \mu } \right)}^2}}}{{2{\sigma ^2}}}} \right),\Theta  = \left\{ {{\boldsymbol{\theta} } = {{(\mu ,\sigma^2 )}^{\rm{T}}}\mid \mu  \in \mathbb{R},\sigma  \in {\mathbb{R}^{* + }}} \right\},\quad m = 2} \right\}.
		\end{equation}}
	\end{ME}

	Clearly, the knowledge of the law followed by $\boldsymbol{X}$ leads to knowing absolutely the probability distribution function $p\left(\boldsymbol{x}; \boldsymbol{\theta}\right)$.  This necessarily implies knowledge of the values of the unknown parameters $\boldsymbol{\theta}$. In this context, estimation theory aims at numerically approximating the value of unknown parameters by using the notion of the estimator.
	\subsubsection{Concept and properties of the estimator}
	\begin{MD}\textbf{(an estimator)}:
		An estimator $\boldsymbol{\theta}_N^{est}:\boldsymbol{X} \to \Theta$ is a random variable from the sample space $\mathcal{S}$ to the parameter space $\Theta$. Their realization, $\boldsymbol{\theta}_N^{est}\left(\boldsymbol{x}\right)$, is an estimated value of the estimating parameter. The sets of values that are reasonable to take are, necessarily, located in the parameter space.
	\end{MD}
	
	Since the estimator $\boldsymbol{\theta}_N^{est}\left(\boldsymbol{X}\right)$ is a random variable, then the estimated value is generally different from the true value of the parameter that will be estimated. In addition, the $\boldsymbol{\theta}_N^{est}\left(\boldsymbol{X}\right)$ is a function of $\boldsymbol{X}$, so it is possible to generate infinities functions from a series of random variables. For this reason, there are many possible estimators in a given estimation problem. Thus, one may ask what criteria we will use to decide between estimators? In this context, we will investigate the criteria for measuring the quality of the estimators constructed by introducing the notions of unbiased and convergent estimators. Among the bases of these criteria, there are the expected value and the covariance of the estimators. These last concepts play an important role in in deciding unbiased and convergent estimators. Therefore, we will define the expected value $E_{\boldsymbol{\theta}}\left( {\boldsymbol{\theta} _N^{est}\left( \boldsymbol{X} \right)} \right)$ and the covariance $\mathtt{Cov}_{\boldsymbol{\theta}}\left( {\boldsymbol{\theta} _N^{est}\left( \boldsymbol{X} \right)} \right)$, which are respectively described by
	\begin{equation}
		E_{\boldsymbol{\theta}}\left(\boldsymbol{\theta}_{N}^{\text {est }}(\boldsymbol{X})\right)=\sum_{\boldsymbol{x} \in \boldsymbol{X}} \boldsymbol{\theta}_{N}^{\text {est }}(\boldsymbol{x}) p\left(\boldsymbol{x}; \boldsymbol{\theta}\right), \quad \text { or } \quad E_{\boldsymbol{\theta}}\left(\boldsymbol{\theta}_{N}^{\text {est }}(\boldsymbol{X})\right)=\int_{\boldsymbol{x}} \boldsymbol{\theta}_{N}^{\text {est }}(\boldsymbol{x}) p\left(\boldsymbol{x}; \boldsymbol{\theta}\right) d \boldsymbol{x},
	\end{equation}
	\begin{equation}
		\operatorname{Cov}_{\boldsymbol{\theta}}\left(\boldsymbol{\theta}_{N}^{\text {est }}(\boldsymbol{X})\right)=\sum_{\boldsymbol{x} \in \boldsymbol{X}} E_{\boldsymbol{\theta}}\left[\left(\boldsymbol{\theta}_{N}^{\text {est }}(\boldsymbol{X})-E_{\boldsymbol{\theta}}\left(\boldsymbol{\theta}_{N}^{\text {est }}(\boldsymbol{X})\right)\right)\left(\boldsymbol{\theta}_{N}^{\text {est }}(\boldsymbol{X})-E_{\boldsymbol{\theta}}\left(\boldsymbol{\theta}_{N}^{\text {est }}(\boldsymbol{X})\right)\right)^{\top}\right].
	\end{equation}
	In particular, the covariance matrix is reduced to the variance when we have a statistical model parameterized by a single unknown parameter, and one can be writing
	\begin{equation}
		\operatorname{Var}_{\theta}\left(\theta_{N}^{\text {est }}(\boldsymbol{X})\right)=\sum_{\boldsymbol{x} \in \boldsymbol{X}} E_{\theta}\left[\left(\theta_{N}^{\text {est }}(\boldsymbol{X})-E_{\theta}\left(\theta_{N}^{\text {est }}(\boldsymbol{X})\right)\right)^{2}\right].
	\end{equation}
	\textbf{ Criteria 1 (Unbiased estimator)}: An unbiased estimator is an estimator satisfying
	\begin{equation}\label{Eqtion}
		E_{\boldsymbol{\theta}}\left(\boldsymbol{\theta}_{N}^{\text {est }}(\boldsymbol{X})\right)=\boldsymbol{\theta}, \quad \forall \boldsymbol{\theta} \in \Theta. 
	\end{equation}
	The unbiased estimator is an influential condition for an estimator. This unbiased condition ensures that, over many independent repetitions of the protocol, the realization of the estimator will fluctuate around the real value of the estimated parameter. Thus, the true value of $\boldsymbol{\theta}$ will be correct on average.
	\begin{MD} \textbf{(Bias of an estimator)}\label{DEFbaias}
		Let $\boldsymbol{\theta}_N^{est}\left(\boldsymbol{X}\right)$ be an estimator of $\boldsymbol{\theta}$. The bias of $\boldsymbol{\theta}_N^{est}\left(\boldsymbol{X}\right)$ is defined by
		\begin{equation}
			\operatorname{Bias}_{\boldsymbol{\theta}}\left(\boldsymbol{\theta}_{N}^{est}(\boldsymbol{X})\right)=E_{\boldsymbol{\theta}}\left(\boldsymbol{\theta}_{N}^{\text {est }}(\boldsymbol{X})\right)-\boldsymbol{\theta}, \quad \forall \boldsymbol{\theta} \in \Theta.
		\end{equation}
	\end{MD}
	If $\operatorname{Bias}_{\boldsymbol{\theta}}\left(\boldsymbol{\theta}_{N}^{est}(\boldsymbol{X})\right)=0$, we conclude that $\boldsymbol{\theta}_{N}^{est}\left(\boldsymbol{X}\right)$ is an unbiased estimator of $\boldsymbol{\theta}$. The bias measures whether the estimator $\boldsymbol{\theta}_{N}^{est}\left(\boldsymbol{X}\right)$ underestimates (negative bias) or overestimates (positive bias) the value of $\boldsymbol{\theta}$. For this reason, we primarily look for unbiased estimators of $\boldsymbol{\theta}$; this means that, on average, they return to the true value of the estimated parameter. In general, to quantify the performance of one estimator over another, it is usual to take as a figure of merit the mean square error (MSE), so that the small MSE of an estimator implies that the considered estimator is more accurate, and one can write
	\begin{equation}
		\operatorname{MSE}_{\boldsymbol{\theta}}\left(\boldsymbol{\theta}_{N}^{\text {est }}(\boldsymbol{X})\right)= 	\mathtt{E}_{\theta}\left[\left(\boldsymbol{\theta}_{N}^{\text {est }}(\boldsymbol{X})-\boldsymbol{\theta}\right)^2\right]=\operatorname{Bias}_{\boldsymbol{\theta}}\left(\boldsymbol{\theta}_{N}^{\text {est }}(\boldsymbol{X})\right)^{2}+\operatorname{Cov}_{\boldsymbol{\theta}}\left(\boldsymbol{\theta}_{N}^{\text {est }}(\boldsymbol{X})\right), \quad \forall \boldsymbol{\theta} \in \Theta. \label{Eq. 1.18}
	\end{equation}
	If the estimator satisfies the unbiased condition. Then the last equation reduces to
	\begin{equation}
		\operatorname{MSE}_{\boldsymbol{\theta}}\left(\boldsymbol{\theta}_{N}^{\text {est }}(\boldsymbol{X})\right)=\operatorname{Cov}_{\boldsymbol{\theta}}\left(\boldsymbol{\theta}_{N}^{\text {est }}(\boldsymbol{X})\right), \quad \forall \boldsymbol{\theta} \in \Theta.
	\end{equation}
	\textbf{Criteria 2 (Convergent estimator)}: Let $\boldsymbol{\theta}_N^{est}\left(\boldsymbol{X}\right)$ be an estimator of $\theta$. $\boldsymbol{\theta}_N^{est}\left(\boldsymbol{X}\right)$ is a convergent estimator if and only if
	\begin{equation}
		\operatorname{MSE}_{\boldsymbol{\theta}}\left(\boldsymbol{\theta}_{N}^{\text {est }}(\boldsymbol{X})\right) \rightarrow 0 \quad \text { when } \quad N \rightarrow \infty.
	\end{equation}
	Thus, repeating the experiment $N$ times independently, with $N$ large,   increases the possibility of obtaining the true value of $\boldsymbol{\theta}$. On the other hand, it refers to the interest that we can give to the variance of the estimator as a second precision criterion that helps us to distinguish a class composed up of unbiased estimators. 
	
	As a conclusion of \textbf{criteria 1} and  \textbf{criteria 2}, for an unbiased estimator, the minimum MSE is equivalent to the convergence of the variance to zero. Therefore, we will try to perform the maximum of experiments to ensure the least MSE. The idea is then to try to construct unbiased estimators that converge in MSE. In what follows, we return to the examples used previously to clarify the utility of \textbf{criteria 1} and \textbf{criteria 2}, which must be satisfied by an estimator for an efficient estimate of the parameters in a given estimation protocol.
	\begin{ME}\label{EX. 3}( \textbf{Estimation of Bernoulli distribution }$\boldsymbol{X}  \sim B\left(N, \theta\right)$)\\
		As previously mentioned in the description of the statistical model associated with a binomial distribution, the estimated parameter is the proportion $\theta$. Then, to estimate the parameter $\theta$, it is necessary to propose an estimator that estimates the most plausible value of the proportion $\theta$ based on the number of the random sorting experiment. In this context, we suggest using the averages of the sorting performed, and we write that
		\begin{equation}\label{Eq. 1.21}
			\theta _N^{{\rm{est }}}({\bf{X}}) = \frac{1}{N}\sum\limits_{i = 1}^N {{X_i}}, \quad \text { the realization of } \quad \theta_{N}^{\text {est }}(\boldsymbol{X}) \quad \text { is } \quad \theta_{N}^{\text {est }}(\boldsymbol{x})=\frac{N_{tails}}{N},
		\end{equation}
		where $N_{tails}$ is the number of tails obtained in $N$ draws.\\
		Now, we are going to verify whether the chosen estimator satisfies both criteria. We start by calculating the bias of $\theta_{N}^{\text {est}}(\boldsymbol{X})$
		\begin{eqnarray}\label{Eq. 1.22}
			\mathtt{Bias}_\theta \left( {\theta _N^{{\rm{est }}}(\boldsymbol{X})} \right) &=& {E_\theta }\left( {\theta _N^{{\rm{est }}}(\boldsymbol{X})} \right) - \theta ,\quad \forall \theta  \in \left] {0,1} \right[\\\notag
			&=&\frac{1}{N}\left[{{E_\theta }\left( {{X_1}} \right),{E_\theta }\left( {{X_2}} \right), \ldots ,{E_\theta }\left( {{X_N}} \right)} \right] - \theta\\\notag
			&=& \frac{N_{tails}}{N}-\theta=0 \notag.
		\end{eqnarray}
		Since $\mathtt{Bias}_\theta \left( {\theta _N^{{\rm{est }}}(\boldsymbol{X})} \right) = 0$, then the estimator $\theta _N^{est}$ is an unbiased estimator of $\theta$.
		In the next step, we are going to compute the variance of $\theta_{N}^{\text {est }}(\boldsymbol{X})$ to confirm the accessibility of the second criteria
		\begin{eqnarray}\label{Eq. 1.23}
			\mathtt{Var}_\theta\left( {\theta _N^{est}(\boldsymbol{X})} \right) &=& E_\theta \left[ {{{\left( {\theta _N^{{\rm{est }}}(\boldsymbol{X}) - E_\theta\left( {\theta _N^{{\rm{est }}}(\boldsymbol{X})} \right)} \right)}^2}} \right]\quad \forall \theta  \in \left] {0,1} \right[\\\notag
			&=&\frac{1}{{{N^2}}}\left( {{\mathtt{Var}_\theta }\left( {{X_1}} \right),{\mathtt{Var}_\theta }\left( {{X_2}} \right), \ldots ,{\mathtt{Var}_\theta}\left( {{X_N}} \right)} \right)\\\notag
			&=&\frac{{\theta \left( {1 - \theta } \right)}}{N}.
		\end{eqnarray}
		If $N\to \infty$, then the variance of $\mathtt{Var}_\theta\left(\theta_N^{est}\right)(\boldsymbol{X})$ tends towards 0. Consequently, the considered estimator is a convergent estimator of the proportion parameter $\theta$. As a result of Eq. (\ref{Eq. 1.22}) and Eq.  (\ref{Eq. 1.23}), we conclude that the estimator proposed to estimate precisely the value of the Bernoulli distribution parameter satisfies both criteria. Therefore,  it is an efficient estimator.
	\end{ME}
	\begin{ME}\label{EX. 4}(\textbf{Estimation of Normal distribution} $\boldsymbol{X}  \sim N\left(\mu, \sigma^2\right)$)\\
		The classical model of this distribution was described in example 2 so that it has parameterized by two unknown parameters, namely the mathematical expectation $\mu$ and the variance $\sigma^2$. To estimate these parameters, it is necessary to construct a two-dimensional estimator vector, such that; the elements of this vector are the estimators of the different parameters
		\begin{equation}
			\boldsymbol{\theta}_{N}^{\text {est }}(\boldsymbol{X})=\left(\mu_{N}^{\text {est }}(\boldsymbol{X}) \quad; \quad\left(\sigma^{2}\right)_{N}^{\text {est }}(\boldsymbol{X})\right)^{\top}, \label{Eq. 1.24}
		\end{equation}
		where the proposed estimators are 
		\begin{equation}
			\mu_{N}^{\mathrm{est}}(\boldsymbol{X})=\frac{1}{N} \sum_{i} X_{i}, \quad\left(\sigma^{2}\right)_{N}^{\text {est }}(\boldsymbol{X})=\frac{1}{N-1} \sum_{i}\left(X_{i}-\mu_{N}^{\text {est }}(\boldsymbol{X})\right)^{2}.\label{Eq. 1.25}
		\end{equation}
		Similarly, one can be obtaining
		\begin{equation}\label{Eq. 1.26}
			E_{\mu}\left(\mu_{N}^{\text {est }}(\boldsymbol{X})\right)=\mu, \quad E_{\sigma^2}\left(\left(\sigma^{2}\right)_{N}^{\text {est }}(\boldsymbol{X})\right)=\sigma^{2}.
		\end{equation}
		\begin{equation}\label{Eq. 1.27}
			\operatorname{Cov}_{\boldsymbol{\theta}}\left[\boldsymbol{\theta}_{N}^{\text {est }}(\boldsymbol{X})\right]=\left(\begin{array}{cc}
				\sigma^{2} / N & 0 \\
				0 & 2 \sigma^{4} /(N-1)
			\end{array}\right).
		\end{equation}
		Therefore, we have 
		\begin{equation} \label{Eq. 1.VAR}
			\mathtt{Var}_{\mu }\left( {\mu _N^{{\rm{est }}}\left( \boldsymbol{X} \right)} \right) = \frac{\sigma ^2}{{{N}}}, \quad \hspace{1cm} \mathtt{Var}_{{\sigma ^2}}\left( {\left( {{\sigma ^2}} \right)_N^{{\rm{est }}}\left( \boldsymbol{X} \right)} \right) = \frac{2\sigma ^4}{{{N-1}}}
		\end{equation}
		From Eq. (\ref{Eq. 1.26}) and Eq. (\ref{Eq. 1.VAR}), we can conclude that the two estimators proposed in Eq. (\ref{Eq. 1.25}) are unbiased and admit converging variances.  Therefore, the vector estimator of Eq. (\ref{Eq. 1.24}) is appropriate but is not necessarily an efficient estimator for estimating the expectation $\mu$ and variance $\sigma^2$ simultaneous.
	\end{ME}
	
	Until now, we've discussed the process of estimating unknown parameters contained in the parametric PDF without knowing the degree of precision that can be achieved.  We say that the appropriate estimator is that will be satisfied both criteria. But it is possible to find more appropriate estimators in the same problem of estimating parameters. In this case, it is reasonable to choose which has a minimum MSE.i.e. which minimizes the covariance  (note that from Eq. (\ref{Eq. 1.18}) that the MSE of an unbiased estimator is just the covariance or the variance in the case of single estimation). Having established what we mean by the best estimate, one might ask whether there is an upper bound on precision that no estimator can violate. The ability to place a lower bound on the variance of any unbiased estimator proves to be extremely useful in practice. In addition, it alerts us to the physical impossibility of finding an unbiased estimator whose variance is less than the bound. This bound has often been useful for the practical aspect of signal processing \cite{kay1993fundamentals}. In fact, there are sets of tractable lower bounds for the MSEs of unbiased estimators, so these sets are derived using the first- and higher-order derivatives of the probability distribution with respect to the parameter of the statistical model. The inequality based on the first-order derivative is called the classical Cramér-Rao bound (CCRB) \cite{lehmann2006theory}.  While those exploited by the higher-order derivatives called Bhattacharyya bound \cite{bhattacharyya1946some}. Any variance of the estimation parameters must always satisfy these inequalities. Therefore, any estimation procedure that saturates the inequality may be considered an efficient estimation. Although many precision bound exist, but the Cramér-Rao lower bound (CRLB) is most widely used on many metrology platforms due to the easier to determine. For these reasons, we will restrain our discussion in what follows to the CRLB.
	\section{Classical CR bound and Fisher information in single estimation models}
	In general, the CRLB has using to determine the minimum MSE of the unbiased estimator so that; if we are able to find the CRLB for an estimator, then it is an efficient estimator. On the other hand, there are minimum variance unbiased estimators that do not satisfy the CRLB, so they are inefficient estimators. Before announcing the CRLB theorem, it is valuable to describe the hidden factors that determine the best estimate of the unknown parameters. Since all our information is performed in the observed data and then in the underlying PDF, it is not surprising that the precision of the estimation parameters depends directly on the parametric PDF of the statistical model $\mathcal{S}$.
	\subsection{Classical Cramér-Rao bound}
	Let $\mathcal{S}$ be a statistical model, and $p\left(\boldsymbol{x};\theta\right)$ is the probability distribution of $\mathcal{S}$.  To estimate a parameters $\theta$, we assume that there exists a lower bound $B\left(\theta\right)$ on the variance of any unbiased estimator of $\theta$. Thus, if we can find an unbiased-estimator $\theta^{est}\left(\boldsymbol{X}\right)$ satisfying that $\mathtt{Var} _{\theta}\left( {{\theta ^{est}}\left( \boldsymbol{X} \right)} \right) = B\left( \theta  \right)$, then we found the efficient estimator. That is the approach we will take, using the concept of CRLB.
	\begin{Mth}\label{TH. 1}\textbf{( Cramér-Rao bound)}:
		Let $X_1, X_2, ...,X_N$ be random variables, and $p\left(\boldsymbol{x};\theta\right)$ the parametric PDF. To estimate the parameter $\theta$, we propose $\theta_N^{est}\left(\boldsymbol{X}\right)$ as an unbiased estimator satisfying
		\begin{equation}
			\frac{\partial}{\partial \theta} \mathrm{E}_{\theta}\left(\theta_N^{\text {est }}(\boldsymbol{X})\right)=\sum_{\boldsymbol{x} \in X} \frac{\partial}{\partial \theta}\left(\theta_N^{\text{est }}(\boldsymbol{x}) p(\boldsymbol{x} ; \theta)\right)=1.\label{Eq. 1.31}
		\end{equation}
		Then, the variance of any unbiased estimator $\theta_N^{est}\left(\boldsymbol{X}\right)$ must satisfy the following inequality
		\begin{equation}\label{Eq. 1.32}
			\mathtt{Var}_{\theta}\left(\theta_N^{\text {est }}(\boldsymbol{X})\right) \geq \frac{1}{\mathrm{E}_{\theta}\left(\left(\frac{\partial}{\partial \theta} \log (p(\boldsymbol{x}; \theta))\right)^{2}\right)}.
		\end{equation}
		An unbiased estimator will be efficient to estimate $\theta$, if and only if
		\begin{equation}\label{Eq. 1.33}
			\mathtt{Var}_{\theta}\left(\theta_N^{\text {est }}(\boldsymbol{X})\right) = \frac{1}{\mathrm{E}_{\theta}\left(\left(\frac{\partial}{\partial \theta} \log (p(\boldsymbol{x}; \theta))\right)^{2}\right)}.
		\end{equation}
		In Eqs. (\ref{Eq. 1.32}) and (\ref{Eq. 1.33}), the expectation value is
		\begin{equation}
			\mathrm{E}_{\theta}\left(\left(\frac{\partial}{\partial \theta} \log (p(\boldsymbol{x} ; \theta))\right)^{2}\right)=\sum_{\boldsymbol{x} \in \boldsymbol{X}}\left(\frac{\partial}{\partial \theta} \log (p(\boldsymbol{x}; \theta))\right)^{2} p(\boldsymbol{x}; \theta).
		\end{equation}
	\end{Mth}
	\begin{MPreuve}:
		To prove this Theorem, we will directly apply the Cauchy-Schwartz inequality. In statistics probabilistic, for any two random variables $\mathbb{X}$ and $\mathbb{Y}$, the Cauchy-Schwartz inequality is given by
		\begin{equation}\label{Eq. 1.35}
			|\mathtt{Cov}(\mathbb{X}, \mathbb{Y})| \leq \sqrt{\mathtt{Var}(\mathbb{X})} \sqrt{\mathtt{Var}(\mathbb{Y})}.
		\end{equation}
		From Eq. (\ref{Eq. 1.35}), we can get a lower bound on the variance of $\mathbb{X}$ as
		\begin{equation}\label{Eq. 1.36}
			\mathtt{Var}(\mathbb{X}) \geq \frac{|\mathtt{Cov}(\mathbb{X}, \mathbb{Y})|^{2}}{\mathtt{Var}(\mathbb{Y})}.
		\end{equation}
		An interesting trick of this proof is to choose $\mathbb{X}$ as the estimator $\theta_N^{est}\left(\boldsymbol{X}\right)$ and $\mathbb{Y}$ to be the quantity $\frac{{\partial}}{{\partial \theta }}\log \left( {p\left( {\boldsymbol{x};\theta } \right)} \right)$ and applying the Cauchy-Schwartz inequality. In first, note that
		\begin{eqnarray}
\frac{d}{{d\theta }}{{\rm{E}}_\theta }\left( {{\theta_N^{{\rm{est }}}}(\boldsymbol{X})} \right) =\sum\limits_{\boldsymbol{x} \in \boldsymbol{X}} {{\theta_N^{{\rm{est}}}}} (\boldsymbol{X})\frac{{\partial p(\boldsymbol{x};\theta )}}{{\partial \theta }},
		\end{eqnarray}
	which implied that
		\begin{eqnarray}\label{Eq. 1.37}
		\begin{aligned}
			\frac{d}{{d\theta }}{{\rm{E}}_\theta }\left( {{\theta_N^{{\rm{est }}}}(\boldsymbol{X})} \right)&=& \sum\limits_{\boldsymbol{x} \in \boldsymbol{X}}{{\theta_N^{{\rm{est }}}}} (\boldsymbol{X})\left( {\frac{1}{{p(\boldsymbol{x};\theta )}}\frac{{\partial p(\boldsymbol{x};\theta )}}{{\partial \theta}}} \right)p(\boldsymbol{x};\theta )\\
			&=& \sum\limits_{\boldsymbol{x} \in \boldsymbol{X}} {{\theta_N^{{\rm{est }}}}} (\boldsymbol{X})\left( {\frac{{\partial \log (p(\boldsymbol{x};\theta ))}}{{\partial \theta }}} \right)p(\boldsymbol{x};\theta )\\
			&=& {{\rm{E}}_\theta }\left( {{\theta_N^{{\rm{est }}}}(\boldsymbol{X})\left( {\frac{{\partial \log (p(\boldsymbol{x};\theta ))}}{{\partial \theta }}} \right)} \right).
		\end{aligned}
		\end{eqnarray}
		Now, we are going to replace $\mathbb{X}$ and $\mathbb{Y}$ in Eq. (\ref{Eq. 1.36}) by $\theta_N^{est}\left(\boldsymbol{X}\right)$ and $\frac{{\partial}}{{\partial \theta }}\log \left( {p\left( {\boldsymbol{x};\theta } \right)} \right)$, one gets
	 \begin{eqnarray}
	\mathtt{Var}_{\theta }\left( {{\theta_N^{est}}(\boldsymbol{X})} \right) \ge \frac{{{{\left| { \mathtt{Cov}_{\theta}\left( {{\theta_N^{{\rm{est }}}}(\boldsymbol{X}),\frac{{\partial \log (p(\boldsymbol{x};\theta ))}}{{\partial \theta }}} \right)} \right|}^2}}}{{\mathtt{Var}_{\theta }\left( {\frac{{\partial \log (p(\boldsymbol{x};\theta ))}}{{\partial \theta }}} \right)}}
	\end{eqnarray}
which is rewriting as
	\begin{eqnarray}\label{Eq. 1.38}
	\mathtt{Var}_{\theta }\left( {{\theta_N^{est}}(\boldsymbol{X})} \right)	\ge\frac{{{{\left| {{{\rm{E}}_\theta }\left[ {\left( {{\theta_N^{{\rm{est }}}}(\boldsymbol{X}) - {{\rm{E}}_\theta }\left( {{\theta_N^{{\rm{est }}}}(\boldsymbol{X})} \right)} \right)\left( {\frac{{\partial \log (p(\boldsymbol{x};\theta ))}}{{\partial \theta }} - {{\rm{E}}_\theta }\left( {\frac{{\partial \log (p(\boldsymbol{x};\theta ))}}{{\partial \theta }}} \right)} \right)} \right]} \right|}^2}}}{{{{\rm{E}}_\theta }\left( {{{\left( {\frac{{\partial \log (p(\boldsymbol{x};\theta ))}}{{\partial \theta }}} \right)}^2}} \right) - {{\left( {{{\rm{E}}_\theta }\left( {\frac{{\partial \log (p(\boldsymbol{x};\theta ))}}{{\partial \theta }}} \right)} \right)}^2}}}.
		\end{eqnarray}
		If we apply the result of Eq. (\ref{Eq. 1.37}) with $\theta_N^{est}\left(\boldsymbol{X}\right)=1$, we have
		\begin{equation}\label{Eq. 1.39}
			\mathrm{E}_{\theta}\left(\frac{\partial \log (p(\boldsymbol{x}; \theta))}{\partial \theta}\right)=\frac{d}{d \theta} \mathrm{E}_{\theta}(1)=0.
		\end{equation}
		Then, the denominator of Eq. (\ref{Eq. 1.38}) becomes
		\begin{equation}\label{Eq. 1.40}
			\mathtt{Var}_{\theta}\left(\frac{\partial \log (p(\boldsymbol{x}; \theta))}{\partial \theta}\right)=\mathrm{E}_{\theta}\left(\left(\frac{\partial \log (p(\boldsymbol{x}; \theta))}{\partial \theta}\right)^{2}\right). 
		\end{equation}
		On the other hand, the numerator of Eq. (\ref{Eq. 1.38}) is equal to the square of expectation of the product, that has derived from Eq. (\ref{Eq. 1.37}) and Eq. (\ref{Eq. 1.39}), and we obtain
		\begin{equation}
			{\left| {\mathtt{Cov}_{\theta}\left( {{\theta_N^{{\rm{est }}}}(\boldsymbol{X}),\frac{{\partial \log (p(\boldsymbol{x};\theta ))}}{{\partial \theta }}} \right)} \right|^2} = {\left| {{{\rm{E}}_\theta }\left( {{\theta_N^{{\rm{est }}}}(\boldsymbol{X})\frac{{\partial \log (p(\boldsymbol{x};\theta ))}}{{\partial \theta }}} \right)} \right|^2} = {\left| {\frac{d}{{d\theta }}{{\rm{E}}_\theta }\left( {{\theta_N^{{\rm{est }}}}(\boldsymbol{X})} \right)} \right|^2}.
		\end{equation}
		Since $\theta_N^{est}\left(\boldsymbol{X}\right)$ is an unbiased estimator, so it satisfies the condition of Eq. (\ref{Eq. 1.31}). Therefore
		\begin{equation}\label{Eq. 1.42}
			{\left| {\mathtt{Cov}_{\theta }\left( {\theta _N^{{\rm{est }}}(\boldsymbol{X}),\frac{{\partial \log (p(\boldsymbol{x};\theta ))}}{{\partial \theta }}} \right)} \right|^2} = 1.
		\end{equation}
		Finally, we substituted the results of Eq. (\ref{Eq. 1.40}) and Eq. (\ref{Eq. 1.42}) together in Eq. (\ref{Eq. 1.38}), we obtain the Cramér-Rao inequality
		{\large \begin{equation}
				\mathtt{Var}_{\theta}\left(\theta_N^{\text {est }}(\boldsymbol{X})\right) \geq \frac{1}{\mathrm{E}_{\theta}\left(\left(\frac{\partial \log (p(\boldsymbol{x}; \theta))}{\partial \theta}\right)^{2}\right)}.
		\end{equation}}
	\end{MPreuve}
	
	In particular, if we assume that the random variables $X_1, X_2, ..., X_N$ are independent and identically distributed (\textbf{iid}) with PDF, then the Cramér-Rao inequality is reduced to the form shown in the following Corollary:
	\begin{MC}\label{PC. 1.3.1}\textbf{ (Cramér-Rao inequality in the case of iid )}: If $X_1, X_2,...,X_N$ be \textbf{iid} random variables and $p(\boldsymbol{x}; \theta)$ the parametric PDF, and $\theta_N^{est}\left(\boldsymbol{X}\right)$ must be an unbiased estimator satisfying the condition proposed in Eq. (\ref{Eq. 1.31}). Then, the Cramér-Rao inequality is reduced to
		{\large \begin{equation}
				\mathtt{Var}_{\theta}\left(\theta_N^{\text {est }}(\mathbf{X})\right) \geq \frac{1}{N\operatorname{E}_{\theta}\left(\left(\frac{\partial \log (p(x; \theta))}{\partial \theta}\right)^{2}\right)}
		\end{equation}}
	\end{MC}
	\begin{MPreuve}
		(\textbf{\textit{Proof of Corollary \ref{PC. 1.3.1}}}): To prove the result of this Corollary, it suffices to show that 
		\begin{equation}
			\mathrm{E}_{\theta}\left(\left(\frac{\partial \log (p(\boldsymbol{x}; \theta))}{\partial \theta}\right)^{2}\right)=N \operatorname{E}_{\theta}\left(\left(\frac{\partial \log (p(x; \theta))}{\partial \theta}\right)^{2}\right).
		\end{equation}
		Since $X_1, X_2, ..., X_N$ are independent, then it is reasonable to write that
		\begin{eqnarray}\label{Eq. 1.46}
			\notag\mathrm{E}_{\theta}\left(\frac{\partial}{\partial \theta} \log p(\boldsymbol{x} ; \theta)\right)^{2}&=&\mathrm{E}_{\theta}\left(\left(\frac{\partial}{\partial \theta} \log \prod_{i=1}^{N} p\left(x_{i}; \theta\right)\right)^{2}\right)\\
			&=&\sum_{i=1}^{N} \mathrm{E}_{\theta}\left(\left(\frac{\partial}{\partial \theta} \log p\left(x_{i}; \theta\right)\right)^{2}\right)\\\notag
			&+&\sum_{i \neq j}^{N} \mathrm{E}_{\theta}\left(\left(\frac{\partial}{\partial \theta} \log p\left(x_{i};\theta\right)\right)\left(\frac{\partial}{\partial \theta} \log p\left(x_{j}; \theta\right)\right)\right).
		\end{eqnarray}
		Using the independent measures of $X_1,X_2,...,X_N$ and applied the result of Eq. (\ref{Eq. 1.39}), one gets that
		\begin{equation}
			\mathrm{E}_{\theta}\left(\left(\frac{\partial}{\partial \theta} \log p\left(x_{i};\theta\right)\right)\left(\frac{\partial}{\partial \theta} \log p\left(x_{j};\theta\right)\right)\right)=\mathrm{E}_{\theta}\left(\frac{\partial}{\partial \theta} \log p\left(x_{i} ; \theta\right)\right) \mathrm{E}_{\theta}\left(\frac{\partial}{\partial \theta} \log p\left(x_{j} ; \theta\right)\right)=0.\label{Eq. 1.47}
		\end{equation}
		On the other hand, since $X_1, X_2, ..., X_N$ are identically distributed, then the first term of Eq. (\ref{Eq. 1.46}) has rewritten as
		\begin{equation}\label{Eq. 1.48}
			\sum_{i=1}^{N} \mathrm{E}_{\theta}\left(\left(\frac{\partial}{\partial \theta} \log p\left(x_{i}; \theta\right)\right)^{2}\right)= N \operatorname{E}_{\theta}\left(\left(\frac{\partial}{\partial \theta} \log p(x ; \theta)\right)^{2}\right).
		\end{equation}
		Finally, to establish the Corollary \ref{PC. 1.3.1}, it suffices to insert the results of Eq. (\ref{Eq. 1.47}) and Eq. (\ref{Eq. 1.48}) into Eq. (\ref{Eq. 1.46}).
	\end{MPreuve}
	
	Before going on, although the CRLB is stated for discrete random variables, it also applies to continuous random variables. The key is simply to switch from summation to integration this switch depends on the nature of the statistical model under consideration; for example, if $p\left(\boldsymbol{x}; \theta\right)$ is a P.F.M, i.e., belonging to the family of discrete laws, then we should use summation as described above. Contrariwise, in the case where the statistical model considered belongs to the family of continuous distributions characterized by P.D.F. In this situation, we must be to use integration instead of summation.
	\subsection{Classical Fisher information}
	The quantity defined in the denominator of Eq. (\ref{Eq. 1.32}) has referred to as the \textbf{\textit{Classical Fisher Information} (CFI)} for the data $\boldsymbol{x}$ and the parameter $\theta$, it is given by
	\begin{equation}\label{Eq. 1.49}
		\mathcal{F}_{C}(\theta)=\mathrm{E}_{\theta}\left(\left(\frac{\partial}{\partial \theta} \log (p(\boldsymbol{x} ; \theta))\right)^{2}\right)=\sum_{\boldsymbol{x} \in \boldsymbol{X}}\left(\frac{\partial \log (p(\boldsymbol{x} ; \theta))}{\partial \theta}\right)^{2} p(\boldsymbol{x} ; \theta). 
	\end{equation}
	The terminology of CFI reflects, in the fact, that the information amount gives a bound on the variance of the best-unbiased estimator of $\theta$.  As well as the {CFI} gets more important, and we have more information about the estimated parameter, then we have a smaller bound on the variance of the best-unbiased estimator; this exactly appears in Eq. (\ref{Eq. 1.32}). In other words, more {CFI} leads to a minimum on the CCRB. Thus, reaching the CRLB. The {CFI} given in Eq. (\ref{Eq. 1.49}) can be expressed in a slightly different form. Although Eq. (\ref{Eq. 1.49})  is generally more compact and has sometimes useful for theoretical work,  the alternative form is more usually convenient for evaluation. This alternative form is that we will confirmed in the next lemma.
	\begin{LM}\label{LM. 1.3} \textbf{(Alternative form of classical Fisher information )}:\\
		Let $X_1, X_2, ..., X_N$ be a random variable and $p\left(\boldsymbol{x}; \theta\right)$ the parametric PDF of the data $\boldsymbol{x}$.  If $p\left(\boldsymbol{x}; \theta\right)$  satisfies the constraint of Eq. (\ref{Eq. 1.39}), then the {CFI} of Eq. (\ref{Eq. 1.49}) becomes
		\begin{equation}\label{Eq. 1.50}
			\mathcal{F}_{C}(\theta)=-\mathrm{E}_{\theta}\left(\frac{\partial^{2}}{\partial \theta^{2}} \log (p(\boldsymbol{x} ; \theta))\right)=-\sum_{\boldsymbol{x} \in \boldsymbol{X}}\left(\frac{\partial^{2}}{\partial \theta^{2}} \log (p(\boldsymbol{x} ; \theta))\right) p(\boldsymbol{x} ; \theta). 
		\end{equation}
	\end{LM}
	\begin{MPreuve}({\textbf{Proof of lemma \ref{LM. 1.3}:}}) One can easily prove this lemma if only the condition of Eq. (\ref{Eq. 1.39}) is satisfied by the probability distribution function of the random variable $\boldsymbol{X}$. We assume that $p\left(\boldsymbol{x}; \theta\right)$ satisfies (\ref{Eq. 1.39}), then we have 
		\begin{equation*}
			\mathrm{E}_{\theta}\left(\frac{\partial \log (p(\boldsymbol{x} ; \theta))}{\partial \theta}\right)=0
		\end{equation*}
		\begin{equation*}
			\sum_{\boldsymbol{x} \in \boldsymbol{X}} \frac{\partial \log (p(\boldsymbol{x} ; \theta))}{\partial \theta} p(\boldsymbol{x} ; \theta)=0.
		\end{equation*}
		We insert the derivative with respect to $\theta$ at both sides of the previous equation
		\begin{equation*}
			\frac{\partial}{\partial \theta}\left(\sum_{x \in X} \frac{\partial \log (p(\boldsymbol{x} ; \theta))}{\partial \theta} p(\boldsymbol{x} ; \theta)\right)=0
		\end{equation*}
		\begin{equation*}
			\sum_{x \in X} \frac{\partial^{2} \log (p(\boldsymbol{x} ; \theta))}{\partial \theta^{2}} p(\boldsymbol{x} ; \theta)+\frac{\partial p(\boldsymbol{x} ; \theta)}{\partial \theta} \frac{\partial \log (p(\boldsymbol{x} ; \theta))}{\partial \theta}=0
		\end{equation*}
		\begin{equation*}
			\sum_{x \in X} \frac{\partial^{2} \log (p(\boldsymbol{x} ; \theta))}{\partial \theta^{2}} p(\boldsymbol{x} ; \theta)=-\sum_{\boldsymbol{x} \in \boldsymbol{X}} \frac{\partial \log (p(\boldsymbol{x} ; \theta))}{\partial \theta} \frac{\partial \log (p(\boldsymbol{x} ; \theta))}{\partial \theta} p(\boldsymbol{x} ; \theta)
		\end{equation*}
		\begin{equation*}
			\mathrm{E}_{\theta}\left(\frac{\partial^{2} \log (p(\boldsymbol{x} ; \theta))}{\partial \theta^{2}}\right)=-\mathrm{E}_{\theta}\left(\left(\frac{\partial \log (p(\boldsymbol{x} ; \theta))}{\partial \theta}\right)^{2}\right)
		\end{equation*}
	\end{MPreuve}
	
	As a consequence, the {CFI}  can be written in both equivalent forms
	\begin{equation}
		\mathcal{F}_{C}(\theta)=\mathrm{E}_{\theta}\left(\left(\frac{\partial \log (p(\boldsymbol{x} ; \theta))}{\partial \theta}\right)^{2}\right)=-\mathrm{E}_{\theta}\left(\frac{\partial^{2} \log (p(\boldsymbol{x} ; \theta))}{\partial \theta^{2}}\right)
	\end{equation}
	Due to the last lemma (\ref{LM. 1.3}), we can see that the {CFI} does not admit negative values. Therefore, we will refer to the \textbf{non-negativity} as the first property of {CFI}. On the other hand, if we assume that $X_1, X_2, ..., X_N$ are \textbf{iid} random variables, and we return to the results presented in {Corollary \ref{PC. 1.3.1}}, we will conclude that
	\begin{equation}\label{Eq. 1.52}
		\mathcal{F}_{C}(\theta)=N f_{C}(\theta),
	\end{equation}
	where $f_{C}(\theta)$ denoted the Fisher information for one sample.  It is writing as
	\begin{equation}
		f_{C}(\theta)=\mathrm{E}_{\theta}\left(\left(\frac{\partial \log (p(x ; \theta))}{\partial \theta}\right)^{2}\right)=-\mathrm{E}_{\theta}\left(\frac{\partial^{2} \log (p(x ; \theta))}{\partial \theta^{2}}\right)
	\end{equation}
	As a remarkable result of Eq. (\ref{Eq. 1.52}), the Fisher information is an additive quantity for independent random variables. This \textbf{additive property} has been indicating as the second property of {CFI}. It is led to the result that the CRLB for $N$ \textbf{iid} observations is $1/N$ times that for one observation, which means that repeating the measurement $N$ times guide to an increase in the possibility of reaching the ultimate value of the estimated parameter (CRLB). Before extending these results to the case of multiparameter estimation, we will clarify the usefulness of using the {CFI} by treating the example of statistical model that have been previously stated.
	\begin{ME} \label{EX. 5}(\textbf{CFI} \textbf{in the  Bernoulli distribution} $\boldsymbol{X}  \sim B\left(N, \theta\right)$)\\
		As mentioned in the description of the statistical model given in Example \ref{EX. 1}, such as the parametric probability of one sample $X_i$ is given by
		\begin{equation}\label{Eq. 1.54}
			p\left( {{x_i};\theta } \right) = \theta^{x_i} (1 - \theta)^{1-x_i}. 
		\end{equation}
		This function $p\left( {x_i; \theta} \right)$ is also known as the likelihood function of Bernoulli distribution. Now, we use the definition of {CFI} introduced in Eq. (\ref{Eq. 1.50})  to evaluate the CR bound for fixing the variance of the estimator of a binomial proportion $\theta$.\\
		The log-likelihood function of Eq. (\ref{Eq. 1.54}) is
		\begin{equation*}
			\log p\left( {{x_i};\theta } \right) = x_i \log \,\theta  +(1-x_i) \log \left( {1 - \theta } \right).
		\end{equation*}
		Differentiating once produces
		\begin{equation*}
			\frac{{\partial \,\log p\left( {{x_i};\theta } \right)}}{{\partial \theta }} = \frac{x_i}{\theta } - \frac{1-x_i}{{1 - \theta }},
		\end{equation*}
		and a second differentiation results in
		\begin{equation*}
			\frac{{{\partial ^2}\log p\left( {{x_i};\theta } \right)}}{{\partial {\theta ^2}}} =  - \frac{x_i}{{{\theta ^2}}} - \frac{1-x_i}{{{{\left( {1 - \theta } \right)}^2}}}.
		\end{equation*}
		Upon taking the negative expectations, the classical Fisher information becomes
		\begin{equation}
			{f_C}(\theta ) =  - {{\rm{E}}_\theta }\left( {\frac{{{\partial ^2}\log (p(x;\theta ))}}{{\partial {\theta ^2}}}} \right) = \frac{1}{{\theta \left( {1 - \theta } \right)}}.
		\end{equation}
		If we considered that $X_1, X_2, ..., X_N$ are \textbf{iid} random variables, thus we have 
		\begin{equation}
			{{\cal F}_C}(\theta ) = \frac{N}{{\theta \left( {1 - \theta } \right)}}. \label{Eq. 1.56}
		\end{equation}
		By inverting the classical Fisher information of Eq (\ref{Eq. 1.56}) and using the result of Theorem (\ref{TH. 1}), we found 
		\begin{equation}\label{Eq. 1.57}
			\mathtt{Var}_{\theta }\left( {\theta _N^{{\rm{est }}}({\bf{X}})} \right) \ge \frac{{\theta \left( {1 - \theta } \right)}}{N}.
		\end{equation}
		The saturation of the last Eq. (\ref{Eq. 1.57}) leads to the CRLB, which is given by
		\begin{equation}\label{Eq. 1.58}
			\mathtt{Var}_{\theta }\left( {\theta _N^{{\rm{est }}}({\bf{X}})} \right) = \frac{{\theta \left( {1 - \theta } \right)}}{N}.
		\end{equation}
		The result of Eq. (\ref{Eq. 1.58}) corresponds to that obtained in Eq. (\ref{Eq. 1.23}), which means that the estimator proposed in Eq. (\ref{Eq. 1.21}) is an efficient estimator of a Bernoulli  proportion $\theta$.
	\end{ME}
	\section{Extension into multiparameter estimation models}
	In this section, we will extend the results discussed in the previous section to the case where we need to estimate a set of unknown parameters. The several parameters that we expect to estimates are constructed in the vector defined in the space parameter, named the vector parameter $\boldsymbol{\theta}=\left(\theta_{1}, \theta_{2}, \ldots, \theta_{m}\right)^{\top} \in \Theta \subset \mathbb{R}^{m}$. Thus, we can define the estimator vector corresponding to it as $\boldsymbol{\theta}^{\text {est }}(\boldsymbol{X})=\left(\theta_{1}^{\text {est }}(\boldsymbol{X}), \theta_{2}^{\text {est }}(\boldsymbol{X}), \ldots, \theta_{m}^{\text {est }}(\boldsymbol{X})\right)^{\top}$, and  the realization of this estimator vector is denoted $\boldsymbol{\theta}^{\text {est }}(\boldsymbol{x})=\left(\theta_{1}^{\text {est }}(\boldsymbol{x}), \theta_{2}^{\text {est }}(\boldsymbol{x}), \ldots, \theta_{m}^{\text {est }}(\boldsymbol{x})\right)^{\top}$. We assume that the $\boldsymbol{\theta}^{\text {est }}(\boldsymbol{X})$ satisfies the unbiased condition mentioned in Eq. (\ref{Eqtion}). The vector parameter will allow us to place a limit on the variance of each parameter, which will be done by generalizing the scalar CCR inequality given in Theorem \ref{TH. 1} to the matrix CR inequality, which we will affirm in the following.
	\subsection{Matrix CR bound and classical Fisher information matrix}
	\begin{Mth} \label{Mth. 1.4}\textbf{(Matrix Cramér-Rao inequality)}
		Let $X_1, X_2,...,X_N$ be a random variable and $p(\boldsymbol{x}; \boldsymbol{\theta})$ the parametric PDF. For an unbiased estimator vector $\boldsymbol{\theta}^{\text {est }}(\boldsymbol{X})$, the following inequality is always satisfied
		\begin{equation}
			{{\mathop{\rm Cov}\nolimits} _{\boldsymbol{\theta} }}\left( {\boldsymbol{\theta} _N^{est}({\bf{X}})} \right)\geq \frac{1}{\mathcal{F}_{c}(\boldsymbol{\theta})},
		\end{equation}
		where $\mathcal{F}_{c}(\boldsymbol{\theta})$ is the $m \times m$ \textbf{\textit{classical Fisher information matrix (CFIM)}}, which defined as
		\begin{equation}\label{CFIM}
			\begin{aligned}
				\left[\mathcal{F}_{C}(\boldsymbol{\theta})\right]_{j k}&=E_{\boldsymbol{\theta}}\left(\frac{\partial \log (p(\boldsymbol{x} ; \boldsymbol{\theta}))}{\partial \theta_{j}} \frac{\partial \log (p(\boldsymbol{x} ; \boldsymbol{\theta}))}{\partial \theta_{k}}\right) \quad \text { for all } \mathrm{j}, \mathrm{k}=1 \ldots \mathrm{m}\\
				&=\sum_{\boldsymbol{x} \in \boldsymbol{X}}\left(\frac{\partial \log (p(\boldsymbol{x} ; \boldsymbol{\theta}))}{\partial \theta_{j}} \frac{\partial \log (p(\boldsymbol{x} ; \boldsymbol{\theta}))}{\partial \theta_{k}}\right) p(\boldsymbol{x} ; \boldsymbol{\theta})\\
				&=- E_{\boldsymbol{\theta}}\left( {\frac{{{\partial ^2}\log p\left( {\boldsymbol{x}; \boldsymbol{\theta}} \right)}}{{\partial {\theta _j}\partial {\theta _k}}}} \right).
			\end{aligned}
		\end{equation}
	\end{Mth}
	Noted in Eq. (\ref{CFIM}), if $j=k$, then we have $\left[\mathcal{F}_{C}(\boldsymbol{\theta})\right]_{jj}=\mathcal{F}_{C}(\theta)$. Thus, the matrix inequality of Eq. (\ref{CFIM}) reduced to the scalar inequality of Eq. (\ref{Eq. 1.49}) that associated with the individual estimation parameter. If the {CFIM} is not invertible, i.e,  the {CFIM} has an eigenvalue 0. In this case, we can define a general inverse for an arbitrary matrix which is the Moore-Penrose pseudo-inverse \cite{ben2003generalized, rao1971further}. Some examples (still in examples introduce previously) are now giving to illustrate the evaluation of the CRLB in the multiparameter case.
	\begin{ME}\label{ME. 1.4}(\textbf{CFIM in Normal distribution} $\boldsymbol{X}  \sim N\left(\mu, \sigma^2\right)$)\\
		In the description of the statistical model mentioned in Example \ref{EX. 2}, the likelihood function of the normal distribution is given in Eq. (\ref{Eq. 1.5}) and the vector of unknown parameters is $\boldsymbol{\theta}=\left(\mu, \sigma^{2}\right)^{\top}$. Hence, the $2\times2$ classical Fisher information matrix is given by
		\begin{equation}
			\mathcal{F}_{C}\left( \boldsymbol{\theta}  \right) = \left( {\begin{array}{*{20}{l}}
					{ - E_{\boldsymbol{\theta}}\left({\frac{{{\partial ^2}\log p\left( {\boldsymbol{x}; \boldsymbol{\theta}} \right)}}{{\partial {\mu ^2}}}} \right)}&{ - E_{\boldsymbol{\theta}}\left( {\frac{{{\partial ^2}\log p\left( {\boldsymbol{x}; \boldsymbol{\theta}} \right)}}{{\partial \mu \partial {\sigma ^2}}}} \right)}\\
					{ - E_{\boldsymbol{\theta}}\left( {\frac{{{\partial ^2}\log p\left( {\boldsymbol{x}; \boldsymbol{\theta}} \right)}}{{\partial {\sigma ^2}\partial \mu }}} \right)}&{ - E_{\boldsymbol{\theta}}\left( {\frac{{{\partial ^2}\log p\left( {\boldsymbol{x}; \boldsymbol{\theta}} \right)}}{{\partial {\sigma ^2}^2}}} \right)}
			\end{array}} \right). \label{Eq. 1.62}
		\end{equation}
		Since the order of partial differential may be interchanged, then the CFIM  is a symmetric matrix. By using the likelihood function given in Eq. (\ref{Eq. 1.5}), one can write the log-likelihood function as
		\begin{equation}
			\log p\left( {\boldsymbol{x}; \boldsymbol{\theta}} \right) =  - \frac{1}{2}\log \left( {2\pi {\sigma ^2}} \right) - \frac{{{{\left( {\boldsymbol{x} - \mu } \right)}^2}}}{{2{\sigma ^2}}}. 
		\end{equation}
		The derivatives are easily found as
		\begin{equation*}
			\frac{{\partial \log p\left( {\boldsymbol{x}; \boldsymbol{\theta}} \right)}}{{\partial {\sigma ^2}}} =  - \frac{1}{{2{\sigma ^2}}} + \frac{{{{\left( {\boldsymbol{x} - \mu } \right)}^2}}}{{2{\sigma ^4}}}, \quad 
			\frac{{\partial \log p\left( {\boldsymbol{x}; \boldsymbol{\theta}} \right)}}{{\partial \mu }} = \frac{{\left( {\boldsymbol{x} - \mu } \right)}}{{{\sigma ^2}}}
		\end{equation*}
		\begin{equation*}
			\frac{{{\partial ^2}\log p\left( {\boldsymbol{x}; \boldsymbol{\theta}} \right)}}{{{\partial ^2}{\sigma ^2}}} = \frac{1}{{2{\sigma ^4}}} - \frac{{{{\left( {\boldsymbol{x} - \mu } \right)}^2}}}{{{\sigma ^6}}}, \quad
			\frac{{{\partial ^2}\log p\left( {\boldsymbol{x};\boldsymbol{\theta} } \right)}}{{{\partial ^2}\mu }} = \frac{{ - 1}}{{{\sigma ^2}}},\quad
			\frac{{{\partial ^2}\log p\left( {x;\boldsymbol{\theta} } \right)}}{{\partial \mu \partial {\sigma ^2}}} =- \frac{{\left( {\boldsymbol{x} - \mu } \right)}}{{{\sigma ^4}}}.
		\end{equation*}
		Upon taking the negative expectations, for $N$ samples, the CFIM becomes
		\begin{equation}
			\mathcal{F}_{C}\left( \theta  \right) = \left( {\begin{array}{*{20}{l}}
					{{\raise0.7ex\hbox{$N$} \!\mathord{\left/
								{\vphantom {N {{\sigma ^2}}}}\right.\kern-\nulldelimiterspace}
							\!\lower0.7ex\hbox{${{\sigma ^2}}$}}}&0\\
					0&{{\raise0.7ex\hbox{$N$} \!\mathord{\left/
								{\vphantom {N {2{\sigma ^4}}}}\right.\kern-\nulldelimiterspace}
							\!\lower0.7ex\hbox{${2{\sigma ^4}}$}}}
			\end{array}} \right)
		\end{equation}
		Although this is not true in general, in this example, the CFIM is diagonal and therefore easily inverted, which lead after using the results of Theorem \ref{Mth. 1.4} to
		\begin{equation}
			\operatorname{Cov}_{\boldsymbol{\theta}}\left[\boldsymbol{\theta}_{N}^{\text {est }}(\boldsymbol{X})\right] \ge \left(\begin{array}{cc}
				\sigma^{2} / N & 0 \\
				0 & 2 \sigma^{4} /N
			\end{array}\right).
		\end{equation}
		Therefore, we have
		\begin{equation}\label{Eq. 1.67}
			\mathtt{Var}_{\mu }\left( {\mu _N^{{\rm{est}}}\left( {\bf{X}} \right)} \right) \ge \frac{{{\sigma ^2}}}{N},\quad  \hspace*{1cm}\mathtt{Var}_{{\sigma ^2}}\left( {\left( {{\sigma ^2}} \right)_N^{{\rm{est}}}\left( {\bf{X}} \right)} \right) \ge \frac{{2{\sigma ^4}}}{N}. 
		\end{equation}
		If we go back to Example \ref{EX. 5} and compare the result of Eq. (\ref{Eq. 1.VAR})  and the result of Eq. (\ref{Eq. 1.67}), we found 
		\begin{equation}\label{Eq. 1.68}
			\mathtt{Var}_{\mu }\left( {\mu _N^{{\rm{est}}}\left( {\bf{X}} \right)} \right) = \frac{{{\sigma ^2}}}{N},\quad  \hspace*{1cm}\mathtt{Var}_{{\sigma ^2}}\left( {\left( {{\sigma ^2}} \right)_N^{{\rm{est}}}\left( {\bf{X}} \right)} \right) \ge \frac{{2{\sigma ^4}}}{N}. 
		\end{equation}
	\end{ME}
	
	As result of this example, the estimator of $\sigma^2$ (proposed in Example (\ref{EX. 4})) does not attain the CRLB. In addition, due to the diagonal nature of CFIM, we note that the CR inequality in Eq. (\ref{Eq. 1.67}) is the same as for the case in when one of the parameters is assume known. So it is natural to ask; there is a better-unbiased estimator of $\sigma^2$ that attains the CRLB, or the CRLB is unattainable?
	\subsubsection{Attainment the CR lower bound}
	The answer to the question, which posed in the last section, is quite simple.  Recall that the CR bound was deriving by applying the Cauchy-Schwartz inequality, so the condition for attaining the CCRB is the equality condition in the Cauchy-Schwartz inequality. In the following Corollary, we will be stating the attainment condition of the CRLB. We can consider this Corollary as a beneficial tool; because it implicitly gives us a way of finding a best-unbiased estimator of estimating the unknown parameter.
	\begin{MC}\label{PC. 1.4}\textbf{(Attainment condition of the CRLB)}:  Let $X_1, X_2,...,X_N$ be \textbf{iid} random variables and $p(\boldsymbol{x}; \boldsymbol{\theta})$ the likelihood function satisfies the regularity condition of Eq. (\ref{Eq. 1.39}), with $p\left({\boldsymbol{x};\boldsymbol{\theta} } \right) = \prod\limits_{i = 1}^N {p\left( {{x_i};\boldsymbol{\theta} } \right)}$. If ${\boldsymbol{\theta} }_N^{{\rm{est }}}\left( \boldsymbol{X} \right)$ is any unbiased estimator of any vector of parameters $\boldsymbol{\theta}$, then  ${\boldsymbol{\theta} }_N^{{\rm{est }}}\left( \boldsymbol{X} \right)$ attains the CRLB if and only if
		\begin{equation}\label{Eq. 1.69}
			\frac{{\partial \,\log p\left( {\boldsymbol{x};\boldsymbol{\theta} } \right)}}{{\partial \boldsymbol{\theta} }} = \mathcal{F}_{C}\left( \boldsymbol{\theta}  \right)\left( {{\boldsymbol{\theta}}_N^{{\rm{est }}}\left( \boldsymbol{X} \right) - \boldsymbol{\theta} } \right), 
		\end{equation}
	\end{MC}
	where the estimator has $m$ dimension and $m\times m$ is the dimension of $\mathcal{F}_{C}\left(\boldsymbol{\theta}\right)$.
	
	\begin{MPreuve}
		(\textbf{Proof of Corollary \ref{PC. 1.4}}): Recall that the CCRB given in Eq. (\ref{Eq. 1.38}) can be also written, in multiparameter case,  as follows
		\begin{equation}\label{Eq. 1.70}
			{\left| {\mathtt{Cov}_{\boldsymbol{\theta} }\left( {{\boldsymbol{\theta} }_N^{{\rm{est }}}\left( \boldsymbol{X} \right),\frac{\partial }{{\partial \boldsymbol{\theta} }}\log \prod\limits_{i = 1}^N {p\left( {{x_i};\boldsymbol{\theta} } \right)} } \right)} \right|^2} \le \mathtt{Var} _{\boldsymbol{\theta} }\left( {\boldsymbol{\theta}_N^{{\rm{est }}}\left( \boldsymbol{X} \right)} \right)\mathtt{Var}_{\theta }\left( {\frac{\partial }{{\partial \boldsymbol{\theta} }}\log \prod\limits_{i = 1}^N {p\left( {{x_i};\boldsymbol{\theta} } \right)} } \right). 
		\end{equation}
		It's clear that  the attainment of this inequality requires the following constraint
		\begin{equation}\label{Eq. 1.71}
			\frac{\partial }{{\partial \boldsymbol{\theta} }}\log \prod\limits_{i = 1}^N {p\left( {{x_i};\boldsymbol{\theta} } \right)} = A\left( \boldsymbol{\theta}  \right)\left( {{\boldsymbol{\theta}}_N^{{\rm{est }}}\left( \boldsymbol{X} \right) + C\left( \boldsymbol{\theta} \right)} \right), 
		\end{equation}
		where $A\left( \boldsymbol{\theta}  \right)$ and $C\left( \boldsymbol{\theta}  \right)$ are functions that depend on $\boldsymbol{\theta}$ but not on $\boldsymbol{X}$. By identification Eq. (\ref{Eq. 1.70}) with Eq. (\ref{Eq. 1.71}), it is sufficient to prove that;
		\begin{equation*}
			C\left( \boldsymbol{\theta}  \right) =  - \boldsymbol{\theta} \quad \text{and} \quad	A\left( \boldsymbol{\theta}  \right) = \mathcal{F}_{C}\left( \boldsymbol{\theta}  \right). 
		\end{equation*}
		For doing this, we are inserting the expected value in Eq. (\ref{Eq. 1.71}), which leads to
		\begin{equation}\label{Eq. 1.72}
			E_{\boldsymbol{\theta} }\left( {\frac{\partial }{{\partial {\boldsymbol{\theta} }}}\log \prod\limits_{i = 1}^N p \left( {{x_i};{\boldsymbol{\theta}}} \right)} \right) = A({\boldsymbol{\theta}})\left( {E_{\boldsymbol{\theta} }\left( {{\boldsymbol{\theta}}_N^{{\rm{est}}}({\boldsymbol{X}})} \right) + C({\boldsymbol{\theta}})} \right). 
		\end{equation}
		Because the likelihood function $p\left(\boldsymbol{x}; \boldsymbol{\theta}\right)$ satisfies the regularity condition of Eq. (\ref{Eq. 1.39}), so we have
		\begin{equation}
			E_{\boldsymbol{\theta} }\left( {{\boldsymbol{\theta}}_N^{{\rm{est}}}({\boldsymbol{X}})} \right) + C({\boldsymbol{\theta}})=0. 
		\end{equation}
		Since the estimator ${{\boldsymbol{\theta}}_N^{{\rm{est}}}({\boldsymbol{X}})}$ satisfies the unbiased condition given in Eq. (\ref{Eqtion}), thus we have
		\begin{equation}
			C\left(\boldsymbol{\theta}\right)=-\boldsymbol{\theta}.
		\end{equation}
		Therefore, the Eq. (\ref{Eq. 1.71}) becomes
		\begin{equation}
			\frac{\partial }{{\partial \boldsymbol{\theta} }}\log \prod\limits_{i = 1}^N {p\left( {{x_i};\boldsymbol{\theta} } \right)} = A\left( \boldsymbol{\theta}  \right)\left( {{\boldsymbol{\theta}}_N^{{\rm{est }}}\left( \boldsymbol{X} \right) -\boldsymbol{\theta}} \right)
		\end{equation}
		For any parameter $\theta_j$, we have
		\begin{equation}
			\frac{\partial }{{\partial {\theta _j}}}\log \prod\limits_{i = 1}^N {p\left( {{x_i};\boldsymbol{\theta} } \right)}  = \sum\limits_{i = 1}^m {{{\left[ {A\left( {\boldsymbol{\theta}} \right)} \right]}_{ji}} \left( {\theta_N^{{\rm{est}}}{{\left( \boldsymbol{X} \right)}_i} - {\theta _i}} \right)},
		\end{equation}
		By differentiating once more the last equation, we found that;
		\begin{eqnarray}
			\frac{{{\partial ^2}}}{{\partial {\theta _j}{\theta _k}}}\log \prod\limits_{i = 1}^N {p\left( {{x_i};\boldsymbol{\theta} } \right)}  &=& \sum\limits_{i = 1}^m {\left( {{{\left[ {A\left( {\boldsymbol{\theta}} \right)} \right]}_{ji}}\left( { - {\delta _{ik}}} \right) + \frac{{\partial \,{{\left[ {A\left( \boldsymbol{\theta}  \right)} \right]}_{ji}}}}{{\partial {\theta _k}}}\left( {\theta _N^{{\rm{est}}}{{\left( \boldsymbol{X} \right)}_i} - {\theta _i}} \right)} \right)}\notag\\
			&=&  - {\left[ {A\left( {\boldsymbol{\theta}} \right)} \right]_{jk}} + \sum\limits_{i = 1}^m {\left( {\frac{{\partial \,{{\left[ {A\left( \boldsymbol{\theta}  \right)} \right]}_{ji}}}}{{\partial {\theta _k}}}\left( {\theta _N^{{\rm{est}}}{{\left( \boldsymbol{X} \right)}_i} - {\theta _i}} \right)} \right)}.
		\end{eqnarray}
		Taking the negative expectations and remembering that the estimator must satisfy the unbiased condition, we found 
		\begin{equation}
			- E_{\boldsymbol{\theta} }\left( {\frac{{{\partial ^2}}}{{\partial {\theta _j}{\theta _k}}}\log \prod\limits_{i = 1}^N {p\left( {{x_i};\theta } \right)} } \right) = {\left[ {A\left( {\boldsymbol{\theta} } \right)} \right]_{jk}}= {\left[ {\mathcal{F}_{C}\left( {\boldsymbol{\theta} } \right)} \right]_{jk}}.
		\end{equation}
		Finally, the Eq. (\ref{Eq. 1.71}) becomes again
		\begin{equation}
			\frac{\partial }{{\partial \boldsymbol{\theta} }}\log \prod\limits_{i = 1}^N {p\left( {{x_i};\boldsymbol{\theta} } \right)} = \mathcal{F}_{C}\left( \boldsymbol{\theta}  \right)\left( {{\boldsymbol{\theta}}_N^{{\rm{est }}}\left( \boldsymbol{X} \right) -\boldsymbol{\theta}} \right),
		\end{equation}
		which is agrees with the result of Corollary \ref{PC. 1.4}.
	\end{MPreuve}
	\begin{ME} \label{EX. 8}(\textbf{ Continuation of Example \ref{ME. 1.4} :}) Recall that the likelihood function of $N$ \textbf{iid} samples can be rewritten as 
	\begin{equation} \label{Eq. 1.80}
			p\left( {\boldsymbol{x};\boldsymbol{\theta} } \right) = \prod\limits_{i = 1}^N {p\left( {{x_i};\boldsymbol{\theta} } \right)}  = \frac{1}{{{{\left( {\sqrt {2\pi {\sigma ^2}} } \right)}^N}}}\exp \left( { - \sum\limits_{i = 1}^N {\frac{{{{\left( {x_{i} - \mu } \right)}^2}}}{{2{\sigma ^2}}}} } \right).
		\end{equation}
		Hence, the derivative of the log-likelihood function with respect to parameter $\sigma^2$ is given by
		\begin{eqnarray}
			\frac{\partial }{{\partial {\sigma ^2}}}\log \prod\limits_{i = 1}^N {p\left( {{x_i};{\bf{\theta }}} \right)}  &=& \frac{N}{{2{\sigma ^4}}}\left( {\sum\limits_{i = 1}^N {\frac{{{{\left( {{x_i} - \mu } \right)}^2}}}{{N}} - {\sigma ^2}} } \right)\notag \\
			&=&\mathcal{F}_{C}\left(\sigma^2\right)\left( {\sum\limits_{i = 1}^N {\frac{{{{\left( {{x_i} - \mu } \right)}^2}}}{{N}} - {\sigma ^2}} } \right).
		\end{eqnarray}
		By using the attainment condition of Corollary \ref{PC. 1.4}, one can be showing that the best-unbiased estimator of  $\sigma^2$ is ${\left( {{\sigma ^2}} \right)_N^{{\rm{est}}}\left( \boldsymbol{X} \right) = \sum\limits_{i = 1}^N {\frac{{{{\left( {{X_i} - \mu } \right)}^2}}}{{N}}} }$, which is suitable to attain the CRLB.
	\end{ME}
	
	Although some successes have been achieving above, this section still leaves some questions unanswered. First, what can we do to finding or constructing an efficient estimator if it does exist? Is there another approach desirable to follow in situations where the efficient estimator does not exist or cannot be founding even if it does exist?\\ One reasonable way to answering these questions is to investigate an alternative approach known as the maximum likelihood estimation, which we will be trying to treat in the next section. 
	\section{Maximum likelihood estimation}
	Finding the best-unbiased estimator, based on the maximum likelihood principle \cite{berger1988likelihood}, is overwhelmingly the most popular approach due to their distinct advantage to implemented complicated estimation problems. It was proposed, in the beginning, by R. A. Fisher \cite{fisher1925theory}. It has become an inherent principle in statistical inference.
	\subsection{Maximum likelihood principle}
	Without losing the previous notation, recall that if  $X_1$, $X_2$, $X_N$ are \textbf{iid} samples random variables and $p\left(\boldsymbol{x}; \boldsymbol{\theta}\right)$ the parametric PDF, the likelihood function is defined as
	\begin{equation}
		p\left( {\boldsymbol{x};\boldsymbol{\theta} } \right) = \prod\limits_{i = 1}^N {p\left( {{x_i};\boldsymbol{\theta} } \right)}.
	\end{equation}
	\begin{MD}\textbf{(maximum likelihood principle)}:
		The maximum likelihood estimator (MLE) is defined to be the value of  $\theta^{MLE}\left(\boldsymbol{x}\right)$ in which the $p\left(\boldsymbol{x}; \theta\right)$ attains its maximum, with $\boldsymbol{x}$ fixed.  More precisely, for each realization $x$, let $\theta^{MLE}\left(\boldsymbol{x}\right)$ be a parameter value in which the $p\left(\boldsymbol{x};\boldsymbol{\theta}\right)$ reaches its maximum as a function of $\theta$. The MLE of a parameter $\theta$ based on sample $\boldsymbol{X}$ is $\theta^{MLE}\left(\boldsymbol{X}\right)$.
	\end{MD}
	
	In general and in the context of determining the maximum likelihood estimation, there is an inherent issue associated with the common problem of finding the maximum of a function, and therefore of maximum likelihood estimation. This problem is that of actually finding and verifying the maximum of this function. As well known, in many cases, the problem of maximizing a function is reducing to a simple differential mathematical calculation.  Notice that the maximization is performing over the allowable range of the parameter $\theta$.  In addition, the range of the MLE coincides with the range of the parameter.
	
	Now, we are going to examine the problem of finding the maximum likelihood estimator. If we considered that the likelihood function is differentiable as needed concerning the parameters $\theta_{i}$, then the reasonable candidates for the  the maximum likelihood estimation are the values of $(\theta_{1}, . . , \theta_{m})$ that solve the following equation
	\begin{equation}\label{Cond. M}
		\frac{{\partial \,p\left( {\boldsymbol{x};\boldsymbol{\theta} } \right)}}{{\partial {\theta _i}}} = 0 \quad \text{where } \quad i=1,...,m. 
	\end{equation}
	Since the first derivative of the likelihood function equal to zero is only a necessary condition for a maximum and doesn't a sufficient condition. Then, the solutions of Eq. (\ref{Cond. M}) are only possible candidates for the MLEs. Moreover, the zeros of the first derivative of a maximum likelihood function find only extreme points in the interior of the definition domain of the likelihood function. If the extreme occurs on the boundary, the first derivative may not be zero. Thus, to find the extreme point, it is necessary to verify the boundary. Indeed, the points in which the first derivative of the likelihood function is equal to zero may be local or global minimum, local or global maximum, also knowing as inflection points. Our purpose is to find a global maximum of the likelihood function. For this purpose, it is necessary to prove that the sign of the second derivative, in the solution of Eq. (\ref{Cond. M}), is negative, i.e.
	\begin{equation}
		{\left. {\frac{{{\partial ^2}{\mkern 1mu} p\left( {\boldsymbol{x};\boldsymbol{\theta} } \right)}}{{\partial^2 {\theta _i}}}} \right|_{{\boldsymbol{\theta} ^{MLE}}}} < 0\quad {\rm{where}}\quad i = 1,...,m.
	\end{equation}
	\begin{ME}\textbf{(MLEs in the Normal distribution }$\boldsymbol{X}  \sim N\left(\mu, \sigma^2\right)$)
		Let $X_1, X_2,...,X_N$ are \textbf{iid} random variables following the normal distribution $\boldsymbol{X}\sim N \left(\mu, \sigma^2\right)$ with the likelihood function $p\left(\boldsymbol{x};\boldsymbol{\theta}\right)$ remembered in Eq. (\ref{Eq. 1.80}).  In this case, the solved of the equation $\frac{{\partial \,p\left( {\boldsymbol{x};\boldsymbol{\theta}} \right)}}{{\partial {\theta_i}}} = 0$, where $\boldsymbol{\theta}=\left(\mu,\sigma^2\right)^T$, reduces to the following equations
		\begin{equation}\label{Sol1.85}
			\sum\limits_{i = 1}^N {\left( {{x_i} - \mu } \right)}  = 0, \quad \hspace*{2cm} \sum\limits_{i = 1}^N {\frac{{{{\left( {{x_i} - \mu } \right)}^2}}}{{{\sigma ^2}}}}  = N, 
		\end{equation}
		which have, respectively, the following solutions
		\begin{equation}
			{\mu_N^{MLE}}\left( \boldsymbol{x} \right) = \frac{1}{N}\sum\limits_{i = 1}^N {{x_i}}, \quad \hspace{2cm} {\left( {{\sigma ^2}} \right)_N^{MEL}}\left( \boldsymbol{x} \right) = \frac{1}{N}\sum\limits_{i = 1}^N {{{\left( {{x_i} - \mu } \right)}^2}}.
		\end{equation}
		Therefore, these solutions are candidates for the MLEs respectively to $\mu$ and $\sigma^2$. To verify that these solutions are global maximums of the likelihood function, we can use the argument of the negativity of second derivatives of Eq. (\ref{Sol1.85}) in respect to $\mu$ and $\sigma^2$, i.e.
		\begin{equation}
			{\left. {\frac{{{\partial ^2}{\mkern 1mu} p\left( {\boldsymbol{x};\boldsymbol{\theta} } \right)}}{{{\partial ^2}\mu }}} \right|_{\mu  = {\mu ^{MLE}}}} =  - 1 < 0, \quad \hspace*{0.3cm} {\left. {\frac{{{\partial ^2}{\mkern 1mu} p\left( {\boldsymbol{x};\boldsymbol{\theta} } \right)}}{{{\partial ^2}{\sigma ^2}}}} \right|_{{\sigma ^2} = {{\left( {{\sigma ^2}} \right)}^{MEL}}}} =  - \frac{1}{{2{\sigma ^4}}}\sum\limits_{i = 1}^N {{{\left( {{x_i} - \mu } \right)}^2}}  < 0.
		\end{equation}
		Hence, these solutions are only the extreme points in the interior of the definition domain of the likelihood function, and they are the maximums.  To finalize this verification that these solutions are the global maximums, it must check the boundaries, $\pm \infty $. By taking the limits of the likelihood function in $\pm \infty$, one can be easy to establish that $\mathop {\lim }\limits_{\mu  \to  \pm \infty } p\left( {\boldsymbol{x};\boldsymbol{\theta} } \right) = 0, \hspace{0.1cm}\mathop {\lim }\limits_{{\sigma ^2} \to  + \infty } p\left( {\boldsymbol{x};\boldsymbol{\theta} } \right) = 0$. Therefore, these solutions are the global maximums, and hence ${\mu_N^{MLE}}\left( \boldsymbol{X} \right) = \frac{1}{N}\sum\limits_{i = 1}^N {{X_i}}$ and ${\left( {{\sigma ^2}} \right)_N^{MEL}}\left( \boldsymbol{X} \right) = \frac{1}{N}\sum\limits_{i = 1}^N {{{\left( {{X_i} - \mu } \right)}^2}}$ are the MLEs. These estimators are precisely those proven above in the previous section to be suitable for attaining the {CRLB}. These estimators are precisely those proven above in the previous section (see Examples \ref{ME. 1.4} and \ref{EX. 8}) to be suitable for achieving the {CRLB}. Thus, they are the best-unbiased estimators and are asymptotically efficient.
	\end{ME}
	
	In some cases, the work directly in the differentiation of likelihood function $p\left(\boldsymbol{x}; \boldsymbol{\theta}\right)$ is hard, so difficult to find the MLEs. In these cases, it's desirable to work with the natural logarithm of likelihood function, $\log p\left(\boldsymbol{x}; \boldsymbol{\theta}\right)$  , which is known as the log-likelihood function. Due to the increasing strictly of the logarithm function on $\left[0,\infty \right]$, this approach does not pose any problem because, on the interval $\left[0,\infty \right]$, the extremes points of $p\left(\boldsymbol{x}; \boldsymbol{\theta}\right)$ and $\log p\left(\boldsymbol{x}; \boldsymbol{\theta}\right)$ coincide.
	\begin{ME}(\textbf{MLE in the Bernoulli distribution} $\boldsymbol{X}  \sim B\left(N, \theta\right)$)\\
		Let be $X_1, X_2,..., X_N$ are \textbf{iid} random variables following the Bernoulli distribution with the following likelihood function
		\begin{equation}\label{LikBD}
			p\left( {\boldsymbol{x}; \theta } \right) = \prod\limits_{i = 1}^N {{\theta^{{x_i}}}\left( {1 - {\theta^{{1-x_i}}}} \right)}={\theta ^{\sum\limits_{i = 1}^N {{x_i}} }}{\left( {1 - \theta } \right)^{N - \sum\limits_{i = 1}^N {{x_i}} }},
		\end{equation}
		Although this function is hard to differentiate directly in respect to the proportion parameter $\theta$, it is much easier to differentiable the log-likelihood function, which is given by
		\begin{equation}
			\log p\left( {\boldsymbol{x};\theta } \right) = \log \left( \theta  \right)\sum\limits_{i = 1}^N {{x_i}}  + \left( {N - \sum\limits_{i = 1}^N {{x_i}} } \right)\log \left( {1 - \theta } \right).
		\end{equation}
		Differentiate the log-likelihood function with respect to $\theta$, and setting the result equal to zero, reduces to the following equation;
		\begin{equation}
			\left( {1 - \theta } \right)\sum\limits_{i = 1}^N {{x_i}}  + \theta \left( {N - \sum\limits_{i = 1}^N {{x_i}} } \right) = 0. \label{MLEBDEQ}
		\end{equation}
		Solved of the last equation leads to finding the following solution;
		\begin{equation}\label{MLEBD}
			{\theta_N^{MEL}}\left( \boldsymbol{x} \right) = \frac{1}{N}\sum\limits_{i = 1}^N {{x_i}}.
		\end{equation}
		To Verify that the $\theta^{MLE}\left(\boldsymbol{x}\right)$ is the global maximum of the log-likelihood function, we will check the sign of the second derivative of Eq. (\ref{MLEBDEQ}) with respect to $\theta$, i.e.
		\begin{equation}
			{\left. {\frac{{{\partial ^2}\log \,p\left( {\boldsymbol{x};\theta } \right)}}{{{\partial ^2}\theta }}} \right|_{\theta  = {\theta ^{MEL}}}} = 2\sum\limits_{i = 1}^N {\left( {\frac{1}{2} - {x_i}} \right)}  \le 0.
		\end{equation}
		Therefore, this solution is an extreme point in the interior of the definition domain of the log-likelihood function and is a maximum.  To finalize this check, we will verify the limits of log-likelihood in the boundaries of the definition domain, which are $\mathop {\lim }\limits_{\theta  \to 0}  \log p\left( {\boldsymbol{x};\theta } \right)$, and $\mathop {\lim }\limits_{\theta  \to 1} \,\log \,p\left( {\boldsymbol{x};\theta } \right)$, since $0 \le \theta  \le 1$. By taking these limits, one can be finding that $\mathop {\lim }\limits_{\theta  \to 0}  \log p\left( {\boldsymbol{x};\theta } \right)=\mathop {\lim }\limits_{\theta  \to 1}  \log p\left( {\boldsymbol{x};\theta } \right)=-\infty < {\theta_N^{MEL}}\left( \boldsymbol{x} \right)$, which implies that the solution given in Eq. (\ref{MLEBD}) is a global maximum, and hence ${\theta_N^{MEL}}\left( \boldsymbol{X} \right) = \frac{1}{N}\sum\limits_{i = 1}^N {{X_i}}$ is an MLE of the proportion parameter $\theta$. We noted that the MLE finding here coincides with that proposed in Example \ref{EX. 3}, which satisfies both \textbf{criteria 1, 2}. Therefore, it is satisfied the attainment condition of CRLB, and hence it is an efficient estimator.
	\end{ME}
	
	As a summary of this section, in some estimation problems, often intuition alone can lead us to very good unbiased estimators for estimating the unknown parameters. For example, the sample mean is a good reasonable estimator for estimating the population means. But in more complicated estimation problems, which often arise in practice. It cannot be denying that we need more well-organized and reasonable methods which help us to find good unbiased estimators. In this context, the maximum likelihood principle is the most extended and most utilizable approach in many models of estimation problems. In fact, there is not only the maximum likelihood principle. But there are also many several alternative approaches to the maximum likelihood principle.  For instance,  Bayes estimators principle, method of moments, and hedged maximum-likelihood estimation, the expectation-maximization algorithm, all these approaches are methods using to finding the best-unbiased estimators in the given estimation models. To learn more about these approaches, we recommend the Ref. \cite{casella2021statistical}.
	\section{Conclusion}
	In this chapter, our philosophy in presenting the main concepts of classical estimation theory is to provide the reader with the basic ideas necessary for determining an efficient estimator. We have included the Bernoulli and Normal distributions as standards examples of the practice in introducing almost all basic concepts. As mentioned previously, our goal is to obtain an efficient estimator that saturates the CCRB by satisfying the attainment condition. We have resorted to the MLE principal to find an efficient estimator. The sequence of this chapter has followed this approach: so that, the proposed estimator has discussed first, followed by testing it to be satisfied both \textbf{criteria 1} and \textbf{criteria 2}, hence saturating the CCRB, and finally, we have explaining how to find an efficient estimator by using the MLE principle.
	
	As a matter of fact, in the classical estimation theory, the CCRB dictates the fundamental limit of precision, that obtained by evaluating the inverse of CFI. However, one may wonder whether it is possible to use other statistical resources than the classical one to go beyond the restriction imposed by the CCRB. Of course, this is possible, but only in the quantum part of the estimation theory, which we will discuss in the next chapter.
	
	\chapter{Quantum estimation theory} \label{Ch. 2}
	\section{Introduction}
	Quantum mechanics has not only remained as fundamental theory restricted at a theoretical level, but it has penetrated almost every corner of modern scientific experiments. Indeed, it is the cornerstone of the current technological revolution \cite{caves1981quantum, dowling2014quantum}. Quantum metrology or quantum estimation theory is a typical example of the advantage of using the resources of quantum mechanics over those of its classical counterpart \cite{paris2009quantum}. More precisely, by using quantum measurement approaches, the precision bound imposed by the fundamental limit theorem in classical estimation parameters can be overcome. But how to generalize the classical estimation theory to quantum estimation theory? \textit{The Copenhagen interpretation of quantum mechanics} should be evoked to answer this question. In the Copenhagen interpretation, Niels Bohr has postulated the so-called \textit{Born rule or correspondence rule} as a heuristic principle \cite{van1928correspondence, miller2007classical}. This principle states that the behavior of systems described by quantum mechanics reproduces classical physics under exceptional conditions. In other terms, the correspondence principle stipulates that a new quantum scientific theory must be able to explain phenomena provided that the earlier classical theory is valid. According to this interpretation, quantum estimation theory is a natural generalization of classical estimation theory. The classical concepts have not changed their meaning, but their application has been restricted. Based on the \textit{Born rule}, Helstrom \cite{helstrom1969quantum} and Holevo \cite{holevo2011probabilistic} successfully found the fundamental concepts of the quantum version of the estimation theory. That last aims to perform high-precision for estimating the parameters specifying a given quantum statistical model.
	
	The quantum statistical model is a reformulation, in quantum mechanical terms, of the classical statistical model. In this reformulation, the density operators take the place of the probability density functions of classical statistics. Thus, the parameters to be estimated are encoded in the density operator, so the quantum estimation theory seeks the best estimators of the parameters of a density operator. In addition to the difference given in the parameterization step, the quantum estimation theory is also different from the classical one in the measurement task. In quantum mechanics terms, the performance measurements are the set of operators satisfied with the self-adjoint property. These measurements are called quantum measurements \cite{braginsky1995quantum}, and the most generally used extensively are \textbf{Positive operator-valued Measures} ({POVM}) \cite{paris2012modern}. Based on the latter, sets of quantum estimators are constructed. Similar to the classical case, a quantum counterpart of the Cramér-Rao inequality of conventional statistics sets a lower bound on the {MSE}s of these estimators. Then, we can ask ourselves how we can generalize this quantum Cramér-Rao bound ({QCRB})? And is it unique? Also, how can we generalize the corresponding Fisher information? Is this {QCRB} always saturated or not? All these questions and others will be addressed in this chapter.
	
	This chapter aims to provide a comprehensive overview of the most crucial concepts and methods in quantum estimation theory that go beyond its classical counterpart on improving estimation precision. We start by reviewing the general quantum measurements and presenting the \textit{Born rule} as a fundamental key link between the density operator of the quantum statistical model and the probability distribution function of the classical one. Next, we will exploit the non-commutative nature of quantum mechanics to derive the different families of quantum Fisher information (QFI) and the corresponding {QCRB}s. We shall discuss the attainment condition of {QCRB}s and extend the results into the multiparameter quantum estimation case. In that last case, we will be introducing the  Holevo Cramér-Rao bound ({HCRB}) as a tighter bound and using it together with the different {QCRB}s to classify the multiparameter quantum statistical models. Finally, we will end this chapter with a conclusion.

	\section{Quantum measurement theory}
	In classical models,  the description of a system has performed by a realization of a random variable, $x_1, x_2,...,x_N$, and the probability distribution PDF. For each measurement performed, $f\left(x_1, x_2,...,x_N\right)$, the required expectations values are given by the following equation
	\begin{equation}\label{Eq. 2.1}
		E\left( {f\left( {{x_1},{x_2},...,{x_N}} \right)} \right) = \sum\limits_{i = 1}^N {p\left( {{x_1},{x_2},...,{x_N}} \right)f\left( {{x_1},{x_2},...,{x_N}} \right)}. 
	\end{equation}
	In quantum models, the quantum system has described by a density operator $\hat \rho$, which is a function of the dynamical variables of a quantum system.  In this case, the expectation value of an observable $\hat A \left(\hat {X}_1, \hat {X}_2,...,\hat {X}_N\right)$, which corresponds to a quantum mechanical operator, is given by 
	\begin{equation}\label{Eq. 2.2}
		E\left( {\hat A\left( {{{\hat X}_1},{{\hat X}_2},...,{{\hat X}_N}} \right)} \right)=\operatorname{Tr}\left[\hat{\rho} \hspace{0.2cm}\hat A \left(\hat {X}_1, \hat {X}_2,...,\hat {X}_N\right) \right]. 
	\end{equation}
	Therefore, the density operator is the quantum counterpart of PDF or PMF. In the orthogonal base constructed by the eigenstates of the operator $\hat A \left(\hat {X}_1, \hat {X}_2,...,\hat {X}_N\right)$, i.e.\\ $\hat A\left( {{{\hat X}_1},{{\hat X}_2},...,{{\hat X}_N}} \right)\left| {{x_1},{x_2},...,{x_N}} \right\rangle  = a\left( {{x_1},{x_2},...,{x_N}} \right)\left| {{x_1},{x_2},...,{x_N}} \right\rangle $, the matrix density is diagonal and written as;
	\begin{equation}\label{Eq. 2.3}
		\hat \rho  = \sum\limits_{i = 1}^N {p\left( {{x_1},{x_2},...,{x_N}} \right)\left| {{x_1},{x_2},...,{x_N}} \right\rangle \left\langle {{x_1},{x_2},...,{x_N}} \right|}.  
	\end{equation}
	The expectations values given in Eq. (\ref{Eq. 2.2}) reduces to 
	\small {\begin{equation}\label{Eq. 2.4}
			E\left( {\hat A\left( {{{\hat X}_1},{{\hat X}_2},...,{{\hat X}_N}} \right)} \right) = \sum\limits_{i = 1}^N {p\left( {{x_1},{x_2},...,{x_N}} \right)\left\langle {{x_1},{x_2},...,{x_N}} \right|} \hat A\left( {{{\hat X}_1},{{\hat X}_2},...,{{\hat X}_N}} \right)\left| {{x_1},{x_2},...,{x_N}} \right\rangle. 
	\end{equation}}
	By identification the last equation with the Eq. (\ref{Eq. 2.1}), we find;
	\begin{eqnarray}\label{Eq. 2.}
		f\left( {{x_1},{x_2},...,{x_N}} \right) &=& \langle {x_1},{x_2},...,{x_N}|\hat A\left( {{{\hat X}_1},{{\hat X}_2},...,{{\hat X}_N}} \right)\left| {{x_1},{x_2},...,{x_N}} \right\rangle \\ 
		&=&a\left( {{x_1},{x_2},...,{x_N}} \right)\notag.
	\end{eqnarray}
	Thus, the average value of quantum observable is the measurable function in the classical counterpart. As a consequence, the quantum statistical theory includes the classical as a particular case.

	\subsection{Born rule as a key postulate of quantum mechanics}
	Quantum statistical aspect, which appears in theory, also has a normative and methodological treatment of estimation. It investigates the best procedures for making statements about the condition of a system under observation data. These statements have based on observational data that are subject to unavoidable random error. Logically, the best methods are those leads to minimize the influence of error by evaluating their quality. This last allowed to determine the ultimate limits imposed by statistical uncertainty on the accuracy of decisions and measurements. In fact, the outcomes of quantum experiment data are probabilistic, which means that there must exist an appropriate probability measurement, $p\left(\boldsymbol{x}\right)$, that appears in the probability space. This latter is almost similar to the classical one introduced in the last chapter. The main difference with the classical counterpart is that the probability function is not arbitrary to depend on the realization of a random variable, but in fact, is a specific function depending on both of the states of quantum models $\hat{\rho}$ and on the measurement performed $\mathcal{\hat M}$, and denoted as $p_{\hat{\rho}}\left(\mathcal{\hat{M}}\right)$. The map $\left( {\hat \rho , \mathcal{\hat M} \left(\boldsymbol{\hat X}\right)} \right) \to {p_{\hat \rho }}\left({\mathcal{\hat M} \left(\boldsymbol{\hat X}\right)}\right)$ is known as the \textit{Born rule} and given by
	\begin{equation}\label{Eq. 2.6}
		{p_{\hat \rho }}\left( {\mathcal{\hat M} \left(\boldsymbol{x}\right)} \right) = \operatorname{Tr}\left[ {\hat \rho \hspace{0.1cm}  \mathcal{\hat M}\left( \boldsymbol{\hat X} \right)} \right]. 
	\end{equation}
	Thus, the probability of obtaining the outcome $x_i$ from the measurement $\mathcal{\hat M} \left(\hat X_i\right)$ of the observable $\hat X_i$ is given by;
	\begin{eqnarray}
		p_{\hat \rho}\left(x_i\right)&=&\operatorname{Tr}\left[ {\hat \rho \hspace{0.1cm}  \mathcal{\hat M}\left(\hat X_i \right)} \right]\\ \notag
		&=& \sum\limits_{j = 1}^N {\left\langle {{x_j}} \right|} \hat \rho \hspace{0.1cm} \mathcal{\hat M}\left( {{\hat X_i}} \right)\left| {{x_j}} \right\rangle\\\notag
		&=&\sum\limits_{j, i = 1}^N {\left\langle {{x_j}} \right|p\left( {{x_i}} \right)\left| {{x_1, x_2,...,x_N}} \right\rangle \left\langle {{x_1, x_2,...,x_N }} \right|\mathcal{\hat M}\left( {{ \hat X_i}} \right)\left| {{x_j}} \right\rangle }\\\notag
		&=& \sum\limits_{j, i = 1}^N {p\left( {{x_i}} \right)\left\langle {{{x _j}}}
			\mathrel{\left | {\vphantom {{{\psi _j}} {{x_1, x_2,...,x_N}}}}
				\right. \kern-\nulldelimiterspace}
			{{{x_1, x_2,...,x_N}}} \right\rangle \left\langle {{x_1, x_2,...,x_N}} \right|\mathcal{\hat M}\left( {{\hat X_i}} \right)\left| {{x _j}} \right\rangle }.
	\end{eqnarray}
	On the orthogonal basis, we have $\left\langle {{{x _j}}}
	\mathrel{\left | {\vphantom {{{x _j}} {{x _i}}}}
		\right. \kern-\nulldelimiterspace}
	{{{x _i}}} \right\rangle  = {\delta _{ji}}$. Hence, the probability of the outcome  $x_i$ becomes
	\begin{equation*}
		{p_{\hat \rho }}\left( {{x_i}} \right) = \sum\limits_{i = 1}^N {p\left( {{x_i}} \right)\left\langle {{x _i}} \right|\mathcal{\hat M}\left( {{\hat X_i}} \right)\left| {{x _i}} \right\rangle }.
	\end{equation*}
	By correspondence, the measurement $\mathcal{\hat M}$ should be equal to $\left| {{x_i}} \right\rangle \left\langle {{x_i}} \right|$,  which is known as the projective measurement. This result is classified as one consequence of the second postulate of quantum mechanics, which we will mention in the following:
	\begin{tcolorbox}
		\textbf{Second postulate of quantum mechanics (Quantum measurements) }: Every observable $X$ in classical mechanics corresponds to a linear Hermitian operator $\hat X$ in quantum mechanics . Any Hermitian operator, $\hat X =\hat X^{\dag}$, admits a spectral decomposition $\hat X = \sum\limits_{i = 1}^N {{x_i}{\hat P_{{x_i}}}} $, in terms of its real eigenvalues $x_i$ and of the projectors, $\hat P_{x_{i}} = {\left| {{x_i}} \right\rangle \left\langle {{x_i}} \right|}$, on its eigenvectors $\hat X\left| x \right\rangle  = x\left| x \right\rangle$. In the Hilbert space, the sets of eigenvectors form a basis with an orthonormal property, i.e. $\left\langle {x}
		\mathrel{\left | {\vphantom {x {x'}}}
			\right. \kern-\nulldelimiterspace}
		{{x'}} \right\rangle  = {\delta _{xx'}}$ and $\sum\limits_{i = 1}^N {\left| {{x_i}} \right\rangle \left\langle {{x_i}} \right| = \mathbb{1}}$. The probability of obtaining the outcome $x$ from the measurement of the observable $\hat X$ is given as $p\left( x \right) = {\left| {\left\langle {\psi } \mathrel{\left | {\vphantom {\psi  x}} \right. \kern-\nulldelimiterspace}{x} \right\rangle } \right|^2}$, with ${\left| \psi  \right\rangle }$  representing the possible states of a physical system. Thus, one can get;
		\begin{eqnarray}
			p\left( x \right) =\left\langle {\psi }\mathrel{\left | {\vphantom {\psi  x}}\right. \kern-\nulldelimiterspace}{x} \right\rangle \left\langle {x}
			\mathrel{\left | {\vphantom {x \psi }}\right. \kern-\nulldelimiterspace}	{\psi } \right\rangle =\left\langle \psi  \right|{\hat P_x}\left| \psi  \right\rangle =\sum\limits_{i = i}^N {\left\langle {{x_i}} \right|{{\hat P}_x}\left| \psi  \right\rangle \left\langle {\psi }
				\mathrel{\left | {\vphantom {\psi  {{x_i}}}}
					\right. \kern-\nulldelimiterspace}
				{{{x_i}}} \right\rangle }=\operatorname{Tr}\left[ {\left| \psi  \right\rangle \left\langle \psi  \right|{{\hat P}_x}} \right]. 
		\end{eqnarray}
	\end{tcolorbox}
	Here, we have $\hat \rho  =  \ket{\psi \bra{\psi} }$, which is the density matrix associated with the pure\footnote{A quantum state is said to be ‘pure’ if it is the projector on a one-dimensional subspace of the Hilbert space, which is equivalent to stating that a quantum state is pure if and only if all of its eigenvalues are 0 except for one where is 1. Analogously, pure states may be seen as the extreme points in the convex set of quantum states. Therefore, any quantum state $\hat \rho$ admits an “ensemble” decomposition into a combination of pure states $\hat \rho  = \sum\limits_{i = 1}^N {{p_i}\left| {\,{\psi _i}} \right\rangle \left\langle {{\psi _i}} \right|}$.  That last expression represent the statistical (density) operator describing the system under investigation.  This system knows as the mixed state.} physical system, and the expectation value of the observable is $\left\langle {\hat X} \right\rangle  = \left\langle \psi  \right|\hat X\left| \psi  \right\rangle  = \operatorname{Tr}\left[ {\left| \psi  \right\rangle \left\langle \psi  \right|\hat X} \right]$. Therefore, the Born rule represents the main recipe that connects the mathematical description of a quantum state to the prediction of quantum mechanics about the results of an experiment.
	
	\subsection{Quantum measurement}
	In the quantum system of the state $\ket{\psi} \in \mathcal{H}$,  the quantum measurement of an observable is characterized by a set of operators $\left\{\boldsymbol{\mathcal{\hat M}}_{i=1,2,...\lambda}\right\}$,  where $\left\{\lambda_{i}\right\}$ are the sets of the possible outcomes of the performed measurement in the quantum system.  Indeed, these outcomes are eigenvalues of the measurement operators (observable) and must be real numbers to allow a physical interpretation of the measurement process. As knows, only self-adjoint operators have real eigenvalues. Thus, the measurement operators, $\boldsymbol{\mathcal{\hat M}}_i$, must be satisfied the self-adjoint property; 
	\begin{equation*}
		{\boldsymbol{\mathcal{\hat M}}}_i^\dag {{\boldsymbol{\mathcal{\hat M}}}_i} = {\boldsymbol{\mathcal{\hat M}}}_i^2 = {{\boldsymbol{\mathcal{\hat M}}}_i} \Leftrightarrow {{\boldsymbol{\mathcal{\hat M}}}_i} = {\boldsymbol{\mathcal{\hat M}}}_i^\dag.
	\end{equation*}
	Using results of the measurement postulate of quantum mechanics, the probability of finding $\lambda_{i}$ as a result of the performed measurement $\boldsymbol{\mathcal{\hat M}}_i$ is
	\begin{equation*}
		p\left(\lambda_{i}\right)=\bra{\psi}\boldsymbol{\mathcal{\hat M}}_i\ket{\psi}.
	\end{equation*}
	On the other hand, the sum of the probabilities of all possible outcomes of the measurement must be satisfied the unitary condition, i.e. 
	\begin{equation*}
		\sum_{i}{p\left(\lambda_{i}\right)}=\sum_{i}{\bra{\psi}\boldsymbol{\mathcal{M}}_i\ket{\psi}}=1.
	\end{equation*}
	That last equation implied that the measurement operators  must satisfy the completeness relation
	\begin{equation*}
		\sum_{i}{\boldsymbol{\mathcal{\hat M}}_i}=\sum_{i}{\ket{x_i}\bra{x_i}}=\mathbb{\hat 1}, 
	\end{equation*}
	where $\ket{x_i}$ are the set of the eigenvectors of $\boldsymbol{\mathcal{\hat M}}_i$. According to quantum mechanics principles, the state of the quantum system after and before the measurement are not coincide. This is due to the influence of  the measurement that causes the system to change its state; if the state of the system immediately before the measurement is $\ket{\psi}$, then the state of the system  after the measurement $\boldsymbol{\mathcal{\hat M}}_i$ is $\ket{\psi_i'}=\boldsymbol{\mathcal{\hat M}}_i\ket{\psi}$. 	The normalized state of the system is given by 
	\begin{equation*}
		\ket{\psi_i'}=\frac{\boldsymbol{\mathcal{\hat M}}_i\ket{\psi}}{\bra{\psi}\boldsymbol{\mathcal{\hat M}}_i\ket{\psi}}.
	\end{equation*}
	This result is often known, in quantum mechanics, as phenomenal of  \textit{\textbf{"wave function collapse"}} because the measurement appears to collapse a complicated quantum state into a state compatible (classical state) with the measurement. The following Fig. (\ref{Fig. 2.1}) is clarified, schematically, this phenomenon.
	\begin{figure}[H]
		\tikzset{every picture/.style={line width=0.75pt}} 
		
		\begin{tikzpicture}[x=0.98pt,y=0.75pt,yscale=-1,xscale=1]
			\draw  [color={rgb, 255:red, 74; green, 144; blue, 226 }  ,draw opacity=1 ][line width=3]  (159.62,92) -- (320.24,92) -- (320.24,189) -- (159.62,189) -- cycle ;
			\draw [color={rgb, 255:red, 74; green, 144; blue, 226 }  ,draw opacity=1 ][line width=2.25]    (17.06,139) -- (158.61,140) ;
			\draw [color={rgb, 255:red, 74; green, 144; blue, 226 }  ,draw opacity=1 ][line width=2.25]    (323,142) -- (449,142) ;
			\draw [color={rgb, 255:red, 86; green, 74; blue, 226 }  ,draw opacity=1 ][line width=1.5]    (187.73,140) .. controls (222.86,86) and (261.01,103) .. (283.1,140)(190.37,129.3) -- (196.73,134.15)(197.01,121.39) -- (202.87,126.83)(204.86,113.92) -- (209.97,120.08)(213.4,107.92) -- (217.41,114.84)(223.17,103.56) -- (225.59,111.18)(234.07,101.62) -- (234.39,109.61)(245.2,102.71) -- (243.35,110.49)(255.71,106.64) -- (252.06,113.75)(264.5,112.23) -- (259.64,118.58)(272.59,119.48) -- (266.85,125.04)(279.17,127.11) -- (272.84,132)(285.14,135.68) -- (278.36,139.93) ;
			\draw [color={rgb, 255:red, 0; green, 0; blue, 0 }  ,draw opacity=1 ][line width=1.5]    (231.9,147) -- (274.99,102.16) ;
			\draw [shift={(277.07,100)}, rotate = 133.87] [color={rgb, 255:red, 0; green, 0; blue, 0 }  ,draw opacity=1 ][line width=1.5]    (14.21,-4.28) .. controls (9.04,-1.82) and (4.3,-0.39) .. (0,0) .. controls (4.3,0.39) and (9.04,1.82) .. (14.21,4.28)   ;
			\draw   (1,74) -- (516,74) -- (516,210) -- (1,210) -- cycle ;
			\draw  [line width=2.25]  (165.14,97) -- (314.72,97) -- (314.72,184) -- (165.14,184) -- cycle ;
			
			\draw (19,90) node [anchor=north west][inner sep=0.75pt]   [align=left] {\textbf{Quantum state $\ket{\psi}$} };
			\draw (198,155) node [anchor=north west][inner sep=0.75pt]   [align=left] {\textbf{Mesurement}};
			\draw (347,99) node [anchor=north west][inner sep=0.75pt]   [align=left] {\textbf{Classical state}};
			\draw  [color={rgb, 255:red, 0; green, 0; blue, 0 }  ,draw opacity=0.95 ][fill={rgb, 255:red, 155; green, 155; blue, 155 }  ,fill opacity=0.14 ]  (430, 177) circle [x radius= 77.78, y radius= 29.7]   ;
			\draw (376,158) node [anchor=north west][inner sep=0.75pt]   [align=left] {\textbf{Quantum states}\\
				\hspace{0.4cm}\textbf{has collapsed}};
			
		\end{tikzpicture}
		\captionsetup{justification=centerlast, singlelinecheck=false}\captionof{figure}{Measuring a quantum state $\ket{\psi}$.} \label{Fig. 2.1}
	\end{figure}
	
	In summary, maybe see the standard quantum measurement as a set of recipes followed to generate the post-measurement and the probability of observing the result after the measurement. We also notice that the number of possible outcomes is limited by the number of measurements of the orthogonal operator. These types of quantum measurements are known as \textbf{Projection-Valued Measures} (PVM). The latter cannot be greater than the dimensionality of the Hilbert space, in which the state of a quantum system lives. However, it would often be desirable to have more results than the dimension of the Hilbert space while keeping the positivity and the normalization of the probability distributions. Formally, This is possible upon relaxing the assumptions on the mathematical describing of the measurement, and replacing them with more precise ones, still obtaining a meaningful prescription that is come by the \textit{Born rule} to generate probabilities. These measurements are known as \textbf{Positive Operator-Valued Measures} (POVM). Then, in the following sections, we will discuss the details of PVM and POVM.
	\subsubsection{Projection-Valued Measures (PVM)}
	In this part, we shall focus entirely on projective measurements or Von Neumann measurements. These measurements are the simplest type that emerges in quantum mechanics, even though they are not perfect measurements of a quantum system\footnote{The perfect and most general measurement type is, in fact, the positive operator-valued measurement, which can be used to describe some other things in which the imperfect PVM occasionally fails. We’ll discuss these later.}. A measurement is a projective or Von Neumann measurement if the measurement operators $\boldsymbol{\mathcal{\hat{M}}_i}$ are orthogonal projectors.  In this case, the projectors $\boldsymbol{\mathcal{\hat{M}}_i}$ form, at self, an orthogonal set, ${{\mathcal{\hat{M}}_{{i}}}{\mathcal{\hat{M}}_{{j}}} = {\delta _{{i}{j}}}{\mathcal{\hat{M}}_{{j}}}}$. Furthermore, since the collection of $\boldsymbol{\mathcal{\hat{M}}_i}$ must be satisfied completeness condition, $\sum_{i}{\boldsymbol{\mathcal{M}}_i}=\mathbb{\hat 1}$, then any observable $\hat{X}$ of the quantum system has a spectral decomposition:
	\begin{equation}
		\hat X = \sum\limits_{i = 1}^N {{x_i}{\boldsymbol{\mathcal{\hat M}}_i}}, 
	\end{equation}
	with $\boldsymbol{\mathcal{\hat M}}_i$ the projector onto the eigenspectra of $\hat X$ with eigenvalue $x_i$. Hence, the measurement operator corresponding to a basis vector $\ket{x_i}$ is $\boldsymbol{\mathcal{\hat M}}_i=\ket{x_i}\bra{x_i}$. This projector operator is self-adjoint and idempotent, i.e.
	\begin{equation}
		{\boldsymbol{\mathcal{\hat M}}_i} = \boldsymbol{\mathcal{\hat M}}_i^\dag \quad {\rm{ and }}\quad \boldsymbol{\mathcal{\hat M}}_i^2 = |{x_i}\rangle \langle {x_i}\left| {{x_i}} \right.\rangle \langle {x_i}| = |{x_i}\rangle \langle {x_i}| = {\boldsymbol{\mathcal{\hat M}}_i}.
	\end{equation}
	We use the concept of projective measurement to answer the question: is the result of a measurement deterministic?
	To answer this question, we consider a quantum system in the state $\ket{\psi}$ immediately before the measurement that is carried out using the projective measurement operator, $\boldsymbol{\mathcal{\hat{M}}}$.  The state $\ket{{\psi '}}$ of the quantum system, immediately after the measurement, is given by 
	\begin{equation}
		\ket{\psi '}= \frac{\boldsymbol{\mathcal{\hat{M}}}_i \ket{\psi}}{\sqrt{p\left(x_i\right)}},
	\end{equation}
	where $p\left(x_i\right)=\bra{\psi} \boldsymbol{\mathcal{\hat{M}}}_i \ket{\psi}$ is the probability of observing the outcome $x_i$. If we apply the operator $\boldsymbol{\mathcal{\hat{M}}}$ again, this time to the state $\ket{\psi '}$, the state immediately after the second measurement is 
	\begin{equation}
		\ket{\psi ''}= \frac{\boldsymbol{\mathcal{\hat{M}}}_i \ket{\psi '}}{\sqrt{p '\left(x_i\right)}}=\frac{\boldsymbol{\mathcal{\hat{M}}}_i}{\sqrt{p '\left(x_i\right)}} \frac{\boldsymbol{\mathcal{\hat{M}}}_i \ket{\psi}}{\sqrt{p\left(x_i\right)}}= \frac{\boldsymbol{\mathcal{\hat{M}}}_i \ket{\psi}}{\sqrt{p '\left(x_i\right) p\left(x_i\right)}},
	\end{equation}
	where $p '\left(x_i\right)=\bra{\psi '} \boldsymbol{\mathcal{\hat{M}}}_i \ket{\psi '}$ is the probability of observing the outcome $x_i$ in the second measurement. On the other hand, one can derive the expression of $p ' \left(x_i\right)$ as; 
	\begin{equation*}
		{p^\prime }(x_i) = \left\langle {{\psi ^\prime }\left| {{\boldsymbol{\mathcal{\hat{M}}}_i}} \right|{\psi ^\prime }} \right\rangle  = \left( {\langle \psi |\frac{{\boldsymbol{\mathcal{\hat{M}}}_i^\dag }}{{\sqrt {p(x_i)} }}} \right){\boldsymbol{\mathcal{\hat{M}}}_i}\left( {\frac{{{\boldsymbol{\mathcal{\hat{M}}}_i}}}{{\sqrt {p(x_i)} }}|\psi \rangle } \right) = \frac{{\left\langle {\psi \left| {{\boldsymbol{\mathcal{\hat{M}}}_i}} \right|\psi } \right\rangle }}{{p(x_i)}}.
	\end{equation*}
	Then, $\sqrt{p '\left(x_i\right) p\left(x_i\right)}=\sqrt{\bra{\psi} \boldsymbol{\mathcal{\hat{M}}}_i \ket{\psi}}= \sqrt{p\left(x_i\right)}$. Finally, we see that the state has not changed by the second measurement
	\begin{equation}
		\ket{\psi ''}= \frac{\boldsymbol{\mathcal{\hat{M}}}_i \ket{\psi}}{\sqrt{p '\left(x_i\right) p\left(x_i\right)}}= \frac{\boldsymbol{\mathcal{\hat{M}}}_i \ket{\psi}}{\sqrt{p\left(x_i\right)}}= \ket{\psi '}. \label{PVMR}
	\end{equation}
	Since repeated measurements yield the same outcome as the first one, thus the result of Eq. (\ref{PVMR})\footnote{This equation can be written in terms of the density operator (mixed states) as $\hat \rho ' = \frac{1}{{{p_{\hat \rho }}({x_i})}} {{\boldsymbol{\mathcal{\hat{M}}}_i}\hat \rho \boldsymbol{\mathcal{\hat{M}}}_i^\dag }$.} proves,  in reality, that the projective measurements are repetitive. This fact has a simple mathematical and physical interpretation. This interpretation is that the projection of a quantum state onto the vector basis of Hilbert space transforms quantum information into classical one, which means that a complicated quantum state collapses into a classical one and thus provides no additional information about the original quantum system. Therefore, applying again in the second time the same {PVM} operator does not lead to changing the original state. Hence,  the outcome observed as an outcome of the projective measurement is deterministic.                               
	\begin{ME}
		Consider a quantum system of a single qubit, photon polarization, or spin of an electron, the state of this system in the basis $\mathcal{{B}}=\left\{\ket{0}, \ket{1}\right\}$ is $\ket{\psi}=\alpha \ket{0}+\beta \ket{1}$, with $\left|\alpha\right|^2+\left|\beta\right|^2=1$. In canonical basis $\mathcal{{B}}$, there are only two possible outcomes measurement, which is we can only observe the system in the states $\ket{0}$ or $\ket{1}$. The corresponding measurements are made, respectively, by 
		\begin{equation*}
			{\boldsymbol{\mathcal{\hat M}}_0} = |0\rangle \langle 0| = \left( {\begin{array}{*{20}{l}}
					1&0\\
					0&0
			\end{array}} \right)\quad {\rm{ and }}\quad {\boldsymbol{\mathcal{\hat M}}_1} = |1\rangle \langle 1| = \left( {\begin{array}{*{20}{l}}
					0&0\\
					0&1
			\end{array}} \right).
		\end{equation*}
		These last measurement operators are self-adjoint, $\boldsymbol{\mathcal{\hat M}}_{i=0,1}=\boldsymbol{\mathcal{\hat M}}_{i=0,1}^{\dag}$, and idempotent, $\boldsymbol{\mathcal{\hat M}}_{i=0,1}^{2}=\boldsymbol{\mathcal{\hat M}}_{i=0,1}$, and the probability of obtaining the outcome $\ket{0}$  is 
		\begin{equation*}
			{p_0} = \left\langle {\psi \left| {{{\boldsymbol{\mathcal{\hat M}}}_0}} \right|\psi } \right\rangle  = {\left| {{\alpha}} \right|^2}.
		\end{equation*}
		The normalized state of the qubit after applying the measurement operator, $\boldsymbol{\mathcal{\hat M}}_0$, is given by
		\begin{equation*}
			\ket{\psi_0^\prime}=\frac{\boldsymbol{\mathcal{\hat M}}_0 \ket{\psi}}{\sqrt{\left\langle {\psi \left| {{{\boldsymbol{\mathcal{\hat M}}}_0}} \right|\psi } \right\rangle }}=\ket{0}.
		\end{equation*}
		Similarly, the probability of outcome corresponding to state  $\ket{1}$ is
		\begin{equation*}
			{p_1} = \left\langle {\psi \left| {{{\boldsymbol{\mathcal{\hat M}}}_1}} \right|\psi } \right\rangle  = {\left| {{\beta}} \right|^2},
		\end{equation*}
		and the normalized state of the qubit after applying the the measurement operator,  $\boldsymbol{\mathcal{\hat M}}_1$, is
		
		\begin{equation*}
			\ket{\psi_1^\prime}=\frac{\boldsymbol{\mathcal{\hat M}}_1 \ket{\psi}}{\sqrt{\left\langle {\psi \left| {{{\boldsymbol{\mathcal{\hat M}}}_1}} \right|\psi } \right\rangle }}=\ket{1}.
		\end{equation*}
		We noted that the completeness relation is satisfied, i.e.  $\boldsymbol{\mathcal{\hat M}}_0+\boldsymbol{\mathcal{\hat M}}_1=\mathbb{1}_{2\times 2}$.
		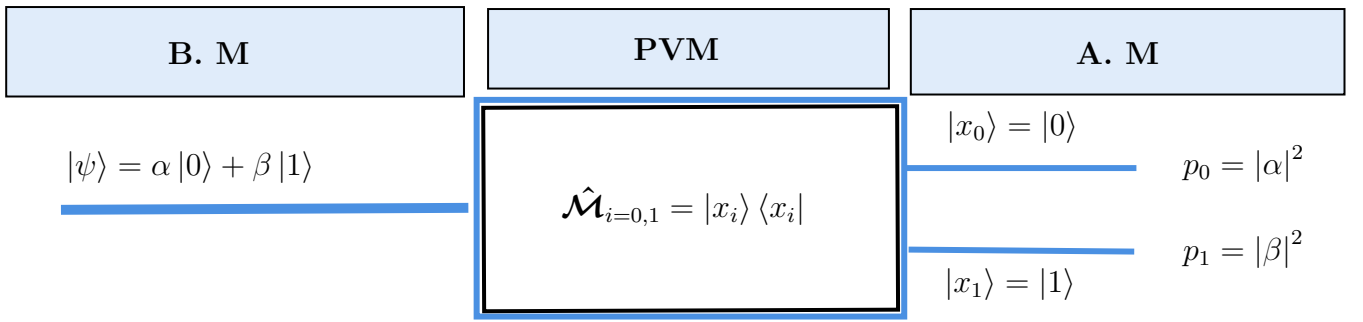
\begin{figure}[H]
			\tikzset{every picture/.style={line width=0.75pt}} 
			
			\begin{tikzpicture}[x=0.82pt,y=0.75pt,yscale=-1,xscale=1]
				
				\draw [color={rgb, 255:red, 74; green, 144; blue, 226 }  ,draw opacity=1 ][line width=3.75]    (58,131) -- (244,130) ;
				\draw  [color={rgb, 255:red, 74; green, 144; blue, 226 }  ,draw opacity=1 ][line width=2.25]  (247.78,75.69) -- (443,75.69) -- (443,184.69) -- (247.78,184.69) -- cycle ;
				\draw  [color={rgb, 255:red, 0; green, 0; blue, 0 }  ,draw opacity=1 ][line width=1.5]  (250.78,79.68) -- (439.75,78.89) -- (440.18,181.22) -- (251.21,182.01) -- cycle ;
				\draw [color={rgb, 255:red, 74; green, 144; blue, 226 }  ,draw opacity=1 ][line width=2.25]    (444,110) -- (549,110) ;
				\draw [color={rgb, 255:red, 74; green, 144; blue, 226 }  ,draw opacity=1 ][line width=2.25]    (445,151) -- (548,152) ;
				\draw  [fill={rgb, 255:red, 74; green, 144; blue, 226 }  ,fill opacity=0.17 ] (33,29) -- (242,29) -- (242,74) -- (33,74) -- cycle ;
				\draw  [fill={rgb, 255:red, 74; green, 144; blue, 226 }  ,fill opacity=0.17 ] (253,30) -- (436,30) -- (436,70) -- (253,70) -- cycle ;
				\draw  [fill={rgb, 255:red, 74; green, 144; blue, 226 }  ,fill opacity=0.17 ] (446,30) -- (644,30) -- (644,73) -- (446,73) -- cycle ;
				
				\draw (100,45) node [anchor=north west][inner sep=0.75pt]   [align=left] {\textbf{ B. M}};
				\draw (59,100) node [anchor=north west][inner sep=0.75pt]   [align=left] {$\ket{\psi}=\alpha \ket{0}+\beta \ket{1}$};
				\draw (318,43) node [anchor=north west][inner sep=0.75pt]   [align=left] {\textbf{PVM}};
				\draw (520,45) node [anchor=north west][inner sep=0.75pt]   [align=left] {\textbf{A. M}};
				\draw (461,80) node [anchor=north west][inner sep=0.75pt]   [align=left] {$\ket{x_0}=\ket{0}$};
				\draw (460,159) node [anchor=north west][inner sep=0.75pt]   [align=left] {$\ket{x_1}=\ket{1}$};
				\draw (569,97) node [anchor=north west][inner sep=0.75pt]   [align=left] {${p_0} = {\left| \alpha  \right|^2}$};
				\draw (569,140) node [anchor=north west][inner sep=0.75pt]   [align=left] {${p_1} = {\left| \beta  \right|^2}$};
				\draw (285,118) node [anchor=north west][inner sep=0.75pt]   [align=left] {$\boldsymbol{\mathcal{\hat M}}_{i=0,1}=\ket{x_i}\bra{x_i}$};

			\end{tikzpicture}
			
			\centering
			
			\captionsetup{justification=centerlast, singlelinecheck=false}\captionof{figure}{\textit{The measurement of a single state $\ket{\psi_{i=1,2}}$ in the state $\ket{\psi}$ using a single state measurement gate $\boldsymbol{\mathcal{\hat M}}_{i=1,2}=\ket{x_i}\bra{x_i}$. Before measurement (B. M),  we have a superposed quantum state formed by some linear combination of $\ket{0}$ and $\ket{1}$. After measurement (A. M), it becomes a classical state $\left(\ket{0} \text{or} \ket{1}\right)$, which means that the superposed quantum state has collapsed due to the measurement influence}.}
			\label{Fig1}
		\end{figure}
	\end{ME}
	
	In overview, the {PVM} are projectors self-adjoint operators and form an orthogonal set. Hence, orthogonal measurement operators commute, which corresponds to simultaneous observables. In addition, the number of such projectors operators is equal to the dimension of the Hilbert space in which lives the state of a quantum system. But there exist, in practice, quantum systems require to generate the quantum measurements having more results than the dimension of Hilbert space, taking into account the preservation of the positivity and normalization of the probability distributions.  For this, it is necessary to generalize the notion of projective measurement to POVM.
	\subsubsection{Positive operator-valued measurement (POVM) }
	As mentioned above, {PVM} is not always possible. Because sometimes, we destroy a quantum particle in the measurement process. Hence the violation of the repeatability of a projective measurement. For instance, a photon may be absorbed by a polarization filter and not be available for measurements again. In such cases where the system is measured only once, the state of a quantum system immediately after the measurement is no longer of interest. Also, the probabilities of the measurement outcomes are the ones that count for the first measure. In this category of system, a generalized type of measurement,  {positive operator-valued measurement}, is most suitable to perform the measurement process. Mathematically, a {POVM} is a set of non-negative Hermitian operators that are not necessarily orthogonal or commute. These sets of positive operators are acted in the Hilbert space and satisfy the completeness condition. In other words, given the sets of measurement operators $\left\{\boldsymbol{\mathcal{\hat{M}}}_i\right\}$ describing a measurement performed on a quantum system in state $\ket{\psi}$, a {POVM} has the elements of $\left\{\boldsymbol{\mathcal{\hat{M}}}_i\right\}$ satisfy
	\begin{equation} \label{POVM}
		\boldsymbol{\hat \Pi}_i \ge 0; \hspace{0.7cm} \boldsymbol{\hat \Pi_i}=\boldsymbol{\mathcal{\hat{M}}}_i^{\dag}\boldsymbol{\mathcal{\hat{M}}}_i \hspace{0.7cm} \text{and} \hspace{0.7cm} \sum_{i} \boldsymbol{\hat \Pi}_i=\mathbb{1}.
	\end{equation}
	This means that is the positive operators still represent a resolution of the identity, as the set of projectors over the eigenstates of a self-adjoint operator. The probability $p\left(x_i\right)$ of outcome $x_i$, if the system is in a state $\ket{\psi}$, is given by\footnote{In the case where the state of a quantum system is described by the matrix density $\hat{\rho}$, the probability of the outcome $x_i$ is given by $p_{\hat{\rho}}\left(x_i\right)=\operatorname{Tr}\left[\hat{\rho}\hspace{0.15cm} \boldsymbol{\hat{\Pi}}_i\right]$.}
	\begin{equation}\label{BOPOVM}
		p\left(x_i\right)= \bra{\psi} \boldsymbol{\hat{\Pi}_i \ket{\psi}}.
	\end{equation}
	The state $\ket{{\psi '}}$ of the quantum system, immediately after the measurement, is given by\footnote{This relation may also be written in terms of the density operator $\hat \rho ' = \frac{1}{{{p_{\hat \rho }}({x_i})}} {{\boldsymbol{\hat{\Pi}}_i}\hat \rho \boldsymbol{\hat{\Pi}}_i^\dag }$.
	}
	\begin{equation}
		\ket{\psi '}= \frac{\boldsymbol{\hat{\Pi}}_i \ket{\psi}}{\sqrt{p\left(x_i\right)}}.
	\end{equation}
	By applying $\boldsymbol{\hat{\Pi}}_i$ again, this time to the state $\ket{\psi '}$, the state immediately after the second measurement is
	\begin{equation} \label{STaPOVM}
		\ket{\psi ''}= \frac{\boldsymbol{\hat{\Pi}}_i }{\sqrt{p '\left(x_i\right)}} \ket{\psi '}=\frac{\boldsymbol{\hat{\Pi}}_i \boldsymbol{\hat{\Pi}}_i}{\sqrt{p '\left(x_i\right) p\left(x_i\right)}}\ket{\psi},
	\end{equation}
	As long as the orthogonality condition is no longer a requirement, then the number of elements of a {POVM} is not restricted, nor is the number of possible outcomes from the measurement. In addition, the above equations (Eqs.(\ref{BOPOVM}, \ref{STaPOVM})) generalized, respectively, \textit{Born's rule} and the post-measurement states of the system immediately after the measurements. The formulation constructed in both equations says that any set of measurement operators satisfying (\ref{POVM}) corresponds to legitimate operations leading to a proper probability distribution and a collective of the post-measurement states. Moreover, the {POVM} allows us to distinguish the cases where the outcome of a measurement identifies, with certainty, the original state from those where the identification is not possible.
	\begin{ME}
		In this example, we will use the {POVM} to distinguish the original state of a photon have two possible non-orthogonal polarization states. For this, we consider a photon in one of two possible states;
		\begin{equation*}
			\ket{\psi_1}=\ket{0} \hspace{0.7cm} \text{or} \hspace{0.7cm} \ket{\psi_2}=\alpha\ket{0}-\beta\ket{1} \hspace{0.7cm} \text{with} \hspace{0.7cm} \left|\alpha\right|^2+\left|\beta\right|^2=1.
		\end{equation*}
		According to Ref. \cite{brandt1999positive}, we can define the set of POVM with three positive operators. The two first are, in general, defined as
		\begin{equation}
			\boldsymbol{\hat{\Pi}_{i=1,2}}=\left(1+\braket{\psi_i| \psi_j}\right)^{-1}\left(\mathbb{1}-\ket{\psi_i}\bra{\psi_i}\right); \quad \text{with} \quad i \ne j,
		\end{equation}
		which mean that 
		\begin{equation*}
			\boldsymbol{\hat{\Pi}_{1}}=\left(1+\alpha\right)^{-1}\ket{1}\bra{1}, \hspace{0.7cm} \text{and} \hspace{0.7cm} \boldsymbol{\hat{\Pi}_{2}}=\left(1+\alpha\right)^{-1}\left(\alpha \ket{0}+\beta\ket{1}\right)\left(\bra{0}\alpha^*+\bra{1}{\beta ^ * }\right).
		\end{equation*}
		The third element of {POVM} is determined by the ensure completeness condition such as; 
		\begin{equation}
			\boldsymbol{\hat{\Pi}_{3}}=\mathbb{1}-\boldsymbol{\hat{\Pi}_{1}}-\boldsymbol{\hat{\Pi}_{2}}.
		\end{equation}
		Using the three possible outcomes of the measurement associated with $\boldsymbol{\hat{\Pi}_{1}}$, $\boldsymbol{\hat{\Pi}_{2}}$,  $\boldsymbol{\hat{\Pi}_{3}}$ lead us to the following conclusions regarding the original state of the system:\\
		$\bullet$ For the first element of {POVM},  $\boldsymbol{\hat{\Pi}_{1}}$, we have; $ \boldsymbol{\hat{\Pi}_{1}}\ket{\psi_1}=0$ and $ \boldsymbol{\hat{\Pi}_{1}}\ket{\psi_2}=\beta \left(1+\alpha\right)^{-1}\ket{1}$, thus the original state was $\ket{\psi_2}$ with the probability \hspace{0.25cm} $p_{\boldsymbol{\hat{\Pi}_{1}}}\left(\psi_2\right)=\left|\beta\right|^2\left(1+\alpha\right)^{-1}$.\\
		$\bullet$ For the second element of {POVM},  $\boldsymbol{\hat{\Pi}_{2}}$, we have; $ \boldsymbol{\hat{\Pi}_{2}}\ket{\psi_2}=0$ and $ \boldsymbol{\hat{\Pi}_{2}}\ket{\psi_1}=\beta \left(1+\alpha\right)^{-1}\ket{1}$, thus the original state was $\ket{\psi_1}$ with the probability \hspace{0.25cm} $p_{\boldsymbol{\hat{\Pi}_{2}}}\left(\psi_1\right)=\left|\alpha\right|^2\left(1+\alpha\right)^{-1}$.\\
		$\bullet$ For the third element of {POVM},  $\boldsymbol{\hat{\Pi}_{3}}$, we have; $ \boldsymbol{\hat{\Pi}_{3}}\ket{\psi_1}=0$ and $ \boldsymbol{\hat{\Pi}_{3}}\ket{\psi_2}=\beta \left(1+\alpha\right)^{-1}\ket{1}$, thus we do not know if the original state was $\ket{\psi_1}$ or $\ket{\psi_1}$.
		
		Finally, we conclude that there is never a mistake in identifying the original state when applying the {POVM} measurement operators $\boldsymbol{\hat{\Pi}_{1}}$ and $\boldsymbol{\hat{\Pi}_{2}}$. But there is not enough information for a reliable identification when using the {POVM} element $\boldsymbol{\hat{\Pi}_{3}}$.
	\end{ME}
	
	We conclude this section with a brief discussion of {PVM}  and {POVM} measurements. The prevalent difference between  {PVM}  and {POVM}  is the number of available preparations of the system and the number of outcomes that may be different from each other and different from the dimension of the Hilbert space. Another difference from the projective measurements is that a {POVM} measurement is, in general, not repeatable.
	\section{Quantum estimation problem}
	After the overview of quantum measurement, in the previous section. Now, we focus on the estimation problem, which is the formulation of statistical and estimation theory in quantum mechanical terms. It involves substituting the probability density functions that appear in the classical estimation theory with quantum-mechanical density operators. Thus, replacing a classical statistical model with its quantum counterpart.
	\subsection{Quantum statistical models}
	We consider a quantum system described by density operators with Hilbert space, $\hat{\rho}\in \mathcal{H}$.  Analogically with the classical case, the quantum statistical model "$\mathcal{S}$" is defined as a family of density operators in $\mathcal{S}\left(\mathcal{H}\right)$ and parameterized by a set of reals parameters $\boldsymbol{\theta} \in \Theta$, where $\Theta$ is an open subset of $R^m$. Formally,
	\begin{equation}
		\mathcal{S}=\left\{\hat{\rho}\left(\boldsymbol{\theta}\right)\in \mathcal{S}\left(\boldsymbol{\theta}\right)/\boldsymbol{\theta}=\boldsymbol{\theta} = \left(\theta_1, \theta_2, ..., \theta_m\right)\in \Theta  \subset \mathbb{R}^m\right\},
	\end{equation}
	where the parameterization map $\Theta \ \to \mathcal{S}\left(\boldsymbol{\theta}\right)$ is injective, and the parameterized density operator, $\hat{\rho}\left(\boldsymbol{\theta}\right)$, can be differentiated as many times as needed with respect to the parameters.
	
	Given a quantum statistical model $\mathcal{S}$,  performing measurements on the quantum system is always required for obtaining information about the system. The performance of a {POVM} measurement on $\hat{\rho}\left(\boldsymbol{\theta }\right)$ leads to the probability of the outcome $x$, $p\left(x; \boldsymbol{\theta }\right)=\operatorname{Tr}\left[\hat{\rho}\left(\boldsymbol{\theta }\right) \boldsymbol{\hat{\Pi}}\left(x\right)\right]$. This means that the probability distribution is parameterized by $\boldsymbol{\theta }$. Therefore, a classical statistical model is a special case arising from its quantum counterparts. Besides, the performance measurement and the state of a quantum system are, together with\textit{ Born’s rule}, model the probabilistic nature of the measurement data.
	\begin{ME}{\textbf{Quantum statistical model of qubit state}.}
		We consider a quantum state of a two-level system, photon polarization, electron position, described by $\ket{\psi}$ and decomposed in the binary base such as; $\ket{\psi}=\cos\left(\frac{\theta}{2}\right)\ket{0}+e^{i\phi}\sin\left(\frac{\theta}{2}\right)\ket{1}$, where $0 \leq \theta \leq \pi$ and $0 \leq \phi \leq 2\pi$. This state specifies, uniquely,  a point on the sphere $\mathbb{R}^{3}$ and is called the Bloch representation of a qubit. 
		\begin{figure}[H]
			\centering \includegraphics[scale=0.7]{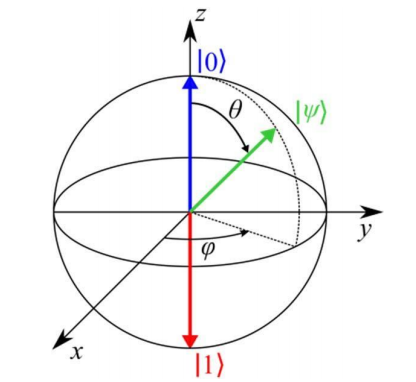}
			\caption{ Representation of the superposed state in the Bloch sphere, the blue color represents the projection of $\ket{\psi}$ into  $\ket{0}$,  and the red color represents the projection of  $\ket{\psi}$ into  $\ket{1}$}
		\end{figure}
		The corresponding density operator\footnote{For an arbitrary qubit, the corresponding density matrix can be written in the Bloch representation as $\hat{\rho}=\frac{1}{2} \sum_{i=0}^{3} r_{i} \hat{\sigma}_{i}, \quad \hat{\sigma}_{0}=\hat{\mathbb{1}}, \quad r_{i}=\operatorname{Tr}\left(\hat{\sigma}_{i} \hat{\rho}\right)$, where $\hat \sigma_{1,2,3}$ are the three Pauli matrices.\\} of this state is
		\begin{equation}
			\hat{\rho}\left(\theta\right)=\frac{1}{2}\left(\begin{array}{cc}
				1+\cos \theta / 2 & 1-i \cos \theta / 2 \sin \theta / 2 \\
				1+i \cos \theta / 2 \sin \theta / 2 & 1-\cos \theta / 2
			\end{array}\right).
		\end{equation}
		The set of {PVM}\footnote{Since the considered system has only one state, then the {POVM} has become projective, thus coinciding with {PVM}.} that we will perform on $\hat{\rho}\left(\theta\right)$ are
		\begin{equation}
			\hat{\Pi}_{\uparrow}=\frac{1}{2}\left(1+\hat{\sigma}_{z}\right), \hspace{0.8cm} \hat{\Pi}_{\downarrow}=\frac{1}{2}\left(1-\hat{\sigma}_{z}\right),
		\end{equation}
		which leads to the probability of the outcomes $\boldsymbol{x}=\left\{\uparrow, \downarrow\right\}$
		\begin{equation}
			p(\uparrow ; {\theta})=\operatorname{Tr}\left[\hat{\Pi}_{\uparrow} \hat{\rho}\right]=\frac{1}{2}(1+\cos \theta / 2), \quad p(\downarrow ; {\theta})=\operatorname{Tr}\left[\hat{\Pi}_{\downarrow} \hat{\rho}\right]=\frac{1}{2}(1-\cos \theta / 2).
		\end{equation}
	\end{ME}
	
	The principal quantities of the classical estimation theory introduced in the last chapter were derived from PDF. Based on \textit{ Born's rule}, we can immediately generalizes theses quantities to the case where the probability distribution outcomes from a quantum mechanical measurement. Related to what was stated in the conclusion of the last chapter, a natural question arises: what is the best possible precision obtained in estimating a set of parameters encoded in the quantum state? By optimizing the overall possible {POVM} measurement, the answer to this question is included in the quantum Cramér-Rao bound ({QCRB}). The last is regarded, in fact, as an intrinsic key of quantum estimation theory. It is determined entirely by the quantum Fisher information ({QFI}), which depends only on the state $\hat{\rho}\left(\boldsymbol{\theta}\right)$. In the next section, we shall introduce the different quantum mechanical versions of the Fisher information.
	\section{Quantum Fisher information}
	
	In this section, we will inspire by the derivation of the classical Fisher information, derived in the first chapter, to discover the quantum Fisher information.  To simplify this task, we first restrict ourselves to the case of a single estimation mode and then extend it to the multi-parameter estimation model. In classical estimation theory, the classical Fisher information is defined as  variance\footnote{he variance of a random variable  ${X}$  is defined as $\mathtt{Var}(X) = {\rm{E}}\left[ {{{\left( {X - E\left( X \right)} \right)}^2}} \right]$} of a random variable defined by 
	\begin{equation}
		{l_\theta }\left( \boldsymbol{x} \right) = \frac{\partial }{{\partial \theta }}\log \,p\left( {\boldsymbol{x},\theta } \right) \quad \text{or} \quad {l_\theta }\left( x \right) = \frac{1}{{p\left( {\boldsymbol{x},\theta } \right)}}\frac{\partial }{{\partial \theta }}\,p\left( {\boldsymbol{x},\theta } \right).
	\end{equation}
	The quantity ${l_\theta }\left( \boldsymbol{x} \right)$ is often called the classical logarithmic derivative ({LD}).  Analogically, to find the quantum version of Fisher information, it is sufficient to replace the classical {LD} with its corresponding one in quantum mechanical terms. The corresponding operator of classical {LD} is evolving by replacing the probability density functions with the quantum-mechanical density operators. Reasonably, any  quantum {LD} operator is admitted as long as it collapses on the {LD} in the classical case. Due to the non-commutativity of quantum mechanics, there are many quantum analogs of a classical {LD}. Therefore, an infinite number of quantum generalizations of the classical Fisher information.  In general, the family of super-operators that generalized the {LD} is defined implicitly by the following differential equation: 
	\begin{equation}\label{SOLD}
		{\partial _\theta }{\hat \rho _\theta } = \left( {q{\mathcal{\hat L}_{\theta ,q}}{\hat \rho _\theta } + (1 - q){\hat \rho _\theta }{\mathcal{\hat L}_{\theta ,q}}} \right),
	\end{equation}
	where $q \in \left[0,1\right]$. In the classical case, we noted that the quantum-mechanical density operator is a diagonal matrix, which implies that operator "$\mathcal{\hat L}_{\theta ,q}$" is also classical for any $q \in \left[0,1\right]$. Thus, the quantum nature of "$\mathcal{\hat L}_{\theta ,q}$" has closed to the classical one "$l_\theta$". Indeed, for each value of $q$, there is an associated operator. But two families of {LD} operators are most frequently used in the literature. The first, called the  Symmetric Logarithmic Derivative ({SLD}) operator, is derived from Eq. (\ref{SOLD}) when $q=1/2$. When $q=0$, the second one is derived and called the Right Logarithmic Derivative ({RLD}) operator. That is, the {SLD} and {RLD} operators are defined, respectively, via the following differential equations:
	\begin{equation} \label{SLD}
		{\partial _\theta }{\hat \rho _\theta } = \frac{1}{2}\left( {{\mathcal{\hat L}_\theta^{(S)} }{\hat \rho _\theta } + {\hat \rho _\theta }{\mathcal{\hat L}_\theta^{(S)} }} \right),
	\end{equation}
	\begin{equation} \label{RLD}
		{\partial _\theta }{\hat \rho _\theta } =  {\hat \rho _\theta }{\mathcal{\hat L}_\theta^{(R)} }= \left({\mathcal{\hat L}_\theta^{(R)} }\right)^{\dag} {\hat \rho _\theta }.
	\end{equation}
	
	Now that we have defined the {SLD} and {RLD} operators, we are in the most adequate position to determine the quantum version of the Fisher information that corresponds to each.
	\subsection{SLD quantum Fisher information}
	As mentioned above, the link between the classical and quantum statistical models is the Borne rule, $p\left(\boldsymbol{x}; {\theta }\right)=\operatorname{Tr}\left[\hat{\rho}\left({\theta }\right) {\hat{\Pi}}\left(\boldsymbol{x}\right)\right]$. If we inserted the derivation operation into the sides of this rule, we have $\partial_\theta p\left(x; {\theta }\right)=\operatorname{Tr}\left[\partial_\theta \hat{\rho}\left({\theta }\right) {\hat{\Pi}}\left(x\right)\right]=\mathfrak{Re}{\left[\rm \operatorname{Tr}\left[\hat{\rho}\left({\theta }\right) {\hat{\Pi}}\left(x\right)\mathcal{\hat L}_\theta^{(S)}\right]\right]}$\footnote{Recall that the real part of a complex number $Z$ is $\mathfrak{Re}\left[Z\right]=1/2 (Z+\bar{Z})$.}. Hence, the classical Fisher information of Eq. (\ref{Eq. 1.49}) becomes
	\begin{equation}
		\mathcal{F}_C\left(\theta\right)=\sum_{\boldsymbol{x}}{\frac{\mathfrak{Re}{\left[\operatorname{Tr}\left[\hat{\rho}\left({\theta }\right) {\hat{\Pi}}\left(\boldsymbol{x}\right)\mathcal{\hat L}_\theta^{(S)}\right]\right]^2}}{\operatorname{Tr}\left[\hat{\rho}\left({\theta }\right) {\hat{\Pi}}\left(\boldsymbol{x}\right)\right]}}.
	\end{equation}
	We maximize, now, this quantity over all {POVM} measurements, $\hat{\Pi}\left(\boldsymbol{x}\right)$. Following\footnote{Let us $Z_1$ and $Z_2$ are the complex number, we have $\mathfrak{Re}{\left[Z_1\right]}\mathfrak{Re}{\left[Z_2\right]} \le \left\|Z_1 Z_2\right\|$}, we have
	\begin{eqnarray}
		\mathcal{F}_C\left(\theta\right) &\le& \sum_{\boldsymbol{x}}{{\left\| {\frac{{\rm Tr\left[ {\hat \rho \left( \theta  \right)\hat \Pi \left( {\boldsymbol{x}} \right)\hat {\cal L}_\theta ^{(S)}} \right]}}{{\sqrt {\rm Tr\left[ {\hat \rho \left( \theta  \right)\hat \Pi \left( {\boldsymbol{x}} \right)} \right]} }}} \right\|^2}}\\\notag
		&=&\sum_{\boldsymbol{x}}{{\left\| {{\mathop{\rm Tr}\nolimits} \left[ {\frac{{\sqrt {\hat \rho (\theta )} \sqrt {\hat \Pi (\boldsymbol{x})} }}{{\sqrt {{\mathop{\rm Tr}\nolimits} \left[ {\hat \Pi (\boldsymbol{x})\hat \rho (\theta )} \right]} }}\sqrt {\hat \Pi (\boldsymbol{x})} {\hat {\cal L}_\theta ^{(S)} }\sqrt {\hat \rho (\theta )} } \right]} \right\|^2}},
	\end{eqnarray}
	where this inequality is saturated if and only if $\rm Tr\left[ {\hat \rho \left( \theta  \right)\hat \Pi \left( {\boldsymbol{x}} \right)\hat {\cal L}_\theta ^{(S)}} \right]$ is a real number. Next, using the Schwartz inequality for the trace\footnote{$\left|\operatorname{Tr}\left[A B^{\dagger}\right]\right|^{2} \leq \operatorname{Tr}\left[A A^{\dagger}\right] \operatorname{Tr}\left[B B^{\dagger}\right]$} leads to
	
	\begin{eqnarray} \label{QFIEX}
		\mathcal{F}_C &\le& \sum\limits_{\boldsymbol{x}} {{\mathop{\rm Tr}\nolimits} \left[ {\sqrt {\hat \Pi \left( \boldsymbol{x} \right)} {\hat{\cal L}_\theta ^{(S)} }\sqrt {\hat \rho \left( \theta  \right)} \sqrt {\hat \Pi \left( \boldsymbol{x} \right)} {\hat{\cal L}_\theta ^{(S)} }\sqrt {\hat \rho \left( \theta  \right)} } \right]}\\\notag
		&=&\sum_{\boldsymbol{x}}{{\rm{Tr}}\left[ {\hat \rho \left( \theta  \right) {\hat{\cal L}}_\theta ^{(S)} \hspace{0.1cm}\hat \Pi \left( \boldsymbol{x} \right)\hat{\cal L}}_\theta ^{(S)}\right]}\\
		&=& {\rm{Tr}}\left[ {\hat \rho \left( \theta  \right){{\left({\mathcal{\hat L}_\theta ^{(S)}} \right)}^2}} \right]=\left\langle{\left({\mathcal{\hat L}_\theta ^{(S)}} \right)}^2 \right\rangle.
	\end{eqnarray}
	In the earlier equality, we used the cyclic permutation propriety of trace\footnote{$\rm Tr\left[ {ABCD} \right] = \rm Tr\left[ {BCDA} \right] = \rm Tr\left[ {CDAB} \right] = \rm Tr\left[ {DABC} \right]$}. In the chain above of inequality, for any quantum measurement, we show that the classical Fisher information is the upper bounded by the so-called \textbf{quantum Fisher information} (QFI), i.e.
	\begin{equation}\label{CQFI}
		\mathcal{	F}_C\left(\theta\right)\le \mathcal{F}_Q^{\left(S\right)}\left(\theta\right)={\rm{Tr}}\left[ {\hat \rho \left( \theta  \right){{\left({\mathcal{\hat L}_\theta ^{(S)}} \right)}^2}} \right]={\rm{Tr}}\left[ {{\partial _\theta }\hat \rho \left( \theta  \right)\hat {\cal L}_\theta ^{(S)}} \right].
	\end{equation}
	Note that the {QFI} is only a function depending on the states of the quantum statistical model and does not depend on the type of measurement performed. As the {CFI},  from Eq. (\ref{QFIEX}),  the {QFI} is also defined as the variance of the {SLD} operator (or the first moment of the {SLD} operator), i.e.
	\begin{equation}
		\mathcal{F}_Q^{\left(S\right)}\left(\theta\right)={\left( {\Delta \hat {\cal L}_\theta ^{(S)}} \right)^2} = \left\langle {{{\left( {\hat {\cal L}_\theta ^{(S)}} \right)}^2}} \right\rangle  - {\left\langle {\left( {\hat {\cal L}_\theta ^{(S)}} \right)} \right\rangle ^2}.
	\end{equation}
	It is straightforward to show this result, using the fact that $\rm Tr\left[\partial_\theta {\hat \rho \left( \theta  \right)}\right]=0$, which implied that $ {\left\langle {\left( {\hat {\cal L}_\theta ^{(S)}} \right)} \right\rangle}=0$. Hence,  $	\mathcal{F}_Q^{\left(S\right)}\left(\theta\right)= \left\langle {{{\left( {\hat {\cal L}_\theta ^{(S)}} \right)}^2}} \right\rangle$. On the other hand, the set of {POVM} measurements that leads to saturated the inequality (\ref{CQFI}), the {CFI} is equal to the {QFI}, are called optimal quantum measurements. This saturation be realized if and only if  
	\begin{equation}
		\frac{{\sqrt {\hat \Pi \left( \boldsymbol{x} \right)} \sqrt {\hat \rho \left( \theta  \right)} }}{{{\mathop{\rm Tr}\nolimits} \left[ {\hat \rho \left( \theta  \right)\hat \Pi \left( \boldsymbol{x} \right)} \right]}} = \frac{{\sqrt {\hat \Pi \left( \boldsymbol{x} \right)} \hat {\cal L}_\theta ^{(S)}\sqrt {\hat \rho \left( \theta  \right)} }}{{{\mathop{\rm Tr}\nolimits} \left[ {\hat \rho \left( \theta  \right)\hat \Pi \left( \boldsymbol{x} \right)\hat {\cal L}_\theta ^{(S)}} \right]}}.
	\end{equation}
	That last condition is satisfied if the set of {POVM} measurement $\left\{\hat \Pi\left(\boldsymbol{x}\right)\right\}$ constructed by the projectors over the eigenstates of $\hat {\cal L}_\theta^{(S)}$, which, in turn, represents the optimal {POVM} to estimate the parameter $\theta$ encoded in $\hat{\rho}\left(\theta\right)$. 
	
	The QFI has many interesting properties. The first, it is additive for independent quantum measurement
	\begin{equation}
		\mathcal{F}_{Q}^{\left(S\right)}\left({\hat \rho _1}\left( \theta  \right) \otimes {\hat \rho _2}\left( \theta  \right)\right)=\mathcal{F}_{Q}^{\left(S\right)}\left({\hat \rho _1}\left( \theta  \right) \right)+\mathcal{F}_{Q}^{\left(S\right)}\left({\hat \rho _2}\left( \theta  \right) \right).
	\end{equation}
	The second one is the convex in the quantum states. This means that for any two states ${\hat \rho _1}\left( \theta  \right)$ with probability $p_1$ and ${\hat \rho _1}\left( \theta  \right)$  with the probability $p_2$ such that $p_1+p_2=1$, we have
	\begin{equation}
		\mathcal{F}_{Q}^{\left(S\right)}\left(p_1{\hat \rho _1}\left( \theta  \right) + p_2 {\hat \rho _1}\left( \theta  \right)\right)\le p_1\mathcal{F}_{Q}^{\left(S\right)}\left({\hat \rho _1}\right)+p_2\mathcal{F}_{Q}^{\left(S\right)}\left({\hat \rho _2}\right).
	\end{equation}
	As well as, for direct sum we have
	\begin{equation}
		{\cal F}_Q^{\left( S \right)}\left( {{p_1}{{\hat \rho }_1}\left( \theta  \right) \oplus {p_2}{{\hat \rho }_2}\left( \theta  \right)} \right) = {p_1}{\cal F}_Q^{\left( S \right)}\left( {{{\hat \rho }_1}\left( \theta  \right)} \right) + {p_2}{\cal F}_Q^{\left( S \right)}\left( {{{\hat \rho }_2}\left( \theta  \right)} \right).
	\end{equation}
	
	In the eigenbasis of $\hat{\rho}\left(\theta\right)$,  we write that:
	\begin{equation}
		\hat \rho \left( \theta  \right) = \sum\limits_{i = 1}^s {{p_i}} \ket{\psi_i}\bra{\psi_i} \hspace{0.5cm} \text{and} \hspace{0.5cm} \hat {\cal L}_\theta^{(S)}=\sum\limits_{i=1}^s{{\left({\hat {\cal L}}_\theta^{(S)}\right)_{jk}}\ket{\psi_j}\bra{\psi_k}},
	\end{equation}
	where $s$ is the number of non vanish eigenstates of $\hat{\rho}\left(\theta\right)$. In the following, we assume that the quantum state of the statistical model is living in the finite-dimensional Hilbert space, which means that "$s$" is finite. In this case, the differential equation of {SLD} operator (\ref{SLD}) is constructed as
	\begin{eqnarray}
		\sum\limits_{jk} {{{\left( {{\partial _\theta }\hat \rho \left( \theta  \right)} \right)}_{jk}}}\ket{\psi_j}\bra{\psi_k} = \frac{1}{2}\sum\limits_{jk} {\left( {{p_j} + {p_k}} \right)} {\left( {\hat {\cal L}_\theta ^{(S)}} \right)_{jk}}\ket{\psi_j}\bra{\psi_k}.
	\end{eqnarray}
	This means that each element of $\partial _{\theta }\hat \rho \left( \theta  \right)$ must correspond to that of $1/2\left( {\hat {\cal L}_\theta ^{(S)}\hat \rho \left( \theta  \right) + \hat \rho \left( \theta  \right)\hat {\cal L}_\theta ^{(S)}} \right)$, and thus we have
	\begin{equation}
		{\left( {\hat {\cal L}_\theta ^{(S)}} \right)_{jk}} = \frac{{2{{\left( {{\partial _\theta }\hat \rho \left( \theta  \right)} \right)}_{jk}}}}{{{p_j} + {p_k}}}.
	\end{equation}
	Inserting this back into ${\hat {\cal L}_\theta ^{(S)}}$, leads to
	\begin{equation}
		{\hat {\cal L}_\theta ^{(S)}}=2\sum_{jk}{\frac{{{{\left( {{\partial _\theta }\hat \rho \left( \theta  \right)} \right)}_{jk}}}}{{{p_j} + {p_k}}}}\ket{\psi_j}\bra{\psi_k}, \label{elem of SLD}
	\end{equation}
	where the sum includes only with $p_j+p_k\ne 0 $. The {QFI} can now be written in the eigenbasis of the density operator as
	\begin{equation}\label{deco QFI}
		\mathcal{F}_Q^{\left(S\right)}\left(\theta\right)= 2\sum\limits_{jk} {\frac{{{{\left| {\bra{\psi_j}\left( {{\partial _\theta }\hat \rho \left( \theta  \right)} \right)\ket{\psi_k} } \right|}^2}}}{{{p_j} + {p_k}}}}. 
	\end{equation}
	Note that the {SLD} operator depends on the density matrix ${\hat \rho \left( \theta  \right)}$, which, in turn, whose the eigenvalues $p_i$ and the eigenstates $\ket{\psi_i}$ may depend on the parameter. For this, we explicitly write the derivation of  ${\hat \rho \left( \theta  \right)}$ as
	\begin{equation}
		{\partial _\theta }\hat \rho \left( \theta  \right) = \sum\limits_i{ {{\partial _\theta }} {p_i}\ket{\psi_i}\bra{\psi_i}+p_i\ket{\partial_\theta{\psi_i}}\bra{\psi_i}+p_i\ket{\psi_i}\bra{\partial_\theta{\psi_i}}}0.\label{elem matrix }
	\end{equation}
	Since $\braket{\psi_j|\psi_k}=\delta_{jk}$, then we have ${\partial _\theta }\braket{{\psi _j}|{\psi _k}}  =\braket{\partial_\theta  \psi_j|\psi_k}  + \braket{\psi_j| \partial_\theta\psi_k} = 0$, and therefore $\braket{\partial_\theta  \psi_j|\psi_k}  = - \braket{\psi_j| \partial_\theta\psi_k}$. Inserting this fact back into Eq. (\ref{elem matrix }) and substituting it into Eq. (\ref{elem of SLD}), we find 
	\begin{equation}
		{\hat {\cal L}_\theta ^{(S)}}=\sum_{i}{\frac{\partial_{\theta }p_i}{p_i}}\ket{\psi_i}\bra{\psi_i}+ 2\sum_{j \ne k}{\frac{p_k-p_j}{p_i+p_j}}\braket{\psi_j|\partial_\theta\psi_k}\ket{\psi_j}\bra{\psi_k},
	\end{equation}
	and also the {QFI} of Eq. (\ref{deco QFI}) become 
	\begin{eqnarray}
		{ {\cal F}_Q^{\left(S\right)}\left(\theta\right)}&=&\sum_{i}{\frac{\left(\partial_{\theta }p_i\right)^2}{p_i}}+ 2\sum_{j \ne k}{\frac{\left(p_k-p_j\right)^2}{p_i+p_j}}\left|\braket{\psi_j|\partial_\theta\psi_k}\right|^2\\
		&=&\sum_{i}{p_i\left(\frac{\partial_{\theta }p_i}{p_i}\right)^2}+ 2\sum_{j \ne k}{\frac{\left(p_k-p_j\right)^2}{p_i+p_j}}\left|\braket{\psi_j|\partial_\theta\psi_k}\right|^2\\
		&=& \mathrm{E}_{\theta}\left(\left(\frac{\partial}{\partial \theta} \log (p(\boldsymbol{x} ; \theta))\right)^{2}\right)+2\sum_{j \ne k}{\frac{\left(p_k-p_j\right)^2}{p_i+p_j}}\left|\braket{\psi_j|\partial_\theta\psi_k}\right|^2. \label{QCFI}
	\end{eqnarray}\\
	The first term of Eq. (\ref{QCFI}) corresponds to the {CFI}, whereas the second one contains the truly quantum contribution. As well as,  if the eigenvectors of $\hat{\rho}\left(\theta\right)$ do not depend on the parameter $\theta$, then the second term vanishes. In this case, we have $\left[\partial_{\theta }\hat{\rho}\left(\theta\right),\hat{\rho}\left(\theta\right)\right]=0$, and the {SLD} operator collapsed to scalar {LD} and the {QFI} be the {CFI}. Hence, the quantum version of Fisher information includes the classical one as a particular case. 
	\subsection{RLD quantum Fisher information}\label{SEC RLD F}
	Similarly, inserting the derivation operation on both sides of the Borne rule based on the definition of RLD leads to ${\partial _\theta }p\left( {\boldsymbol{x};\theta } \right) = \operatorname{Tr}\left[ {{\partial _\theta }\hat \rho \left( \theta  \right)\hat \Pi \left( \boldsymbol{x} \right)} \right] = {\rm Tr\left[ {\hat \rho \left( \theta  \right)\hat {\cal L}_\theta ^{(R)}\hat \Pi \left( x \right)} \right]}= {\rm{Tr}}\left[ {{{\left( {\hat {\cal L}_\theta ^{(R)}} \right)}^\dag }\hat \rho \left( \theta  \right)\hat \Pi \left( x \right)} \right] = \mathfrak{Re}\left[ \rm Tr\left[ {\hat \rho \left( \theta  \right)\hat {\cal L}_\theta ^{(R)}\hat \Pi \left( x \right)} \right]\right]$. If we entered this in the definition of {CFI} (\ref{Eq. 1.49}), then we have 
	\begin{equation}
		{{\cal F}_C}\left( \theta  \right) = \sum\limits_x {\frac{{\mathfrak{Re}{{\left[ {\operatorname{Tr}\left[ {\hat \rho \left( \theta  \right)\hat {\cal L}_\theta ^{(R)} \hspace{0.08cm}\hat \Pi \left( x \right)} \right]} \right]}^2}}}{{\operatorname{Tr}\left[ {\hat \rho \left( \theta  \right)\hat \Pi \left( x \right)} \right]}}}.
	\end{equation}
	Next, we maximize that last quantity over all {POVM}, which leads to 
	\begin{eqnarray}
		{{\cal F}_C}\left( \theta  \right)&\le & \sum\limits_{\boldsymbol{x}} {{{\left\| {\frac{{\operatorname{Tr}\left[ {\hat \rho \left( \theta  \right)\hat {\cal L}_\theta ^{(R)} \hspace{0.08cm}\hat \Pi \left( \boldsymbol{x} \right)} \right]}}{{\sqrt {{\rm{Tr}}\left[ {\hat \rho \left( \theta  \right)\hat \Pi \left( \boldsymbol{x} \right)} \right]} }}} \right\|}^2}}\\ \notag
		&=&\sum\limits_x {{{\left\| {\rm Tr\left[ {\frac{{\sqrt {\hat \Pi \left( \boldsymbol{x} \right)} \sqrt {\hat \rho \left( \theta  \right)} }}{{\sqrt {{\rm{Tr}}\left[ {\hat \rho \left( \theta  \right)\hat \Pi \left( \boldsymbol{x} \right)} \right]} }}\sqrt {\hat \rho \left( \theta  \right)} \hat {\cal L}_\theta ^{(R)}\sqrt {\hat \Pi \left( \boldsymbol{x} \right)} } \right]} \right\|}^2}}. 
	\end{eqnarray}
	That last inequality is saturated if and only if  $\mathfrak{Im}\left[{\operatorname{Tr}\left[ {\hat \rho \left( \theta  \right)\hat {\cal L}_\theta ^{(R)}\hat \Pi \left( \boldsymbol{x} \right)} \right]}\right]$, i.e. if the quantity $\operatorname{Tr}\left[ {\hat \rho \left( \theta  \right)\hat {\cal L}_\theta ^{(R)}\hat \Pi \left( \boldsymbol{x} \right)} \right]$ is a real number for all $\theta$. This requires the  {RLD} to be Hermitian operator, $\hat {\cal L}_\theta ^{(R)} = {\left( {\hat {\cal L}_\theta ^{(R)}} \right)^\dag }$, but, in general, the {RLD} is not Hermitian. Consequently, the set of projectors over the eigenstates of {RLD} does not represent, in general, the optimal {POVM} measurement.  We, now, using the Schwartz inequality of trace 
	\begin{eqnarray}
		{{\cal F}_C}\left(\theta\right)&\le& \sum\limits_{\boldsymbol{x}} {{\rm{Tr}}\left[ {\sqrt {\hat \rho \left( \theta  \right)} \hat {\cal L}_\theta ^{(R)}\sqrt {\hat \Pi \left( \boldsymbol{x} \right)} \sqrt {\hat \Pi \left( \boldsymbol{x} \right)} {{\left( {\hat {\cal L}_\theta ^{(R)}} \right)}^\dag }\sqrt {\hat \rho \left( \theta  \right)} } \right]} \\\notag
		&=&\sum\limits_{\boldsymbol{x}} {{\rm{Tr}}\left[ {\hat \rho \left( \theta  \right)\hat {\cal L}_\theta ^{(R)}\hat \Pi \left( \boldsymbol{x} \right){{\left( {\hat {\cal L}_\theta ^{(R)}} \right)}^\dag }} \right]}\\
		&=&{{\rm{Tr}}\left[ {\hat \rho \left( \theta  \right)\hat {\cal L}_\theta ^{(R)}{{\left( {\hat {\cal L}_\theta ^{(R)}} \right)}^\dag }} \right]}= \mathcal{	F}_Q^{\left(R\right)}\left(\theta\right). \label{RLD FI}
	\end{eqnarray}
	Note that the {RLD} quantum Fisher information developed in Eq. (\ref{RLD FI}) will coincide with the {SLD} quantum Fisher information evolved in Eq. (\ref{CQFI}) if and only if the {RLD} operator is self-adjoint or Hermitian.
	
	In the diagonal basis of the matrix density operator, we write that;
	\begin{equation}\label{DE RLD}
		\hat \rho \left( \theta  \right) = \sum\limits_{i = 1}^s {{p_i}} \ket{\psi_i}\bra{\psi_i} \hspace{0.5cm} \text{and} \hspace{0.5cm} \hat {\cal L}_\theta^{(R)}=\sum\limits_{i=1}^s{{\left({\hat {\cal L}}_\theta^{(R)}\right)_{jk}}\ket{\psi_j}\bra{\psi_k}},
	\end{equation}
	Thus, the differential equation of the {RLD} operator (\ref{RLD}) is constructed on this basis as
	\begin{eqnarray}
		\sum\limits_{jk} {{{\left( {{\partial _\theta }\hat \rho \left( \theta  \right)} \right)}_{jk}}}\ket{\psi_j}\bra{\psi_k}&=&\sum_{ijk}{p_i\left(\mathcal{\hat L}_{\theta}^{\left(R\right)}\right)_{jk}}\ket{\psi_i}\braket{\psi_i|\psi_j}\bra{\psi_k}\\\notag
		&=&\sum_{jk}{p_j\left(\mathcal{\hat L}_{\theta}^{\left(R\right)}\right)_{jk}}\ket{\psi_j}\bra{\psi_k},
	\end{eqnarray}
	so each element of ${{\partial _\theta }\hat \rho \left( \theta  \right)}$ must correspond to that of ${\hat \rho \left( \theta  \right)\hat {\cal L}_\theta ^{(R)}}$, and thus we have 
	\begin{equation}
		{\left( {\hat {\cal L}_\theta ^{\left(R\right)}} \right)_{jk}} = \frac{{{{\left( {{\partial _\theta }\hat \rho \left( \theta  \right)} \right)}_{jk}}}}{{{p_j}}}0.
	\end{equation}
	Inserting this results back into ${\hat {\cal L}_\theta ^{(R)}}$ (\ref{DE RLD}), leads to
	\begin{equation}
		\hat {\cal L}_\theta ^{\left(R\right)} = \sum\limits_{jk} {\frac{{{{\left( {{\partial _\theta }\hat \rho \left( \theta  \right)} \right)}_{jk}}}}{{{p_j}}}}\ket{\psi_j}\bra{\psi_k},
	\end{equation}
	and the conjugate corresponding is 
	\begin{equation}
		\left(\hat {\cal L}_\theta ^{\left(R\right)}\right)^{\dag} = \sum\limits_{jk} {\frac{{{{\left( {{\partial _\theta }\hat \rho \left( \theta  \right)} \right)}_{kj}^{*}}}}{{{p_j}}}}\ket{\psi_k}\bra{\psi_j}.
	\end{equation}
	Thus, in the diagonal basis of $\hat{\rho}\left(\theta\right)$, the {RLD} quantum Fisher information given in Eq. (\ref{RLD FI}) is written as
	\begin{eqnarray}
		\mathcal{F}_Q^{(R)}\left(\theta\right)&=&\rm Tr\left[\hat{\rho}\left(\theta\right)\hat {\cal L}_\theta ^{\left(R\right)}\left(\hat {\cal L}_\theta ^{\left(R\right)}\right)^{\dag}\right]\\\notag
		&=&\rm Tr\left[\partial_\theta \hat{\rho}\left(\theta\right)\left(\hat {\cal L}_\theta ^{\left(R\right)}\right)^{\dag}\right]\\
		&=&\sum_{jk}{\frac{\left|\bra{\psi_j}\partial_\theta \hat{\rho}\left(\theta\right)\ket{\psi_k}\right|^2}{p_j}}.
	\end{eqnarray}
	Using the result of Eq. (\ref{elem matrix }), the last equation become
	\begin{eqnarray}
		\mathcal{F}_Q^{(R)}\left(\theta\right)&=&\sum_{i}{\frac{\left(\partial_{\theta }p_i\right)^2}{p_i}}+ 2\sum_{jk}{\frac{p_j p_k}{p_j}}\left|\braket{\psi_j|\partial_\theta\psi_k}\right|^2\\
		&=& \mathrm{E}_{\theta}\left(\left(\frac{\partial}{\partial \theta} \log (p(\boldsymbol{x} ; \theta))\right)^{2}\right)+2\sum_{jk}{\frac{p_j p_k}{p_j}}\left|\braket{\psi_j|\partial_\theta\psi_k}\right|^2. \label{RLDQFI}
	\end{eqnarray}
	The first term in Eq. (\ref{RLDQFI})  represents the classical Fisher information of the distribution probability density function $p\left(\boldsymbol{x}; \theta\right)$, while the second term represents the truly quantum contribution. Also, note that when the eigenvectors of matrix density are not dependent on parameter $\theta$, the second term will vanish. Therefore, the {RLD} operator collapsed to classical scalar {LD}, and the {RLD} quantum Fisher information coincides with the classical Fisher information.
	\section{Quantum Cramér-Rao bound}
	Now that we have the quantum version of Fisher information in terms of {SLD} and {RLD} operators, we can, therefore, derive the quantum version of Cramér-Rao bound. That last is known as the {"Quantum Cramér-Rao bound"(QCRB)}. On the other hand, as the {QFI} is not uniquely family due to the non-commutativity of quantum mechanics, the {QCRB} is also not uniquely. Then, it is reasonable that each quantum family of Fisher information has the {QCRB} associated. Thus, in the next section, we will try to discuss the various families of {QCRBs}, which are the {SLD} and the {RLD-QCRBs}.
	\subsection{SLD quantum Cramér-Rao bound}
	In this subsection, we will derive the {SLD}-quantum Cramér-Rao inequality.  This inequality was derived originally by Helstrom, and it is occasionally referred to as the Helstrom bound. In order to construct the SLD-QCRB, we consider first a {POVM} measurement on a quantum system in state $\hat \rho \left( \theta  \right)$ that acts as an estimator for $\theta$. In other words, the estimator of $\theta$ is described, in the quantum case,  as a function of the quantum measurement, $\theta^{est}\left(\hat{\Pi}\left(\boldsymbol{X}\right)\right)$. On the other hand,  the expectation value of a measurement operator in quantum mechanics is given by the Born rule, $\mathtt{E}_{\theta}\left(\theta^{est}\left(\hat{\Pi}\left(\boldsymbol{X}\right)\right)\right)=\rm Tr\left[\hat\rho\left(\theta\right)\theta^{est}\left(\hat{\Pi}\left(\boldsymbol{X}\right)\right)\right]$. Thus and by analogical, the classical definition of Biased estimator\footnote{see the Bias definition of an estimator given in Def. (\ref{DEFbaias})} becomes; 
	\begin{eqnarray}
		{{\mathop{\rm Bias}\nolimits} _\theta }\left(\theta^{est}\left(\hat{\Pi}\left(\boldsymbol{X}\right)\right) \right) &=& {\mathtt{E}_\theta }\left( \theta^{est}\left(\hat{\Pi}\left(\boldsymbol{X}\right)\right)\right) - \theta ,\quad \forall \theta  \in \Theta\\\notag
		&=&\rm Tr\left[\hat\rho\left(\theta\right)\theta^{est}\left(\hat{\Pi}\left(\boldsymbol{X}\right)\right)\right]-\theta\hspace{0.08cm} \rm Tr\left(\hat\rho\left(\theta\right)\right)\\
		&=&\rm Tr\left[\hat\rho\left(\theta\right)\left(\theta^{est}\left(\hat{\Pi}\left(\boldsymbol{X}\right)\right)-\theta\right)\right].
	\end{eqnarray}
	Now,  we take the derivation of $\mathtt{Bias}_{\theta}$ with respect to $\theta$  we have then 
	\begin{equation}
		\partial_{\theta }\mathtt{Bias}_{\theta}\left(\theta^{est}\left(\hat{\Pi}\left(\boldsymbol{X}\right)\right) \right)=\rm Tr\left[\partial_\theta\hat\rho\left(\theta\right)\left(\theta^{est}\left(\hat{\Pi}\left(\boldsymbol{X}\right)\right)-\theta\right)\right]-1. \label{Bias Derv}
	\end{equation}
	By inserting the {SLD} \ref{SLD} in the last equation, we get 
	\begin{eqnarray}
		\partial_\theta\mathtt{Bias}_{\theta}\left(\theta^{est}\left(\hat{\Pi}\left(\boldsymbol{X}\right)\right) \right)+1&=&\rm Tr\left[\frac{1}{2}\left( {{\mathcal{\hat L}_\theta^{(S)} }{\hat\rho\left(\theta\right) } + {\hat\rho\left(\theta\right) }{\mathcal{\hat L}_\theta^{(S)} }} \right)\left(\theta^{est}\left(\hat{\Pi}\left(\boldsymbol{X}\right)\right)-\theta\right)\right]\\
		&=&\mathfrak{Re} \hspace{0.08cm}{\rm Tr\left[ {\mathcal{\hat L}_\theta^{(S)} }{\hat\rho\left(\theta\right)}\left(\theta^{est}\left(\hat{\Pi}\left(\boldsymbol{X}\right)\right)-\theta\right)\right]}.
	\end{eqnarray}
	Next, we square the last equation and use together the Schwartz inequality for the trace, we have then 
	\vspace{-0.5cm}
	\begin{eqnarray}
		\notag\left(\partial_\theta\mathtt{Bias}_{\theta}\left(\theta^{est}\left(\hat{\Pi}\left(\boldsymbol{X}\right)\right) \right)+1\right)^2&=&\left(\mathfrak{Re} \hspace{0.08cm}{\rm Tr\left[ {\mathcal{\hat L}_\theta^{(S)} }{\hat\rho\left(\theta\right)}\left(\theta^{est}\left(\hat{\Pi}\left(\boldsymbol{X}\right)\right)-\theta\right)\right]}\right)^2\\\notag
		&\le&\left\|\rm Tr\left[ {\mathcal{\hat L}_\theta^{(S)} }{\hat\rho\left(\theta\right)}\left(\theta^{est}\left(\hat{\Pi}\left(\boldsymbol{X}\right)\right)-\theta\right)\right]\right\|^2\\
		&=&\left\|\rm Tr\left[ {\mathcal{\hat L}_\theta^{(S)} }{\sqrt{\hat\rho\left(\theta\right)} \sqrt{\hat\rho\left(\theta\right)}}\left(\theta^{est}\left(\hat{\Pi}\left(\boldsymbol{X}\right)\right)-\theta\right)\right]\right\|^2\\\notag
		&\le&\rm Tr\left[ \left({\mathcal{\hat L}_\theta^{(S)} }\right)^2{\hat\rho\left(\theta\right)}\right] \rm Tr\left[\hat\rho\left(\theta\right)\left(\theta^{est}\left(\hat{\Pi}\left(\boldsymbol{X}\right)\right)-\theta\right)^2\right].
	\end{eqnarray}
	By identifying with the definition of {MSE}\footnote{see Eq. (\ref{Eq. 1.18}) in the last chapter.} and the {QFI} associated with {SLD} (\ref{CQFI}), we get 
	\begin{equation}
		\left(\partial_\theta\mathtt{Bias}_{\theta}\left(\theta^{est}\left(\hat{\Pi}\left(\boldsymbol{X}\right)\right) \right)+1\right)^2\le \mathcal{F}_Q^{(S)}\left(\theta\right)\left(\mathtt{Bias}_{\theta}\left(\theta^{est}\left(\hat{\Pi}\left(\boldsymbol{X}\right)\right) \right)^2+\mathtt{Var}\left(\theta^{est}\left(\hat{\Pi}\left(\boldsymbol{X}\right)\right)\right)\right).
	\end{equation}
	The unbiased condition, $\mathtt{Bias}_{\theta}=0$, which is referred to as the first \textbf{criteria 1} of an efficient estimator, leads to 
	\begin{equation}
		\mathtt{Var}\left(\theta^{est}\left(\hat{\Pi}\left(\boldsymbol{X}\right)\right)\right)\ge\frac{1}{\mathcal{F}_Q^{(S)}\left(\theta\right)}.
	\end{equation}
	That last inequality is the so-called quantum Cramér-Rao bound or quantum Cramér-Rao inequality (SLD-QCRB). It depends only on the state of quantum statistical models and does not depend on types of quantum measurement. For $N$ independent quantum measurements, the additivity property of {QFI} implies that
	\begin{equation}
		\mathtt{Var}\left(\theta^{est}\left(\hat{\Pi}\left({X}\right)\right)\right)\ge\frac{1}{N\mathcal{F}_Q^{(S)}\left(\theta\right)}.
	\end{equation}
	That last form of SLD-QCRB is the most used in the many protocols of quantum estimation theory.
	
	Now that we have the quantum version of the {QCRB}, the question addressed in the following is how to find the optimal estimator that saturates the {QCRB}? Of course, the response to this question includes the optimal measurement for which the {QFI} equals the classical one, which means that the {QCRB} is equal to the classical CRB, i.e.
	
	\begin{equation}
		\mathtt{Var}\left(\theta^{est}\left(\hat{\Pi}\left({\textbf{X}}\right)\right)\right)=\mathtt{Var}\left(\theta^{est}\left(\textbf{X}\right)\right).
	\end{equation}
	As mentioned above, this saturation is realized when the $\mathcal{	F}_C\left(\theta\right)=\mathcal{F}_{Q}^{\left(S\right)}\left(\theta\right)$. That last condition is satisfied only if the measurements performed are constructed by the projectors over the eigenstates of $\mathcal{\hat L}_\theta^{(S)}$, which generally depends on the unknown value of parameter $\theta$ and written as
	\begin{equation}\label{EQ. 2.71}
		\mathcal{\hat L}_\theta^{(S)}=g\left(\theta\right)\left(\theta^{est}\left(\hat{\Pi}\left(\boldsymbol{X}\right)\right)+f\left(\theta\right)\right), 
	\end{equation}
	where $g(\theta)$ and $f(\theta)$ are functions that depend on $\theta$ and not on $\hat{\Pi}\left(\boldsymbol{X}\right)$.  We noted that the last equation corresponds to the quantum version of  Eq. (\ref{Eq. 1.69}), which was mentioned in the classical estimation case. From Eq. (\ref{EQ. 2.71}), we can write the optimal estimator such that
	\begin{equation}\label{EQ. 2.72}
		\theta^{est}\left(\hat{\Pi}\left(\boldsymbol{X}\right)\right)=\frac{\mathcal{\hat L}_\theta^{(S)}}{g\left(\theta\right)}+f\left(\theta\right).
	\end{equation}
	It only remains to determine the functions $g\left(\theta\right)$ and $f\left(\theta\right)$ using the unbiased condition.  Then, we insert the expectation value in both sides of (\ref{EQ. 2.72}), which leads to 
	\begin{equation}\label{EQ. 2.73}
		\operatorname{Tr}\left[\hat \rho\left(\theta\right) \theta^{est}\left(\hat{\Pi}\left(\boldsymbol{X}\right)\right)\right]=\frac{	\operatorname{Tr}\left[\hat \rho\left(\theta\right)\mathcal{\hat L}_\theta^{(S)}\right]}{g\left(\theta\right)}+f\left(\theta\right).
	\end{equation}
	Using the fact that $\operatorname{Tr}\left[\hat \rho\left(\theta\right)\mathcal{\hat L}_\theta^{(S)}\right]=0$ and the  unbiased condition, we gets that $f\left(\theta\right)=\theta$ and $g\left(\theta\right)=\mathcal{F}_Q^{(S)}\left(\theta\right)$. Thus, the explicit form for the optimal quantum estimator is
	\begin{equation}\label{EQ. 2.74}
		\theta^{est}\left(\hat{\Pi}\left(\boldsymbol{X}\right)\right)=\theta +\frac{\mathcal{\hat L}_\theta^{(S)}}{\mathcal{F}_Q^{(S)}\left(\theta\right)}.
	\end{equation}
	Consequently, there is an estimator that saturates the {SLD-QCRB}. It generally depends on the unknown parameter $\theta$.
	\subsection{RLD quantum Cramér-Rao bound}
	Based on the same approaches followed for deriving the {SLD}-QCRB in the last subsection, we will derive, in this subsection, the {RLD}-QCRB associated with the {RLD}-quantum Fisher information. For this, we inserted the {RLD}-operator defined in Eq. (\ref{RLD}) into (\ref{Bias Derv}), we have 
	\begin{eqnarray}
	\notag	\partial_\theta\mathtt{Bias}_{\theta}\left(\theta^{est}\left(\hat{\Pi}\left(\boldsymbol{X}\right)\right) \right)+1&=&\rm Tr\left[{\hat\rho\left(\theta\right) }{\mathcal{\hat L}_\theta^{(R)} } \left(\theta^{est}\left(\hat{\Pi}\left(\boldsymbol{X}\right)\right)-\theta\right)\right]\\
		&=&\rm Tr\left[\frac{1}{2}\left({\hat\rho\left(\theta\right)}{\mathcal{\hat L}_\theta^{(R)}}+\left({\mathcal{\hat L}_\theta^{(R)}}\right)^{\dag}{\hat\rho\left(\theta\right)}\right) \left(\theta^{est}\left(\hat{\Pi}\left(\boldsymbol{X}\right)\right)-\theta\right)\right]\\\notag
		&=&\mathfrak{Re} \hspace{0.08cm}{\rm Tr\left[{\hat\rho\left(\theta\right)}{\mathcal{\hat L}_\theta^{(R)}}\left(\theta^{est}\left(\hat{\Pi}\left(\boldsymbol{X}\right)\right)-\theta\right)\right]}.
	\end{eqnarray}
	Squaring the last equation and using the Schwartz inequality for trace,  we gets
	\begin{eqnarray}
		\notag\left(\partial_\theta\mathtt{Bias}_{\theta}\left(\theta^{est}\left(\hat{\Pi}\left(\boldsymbol{X}\right)\right) \right)+1\right)^2&=&\left(\mathfrak{Re} \hspace{0.08cm}{\rm Tr\left[ {\hat\rho\left(\theta\right)}{\mathcal{\hat L}_\theta^{(R)}}\left(\theta^{est}\left(\hat{\Pi}\left(\boldsymbol{X}\right)\right)-\theta\right)\right]}\right)^2\\\notag
		&\le&\left\|\rm Tr\left[{\hat\rho\left(\theta\right)}{\mathcal{\hat L}_\theta^{(R)}}\left(\theta^{est}\left(\hat{\Pi}\left(\boldsymbol{X}\right)\right)-\theta\right)\right]\right\|^2\\
		&=&\left\|\rm Tr\left[ \sqrt{{\hat\rho\left(\theta\right)}}{\mathcal{\hat L}_\theta^{(R)}}\left(\theta^{est}\left(\hat{\Pi}\left(\boldsymbol{X}\right)\right)-\theta\right)\sqrt{{\hat\rho\left(\theta\right)}}\right]\right\|^2\\\notag
		&\le&\rm Tr\left[{\hat\rho\left(\theta\right)}{\mathcal{\hat L}_\theta^{(R)}}\left({\mathcal{\hat L}_\theta^{(R)}}\right)^{\dag}\right] \rm Tr\left[\hat\rho\left(\theta\right)\left(\theta^{est}\left(\hat{\Pi}\left(\boldsymbol{X}\right)\right)-\theta\right)^2\right].
	\end{eqnarray}
	Identifying both with the {MSE} and the {RLD}-quantum Fisher information, we have 
	\begin{equation}
		\left(\partial_\theta\mathtt{Bias}_{\theta}\left(\theta^{est}\left(\hat{\Pi}\left(\boldsymbol{X}\right)\right) \right)+1\right)^2\le \mathcal{F}_Q^{(R)}\left(\theta\right)\left(\mathtt{Bias}_{\theta}\left(\theta^{est}\left(\hat{\Pi}\left(\boldsymbol{X}\right)\right) \right)^2+\mathtt{Var}\left(\theta^{est}\left(\hat{\Pi}\left(\boldsymbol{X}\right)\right)\right)\right).
	\end{equation}
	For the unbiased estimator, $\mathtt{Bias}_{\theta}=0$, we have finally
	\begin{equation} \label{RLD-QCRB}
		\mathtt{Var}\left(\theta^{est}\left(\hat{\Pi}\left(\boldsymbol{X}\right)\right)\right)\ge\frac{1}{\mathcal{F}_Q^{(R)}\left(\theta\right)}.
	\end{equation}
	As mentioned in Sec. (\ref{SEC RLD F}), the {RLD}-operator is not, in general, Hermitian, which implies that the set of projectors constructed by its eigenvectors does not always correspond to the optimal measurement and thus does not match the {POVM}. Therefore, the {RLD}-quantum Cramér-Rao inequality derived in Eq. (\ref{RLD-QCRB}) is not always achievable. 
	
	Indeed, in the case of single-parameter estimation, the unattainable {QCRB} based on the {RLD}-operator is not an issue because the {SLD}-{QCRB} is always attainable and higher than the {RLD}-{QCRB}. To show this, we return to comparing Eq. (\ref{deco QFI}) with the Eq. (\ref{RLDQFI}). This comparison is straightforward only we noted that 
	\begin{equation}
		\frac{2}{p_j+p_k}\le\frac{1}{p_j} \quad \text{for each} \quad p_{j}\in \left[0,1\right].
	\end{equation}
	This leads to conclude that $\mathcal{F}_{Q}^{(R)}\ge \mathcal{F}_{Q}^{(S)}$. Thus, the {QCRB} derived using {SLD}-quantum Fisher information is never less than that derived by using the {RLD}-quantum Fisher information.  Consequently, the {RLD}-quantum Cramér-Rao bound is appropriate only in the case of multiparameter estimation models, which we will analyze extensively in the next section.
	\section{Multiparameter quantum estimation }
	Many applications of quantum estimation theory require precision estimation involving several unknown parameters, for example, thermometry \cite{correa2015individual}, multiple phases estimation \cite{humphreys2013quantum}, super-resolution quantum imaging \cite{tsang2016quantum}, magnetic field detection \cite{zhang2014fitting, bakmou2019quantum}. For this, we will extend the result discussed above to the case in which we need to improve quantum parameter estimation of multiparameter simultaneously. We will, firstly, derive the SLD-quantum Fisher information matrix (SLD-QFIM) and the corresponding {SLD-QCRB}. Next, we shall derive the {RLD}-{QFIM} and corresponding {RLD-QCRB}. We will end this section with a discussion of the Holevo Cramér-Rao bound (HCRB) and the general attainability condition of {QCRB}.
	\subsection{SLD quantum Fisher information matrix } \label{Sub SLDQFIM}
	Consider a family of quantum states, $\hat \rho\left(\boldsymbol{\theta}\right)$, with coded values of $m$-real parameters, which we will represent as a vector in the parameter space $\boldsymbol{\theta}=\left(\theta_{1}, \theta_{2}, \ldots, \theta_{m}\right)^{T} \in \Theta \subset \mathbb{R}^{m}$.  Since the generalization to the $m$-dimensional parameter space creates a natural matrix quantity, then the {QFI} becomes a matrix of dimension $m\times m$ and written as 
	\begin{equation}\label{QFIMSLD1}
		\left[\mathcal{F}_Q^{(S)}\left(\boldsymbol{\theta}\right)\right]_{jk}=\rm Tr\left[\partial_{{\theta _j}}\hat \rho\left(\boldsymbol{\theta}\right)\mathcal{\hat L}_{\theta_{k}}^{(S)}\right] \quad \text{for all} \quad j,k =1,2,...,m. 
	\end{equation}
	In terms of the different {SLD}-operators, the {QFIM} of Eq. (\ref{QFIMSLD1})  becomes
	\begin{equation}\label{SLDQFIM}
		\left[\mathcal{F}_Q^{(S)}\left(\boldsymbol{\theta}\right)\right]_{jk}=\frac{1}{2}\rm Tr\left[\left\{\mathcal{\hat L}_{\theta_{j}}^{(S)}, \mathcal{\hat L}_{\theta_{k}}^{(S)}\right\}\hat \rho\left(\boldsymbol{\theta}\right)\right],
	\end{equation}
	where $\left\{\right\}$ denotes the anti-commutator between $\mathcal{\hat L}_{\theta_{j}}^{(S)}$ and $\mathcal{\hat L}_{\theta_{k}}^{(S)}$. If $j=k$, then we have $\left[\mathcal{F}^{(S)}\left(\boldsymbol{\theta}\right)\right]_{ii}=\mathcal{F}^{(S)}\left({\theta_{i}}\right)$, which means that the {QFIM} is reduced to the {QFI} scalar. Thus, the issue of multiparameter estimation includes the single as a special case. 
	
	In the diagonal basis of the matrix density, $\hat \rho \left( \boldsymbol{\theta}  \right) = \sum\limits_{i = 1}^s {{p_i}} \ket{\psi_i}\bra{\psi_i}$, we can write the matrix elements of {QFIM} as follows
	\begin{equation}
		\left[\mathcal{F}_Q^{(S)}\left(\boldsymbol{\theta}\right)\right]_{jk}=\sum_{jk}{\frac{2}{p_j+p_k}\bra{\psi_j}\partial_{\theta_j}\hat \rho \left( \boldsymbol{\theta}  \right)\ket{\psi_k}\bra{\psi_k}\partial_{\theta_k}\hat \rho \left( \boldsymbol{\theta}  \right)\ket{\psi_j}}.
	\end{equation}
	Understanding the {QFIM} form seems quite complicated, and the most general analytical expression requires diagonalizing the density matrix. This diagonalization is, in turn,  hard to evaluate for any high dimensional density matrix. However, this problem has been avoided by exploiting some mathematical techniques and derivations, which we will address in next paragraph.
	
	In order to compute, analytically, the {QFIM} without diagonalization of the density matrix, we should first obtain the {SLD}-operator $\mathcal{\hat L}_{\theta_{j}}^{(S)}$ satisfying the continuous Lyapunov equation of the following form
	\begin{equation}\label{VecSLD}
		\frac{\partial\hat \rho \left( \boldsymbol{\theta}  \right)}{\partial{{\theta _j}}}=\frac{1}{2}\left(\mathcal{\hat L}_{\theta_{j}}^{(S)}\hat \rho \left( \boldsymbol{\theta}  \right)+\hat \rho \left( \boldsymbol{\theta}  \right)\mathcal{\hat L}_{\theta_{j}}^{(S)}\right). 
	\end{equation}
	To solve this equation, we will introduce some interesting properties of the concept of vectorization of matrices\footnote{The vectorization of a matrix $A_{n\times n}$, noted $\mathtt{vec}[\rm A]$, is an operation that transforms a matrix $A_{n\times n}$ into a column vector, i.e. the columns of a matrix are put one below the other in a single column. Formally, the vectorization of $A_{n\times n}$ is defined as $\operatorname{vec}[A]=\left(\mathbb{1}_{n \times n} \otimes A\right) \sum_{i=1}^{n} e_{i} \otimes e_{i}$.}, that are
	\begin{equation}\label{vecpro1}
		\operatorname{vec}[A X B]=\left(B^{\top} \otimes A\right) \operatorname{vec}[X]. 
	\end{equation}
	\begin{equation}\label{vec tr}
		\rm Tr\left(A^{\dagger} B\right)=\operatorname{vec}[A]^{\dagger} \operatorname{vec}[B]. 
	\end{equation}
	By inserting the vec-operator into both sides of Eq. (\ref{VecSLD}), we have
	\begin{eqnarray}
		\mathtt{vec}\left[\partial_{{\theta _j}}\hat \rho \left( \boldsymbol{\theta}  \right)\right]&=&\frac{1}{2}\mathtt{vec}\left[\mathcal{\hat L}_{\theta_{j}}^{(S)}\hat \rho \left( \boldsymbol{\theta}  \right)+\hat \rho \left( \boldsymbol{\theta}  \right)\mathcal{\hat L}_{\theta_{j}}^{(S)}\right]\\\notag
		&=&\frac{1}{2}\mathtt{vec}\left[\mathbb{1}\mathcal{\hat L}_{\theta_{j}}^{(S)}\hat \rho \left( \boldsymbol{\theta}  \right)+\hat \rho \left( \boldsymbol{\theta}  \right)\mathcal{\hat L}_{\theta_{j}}^{(S)}\mathbb{1}\right]\\\notag
		&=&\frac{1}{2}\left(\left(\hat \rho \left( \boldsymbol{\theta}  \right)^{\top}\otimes \mathbb{1}\right)\mathtt{vec}\left[\mathcal{\hat L}_{\theta_{j}}^{(S)}\right]+\left(\mathbb{1}\otimes\hat \rho \left( \boldsymbol{\theta}  \right)\right)\mathtt{vec}\left[\mathcal{\hat L}_{\theta_{j}}^{(S)}\right]\right)\\\notag
		&=&\frac{1}{2}\left(\left(\hat \rho \left( \boldsymbol{\theta}  \right)^{\top}\otimes \mathbb{1}\right)+\left(\mathbb{1}\otimes\hat \rho \left( \boldsymbol{\theta}  \right)\right)\right)\mathtt{vec}\left[\mathcal{\hat L}_{\theta_{j}}^{(S)}\right].\label{solve SLD}
	\end{eqnarray}
	Assuming that $\hat \rho \left( \boldsymbol{\theta}  \right)$ is invertible, then the solution of this equation is written as
	\begin{equation}
		\mathtt{vec}\left[\mathcal{\hat L}_{\theta_{j}}^{(S)}\right]=2\left(\left(\hat \rho \left( \boldsymbol{\theta}  \right)^{\top}\otimes \mathbb{1}\right)+\left(\mathbb{1}\otimes\hat \rho \left( \boldsymbol{\theta}  \right)\right)\right)^{-1}\mathtt{vec}\left[\partial_{{\theta _j}}\hat \rho \left( \boldsymbol{\theta}  \right)\right].
	\end{equation}
	If we insert the above solution together with the use of the property (\ref{vec tr}) into the definition of {QFIM} given in (\ref{QFIMSLD1}), we get
	\begin{equation}
		\left[\mathcal{F}_Q^{(S)}\left(\boldsymbol{\theta}\right)\right]_{jk}=2 \mathtt{vec}\left[\partial_{{\theta _j}}\hat \rho \left( \boldsymbol{\theta}  \right)\right]^{\dag} \left(\left(\hat \rho \left( \boldsymbol{\theta}  \right)^{\top}\otimes \mathbb{1}\right)+\left(\mathbb{1}\otimes\hat \rho \left( \boldsymbol{\theta}  \right)\right)\right)^{-1}\mathtt{vec}\left[\partial_{{\theta _k}}\hat \rho \left( \boldsymbol{\theta}  \right)\right].
	\end{equation}
	In the case where the $\hat \rho \left( \boldsymbol{\theta}  \right)$ is not invertible, Eq. (\ref{solve SLD}) is solved using the "\textit{Moore-Penrose pseudo-inverse}" \footnote{Moore-Penrose pseudo-inverse is a generalization of the inverse, which we can calculate by using the Tikhonov regularization: $A^{+}=\lim _{\delta \searrow 0}\left(A^{\dagger}\left(A A^{\dagger}+\delta I\right)^{-1}\right)=$ $\lim _{\delta\searrow 0}\left(\left(A^{\dagger} A+\delta I\right)^{-1} A^{\dagger}\right)$. These limits exist even if $A^{-1}$ does not exist (for more details see Ref. \cite{ben2003generalized, campbell2009generalized}  )}and can be written its solution as;
	\begin{equation}
		\mathtt{vec}\left[\mathcal{\hat L}_{\theta_{j}}^{(S)}\right]=2\left(\left(\hat \rho \left( \boldsymbol{\theta}  \right)^{\top}\otimes \mathbb{1}\right)+\left(\mathbb{1}\otimes\hat \rho \left( \boldsymbol{\theta}  \right)\right)\right)^{+}\mathtt{vec}\left[\partial_{{\theta _j}}\hat \rho \left( \boldsymbol{\theta}  \right)\right].
	\end{equation}
	Thus, the {QFIM} elements are \cite{vsafranek2018simple, liu2019quantum}
	\begin{equation}
		\left[\mathcal{F}_Q^{(S)}\left(\boldsymbol{\theta}\right)\right]_{jk}=2 \mathtt{vec}\left[\partial_{{\theta _j}}\hat \rho \left( \boldsymbol{\theta}  \right)\right]^{\dag} \left(\left(\hat \rho \left( \boldsymbol{\theta}  \right)^{\top}\otimes \mathbb{1}\right)+\left(\mathbb{1}\otimes\hat \rho \left( \boldsymbol{\theta}  \right)\right)\right)^{+}\mathtt{vec}\left[\partial_{{\theta _k}}\hat \rho \left( \boldsymbol{\theta}  \right)\right].
	\end{equation}
	Note that this expression is valid for finite-dimensional systems, and can be evaluated directly based on matrix forms of the density matrix $\hat \rho \left( \boldsymbol{\theta}  \right)$ and its derivatives $\partial_{{\theta _j}}\hat \rho \left( \boldsymbol{\theta}  \right)$.
	\subsection{ Multiparameter SLD-quantum Cramér-Rao bound } \label{Sub. SLD-QCRB}
	Now that we have the {SLD-QFIM}, we will derive the {QCRB} corresponding. We start by considering the derivation of an unbiased estimator vector with respect to $\theta_j$  such that
	\begin{equation}\label{Ubsaid vect}
		\partial_{{\theta_j} }\mathtt{Bias}_{\boldsymbol{\theta}}\left(\boldsymbol{\theta}^{est}\left(\hat{\Pi}\left(\boldsymbol{X}\right)\right) \right)=\rm Tr\left[\partial_{\theta_j}\hat\rho\left(\boldsymbol{\theta}\right)\left(\theta^{est}\left(\hat{\Pi}\left(\boldsymbol{X}\right)\right)_k-\theta_k\right)\right]=\delta_{jk}. 
	\end{equation}
	Inserting {SLD} into the last equation leads to 
	\begin{equation}
		\frac{1}{2}\rm Tr\left[\left(\mathcal{\hat L}_{\theta_{j}}^{(S)}\hat \rho \left( \boldsymbol{\theta}  \right)+\hat \rho \left( \boldsymbol{\theta}  \right)\mathcal{\hat L}_{\theta_{j}}^{(S)}\right)\left(\theta^{est}\left(\hat{\Pi}\left(\boldsymbol{X}\right)\right)_k-\theta_k\right)\right]=\delta_{jk}. 
	\end{equation}
	Next, we introduce two arbitrary real vectors, $\boldsymbol{a}$ and $\boldsymbol{a}$ . The dot product or scalar product between $\boldsymbol{a}$  and $\boldsymbol{b}$  is defined as
	\begin{equation}
		\boldsymbol{a}^{\top}\boldsymbol{b}=\sum_{i}{a_i b_i} =\sum_{jk}{a_j b_k \delta_{jk}}.
	\end{equation}
	If we insert the result of (\ref{Ubsaid vect}) into the definition of the dot product defined above, then we have 
	\begin{eqnarray}
		\boldsymbol{a}^{\top}\boldsymbol{b}&=&\frac{1}{2}\sum_{jk}{a_j b_k  \operatorname{Tr}\left[\left(\mathcal{\hat L}_{\theta_{j}}^{(S)}\hat \rho \left( \boldsymbol{\theta}  \right)+\hat \rho \left( \boldsymbol{\theta}  \right)\mathcal{\hat L}_{\theta_{j}}^{(S)}\right)\left(\theta^{est}\left(\hat{\Pi}\left(\boldsymbol{X}\right)\right)_k-\theta_k\right)\right]}\\\notag
		&=&\frac{1}{2}\sum_{jk}{ \operatorname{Tr}\left[a_j\left( \mathcal{\hat L}_{\theta_{j}}^{(S)}\hat \rho \left( \boldsymbol{\theta}  \right)+\hat \rho \left( \boldsymbol{\theta}  \right)\mathcal{\hat L}_{\theta_{j}}^{(S)}\right)\left(\theta^{est}\left(\hat{\Pi}\left(\boldsymbol{X}\right)\right)_k-\theta_k\right)b_k\right]}\\\notag
		&=& \mathfrak{Re}{\operatorname{Tr}\left[\left(\sum_j{a_j \mathcal{\hat L}_{\theta_{j}}^{(S)}}\right)\hat \rho \left( \boldsymbol{\theta}\right)\left(\sum_k{\left(\theta^{est}\left(\hat{\Pi}\left(\boldsymbol{X}\right)\right)_k-\theta_k\right)b_k}\right)\right]}\\
		&\le&\left\|\operatorname{Tr}\left[\left(\sum_j{a_j \mathcal{\hat L}_{\theta_{j}}^{(S)}}\right)\hat \rho \left( \boldsymbol{\theta}\right)\left(\sum_k{\left(\theta^{est}\left(\hat{\Pi}\left(\boldsymbol{X}\right)\right)_k-\theta_k\right)b_k}\right)\right]\right\|.\label{Eq. 2.91}
	\end{eqnarray}
	Squaring the last inequality and using the Schwartz inequality for traces, we gets
	\begin{eqnarray}
		\left(\boldsymbol{a^{\top}} \boldsymbol{b}\right)^2\notag &\le&\left\|\operatorname{Tr}\left[\left(\sum_j{a_j \mathcal{\hat L}_{\theta_{j}}^{(S)}}\right)\sqrt{\hat \rho \left( \boldsymbol{\theta}\right)}\sqrt{\hat \rho \left( \boldsymbol{\theta}\right)}\left(\sum_k{\left(\theta^{est}\left(\hat{\Pi}\left(\boldsymbol{X}\right)\right)_k-\theta_k\right)b_k}\right)\right]\right\|^2\\\notag
		&\le&\operatorname{Tr}\left[\hat \rho \left( \boldsymbol{\theta}\right)\left(\sum_j{a_j \mathcal{\hat L}_{\theta_{j}}^{(S)}}\right)^2\right]\operatorname{Tr}\left[\hat \rho \left( \boldsymbol{\theta}\right)\left(\sum_k{\left(\theta^{est}\left(\hat{\Pi}\left(\boldsymbol{X}\right)\right)_k-\theta_k\right)b_k}\right)^2\right]\\
		&=& \left(\boldsymbol{a}^{\top}\mathcal{F}_Q^{(S)}\left(\boldsymbol{\theta}\right)\boldsymbol{a}\right)\left(\boldsymbol{b}^{\top}\mathtt{Cov}_{\boldsymbol{\theta}}\left[\boldsymbol{\theta}^{est}\left(\hat{\Pi}\left(\boldsymbol{X}\right)\right)\right]\boldsymbol{b}\right). \label{EQ. 2.92}
	\end{eqnarray}
	Since $\boldsymbol{a}$ and  $\boldsymbol{b}$  are arbitrary vectors, then we can assume that $\boldsymbol{a}=\mathcal{F}_Q^{(S)}\left(\boldsymbol{\theta}\right)^{-1}\boldsymbol{b}$. Hence, the inequality (\ref{EQ. 2.92}) becomes
	\begin{equation}
		\left(\boldsymbol{b^{\top}} \left(\mathcal{F}_Q^{(S)}\left(\boldsymbol{\theta}\right)^{-1}\right)^{\top}\boldsymbol{b}\right)^2\le	\left(\boldsymbol{b^{\top}} \left(\mathcal{F}_Q^{(S)}\left(\boldsymbol{\theta}\right)^{-1}\right)^{\top}\boldsymbol{b}\right)\left(\boldsymbol{b}^{\top}\mathtt{Cov}_{\boldsymbol{\theta}}\left[\boldsymbol{\theta}^{est}\left(\hat{\Pi}\left(\boldsymbol{X}\right)\right)\right]\boldsymbol{b}\right).
	\end{equation}
	Therefore, we have
	\begin{equation}
		\boldsymbol{b^{\top}}\left(\mathcal{F}_Q^{(S)}\left(\boldsymbol{\theta}\right)^{-1}\right)^{\top}\boldsymbol{b}\le\left(\boldsymbol{b}^{\top}\mathtt{Cov}_{\boldsymbol{\theta}}\left[\boldsymbol{\theta}^{est}\left(\hat{\Pi}\left(\boldsymbol{X}\right)\right)\right]\boldsymbol{b}\right).
	\end{equation}
	This matrix-bound is valid for any vector $\boldsymbol{b}$,  as well as the {QFIM} is symmetric, which leads to   
	\begin{equation}
		\mathtt{Cov}_{\boldsymbol{\theta}}\left[\boldsymbol{\theta}^{est}\left(\hat{\Pi}\left(\boldsymbol{X}\right)\right)\right]\ge \frac{1}{\mathcal{F}_Q^{(S)}\left(\boldsymbol{\theta}\right)}. \label{QCRBM}
	\end{equation}
	That last inequality is so-called the multi-parameter quantum Cramér-Rao inequality associated with the {SLD}-quantum Fisher information matrix. In the case where the sets of the estimate parameters are independent, we have   $\mathtt{Cov}\left[{\theta}_j^{est}\left(\hat{\Pi}\left(\boldsymbol{X}\right)\right),{\theta}_k^{est}\left(\hat{\Pi}\left(\boldsymbol{X}\right)\right)\right]=0$ for $j \ne k$. Thus, the inequality (\ref{QCRBM}) is reduced to
	\begin{equation}
		\mathtt{Var}_{{\theta_j}}\left[{\theta}_j^{est}\left(\hat{\Pi}\left(\boldsymbol{X}\right)\right)\right]\ge \frac{1}{\left[\mathcal{F}_Q^{(S)}\left(\boldsymbol{\theta}\right)\right]_{jj}}.
	\end{equation}
	Given a positive definite matrix $\boldsymbol{W}$, we can weigh the uncertainty of different estimating parameters. This leads to 
	\begin{equation}
		\operatorname{Tr}\left[\boldsymbol{W} \mathtt{Cov}_{\boldsymbol{\theta}}\left[\boldsymbol{\theta}^{est}\left(\hat{\Pi}\left(\boldsymbol{X}\right)\right)\right]\right]\ge \operatorname{Tr}\left[\boldsymbol{W} \left(\mathcal{F}_Q^{(S)}\left(\boldsymbol{\theta}\right)\right)^{-1}\right]. 
	\end{equation}
	If we choose $\boldsymbol{W} = \mathbb{1}$, we find that the  bound on the sum of the variances of the estimators of the estimated parameters
	\begin{equation}\label{BS}
		\sum\limits_{j = 1}^m {{{\mathtt{Var}_{\theta_j}}} } \left( {\theta _j^{est}\left(\hat{\Pi}\left(\boldsymbol{X}\right)\right)} \right) \ge  {\mathop{\rm Tr}\nolimits} \left[\left(\mathcal{F}_Q^{(S)}\left(\boldsymbol{\theta}\right)\right)^{-1} \right]={B_S}. 
	\end{equation}
	In the case of a single parameter estimation problem, the {QCRB} can, in principle, be achieved asymptotically by a suitable measurement that is the optimal measurement. The natural question that arises is whether {QCRB} can be achievable in the case of multiparameter estimation. This question that we will address in the following subsection.           
	\subsection{Attainment of the lower bound of QCRB} \label{subsect2.6.3}
	As derived in Eq. (\ref{EQ. 2.74}), the optimal quantum estimator that attainable the the lower bound  of the {QCRB} is, for each parameter individually $\theta_j$, given by 
	\begin{equation}\label{EQ. 2.103}
		\theta_j^{est}\left(\hat{\Pi}\left(\boldsymbol{X}\right)\right)=\theta_j +\frac{\mathcal{\hat L}_{\theta_j}^{(S)}}{\left[\mathcal{F}_Q^{(S)}\left(\boldsymbol{\theta}\right)\right]_{jj}}.
	\end{equation}
	In general, the optimal observables in Eq. (\ref{EQ. 2.103}) may not be compatible, which means that the optimal estimation vector $\boldsymbol{\theta}^{est}\left(\hat{\Pi}\left(\boldsymbol{X}\right)\right)$ is, in general, a difficult task to determine. Therefore, the {QCRB} for multiple parameters is generally not saturable. That fact is an issue major in multiparameter quantum estimation theory.  However, this problem has been overcome in exceptional estimation cases, where the different {SLD}-operators are satisfied with the \textit{compatible condition}
	\begin{equation}\label{EQ. 2.104}
		U_{\theta_j \theta_k}=-\frac{i}{2}\operatorname{Tr}\left[\hat \rho \left( \boldsymbol{\theta}\right)\left[\mathcal{\hat L}_{\theta_j}^{((S))}, \mathcal{\hat L}_{\theta_k}^{((S))}\right]\right]=0, \quad \text{with}\quad j,k=1,2,...,m. 
	\end{equation}
	The matrix $\boldsymbol{U}$ is noted as a measure of \textit{incompatibility} between $\theta_j$ and $\theta_{k}$. This \textit{incompatibility} arises from the inherent non-commutativity nature of quantum mechanics, which, in turn, emerges in this case from the non-commuting observables $\left[\mathcal{\hat L}_{\theta_j}^{(S)}, \mathcal{\hat L}_{\theta_k}^{(S)}\right]$. When the compatible condition is satisfied, we have immediately  $\left[\mathcal{\hat L}_{\theta_j}^{(S)}, \mathcal{\hat L}_{\theta_k}^{(S)}\right]=0$. That means there is a common eigenbasis between $\mathcal{\hat L}_{\theta_j}^{(S)}$ and $\mathcal{\hat L}_{\theta_k}^{(S)}$ and then a single optimal measurement attaining the lower bound of {QCRB}. Except for the quantum statistical models that satisfied condition (\ref{EQ. 2.104}),  the lower bound of {QCRB} for multiparameter estimation is never attainable, so the maximal precision of the several parameters is not necessarily the trace of the inverse of {QFIM}. Since,  in this case,  we cannot attain the lower bound of {QCRB}, then the question arises as to how we can derive a tight bound on precision? In other words, is there an upper bound that is high and then tight that allows us to predict where the precision limit is? Reasonably, this question immediately leads us to guess at the validity of {QCRB} derived from using the {RLD}-QFIM.
	\subsection{RLD-quantum Fisher information matrix}
	Following the same procedure as in the Sub. (\ref{Sub SLDQFIM}) for deriving the {SLD}-{QFIM}, the {RLD}-operators lead to derive the {RLD}-quantum Fisher information matrix such as
	\begin{equation}\label{QFIMRLD1}
		\left[\mathcal{F}_Q^{(R)}\left(\boldsymbol{\theta}\right)\right]_{jk}=\rm Tr\left[\partial_{{\theta _j}}\hat \rho\left(\boldsymbol{\theta}\right)\left(\mathcal{\hat L}_{\theta_{k}}^{(R)}\right)^{\dag}\right], \quad \text{for all} \quad j,k =1,2,...,m 
	\end{equation}
	In terms of {RLD}-operators, the {RLD}-{QFIM} is written as
	\begin{equation}
		\left[\mathcal{F}_Q^{(R)}\left(\boldsymbol{\theta}\right)\right]_{jk}=\rm Tr\left[\hat \rho\left(\boldsymbol{\theta}\right)\mathcal{\hat L}_{\theta_{j}}^{(R)}\left(\mathcal{\hat L}_{\theta_{k}}^{(R)}\right)^{\dag}\right].
	\end{equation}
	In the standard diagonal basic of the density operator, we can write the matrix elements of the {RLD}-{QFIM}  as follows
	\begin{equation}
		\left[\mathcal{F}_Q^{(R)}\left(\boldsymbol{\theta}\right)\right]_{jk}=\frac{1}{2}\sum_{jk}{\left(\frac{1}{p_j}+\frac{1}{p_k}\right)\bra{\psi_j}\partial_{\theta_j}\hat \rho \left( \boldsymbol{\theta}  \right)\ket{\psi_k}\bra{\psi_k}\partial_{\theta_k}\hat \rho \left( \boldsymbol{\theta}  \right)\ket{\psi_j}}.
	\end{equation}
	That last explicit expression of {RLD}-{QFIM} requires the diagonalization of the density matrix, which is, in turn, a task most hard to do in the statistical model of a high dimensional density operator.To avoid this challenging approach, we use the \textit{vectorization method} to derive a new explicit expression of the {RLD}-{QFIM}. For this, we insert the vec-operator into the differential equation of the {RLD}-operator (\ref{RLD}), which lead to
	\begin{eqnarray}
		\mathtt{vec}\left[{\partial _{\theta_j} }{\hat \rho \left( \boldsymbol{\theta}  \right) } \right]&=& \mathtt{vec} \left[{\hat \rho \left( \boldsymbol{\theta}  \right)}{\mathcal{\hat L}_{\theta_j}^{(R)} }\right].\\
		&=& \mathtt{vec} \left[{\hat \rho \left( \boldsymbol{\theta}  \right)}{\mathcal{\hat L}_{\theta_j}^{(R)} \mathbb{1} }\right].
	\end{eqnarray}
	By using the property of Eq. (\ref{vecpro1}), we get  
	\begin{equation}
		\mathtt{vec}\left[\mathcal{\hat L}_{\theta_j}^{(R)} \right]=\left(\mathbb{\hat 1} \otimes \hat \rho \left( \boldsymbol{\theta}  \right) \right)^{-1}\mathtt{vec}\left[{\partial _{\theta_j} }{\hat \rho \left( \boldsymbol{\theta}  \right) } \right].
	\end{equation}
	Inserting the above solution together with the use of the property (\ref{vec tr}) into (\ref{QFIMRLD1}) leads us to find the explicit expression of {RLD}-{QFIM} as follows 
	\begin{equation}
		\left[\mathcal{F}_Q^{(R)}\left(\boldsymbol{\theta}\right)\right]_{jk}=\mathtt{vec}\left[{\partial _{\theta_j} }{\hat \rho \left( \boldsymbol{\theta}  \right) } \right]^{\dag}\left(\mathbb{\hat 1} \otimes \hat \rho \left( \boldsymbol{\theta}  \right) \right)^{-1}\mathtt{vec}\left[{\partial _{\theta_k} }{\hat \rho \left( \boldsymbol{\theta}  \right) } \right].
	\end{equation}
	Note that the conjugate transpose distributes over tensor products\footnote{Let us $A$ and $B$ are two complex matrices, then we have $\left(A\otimes B\right)^{\dag}=A^{\dag}\otimes B^{\dag}$.}. In the case where the $\hat \rho \left( \boldsymbol{\theta}  \right)$ is not invertible, Eq. (\ref{RLD}) is solved using the "\textit{Moore-Penrose pseudo-inverse}", and then the elements of {RLD}-{QFIM} are
	\begin{equation}
		\left[\mathcal{F}_Q^{(R)}\left(\boldsymbol{\theta}\right)\right]_{jk}=\mathtt{vec}\left[{\partial _{\theta_j} }{\hat \rho \left( \boldsymbol{\theta}  \right) } \right]^{\dag}\left(\mathbb{\hat 1} \otimes \hat \rho \left( \boldsymbol{\theta}  \right) \right)^{+}\mathtt{vec}\left[{\partial _{\theta_k} }{\hat \rho \left( \boldsymbol{\theta}  \right) } \right].
	\end{equation}
As for {SLD}-{QFIM}, also the {RLD}-{QFIM} is evaluated directly based on the matrix density elements and their derivation. We noted that this approach stays useful as long as the density matrix of the statistical model lives in the Hilbert space of finite-dimensional.
	\subsection{ Multiparameter RLD-quantum Cramér-Rao bound}
	Now that we have the {RLD}-quantum Fisher information matrix, we can look at deriving the corresponding {QCRB}. Similar to what we did for the derived {SLD}-{QCRB} in Sub. (\ref{Sub. SLD-QCRB}), we will follow the same approach to derive the {RLD}-{QCRB} based on {RLD}-operators. The unbiased condition of the estimator vector $\boldsymbol{\theta}^{est}\left(\hat{\Pi}\left(\boldsymbol{X}\right)\right)$ obey to 
	\begin{eqnarray}
		\operatorname{Tr}\left[\partial_{\theta_j}\hat\rho\left(\boldsymbol{\theta}\right)\left(\theta^{est}\left(\hat{\Pi}\left(\boldsymbol{X}\right)\right)_k-\theta_k\right)\right]&=&\operatorname{Tr}\left[\hat\rho\left(\boldsymbol{\theta}\right)\mathcal{\hat L}_{\theta_j}^{(R)}\left(\theta^{est}\left(\hat{\Pi}\left(\boldsymbol{X}\right)\right)_k-\theta_k\right)\right]\\\notag \label{Ubsaid co}
		&=&\operatorname{Tr}\left[\left(\mathcal{\hat L}_{\theta_j}^{(R)}\right)^{\dag}\hat\rho\left(\boldsymbol{\theta}\right)\left(\theta^{est}\left(\hat{\Pi}\left(\boldsymbol{X}\right)\right)_k-\theta_k\right)\right]=\delta_{jk}.
	\end{eqnarray}
	Now, we consider two arbitrary complex vectors, $\boldsymbol{a}=(a_1,a_2,...,a_m)^{\top}$ and $\boldsymbol{b}=(b_1,b_2,...,b_m)^{\top}$, the scalar product between $\boldsymbol{a}$ and  $\boldsymbol{b}$ is defined as \cite{arfken1999mathematical, brand2020vector}
	\begin{equation}
		\boldsymbol{a}^{\dag}\boldsymbol{b}=\sum_{i}{a_i^{*} b_i} =\sum_{jk}{a_j^{*} b_k \delta_{jk}}.
	\end{equation}
	By inserting the result of Eq. (\ref{Ubsaid co}) into the scalar product definition above, we get
	\begin{eqnarray}
		\boldsymbol{a}^{\dag}\boldsymbol{b}&=&\sum_{jk}{a_j^{*} b_k}\operatorname{Tr}\left[\hat\rho\left(\boldsymbol{\theta}\right)\mathcal{\hat L}_{\theta_j}^{(R)}\left(\theta^{est}\left(\hat{\Pi}\left(\boldsymbol{X}\right)\right)_k-\theta_k\right)\right]\\\notag
		&=&\mathfrak{Re}{\operatorname{Tr}\left[\left(\sum_{j}{{a_j^{*}\left(\mathcal{\hat L}_{\theta_j}^{(R)}\right)^{\dag}}}\right)\hat\rho\left(\boldsymbol{\theta}\right)\left(\sum_{k}{\left(\theta^{est}\left(\hat{\Pi}\left(\boldsymbol{X}\right)\right)_k-\theta_k\right) b_k}\right)\right]}\\\notag
		&\le&\left\|\operatorname{Tr}\left[\left(\sum_{j}{{a_j^{*}\left(\mathcal{\hat L}_{\theta_j}^{(R)}\right)^{\dag}}}\right)\hat\rho\left(\boldsymbol{\theta}\right)\left(\sum_{k}{\left(\theta^{est}\left(\hat{\Pi}\left(\boldsymbol{X}\right)\right)_k-\theta_k\right) b_k}\right)\right]\right\|.
	\end{eqnarray}
	Squaring the last inequality and using together the Schwartz inequality of trace, leads to
	\begin{eqnarray}
		\left(\boldsymbol{a}^{\dag}\boldsymbol{b}\right)^2&\le&\notag\left\|\operatorname{Tr}\left[\left(\sum_{j}{{a_j^{*}\left(\mathcal{\hat L}_{\theta_j}^{(R)}\right)^{\dag}}}\right)\sqrt{\hat\rho\left(\boldsymbol{\theta}\right)}\sqrt{\hat\rho\left(\boldsymbol{\theta}\right)}\left(\sum_{k}{\left(\theta^{est}\left(\hat{\Pi}\left(\boldsymbol{X}\right)\right)_k-\theta_k\right) b_k}\right)\right]\right\|^2\\\notag
		&\le&\left(\sum_{jk}a_j^{*}\operatorname{Tr}\left[\hat\rho\left(\boldsymbol{\theta}\right)\mathcal{\hat L}_{\theta_j}^{(R)}\left(\mathcal{\hat L}_{\theta_k}^{(R)}\right)^{\dag}\right]a_k\right)\left(\sum_{jk}b_j^{*}\left[\mathtt{Cov}_{\boldsymbol{\theta }}\left[\boldsymbol{\theta}^{est}\left(\hat{\Pi}\left(\boldsymbol{X}\right)\right)\right]\right]_{jk}b_k\right)\\
		&=&\left(\boldsymbol{a}^{\dag}\mathcal{	F}_Q^{(R)}\left(\boldsymbol{\theta}\right)\boldsymbol{a}\right)
		\left(\boldsymbol{b}^{\dag}\mathtt{Cov}_{\boldsymbol{\theta }}\left[\boldsymbol{\theta}^{est}\left(\hat{\Pi}\left(\boldsymbol{X}\right)\right)\right]\boldsymbol{b}\right).
	\end{eqnarray}
	Since $\boldsymbol{a}$ and  $\boldsymbol{b}$  are arbitrary complex vectors, then we can assume that $\boldsymbol{a}=\mathcal{F}_Q^{(R)}\left(\boldsymbol{\theta}\right)^{-1}\boldsymbol{b}$. Hence, the inequality (\ref{EQ. 2.116}) becomes
	\begin{equation}
		\left(\boldsymbol{b^{\dag}} \left(\mathcal{F}_Q^{(R)}\left(\boldsymbol{\theta}\right)^{-1}\right)^{\dag}\boldsymbol{b}\right)^2\le	\left(\boldsymbol{b^{\dag}} \left(\mathcal{F}_Q^{(S)}\left(\boldsymbol{\theta}\right)^{-1}\right)^{\dag}\boldsymbol{b}\right)\left(\boldsymbol{b}^{\dag}\mathtt{Cov}_{\boldsymbol{\theta}}\left[\boldsymbol{\theta}^{est}\left(\hat{\Pi}\left(\boldsymbol{X}\right)\right)\right]\boldsymbol{b}\right).
	\end{equation}
	Thus, we have
	\begin{equation}
		\boldsymbol{b^{\dag}} \left(\mathcal{F}_Q^{(R)}\left(\boldsymbol{\theta}\right)^{-1}\right)^{\dag}\boldsymbol{b}\le \left(\boldsymbol{b}^{\dag}\mathtt{Cov}_{\boldsymbol{\theta}}\left[\boldsymbol{\theta}^{est}\left(\hat{\Pi}\left(\boldsymbol{X}\right)\right)\right]\boldsymbol{b}\right).
	\end{equation}
	Again, since this must be true for any complex vector $\boldsymbol{b}$, then we obtain the following  inequality
	\begin{equation}
		\mathtt{Cov}_{\boldsymbol{\theta}}\left[\boldsymbol{\theta}^{est}\left(\hat{\Pi}\left(\boldsymbol{X}\right)\right)\right]\ge \frac{1}{\left(\mathcal{F}_Q^{(R)}\left(\boldsymbol{\theta}\right)\right)}.
	\end{equation}
	Therefore, the {RLD}-quantum Fisher information matrix has a similar role to that played by {SLD}-quantum Fisher information matrix. Given a positive weight matrix, $\boldsymbol{W}$,  we can weigh the uncertainty of different estimating parameters based on the {RLD}-quantum Fisher information matrix, and we get 
	\begin{equation}
		\operatorname{Tr}\left[\boldsymbol{W} \mathtt{Cov}_{\boldsymbol{\theta}}\left[\boldsymbol{\theta}^{est}\left(\hat{\Pi}\left(\boldsymbol{X}\right)\right)\right]\right]\ge \operatorname{Tr}\left[\boldsymbol{W} \left(\mathcal{F}_Q^{(R)}\left(\boldsymbol{\theta}\right)\right)^{-1}\right]. \label{}
	\end{equation}
	On the other hand, since the {RLD}-quantum Fisher information matrix has been defined as a complex matrix, then we rewrite the last inequality such that;
	\begin{equation}
		\operatorname{Tr}\left[\boldsymbol{W} \mathtt{Cov}_{\boldsymbol{\theta}}\left[\boldsymbol{\theta}^{est}\left(\hat{\Pi}\left(\boldsymbol{X}\right)\right)\right]\right]\ge \operatorname{Tr}\left[\boldsymbol{W} \mathfrak{Re} \left[\mathcal{F}_Q^{(R)}\left(\boldsymbol{\theta}\right)\right]^{-1}\right]+\operatorname{Tr}\left[\boldsymbol{W}\mathfrak{Im} \left[\mathcal{F}_Q^{(R)}\left(\boldsymbol{\theta}\right)\right]^{-1}\right]. 
	\end{equation}
	If we choose $\boldsymbol{W} = \mathbb{1}$, we find that the bound on the sum of the variances of the estimators of the estimated parameters based on the {RLD}-quantum Fisher information matrix
	\begin{equation}\label{BR}
		\sum\limits_{j = 1}^m {{{\mathtt{Var}_{\theta_j}}} } \left( {\theta _j^{est}\left(\hat{\Pi}\left(\boldsymbol{X}\right)\right)} \right) \ge \operatorname{Tr}\left[\mathfrak{Re} \left[\mathcal{F}_Q^{(R)}\left(\boldsymbol{\theta}\right)\right]^{-1}\right]+\operatorname{Tr}\operatorname{Abs}\left[\mathfrak{Im} \left[\mathcal{F}_Q^{(R)}\left(\boldsymbol{\theta}\right)\right]^{-1}\right]=B_R. 
	\end{equation}
	Here, for a complex matrix $\boldsymbol{A}$, $\mathfrak{Re}\left[\boldsymbol{A}\right]=\frac{\boldsymbol{A}+{\boldsymbol{\bar A}}}{2}$,  $\mathfrak{Im}\left[\boldsymbol{A}\right]=\frac{\boldsymbol{A}-{\boldsymbol{\bar A}}}{2i}$ and the $\operatorname{TrAbs}$ denotes the absolute sum of the eigenvalues of $\boldsymbol{A}$.
	
	As demonstrated in the single estimation problem, the {RLD}-operators is not, generally, Hermitian. Consequently, the optimal estimator derived by the {RLD}-{QFI} may not correspond to a physical {POVM} measurement. Despite this, there are instances, in the multiparameter estimation case, in which the {RLD}-{QCRB} could be tighter and therefore becomes more important than the {SLD}-{QCRB}. Generally, there is no hierarchy between {SLD} and {RLD}-{QCRB}s, which leads us to ask which one of these bounds is more informative? In other words, which one of these bounds is higher and then tighter? Holevo answered this question by introducing a tighter bound called "{Holevo Cramér-Rao bound (HCRB)}" and defined it as the most general quantum extension of the classical Cramér-Rao bound \cite{holevo2011probabilistic, sidhu2021tight, albarelli2019evaluating}. 
	\subsection{Holevo Cramér-Rao bound}
	In the 1970s, Holevo proposed a bound named {HCRB} aiming to derive a fundamental precision limit for the quantum parameter estimation problem. At that time, it was not nonetheless evident whether this bound was tight or not. Over the last decade, there have been many meaningful accomplish on asymptotic analysis of quantum parameter estimation theory showing that the Holevo bound is assuredly the best asymptotically attainable bound \cite{ albarelli2019evaluating, sidhu2021tight, carollo2019quantumness, li2022geometric}. These accomplished progress results confirmed that the {HCRB} plays a crucial role in the asymptotic theory of quantum estimation problems, particularly in the multiparameter quantum estimation case. Although we now have the fundamental asymptotic limit attainable, {HCRB} has a major obstacle. It is not an explicit form in terms of a given statistical model such as {QCRB}, but rather an optimization of some non-trivial function. In other words, unlike the {QCRB}, which evaluate directly from the quantum Fisher information that, in turn, depends strongly on the matrix density of the statistical model, the structure of {HCRB}  is not expressed in terms of the given statistical model.
	
	After these introductory remarks, we wish to gain a deeper insight into the structure of the {HCRB} that reflects the statistical properties of a given statistical model. To achieve this desire, we will follow the most malleable approach to stating the formulation of {HCRB}. This approach is based on the so-called "\textbf{Weight Mean Square Error (WMSE)}".  In multiparameter estimation protocols, a fundamental problem arises as to how the set of unknown parameters can be estimated simultaneously. Reasonably, a meaningful measurement strategy minimizes the weighted sum of parameter estimate variances. For this,  a $m\times m$-positive definite square weight matrix, $\boldsymbol{W}$, is chosen to define the weighted mean square error $\operatorname{Tr}\left[\boldsymbol{W} \mathtt{Cov}_{\boldsymbol{\theta}}\left[\boldsymbol{\theta}^{est}\left(\hat{\Pi}\left(\boldsymbol{X}\right)\right)\right]\right]$. The idea of Holevo is to prove an equivalence between a matrix inequality and its corresponding scalar inequality, which allows minimizing, optimally, the {WMSE}. To prove this, we start with a essential lemma.
	\begin{LM}
		Given a weight matrix $\boldsymbol{W}$, for any real symmetric matrix $\boldsymbol{V}$ and a Hermitian matrix $\boldsymbol{M}$, the inequality $\boldsymbol{V}>\boldsymbol{M}$ implies that 
		\begin{equation}
			\operatorname{Tr}\left[\boldsymbol{W} \boldsymbol{V}\right]\ge \operatorname{Tr}\left[\boldsymbol{W} \mathfrak{Re}\left[\boldsymbol{M}\right]\right]+\operatorname{TrAbs}\left[\sqrt{\boldsymbol{W}} \mathfrak{Im}\left[\boldsymbol{M}\right]\sqrt{\boldsymbol{W}}\right].
		\end{equation}
		where $\operatorname{TrAbs}[.]$ denotes the sum of the absolute values of the eigenvalues of a matrix. It is equivalent to the trace norm $\left\|.\right\|_1$, which is more commonly used in the literature.
	\end{LM}
	Since the covariance matrix $\mathtt{Cov}_{\boldsymbol{\theta}}\left[\boldsymbol{\theta}^{est}\left(\hat{\Pi}\left(\boldsymbol{X}\right)\right)\right]$ is always real and symmetric, then we can identify the matrix $\boldsymbol{V}$ with the $\mathtt{Cov}_{\boldsymbol{\theta}}\left[\boldsymbol{\theta}^{est}\left(\hat{\Pi}\left(\boldsymbol{X}\right)\right)\right]$. Thus, the \textbf{WMSE} is written as
	\begin{equation}\label{Eq. 2.123}
		\operatorname{Tr}\left[\boldsymbol{W} \mathtt{Cov}_{\boldsymbol{\theta}}\left[\boldsymbol{\theta}^{est}\left(\hat{\Pi}\left(\boldsymbol{X}\right)\right)\right]\right]\ge \operatorname{Tr}\left[\boldsymbol{W} \mathfrak{Re}\left[\boldsymbol{M}\right]\right]+\left\|\sqrt{\boldsymbol{W}} \mathfrak{Im}\left[\boldsymbol{M}\right]\sqrt{\boldsymbol{W}}\right\|_1.
	\end{equation}
	For a given weight matrix $\boldsymbol{W}$ and Hermitian matrix $\boldsymbol{M}$, we want to minimize the scalar {WMSE} to obtain better estimates simultaneously of unknown parameters\footnote{As a remark, if we identify the matrix $\boldsymbol{M}$ with the inverse of different families of {QFIM}s, then we have some lower bound derived in Eqs. (\ref{BS}, \ref{BR}).}. In order to attain the saturation of inequality (\ref{Eq. 2.123}), Holevo considered a set of Hermitian observables collected in a vector $\boldsymbol{\hat X}=\left(\hat{X}_1,\hat{X}_2,...,\hat{X}_m\right)$ and satisfying the locally unbiased conditions
	\begin{equation}
		\operatorname{Tr}\left[\hat \rho\left(\boldsymbol{\theta }\right)\boldsymbol{\hat X}\right]=0, \hspace{1cm} \operatorname{Tr}\left[\partial_{{\theta _j}}\hat \rho\left(\boldsymbol{\theta }\right)\hat X_k\right]=\delta_{jk}.
	\end{equation}
	The covariance matrix of $\boldsymbol{\hat X}$, denoted $\boldsymbol{Z}_{\boldsymbol{\theta}}\left[\boldsymbol{\hat X}\right]$, and its matrix elements are defined such as $\left[\boldsymbol{Z}_{\boldsymbol{\theta}}\left[\boldsymbol{\hat X}\right]\right]_{jk}=\operatorname{Tr}\left[\hat \rho\left(\boldsymbol{\theta }\right){\hat X}_j {\hat X}_j\right]$ and satisfies the inequalities 
	\begin{equation}\label{Eq. 2.126}
		\boldsymbol{Z}_{\boldsymbol{\theta}}\left[\boldsymbol{\hat X}\right]\ge \left[\mathcal{F}_Q^{(S)}\left(\boldsymbol{\theta}\right)\right]^{-1}, \hspace{1cm} \boldsymbol{Z}_{\boldsymbol{\theta}}\left[\boldsymbol{\hat X}\right]\ge \left[\mathcal{F}_Q^{(R)}\left(\boldsymbol{\theta}\right)\right]^{-1}. 
	\end{equation}
	By identifying $\boldsymbol{M}$ in ((\ref{Eq. 2.123})) with the Hermitian matrix $\boldsymbol{Z}\left[\boldsymbol{\hat X}\right]$, we get 
	\begin{equation}\label{Eq. 2.127}
		\operatorname{Tr}\left[\boldsymbol{W} \mathtt{Cov}_{\boldsymbol{\theta}}\left[\boldsymbol{\theta}^{est}\left(\hat{\Pi}\left(\boldsymbol{X}\right)\right)\right]\right]\ge \operatorname{Tr}\left[\boldsymbol{W} \mathfrak{Re}\left[\boldsymbol{Z}_{\boldsymbol{\theta}}\left[\boldsymbol{\hat X}\right]\right]\right]+\left\|\sqrt{\boldsymbol{W}} \mathfrak{Im}\left[\boldsymbol{Z}_{\boldsymbol{\theta}}\left[\boldsymbol{\hat X}\right]\right]\sqrt{\boldsymbol{W}}\right\|_1. 
	\end{equation}
	Given the result of Eq. (\ref{Eq. 2.126}), the optimization of inequality (\ref{Eq. 2.127}) under the appropriate local unbiased condition on $\boldsymbol{\hat{X}}$ leads to obtaining the tightest bound on the {WMSE}.  This optimization is the one that defines the {HCRB}, which is explicitly the minimum of the following minimization problem 
	\begin{eqnarray}
		{B_H}\left(\boldsymbol{\theta}\right)=&\mathop{min}\limits_{\boldsymbol{\hat X}}&\left\{\operatorname{Tr}\left[\boldsymbol{W} \mathfrak{Re}\left[\boldsymbol{Z}_{\boldsymbol{\theta}}\left[\boldsymbol{\hat X}\right]\right]\right]+\left\|\sqrt{\boldsymbol{W}} \mathfrak{Im}\left[\boldsymbol{Z}_{\boldsymbol{\theta}}\left[\boldsymbol{\hat X}\right]\right]\sqrt{\boldsymbol{W}}\right\|_1\right\},\\\notag
		&\text{subject to}& \operatorname{Tr}\left[\hat \rho\left(\boldsymbol{\theta }\right)\boldsymbol{\hat X}\right]=0, \hspace{0.5cm} \operatorname{Tr}\left[\partial_{{\theta _j}}\hat \rho\left(\boldsymbol{\theta }\right)\hat X_k\right]=\delta_{jk}.\label{Eq. 2.128}
	\end{eqnarray}
	That last equation is the {HCRB} that defines a scalar lower bound on the {WMSE} and represents the best precision attainable asymptotically with global measurements. If we take the minimization given in (\ref{Eq. 2.128}) only on the first term, then we get the {SLD}-{QCRB}. Formally, we write;
	\begin{equation}
		B_S\left(\boldsymbol{\theta}\right)=\mathop{\min }\limits_{\boldsymbol{\hat X}}\left\{\operatorname{Tr}\left[\boldsymbol{W} \mathfrak{Re}\left[\boldsymbol{Z}_{\boldsymbol{\theta}}\left[\boldsymbol{\hat X}\right]\right]\right]\right\}=\operatorname{Tr}\left[\boldsymbol{W} \left[\mathcal{F}_Q^{(S)}\left(\boldsymbol{\theta}\right)\right]^{-1}\right]. \label{Eq. 1.129}
	\end{equation}
	Since the second term in Eq. (\ref{Eq. 2.128}) is never get to be negative, thus the result of (\ref{Eq. 1.129}) shows that the {HCRB} is a tighter bound than the {SLD}-{QCRB}. On the other hand, if we multiply both sides of the second inequality of (\ref{Eq. 2.126}) by $\boldsymbol{W}$ and take the trace, we have  
	\begin{eqnarray}
		\operatorname{Tr}\left[\boldsymbol{W}\boldsymbol{Z}_{\boldsymbol{\theta}}\left[\boldsymbol{\hat X}\right]\right]\ge \operatorname{Tr}\left[\boldsymbol{W} \left[\mathcal{F}_Q^{(R)}\left(\boldsymbol{\theta}\right)\right]^{-1}\right].
	\end{eqnarray}
	Taking the minimization over $\boldsymbol{\hat X}$ in the last inequality, we get the {RLD}-{QCRB}
	\begin{eqnarray}
		B_R\left(\boldsymbol{\theta}\right)&=&\mathop{\min }\limits_{\boldsymbol{\hat X}}\left\{\operatorname{Tr}\left[\boldsymbol{W} \left[\boldsymbol{Z}_{\boldsymbol{\theta}}\left[\boldsymbol{\hat X}\right]\right]\right]\right\}=\operatorname{Tr}\left[\boldsymbol{W} \left[\mathcal{F}_Q^{(R)}\left(\boldsymbol{\theta}\right)\right]^{-1}\right]\\\notag&=&\operatorname{Tr}\left[\boldsymbol{W} \mathfrak{Re}\left[\mathcal{F}_Q^{(R)}\left(\boldsymbol{\theta}\right)\right]^{-1}\right]+\left\|\sqrt{\boldsymbol{W}} \mathfrak{Im}\left[\mathcal{F}_Q^{(R)}\left(\boldsymbol{\theta}\right)\right]^{-1}\sqrt{\boldsymbol{W}}\right\|_1. \label{Eq. 2.131}
	\end{eqnarray}
	This result shows that the {HCRB} is a tighter bound than the {RLD}-{QCRB}. Therefore, from the results of Eq. (\ref{Eq. 1.129}) and Eq. (\ref{Eq. 2.131}), we can conclude that the {HCRB}  is more informative and then tighter than both the scalar {SLD} and {RLD}-{QCRB}s and satisfies the following chain of inequalities 
	\begin{equation}
		\operatorname{Tr}\left[\boldsymbol{W} \mathtt{Cov}_{\boldsymbol{\theta}}\left[\boldsymbol{\theta}^{est}\left(\hat{\Pi}\left(\boldsymbol{X}\right)\right)\right]\right]\ge B_H\left(\boldsymbol{\theta}\right)\ge \mathop{max}\left\{B_R\left(\boldsymbol{\theta}\right), B_S\left(\boldsymbol{\theta}\right)\right\}. \label{Eq. 2.132}
	\end{equation}
	
	Despite its importance for attainable optimal precision, the {HCRB} has not been used extensively in quantum metrology so far. The main reason for this is relevant to its evaluation. More precisely, from Eq. (\ref{Eq. 2.128}), the evaluation of {HCRB} requires performing an optimization on the sets of Hermitian operators that are not typically known. These unknown observations make this optimization very difficult, except in some non-trivial cases \cite{bradshaw2017tight, suzuki2019information, yang2019attaining, albarelli2019evaluating, albarelli2019upper}. Thus, the direct evaluation task of {HCRB} remains a foremost obstacle to developing multiparameter quantum metrology.  More recently, most works have focused on determining the upper bounds of {HCRB} so that we can predict the behavior of {HCRB} in cases where its evaluation directly becomes difficult \cite{carollo2019quantumness, albarelli2019upper, tsang2019holevo}. These works established that the {HCRB} is upper bounded and satisfies the following inequalities
	\begin{equation}\label{Eq. tight}
		\begin{aligned}
			\mathop{max}\left\{B_R\left(\boldsymbol{\theta}\right), B_S\left(\boldsymbol{\theta}\right)\right\}&\le B_H\left(\boldsymbol{\theta}\right)\\
			&\le B_S\left(\boldsymbol{\theta}\right)+\left\|\sqrt{\boldsymbol{W}} \left[\mathcal{F}_Q^{(S)}\left(\boldsymbol{\theta}\right)\right]^{-1}\boldsymbol{U}\left[\mathcal{F}_Q^{(S)}\left(\boldsymbol{\theta}\right)\right]^{-1}\sqrt{\boldsymbol{W}}\right\|_1\\
			&\le \left(1+\mathcal{R}_{Q}\right)B_S\left(\boldsymbol{\theta}\right).
		\end{aligned}
	\end{equation}
	where $\boldsymbol{U}$ is the asymptotic incompatibility matrix given in Eq. (\ref{EQ. 2.104}), and the quantity $\mathcal{R}_{Q}$ is a measure of \textit{\textbf{quantumness}} in quantum multi-parameter estimation problems. Its expression is given by
	\begin{equation}
		\mathcal{R}_{Q}=\left\|i\left[\mathcal{F}_Q^{(S)}\left(\boldsymbol{\theta}\right)\right]^{-1}\boldsymbol{U}\right\|_{\infty}. \label{Eq. 2.134}
	\end{equation}
	where $\left\|.\right\|_{\infty}$ denoted the largest eigenvalue of a matrix. It has proved, in Ref. \cite{carollo2019quantumness}, that $0\le \mathcal{R}_{Q} \le 1$. Thus, we can tighten the {HCRB} by the {SLD}-{QCRB} and write
	\begin{equation}\label{Eq. 2.135}
		B_S\left(\boldsymbol{\theta}\right)\le 	B_H\left(\boldsymbol{\theta}\right)\le\left(1+\mathcal{R}_{Q}\right)B_S\left(\boldsymbol{\theta}\right)\le 2 B_S\left(\boldsymbol{\theta}\right). 
	\end{equation}
	From this result, we can conclude that the evaluation together of {SLD}-{QCRB} and the \textit{\textbf{quantumness}} parameter gives, in fact, an estimate of the {HCRB}. Furthermore, we noted that the parameter $\mathcal{R}_{Q}$ is introduced as a figure of merit that measures the amount of incompatibility within a multiparameter estimation model. The saturation of the upper bound, $\mathcal{R}_{Q}=1$, is equivalent to the maximal incompatibility between the simultaneously estimated parameters. In the opposite case, when $\mathcal{R}_{Q}=0$, the parameterization model is compatible.
	\section{Classification of multiparameter quantum statistical models}
	In this section, based on the relationship between the {HCRB} and the different {QCRB}s,  we will classify the multiparameter quantum statistical models. This classification has been originally discussed in Refs. \cite{suzuki2019information, albarelli2020perspective}, where the authors show that multiparameter quantum models can be classified into four different classes.
	\subsection{ Classical quantum statistical model}
	Given a quantum statistical model, $\mathcal{S}$, of a matrix density $\hat \rho\left({\boldsymbol{\theta}}\right)$, for each parameter $\boldsymbol{\theta}\in \Theta$, can be diagonalized the quantum state $\hat \rho\left({\boldsymbol{\theta}}\right)$  with a unitary as
	\begin{equation}
		\rho\left({\boldsymbol{\theta}}\right)=U\left(\boldsymbol{\theta}\right)\boldsymbol{\Lambda}\left(\boldsymbol{\theta}\right)U\left(\boldsymbol{\theta}\right)^{\dag}.
	\end{equation}
	where $\boldsymbol{\Lambda}\left(\boldsymbol{\theta}\right)$ begin a diagonal matrix with the eigenvalues of the state $\hat \rho\left({\boldsymbol{\theta}}\right)$, formally, it is written as follows
	\begin{equation}
		\boldsymbol{\Lambda}\left(\boldsymbol{\theta}\right)=\left(\begin{array}{cccc}
			p\left(\boldsymbol{x}; \theta_1\right) & 0 & \cdots & 0 \\
			0 & p\left(\boldsymbol{x}; \theta_2\right) & \cdots & 0 \\
			\vdots & \vdots & \ddots & \vdots \\
			0 & 0 & \cdots & p\left(\boldsymbol{x}; \theta_m\right)
		\end{array}\right).
	\end{equation}
	When the unitary $U$ begins independent of $\boldsymbol{\theta}$, for all $\boldsymbol{\theta }\in\Theta$, the quantum statistical model is reduced to the classical one. With this identification, we can define the classical-quantum statistical model as following
	\begin{MD}
		For a given quantum statistical model $\mathcal{S}$, the model is classical if and only if the matrix density $\hat \rho\left({\boldsymbol{\theta}}\right)$ can be diagonalized with a $\boldsymbol{\theta}$-independent unitary $U$ as
		\begin{equation}
			\hat \rho\left({\boldsymbol{\theta}}\right)=U\boldsymbol{\Lambda}\left(\boldsymbol{\theta}\right)U^{\dag}, \hspace{1cm} \text{for all} \hspace{1cm} \boldsymbol{\theta } \in \Theta.
		\end{equation}
	\end{MD}
	These quantum statistical models are named the classical models because; they are entirely described in terms of the classical statistical models, in where the quantum density operators are replaced by the classical probability distribution. In this case, the {SLD}-{QCRB}s scalars or matrices are always achievable and constructing the optimal measurements by the projectors over the eigenstates of {SLD}-operators, which implied that the classical Fisher information is equal to the quantum Fisher information at the optimal values. Mathematically, we can write $\mathcal{F}_C\left({\boldsymbol{\theta}_{opt}}\right)=\mathcal{F}_Q^{(S)}\left({\boldsymbol{\theta}}\right)=\mathcal{F}_Q^{(R)}\left({\boldsymbol{\theta}}\right)$ and thus we have
	\begin{equation}
		B_H\left({\boldsymbol{\theta}}\right)=B_S\left({\boldsymbol{\theta}}\right)=B_R\left({\boldsymbol{\theta}}\right).
	\end{equation}
	\subsection{Quasi-classical quantum statistical model}
	As discussed above,  the saturation of {QCRB}s does not require the diagonalization of density operators. But this saturation requires, in fact, a sufficient condition that is
	\begin{equation}\label{Eq. 2.140}
		\left[\mathcal{\hat L}_{\theta_j}^{(S)}, \mathcal{\hat L}_{\theta_k}^{(S)}\right]=0,  \hspace{1cm} \text{for all} \hspace{0.5cm} j,k=1,2,...,m. 
	\end{equation}
	With satisfied this condition, we can define the \textit{\textbf{quasi-classical}} quantum statistical model as follows
	\begin{MD}
		A quantum statistical model, $\mathcal{S}$, of a matrix density $\hat \rho\left(\boldsymbol{\theta}\right)$ is said \textit{\textbf{quasi-classical}} if all {SLD}-operators commute with each other at all $\boldsymbol{\theta }\in \Theta$.
	\end{MD}
	Although {SLD} operators commute, this does not generally mean that estimation problems are reformulated as classical problems. These kinds of models are dubbed \textit{\textbf{quasi-classical}}. Since the {SLD}-operators admit the same eigenstates, thus it is always possible to perform an optimal measurement that saturates the {SLD}-{QCRB} scalars or matrices. In this case,  we can write that $\mathcal{F}_C\left({\boldsymbol{\theta}_{opt}}\right)=\mathcal{F}_Q^{(S)}\left({\boldsymbol{\theta}}\right)$ and therefore we get
	\begin{equation}
		B_H\left({\boldsymbol{\theta}}\right)=B_S\left({\boldsymbol{\theta}}\right). \label{Eq. 2.141}
	\end{equation}
	Clearly, if the model is classical, then it is also quasi-classical. However, the converse statement does not hold in general. Hence, the classical models are considered as a particular case of \textit{\textbf{quasi-classical}} one. We note, from Eq. (\ref{Eq. 2.140}) and Eq. (\ref{Eq. tight}), that the \textit{\textbf{quantumness}} parameter defined in Eq. (\ref{Eq. 2.134}) is equal to zero, $\mathcal{R}_{Q}=0$, which agree with the classical compatibility parameterization of estimation models.
	\subsection{ Asymptotically quantum classical statistical model}
	Indeed, the {SLD}-operators commute with each other is a sufficient condition but does not necessary to saturate the {QCRB} in multiparameter estimation models. As examined in Sub. (\ref{subsect2.6.3}), the most general condition of attainment of the {SLD}-{QCRB} scalars or matrices is that given in  Eq. (\ref{EQ. 2.104}). With fulfilled this \textit{compatible condition}, we can define the asymptotically classical-quantum statistical models as follows
	\begin{MD}
		A quantum statistical model, $\mathcal{S}$, with density operator $\hat \rho\left(\boldsymbol{\theta}\right)$ is called an asymptotically quantum-classical statistical model if and only if all the {SLD}-operators commute on average on $\hat \rho\left(\boldsymbol{\theta}\right)$. In other words, if the \textit{compatible condition} is fulfilled, then we have an asymptotically quantum-classical statistical model.
	\end{MD}
	Given the definition of an asymptotically quantum-classical statistical model introduced above, it is clear that the \textit{\textbf{quantumness}} parameter defined in Eq. (\ref{Eq. 2.134}) has been introduce as a measure of asymptotic incompatibility. It is zero, in this case, since the incompatible matrix $\boldsymbol{U}$ is equal to a zero matrix. Thus, the chain of inequality given in Eq. (\ref{Eq. 2.135}) is saturated, and we get
	\begin{equation}\label{Eq. 2.142}
		B_H\left({\boldsymbol{\theta}}\right)=B_S\left({\boldsymbol{\theta}}\right). 
	\end{equation}
	Consequently, all \textit{\textbf{quasi-classical}} models are also asymptotically quantum-classical. However, the converse statement is not generally valid. Thus, the \textit{\textbf{quasi-classical}} model is considered as a particular case of the most general that is an asymptotically quantum-classical statistical model.
	\subsection{ D-invariant quantum statistical model}
	The D-invariant quantum statistical models were introduced originally by Holevo based on the commutation operator $\mathcal{\hat D}_{\hat \rho_{\boldsymbol{\theta}}}$ called super-operators, and whose action on $\boldsymbol{\hat X}$ is determined by 
	\begin{equation}
		\left[\hat \rho_{\boldsymbol{\theta}}, \boldsymbol{\hat X}\right]=\hat \rho_{\boldsymbol{\theta}}\boldsymbol{\hat X}-\boldsymbol{\hat X}\hat \rho_{\boldsymbol{\theta}}=i\rho_{\boldsymbol{\theta}}\mathcal{\hat D}_{\hat \rho_{\boldsymbol{\theta}}}\left(\boldsymbol{\hat X}\right)+i\mathcal{\hat D}_{\hat \rho_{\boldsymbol{\theta}}}\left(\boldsymbol{\hat X}\right)\rho_{\boldsymbol{\theta}}.
	\end{equation}
	where $\boldsymbol{\hat X}$ are the set of collective Hermitian observables\footnote{They are the same that defined the {HCRB} given in Eq. (\ref{Eq. 2.128}).}. Before introducing the definition of D -invariant, it is most suitable, at first, to clarify some necessary notion, which is the {SLD} tangent space. The {SLD} tangent space is defined by the linear span of {SLD}-operators \footnote{ We can also define the {RLD} tangent space by the linear span of {RLD} operators with complex coefficients $\mathcal{T}_{\hat \rho_{\boldsymbol{\theta}}}\left(\mathcal{S}\right)=\text{span}_{\mathbb{C}}\left\{\mathcal{\hat L}_{\theta_{j}}^{(R)}\right\}$.}such as
	\begin{equation}
		\mathcal{T}_{\hat \rho_{\boldsymbol{\theta}}}\left(\mathcal{S}\right)=\text{span}_{\mathbb{R}}\left\{\mathcal{\hat L}_{\theta_{j}}^{(S)}\right\}, \hspace{0.7cm} \text{where} \quad j=1,2,...,m.
	\end{equation}
	Now that we have both definitions of the commutation super-operator and the {SLD} tangent space, we can then define the D-invariant quantum statistical model as follows;
	\begin{MD}
		A quantum statistical model, $\mathcal{S}$, with a density operator $\hat \rho\left(\boldsymbol{\theta}\right)$ is called D-invariant quantum statistical model if and only if the {SLD} tangent space at $\boldsymbol{\theta}$ is an invariant subspace of the commutation super-operator. Mathematically, we can express this condition as
		\begin{equation}
			\forall \boldsymbol{\hat X}\in \mathcal{T}_{\hat \rho_{\boldsymbol{\theta}}}\left(\mathcal{S}\right),\quad \mathcal{\hat D}_{\hat \rho_{\boldsymbol{\theta}}}\in \mathcal{T}_{\hat \rho_{\boldsymbol{\theta}}}\left(\mathcal{S}\right).
		\end{equation}
	\end{MD}
	As an advantage, in the D-invariant quantum statistical models, the {HCRB} can be expressed analytical and coincides with the {RLD}-{QCRB}. Thus, we get
	\begin{equation}
		B_H\left({\boldsymbol{\theta}}\right)=B_R\left({\boldsymbol{\theta}}\right).
	\end{equation}
	
	Based on the different definitions introduced previously, in the following figure, we will summarize the set of all classification quantum statistical models.
	\begin{figure}[H]      
		\tikzset{every picture/.style={line width=0.5pt}} 
		
		\begin{tikzpicture}[x=0.82pt,y=0.99pt,yscale=-1,xscale=1]
			
			\draw  [color={rgb, 255:red, 208; green, 2; blue, 27 }  ,draw opacity=1 ][line width=2.25]  (48,242) .. controls (48,176.28) and (115.6,123) .. (199,123) .. controls (282.4,123) and (350,176.28) .. (350,242) .. controls (350,307.72) and (282.4,361) .. (199,361) .. controls (115.6,361) and (48,307.72) .. (48,242) -- cycle ;
			\draw  [color={rgb, 255:red, 0; green, 0; blue, 0 }  ,draw opacity=1 ][line width=2.25]  (238,240) .. controls (238,174.83) and (329.11,122) .. (441.5,122) .. controls (553.89,122) and (645,174.83) .. (645,240) .. controls (645,305.17) and (553.89,358) .. (441.5,358) .. controls (329.11,358) and (238,305.17) .. (238,240) -- cycle ;
			\draw  [color={rgb, 255:red, 74; green, 74; blue, 74 }  ,draw opacity=1 ][line width=2.25]  (238,240) .. controls (238,186.98) and (290.83,144) .. (356,144) .. controls (421.17,144) and (474,186.98) .. (474,240) .. controls (474,293.02) and (421.17,336) .. (356,336) .. controls (290.83,336) and (238,293.02) .. (238,240) -- cycle ;
			\draw  [color={rgb, 255:red, 74; green, 144; blue, 226 }  ,draw opacity=1 ][line width=2.25]  (186,60) -- (276,60) -- (276,100) -- (186,100) -- cycle ;
			\draw  [color={rgb, 255:red, 74; green, 144; blue, 226 }  ,draw opacity=1 ][line width=2.25]  (455,59) -- (649,59) -- (649,99) -- (455,99) -- cycle ;
			\draw  [color={rgb, 255:red, 74; green, 144; blue, 226 }  ,draw opacity=1 ][line width=2.25]  (304,60) -- (425,60) -- (425,100) -- (304,100) -- cycle ;
			\draw  [color={rgb, 255:red, 74; green, 144; blue, 226 }  ,draw opacity=1 ][line width=2.25]  (52,61) -- (157,61) -- (157,101) -- (52,101) -- cycle ;
			\draw [color={rgb, 255:red, 74; green, 144; blue, 226 }  ,draw opacity=1 ][line width=2.25]    (118,101) -- (140.68,132.75) ;
			\draw [shift={(143,136)}, rotate = 234.46] [color={rgb, 255:red, 74; green, 144; blue, 226 }  ,draw opacity=1 ][line width=2.25]    (17.49,-5.26) .. controls (11.12,-2.23) and (5.29,-0.48) .. (0,0) .. controls (5.29,0.48) and (11.12,2.23) .. (17.49,5.26)   ;
			\draw [color={rgb, 255:red, 74; green, 144; blue, 226 }  ,draw opacity=1 ][line width=2.25]    (246,100) -- (282.09,166.48) ;
			\draw [shift={(284,170)}, rotate = 241.5] [color={rgb, 255:red, 74; green, 144; blue, 226 }  ,draw opacity=1 ][line width=2.25]    (17.49,-5.26) .. controls (11.12,-2.23) and (5.29,-0.48) .. (0,0) .. controls (5.29,0.48) and (11.12,2.23) .. (17.49,5.26)   ;
			\draw [color={rgb, 255:red, 74; green, 144; blue, 226 }  ,draw opacity=1 ][line width=2.25]    (404,100) -- (389.12,151.16) ;
			\draw [shift={(388,155)}, rotate = 286.22] [color={rgb, 255:red, 74; green, 144; blue, 226 }  ,draw opacity=1 ][line width=2.25]    (17.49,-5.26) .. controls (11.12,-2.23) and (5.29,-0.48) .. (0,0) .. controls (5.29,0.48) and (11.12,2.23) .. (17.49,5.26)   ;
			\draw [color={rgb, 255:red, 74; green, 144; blue, 226 }  ,draw opacity=1 ][line width=2.25]    (564,100) -- (540.94,141.5) ;
			\draw [shift={(539,145)}, rotate = 299.05] [color={rgb, 255:red, 74; green, 144; blue, 226 }  ,draw opacity=1 ][line width=2.25]    (17.49,-5.26) .. controls (11.12,-2.23) and (5.29,-0.48) .. (0,0) .. controls (5.29,0.48) and (11.12,2.23) .. (17.49,5.26)   ;
			\draw   (40,50) -- (658,50) -- (658,366) -- (40,366) -- cycle ;
			
			\draw (87,230) node [anchor=north west][inner sep=0.75pt]   [align=left] {$B_H=B_R$};
			\draw (197,75) node [anchor=north west][inner sep=0.75pt]   [align=left] {\textbf{Classical}};
			\draw (57,75) node [anchor=north west][inner sep=0.75pt]   [align=left] {\textbf{D-iinvariant}};
			\draw (309,75) node [anchor=north west][inner sep=0.75pt]   [align=left] {\textbf{quasi-classical}};
			\draw (461,75) node [anchor=north west][inner sep=0.75pt]   [align=left] {\textbf{asymptotically classical}};
			\draw (278,313) node [anchor=north west][inner sep=0.75pt]  [rotate=-266.81] [align=left]{$\mathcal{F}_Q^{(S)}=\mathcal{F}_Q^{(R)}=\mathcal{F}_C$};
			\draw (353,210) node [anchor=north west][inner sep=0.75pt]   [align=left] {$\left[\mathcal{\hat L}_{\theta_j}^{(S)},\mathcal{\hat L}_{\theta_k}^{(S)}\right]=0$};
			\draw (355,250) node [anchor=north west][inner sep=0.75pt]   [align=left] {$B_H=B_S$};
			\draw (500,250) node [anchor=north west][inner sep=0.75pt]   [align=left] {$B_H=B_S$};
			\draw (510,210) node [anchor=north west][inner sep=0.75pt]   [align=left] {$U_{\theta_{j}\theta_{k}}=0$};
			\draw (315,301) node [anchor=north west][inner sep=0.75pt]  [rotate=-266.81] [align=left] {$B_S=B_R=B_H$};
			
		\end{tikzpicture}
		
		\centering
		\captionsetup{justification=centerlast, singlelinecheck=false}\captionof{figure}{A schematic diagram represents the classification of quantum statistical models. Originally, this classification was introduced in Ref. \cite{suzuki2019information}, also has been reviewed in-depth in \cite{albarelli2020perspective}.}\label{Fig}
	\end{figure}
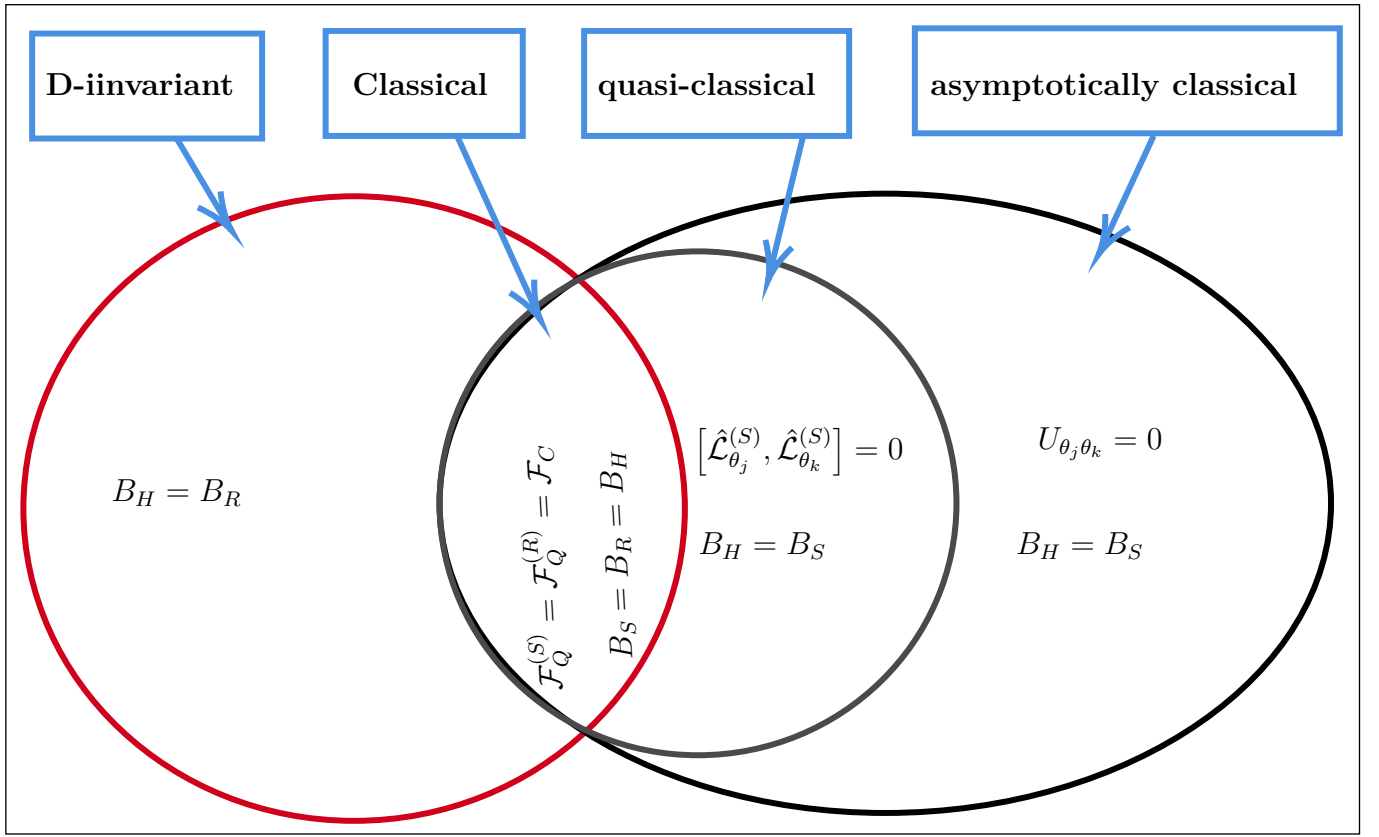
	\section{Conclusion}
	\vspace{-0.3cm}
	Without exaggeration, due to the increased need for more accurate and sensitive detectors, the ability to dive into new approaches for achieving high precision in estimation protocols has been one of the most fundamental drivers of scientific and technological discovery. In this context, the quantum estimation theory seeks scenarios where non-classical resources can realize progress beyond the classical ones. And for this, it tries to identify the measurements that achieve quantum-enhanced precision,  which can be attainment after suppressing all classical sources of measurements. Of course, this has been achieved by using standard quantum measurement, which is, in turn, the {POVM}. Thus,  quantum metrology or quantum estimation theory provides a natural extension of the classical methods of advancing performance improvements using the {POVM}. More precisely, quantum metrology looks for the {POVM} to maximizing the classical Fisher information, thus minimizing via {CCRB} the variance of the estimator.
	
	In this chapter, upon exploiting the  {POVM}, we have described a general approach to solving a quantum statistical model. To this purpose, we have derived the quantum versions of  {CRB}s and then evaluated the bound of precision possible attainable. At first, we focus on single-estimation models. Due to the non-commutativity of quantum mechanics, we have derived two families of  {LD} operators, which are the  {SLD} and  {RLD} operators. Inspired by these families, we derived two versions of the quantum Fisher information, namely the  {SLD} and  {RLD}-quantum Fisher information. These quantities of Fisher information were used to derive the variance bounds for the estimator via the different versions of the quantum  {CRB}s, which are  {SLD} and  {RLD}- {QCRB}s. We have shown that, particularly in the single parameter estimation,  the  {SLD}- {QCRB} is always attaining and more informative than the  {RLD}- {QCRB}. This means that the  {RLD}- {QCRB} is only interesting when looking at multi-parameter estimation cases. Having accomplished this, we extended these results to multiparameter quantum estimation models. In this case, we review the  {SLD}  and  {RLD}-quantum Fisher information matrices. These information matrices were devoted to deriving bounds for elements of the covariance matrix by employing the Schwartz inequality. We have provided a comparison of these bounds and discussed their attainability. Specifically, we report that the commonly used  {SLD} and  {RLD}- {QCRB}s are generally not attainable in multiparameter estimation cases. This fact leads to the difficulty of extracting, directly, the optimal simultaneous estimators. In this context, we have discussed the  {HCRB}, which provides the most fundamental, simultaneously attainable bound for multiparameter estimation problems. Despite its importance, the evaluation of  {HCRB} is still a task most hardly for arbitrary estimation models. This task remains an area of current research interest in multiparameter estimation problems. We have ended this chapter by discussing the different classifications of quantum statistical models, which are; classical, \textit{\textbf{quasi-classical}}, asymptotically classical, and D-invariant models.
	
	Indeed,  due to the assumption restricting our analysis only to finite-dimensional quantum systems, the progress made in this chapter to extend the classical estimate to its quantum counterpart is still incomplete. More precisely, all explicit formulas expressed in this chapter have been discussed only for the quantum statistical models of density operators living in the finite-dimensional Hilbert space. Therefore, one can ask a question that is; how can do to extend this analysis into a quantum statistical model of a matrix density living in the infinite-dimensional Hilbert space? This question we will address in the next chapter.
	\chapter{ Quantum metrology with continuous variables}\label{Ch. 3}
	\section{Introduction}
	In the past decades, quantum metrology has exploited quantum mechanics effects to accomplish many advancements. These advancements were mainly motivated by the current quantum technology revolution \cite{toth2014quantum, braun2018quantum, pezze2018quantum, huang2014quantum}. The applications of quantum metrology range from fundamental science, such as improving time and frequency standards \cite{boss2017quantum, albarelli2018restoring}, to advancing the sensitivity of gravitational wave interferometry \cite{abbott2009ligo}, to more applied scenarios, such as navigation \cite{cai2013chemical, komar2014quantum}, and super-resolution imaging \cite{yang2016far}, magnetic field detection for biomedical diagnostics \cite{taylor2016quantum, bonato2016optimized}. In all these applications, one or more parameters of the system under study are encoded in the states of light, which are the probe fields initialized in continuous and infinite spectrum systems. Then, an attempt is made to recover this value by detecting the light appropriately.  The continuous spectrum of the quantum state of light always requires the adoption of infinite-dimensional Hilbert spaces, even when considering a finite number of degrees of freedom. This fact implies that the analysis of properties related to quantum information, including the quantum metrology of systems with continuous variables, is generally most hard. A widely known way to make continuous-variable systems easier to tractable is to restrict them into  Gaussian states \cite{wang2007quantum, olivares2012quantum, ferraro2005gaussian, weedbrook2012gaussian}.
	
	In quantum optics literature, the study of Gaussian states has more recently become a central theme of quantum information with continuous variable systems. This family of states is relatively easy to generate and manipulate experimentally. From a theoretical viewpoint, it provides beneficial tools for encoded and processing information due to the limit of their degrees of freedom which is limited only to the displacement vector and the covariance matrix. The formalism of the Gaussian state has already been proven and serves as an invaluable tool for describing the quantum states of light and atomic ensembles \cite{hammerer2010quantum}, as well as providing appropriate insight and intuition. They have been successfully used to make incredible advances over the past decade in several areas of quantum physics, such as current quantum optical technology \cite{xiao2019continuous, dowling2014quantum}, description of optomechanical oscillators \cite{tian2010optical, nunnenkamp2011single, knill2001scheme}.
	
	Given the simplicity of Gaussian states in describing quantum states of light as well as the importance of quantum metrology in improving detection sensitivity, it would be preferable to integrate both into a common framework. The goal of this chapter goes in this direction. We will provide the analytical expression of the central quantities of quantum metrology, which are the {SLD} and {RLD-QFIM}s. Indeed, the most efficient way to achieve this goal is to use a finite-dimensional phase-space analysis. To achieve this goal, we must first provide a comprehensive review of the basic concepts of the terminology of continuous variables, specifically, continuous variables that have Gaussian characteristic functions. Next, we present, in Sec. (\ref{Sec. 3.5}), the general framework of quantum metrology for Gaussian states. Specifically, we derive the different families of {QFIM}s that are {SLD} and {RLD-QFIM}s in terms of the first and second moments of Gaussian states. We also discuss the condition of attainment of the lower bound of {SLD-QCRB} since the {RLD-QCRB} is not always saturated. In Sec. (\ref{Sec. 3.6}), we treat quantum metrology with single-mode Gaussian states, such that we deal with the problem of optical phase estimation. This problem is especially applicable experimentally in quantum optics. We clarify, in this problem, how to use the result developed in this chapter, and we will also discuss the different precision limits that are possible in non-classical resources. Finally, we end this chapter with concluding remarks. Technical proof of calculation is provided in the appendices of the supplementary \ref{Ch. 5}.
	\section{Introduction to continuous variables} \label{Sec 3.2}
	It propagates in space and interferes with itself, disperses in optical media such as prisms, and displays polarization effects. All these properties are regarded, commonly, as wave characteristics of light. Moreover,  light behaves like moving particles in motion. These particles appear as a distinct detector named photons. Thus, we can say, at the same time, that light has both wave and particle aspects. This strange property has intrigued countless physicists over the last century. Indeed,  it has not been settled the controversy surrounding yet, but rather it has been formulated more precisely in the quantum theory of light and is known as \textbf{Wave-Particle Duality} \cite{selleri2012wave, dimitrova2008wave}. According to this theory, the wave features of light are regarded as classical aspects, which does not necessarily mean that the particle aspects are entirely quantum. The electromagnetic oscillator is the most relevant model for both the classical and quantum aspects of lights. All classical wave aspects, including polarization, are compressing in one complex vector called a spatial-temporal wave. A simple example of this vector is a plane wave given by 
	\begin{equation}\label{Eq. 3.1}
		u\left(r_i,t\right)=\alpha\exp\left(i\left(\vec k.\vec r - \omega t\right)\right)=\alpha\exp\left(-i\left(\boldsymbol{\vec k}. \boldsymbol{\vec r}\right)\right), 
	\end{equation}
	where $\alpha$ is the complex amplitude of wave, $\omega$ and $k$ are, respectively, the frequency and wave vector. Of course, the spatial-temporal function given in Eq. (\ref{Eq. 3.1}) obeys the laws of classical waves, that is, Maxwell's equations. In the quantum aspect, which is the quantum field theory, the choice of $u\left(r_i,t\right)$ is made by the observer. This observer singles out one mode, one quantum object, from the rest of the world. This object turns out to be a harmonic oscillator described by the annihilation operator $\hat{a}$. The operator $\hat{a}$ allows quantifying the amplitude with which the spatial-temporal mode can be excited.  In classical optics, it would be just a complex number $\alpha=q+ip$ of magnitude $|\alpha|$ and phase $\arg\alpha$. To make all these fuzzy words more precise, we take the electric field strength $\hat{E}$ of the light field as follows;
	\begin{equation}
		\hat{E}=\hat{a}u\left(r_i,t\right)^*+u\left(r_i,t\right)\hat{a}^{\dag},
	\end{equation}
	$\hat{a}$ is the amplitude operator that is, in fact, a bosonic annihilation operator that obeys the following commutation relation
	\begin{equation}\label{Eq. 3.3}
		\left[\hat{a}, \hat{a}^{\dag}\right]=\mathbb{1}.
	\end{equation}
	Throughout this chapter, we set Planck's constant $h=2$. In the following, we introduce the key elements of quantum oscillator harmonic. We define a pair of self-adjoint operators, $\hat{Q}$ and $\hat{P}$, called the quadrature operators. They appear as the "real" and the "imaginary" part, respectively, of the "complex" amplitude $\hat{a}$ multiplied by 2:
	\begin{equation}
		\hat{Q}=\hat{a}+\hat{a}^{\dag}, \quad \quad \hat{P}=i\left(\hat{a}^{\dag}-\hat{a}\right).
	\end{equation}
	In physical optics or wave optics, $\hat{Q}$ and $\hat{P}$ correspond to the in-phase and the out-of-phase component of the electric field amplitude of the spatial-temporal mode with respect to a reference phase. From the bosonic commutation relation (\ref{Eq. 3.3}), it is easy to see that $\hat{Q}$ and $\hat{P}$ are conjugate observables that satisfy the canonical commutation relation
	\begin{equation}\label{Eq. 3.5}
		\left[\hat{Q}, \hat{P}\right]=2i\mathbb{1}. 
	\end{equation}
	The quadrature $\hat{Q}$ and $\hat{P}$ can be regarded as the position and the momentum of the electromagnetic oscillator. They are referred to as a pair of ‘canonical’ operators, in analogy with the terminology of classical Hamiltonian dynamics, where commutators would be replaced with Poisson brackets \footnote{The Poisson bracket is a binary operation in Hamiltonian mechanics. It is played a central role in Hamilton's equations of motion, which govern the time evolution of a Hamiltonian dynamical system. In canonical coordinates  $(q, p)$ on the phase space, given two functions, $f$ and $g$, the Poisson bracket takes the form $\{ f,g\}  = \frac{{\partial f}}{{\partial q}}\frac{{\partial g}}{{\partial p}} - \frac{{\partial f}}{{\partial p}}\frac{{\partial g}}{{\partial q}}$.}. Of course, they do not appear in real space but in the phase space spanned by the complex vibration amplitude $\hat{a}$ of the electromagnetic oscillator. By using the commutation relation (\ref{Eq. 3.5}), we express the photon-number operator $\hat{n}$  that accounts for the photons number in the chosen spatial-temporal wave function and is given by the counterpart of a classical modulus-squared amplitude $\hat{n}=\hat{a}^{\dag}\hat{a}$, and we get 
	\begin{equation}
		\hat H = 2\hat n + 1 = \hat Q^{2}  + \hat P^{2}.
	\end{equation}
	The right-hand side of this equation stands for the energy of a harmonic oscillator, which means that the  double photon number plus $1$ gives the energy of the electromagnetic oscillator with unity mass and frequency $m= \omega=1$. The additional $1$ is called vacuum energy.
	
	As a matter of fact, the representation through finite-dimensional matrices is not appropriate to represent the conjugate operators $\hat Q$ and $\hat P$. This fact will be apparent if we take the trace of the left- and right-hand sides of Eq. (\ref{Eq. 3.5}), i.e.
	\begin{equation}
		\operatorname{Tr}\left[\hat Q \hat P- \hat P \hat Q\right]=	\operatorname{Tr}\left[\hat Q \hat P\right]-	\operatorname{Tr}\left[ \hat P \hat Q\right]=0,
	\end{equation}
	which contradicts with $\operatorname{Tr}\left[\mathbb{1}\right]=\text{dim} \mathcal{H}$. Therefore, the infinitely dimensional representations of the quadratic operators  $\hat Q$ and $\hat P$  do exist, and we can say that they have a continuous spectrum of eigenvalues that satisfies
	\begin{equation}\label{Eq. 3.8}
		\hat{Q} \ket{q}=q \ket{q}, \quad \hat{P}\ket{p}=p\ket{p}. 
	\end{equation}
	In quantum mechanics, the usual representation of the canonical commutation relation is the Schrodinger representation on the Hilbert space $\mathcal{H} = \mathcal{L}^{2}(\mathbb{R})$ of square-integrable functions, where $\hat{Q}$ and $\hat{P}$ acts as 
	\begin{equation}\label{Eq. 3.9}
		\hat{Q}\Ket{f}=q f\left(q\right), \quad \hat{P}\Ket{f}=-i\frac{\partial}{\partial q}f\left(q\right),
	\end{equation}
	where $f\left(q\right)=\braket{q|f}$. According to Eqs. (\ref{Eq. 3.8}), (\ref{Eq. 3.9}), the operators $\hat{Q}$ and $\hat{P}$ admit a spectral decomposition in terms of projectors on these improper eigenvectors. Their eigenvalues form a continuous set covering the whole real line, hence the terminology "\textbf{quantum continuous variables}" for systems described by pairs of canonical operators.
	
	The quantum field theory allows us to include relativistic systems with a certain number of particles through the formalism of second quantization. In this formalism, the commutation relation above (\ref{Eq. 3.5}) applies to pairs of bosonic field operators $\hat{Q}$ and $\hat{P}$. A prominent example of such a quantum field is the electromagnetic one, where $\hat{Q}$  and $\hat{P}$ are the quantum counterpart of the magnetic and electric fields along one polarization direction. For this reason, the systems described by quantum continuous variables are often referred to, in the literature, as “bosonic” systems. 
	\subsection{Canonical commutation relations and quantized fields}
	In quantum mechanics, the \textbf{CV} system was introduced as a system in which the degrees of freedom associated with the canonical operators have a continuous spectrum. The eigenstates of such operators form bases for the infinite-dimensional Hilbert space $\mathcal{H}$ of the system. Now, we consider a canonical infinite-dimensional system composed of $N$-bosonic modes. Each mode, $k$, is described by a pair of quadrature field operators, $\hat{Q}_k$, and $\hat{P}_k$, acting on a Hilbert space $\mathcal{H}_k$. The space  $\mathcal{H}_k$ is spanned by a number basis $\left\{\ket{n}_k\right\}$ of eigenstates of the number operator $\hat{n}_k=\hat{a}_k^{\dag} \hat{a}_k$. The Hilbert space for the whole system is the tensor product of infinite-dimensional Hilbert spaces $\mathcal{H}_k$ of each single-mode, $\mathcal{H}= \mathop\otimes \limits_{k = 1}^N {\mathcal{H}_k}$. In the case of the electromagnetic field, the Hamiltonian of the whole system is the sum of the Hamiltonian's of the single harmonic oscillator. Mathematically, we write
	\begin{equation}
		\hat{H}=\sum_{k=1}^{N}{\left(2 \hat{a}_k^{\dag} \hat{a}_k. +1\right)}.
	\end{equation}
	Here, $\hat{a}_k$ and $\hat{a}_k^{\dag} $ are the annihilation and creation operators associated with mode $k$, and satisfy the following canonical commutation relations
	\begin{equation}
		\left[\hat{a}_{k}, \hat{a}_{l}^{\dagger}\right]=\delta_{k l}, \hspace{1cm} \left[\hat{a}_{k}, \hat{a}_{l}\right]=\left[\hat{a}_{k}^{\dagger}, \hat{a}_{l}^{\dagger}\right]=0. \label{Eq. 3.11}
	\end{equation}
	By adapting that $\hbar=2$, the corresponding quadrature-phase operators for each mode are defined as
	\begin{equation}\label{Eq. 3.12}
		\hat{Q}_k=\hat{a}_k+\hat{a}_k^{\dag}, \quad \quad \hat{P}_k=i\left(\hat{a}_k^{\dag}-\hat{a}_k\right). 
	\end{equation}
	In the phase-space analysis, the quadrature operators $\hat{Q}_k$ and $\hat{P}_k$ are collected in a vector\\ $\hat{\boldsymbol{R}}=\left(\hat{Q}_{1}, \hat{P}_{1}, \ldots, \hat{Q}_{N}, \hat{P}_{N}\right)^{\top}$, which enables us to write the bosonic canonical commutation relations  in the following compact form 
	\begin{equation}\label{Eq. 3.13}
		\left[\hat{R}_{k}, \hat{R}_{l}\right]= 2 i \Omega_{k l}, 
	\end{equation}
	where $\boldsymbol{\Omega}$ is the $2N \times 2N$ symplectic matrix that takes on a standard form:
	\begin{equation}
		\boldsymbol{\Omega}=\bigoplus_{k=1}^{N} \boldsymbol{\omega}, \quad \boldsymbol{\omega}=\left(\begin{array}{cc}
			0 & 1 \\
			-1 & 0
		\end{array}\right).
	\end{equation}
	Notice that $\boldsymbol{\Omega}=-\boldsymbol{\Omega}^{\top}$ and $\boldsymbol{\Omega}^2=-\mathbb{1}_{2N\times 2N}$. Also, $\boldsymbol{\Omega}$ is a real orthogonal transformation
	\begin{equation}
		\boldsymbol{\Omega}^{\top}\boldsymbol{ \Omega}=-\boldsymbol{\Omega}^{2}=\mathbb{1}_{2N \times 2N}.
	\end{equation}
	
	Often, in the mathematical physics literature, the canonical commutation relations are expressed by exponentiating the canonical operators, which has the advantage of limiting the operators involved\footnote{The equivalence between this expression and Eq. (\ref{Eq. 3.5}) is a straightforward consequence the Baker–Campbell–Hausdorff formula: $\mathrm{e}^{\hat{A}+\hat{B}}=\mathrm{e}^{\hat{A}} \mathrm{e}^{\hat{B}} \mathrm{e}^{-[\hat{A}, \hat{B}] / 2}$.}
	\begin{equation}\label{Eq. 3.16}
		\mathrm{e}^{i\left(\hat{Q}_{k}-\hat{P}_{l}\right)}=\mathrm{e}^{i \hat{Q}_{k}} \mathrm{e}^{-i \hat{P}_{l}} \mathrm{e}^{-i \delta_{kl}}=\mathrm{e}^{-i \hat{P}_{l}} \mathrm{e}^{i \hat{Q}_{k}} \mathrm{e}^{i \delta_{kl}}. 
	\end{equation}
	The final phase factors in the previous equation, $\mathrm{e}^{i \delta_{kl}}$, imply the non-commutativity of position and momentum operators and is a typical signature of quantum mechanics. Eq. (\ref{Eq. 3.16}) may be generalized to consider arbitrary shift operators, known as Weyl operators
	\begin{equation}\label{Eq. 3.17}
		\hat{D}(\boldsymbol{R})=e^{i \boldsymbol{R}^{\top} \boldsymbol{\Omega} \hat{\boldsymbol{R}}}, \quad \text { with } \quad \boldsymbol{R}=\left(q_1, p_1, \ldots, q_N, p_N\right)^{\top} \in \mathbb{R}^{2 N}. 
	\end{equation}
	This equation is key central to the construction of the general formalism of \textbf{CV} systems and will be frequently applied throughout this chapter.
	\subsection{Phase space description and  Wigner representation}
	In classical optics, the state of an electromagnetic oscillator is perfectly described by the statistics of the classical amplitude $\alpha$. In the case where the field is coherent, the amplitude is fixed. But, it can also fluctuate when the field is partially coherent or incoherent. As explained in Sec. (\ref{Sec 3.2}), the real and the imaginary part of the complex amplitude ($\alpha=q+ip$) are the position and momentum of the electromagnetic oscillator. Thus, we can characterize the statistics of the complex amplitude by the statistics of the position $q$ and momentum $p$ that represent the components of the classical phase-space distribution. In what follows, we will refer to this distribution as $W\left(q,p\right)$. It quantifies the probability of finding a particular pair of $q$ and $p$ values in their simultaneous measurement. Knowing $W\left(q,p\right)$, all statistical quantities of the electromagnetic oscillator may be predicted by calculation. Thus, the phase-space distribution $W\left(q,p\right)$ describes the state in classical physics. All this is much evident in classical physics. But what will happen when we extend this description to quantum mechanics? In quantum mechanics, the first obstacle is\textbf{ Heisenberg's uncertainty principle} that prevents us from observing position and momentum simultaneously and precisely. So it seems pointless to think about quantum phase space. But wait! in quantum mechanics, we cannot directly observe quantum states either. However, we are legitimately entitled to use the concept of states as if they were existing entities. We use their properties to predict the statistics of observations. But, why not use a quantum phase-space distribution $W(q, p)$ solely to calculate observable quantities in a classical-like fashion? Undoubtedly, the concept of quantum phase space must contain a specific flaw. This flaw is that the probability distribution $W(q, p)$ could become negative or ill-behaved. For this very reason, $W(q, p)$ has been called in quantum mechanics as a \textit{ quasi-probability }distribution. Furthermore, there are certainly infinitely many ways to make up \textit{\textbf{quasi-probability}} distributions. Which one shall we choose? 
	
	Indeed, the first \textit{\textbf{quasi-probability}} distribution introduced in quantum mechanics was proposed by Eugene Paul Wigner. This \textit{quasi-probability} distribution is known as the Wigner function and may be defined as the Fourier transform of the specific function 
	\begin{equation} \label{Eq. 3.18}
		{W_{\hat \rho }}\left( \boldsymbol{R}  \right) = \int_{{\mathbb{R}^{2N}}} {\frac{{{d^{2N}}{\boldsymbol{\xi }}}}{{{{(2\pi )}^{2N}}}}} \exp \left( { i{{\boldsymbol{\xi }}^T}{\boldsymbol{\Omega} \boldsymbol{R} }} \right){\chi _{\hat \rho }}({\boldsymbol{\xi }}).
	\end{equation}
	which is normalized to 1 but generally non-positive. Here, the function ${\chi _{\hat \rho }}({\boldsymbol{\xi }})$ is known as the symmetrically ordered characteristic function associated with the quantum state $\hat \rho$
	\begin{equation}\label{Eq. 3.19}
		\chi_{\hat \rho}(\boldsymbol{\xi})=\operatorname{Tr}\left[\hat{\rho} \hat D(\boldsymbol{\xi})\right], \quad \text{with} \quad \boldsymbol{\xi}\in \mathbb{R}^{2N}. 
	\end{equation}
	In Eq. (\ref{Eq. 3.18}), the continuous variables $\boldsymbol{R}$, constructed in the vector of $2N$-dimensional\\$\boldsymbol{R}=\left(q_1, p_1, \ldots, q_N, p_N\right)^{\top}$,  are the eigenvalues of quadrature operators $\boldsymbol{\hat R}$. This vector belongs to a real $2N$-dimensional space $\mathcal{K}:=\left(\mathbb{R}^{2 N}, \boldsymbol{\Omega}\right)$ equipped with a symplectic form $\boldsymbol{\Omega}$, called quantum phase space, in analogy with the Liouville phase space of classical Hamiltonian mechanics. The single formula of Eq. (\ref{Eq. 3.18}) allows to marries the \textit{quasi-probability} distribution  ${W_{\hat \rho }}\left( \boldsymbol{\xi}  \right)$ with quantum mechanics. The same formula ties ${W_{\hat \rho }}\left( \boldsymbol{\xi}  \right)$ to observable quantities. And, even more remarkably, it links quantum states to observations. Thus, an arbitrary quantum state $\hat{\rho}$ of an $N$-mode bosonic system is equivalent to a Wigner function $W_{\hat{\rho}}$, defined via a Fourier transform of the characteristic function $\chi_{\hat \rho}(\boldsymbol{\xi})$, over a $2N$-dimensional quantum phase space $\mathcal{K}$. Whereas, Eq. (\ref{Eq. 3.19}) is a fundamental key that links the quantum states, represented by a density operator $\hat{\rho}$ acting on the corresponding infinitely-dimensional Hilbert space, to the characteristic function acting on the $2N$ dimensional quantum phase space. Evidently, complete knowledge of ${\chi _{\hat \rho }}({\boldsymbol{\xi }})$ provides one with comprehensive information about the quantum state. Therefore, Eq. (\ref{Eq. 3.19}) allows deriving the properties of the characteristic function ${\chi _{\hat \rho }}({\boldsymbol{\xi }})$ from those of the density operator $\hat{\rho}$. We will summarize the most relevant essential properties in the following table;
	\begin{table}[h!]
		\begin{tabular}{|R{4cm}||C{6cm}||C{6.2cm}|}
			\hline \centering{properties} & density operator $\hat{\rho}$ &   characteristic function ${\chi _{\hat \rho }}({\boldsymbol{\xi }})$  \\
			\hline
			\hline \centering{Dimension} & infinitely-dimensional Hilbert space $\mathcal{H}$ &  $2N$-dimensional quantum phase space $\mathcal{K}$\\
			\hline
			\hline \centering{Normality} & $\operatorname{Tr}\left[\hat{\rho}\right]=1$ &  ${\chi _{\hat \rho }}({{0 }})=\operatorname{Tr}\left[\hat{\rho}\right]=1$\\
			\hline
			\hline  \centering{Purity}  & $\hat{\rho}^2=\hat{\rho}$ \hspace{0.2cm} or \hspace{0.2cm} $\operatorname{Tr}\left[\hat{\rho}\right]=1$ & $\int_{{\mathbb{R}^{2N}}} {{\left| {{\chi _{\hat \rho }}\left( {\boldsymbol{\xi }} \right)} \right|^2}{{d^{2N}}{\boldsymbol{\xi }}}} = {(2\pi )^N}$\\
			\hline
			\hline \centering{Symmetry} & $\hat{\rho}^{\dag}=\hat{\rho}$ & ${\chi _{\hat \rho }}({\boldsymbol{-\xi }})={\chi _{\hat \rho }}({\boldsymbol{\xi }})$\\
			\hline 
		\end{tabular}
		\captionsetup{justification=centering}
		\caption{\label{tab:table-name} A schematic table summarizes the equivalent properties of the density operator and the characteristic function.}
	\end{table}
	\section{Characteristic function of Gaussian states}
	The notion of Gaussian function brings us back to what was presented in the first chapter, more precisely, in our learning of probability theory, often under the name of normal distributions. This function appears everywhere in the study of probability and statistics theories. For this reason, any mathematician or physicist would be well advised to know them. But wait! the quantum Gaussian state is not a familiar term as introduced in the first chapter. In fact, the notion of Gaussian is still reserved, but in this case, with the resources of quantum mechanics. Gaussian states are archetypes of quantum physical states that ubiquitous in laboratories. For instance: coherent states, such as those from a laser; thermal states, as from a black body source; and even the vacuum state. Importantly, Gaussian states are very closely related to the characteristics and Wigner functions.
	\subsection{Gaussian state}
	A quantum state $\hat \rho$ of a continuous variable system with $N$-bosonic modes is said to be Gaussian if and only if its characteristic function, or equivalently its Wigner function is Gaussian. Mathematically, it is given by
	\begin{equation} \label{Eq. 3.20}
	\chi_{\hat\rho}(\boldsymbol{\xi})=\exp\left(-\frac{1}{4} \boldsymbol{\xi}^{\top}\left(\boldsymbol{\Omega} {\boldsymbol{\mathrm V}} \boldsymbol{\Omega}^{\top}\right) \boldsymbol{\xi}-i(\boldsymbol{\Omega} \left\langle {\boldsymbol{\hat{R}}} \right\rangle)^{\top} \boldsymbol{\xi}\right).
	\end{equation} 
	\begin{equation}
		{W_{\hat \rho }}\left( \boldsymbol{R} \right) = \frac{{\exp\left(  - \frac{1}{2}{{\left( {\boldsymbol{R} - \langle \widehat {\bf{R}}\rangle } \right)}^T}{{\boldsymbol{\mathrm V}}^{ - 1}}\left( {\boldsymbol{R} - \langle \widehat {\bf{R}}\rangle } \right)\right)}}{{{\pi ^n}\sqrt {\det \left[{\boldsymbol{\mathrm V}}\right]} }}.
	\end{equation}
	These equations are fully described by the two specific variables, $\left\langle {\boldsymbol{\hat{R}}} \right\rangle$ and ${\boldsymbol{\mathrm V}}$. These variables are known respectively as the first and second canonical moments. The first moment is called the displacement vector or, simply, the mean value of the quadrature operators and expressed by
	\begin{equation}
		\mathbf{d}=\langle\hat{\mathbf{R}}\rangle=\operatorname{Tr}[\hat{\rho} \hat{\mathbf{R}}],
	\end{equation}
	and the second moment is called the covariance matrix ${\boldsymbol{\mathrm V}}$, whose arbitrary element is defined by
	\begin{equation}
		\mathrm{V}_{j k}=\frac{1}{2}\left\langle\hat{R}_{j} \hat{R}_{k}+\hat{R}_{k} \hat{R}_{j}\right\rangle-\left\langle\hat{R}_{j}\right\rangle\left\langle\hat{R}_{k}\right\rangle.
	\end{equation}
	Note that the diagonal elements of the covariance matrix provide variances of the position and momentum operators, i.e.
	\begin{equation}
		\mathrm{V}_{ii} = \left\langle {{{\hat R}_i}^2} \right\rangle  - {\left\langle {{{\hat R_i}^2}} \right\rangle ^2} = \left\langle {{{\left( {\Delta {{\hat R}_i}} \right)}^2}} \right\rangle = \mathtt{Var}_{\hat\rho}\left({\hat R}_i\right).
	\end{equation}
	Also, we note that the covariance matrix ${\boldsymbol{\mathrm V}}$ is $2N \times 2N$-real, symmetric, positive definite matrix must satisfy the following uncertainty relation, reflecting the positivity of the density matrix,
	\begin{equation} \label{Eq. 3.25}
		\boldsymbol{\mathrm V}+i \boldsymbol{\Omega} \geq 0.
	\end{equation}
	The uncertainty principle given in the last equation is directly coming from the commutation relations of Eq. (\ref{Eq. 3.13}) and implies, in turn, the positive definiteness ${\boldsymbol{\mathrm V}}> 0$. From the diagonal elements of in Eq. (\ref{Eq. 3.25}), we can easily derive the usual Heisenberg uncertainty for position and momentum\footnote{We note that we chose the natural unit convention $\hbar=2$, which other authors may have chosen differently.}
	\begin{equation}
		\mathtt{Var}_{\hat\rho}\left({\hat Q}_i\right)\mathtt{Var}_{\hat\rho}\left({\hat P}_i\right)\ge 1.
	\end{equation}
	
	Of course, the Gaussian states of the $\hat{\rho}$ matrix density may be pure or mixed. As the purity is defined by the matrix density, it can also be determined through the covariance matrix ${\boldsymbol{\mathrm V}}$
	\begin{equation}
		\mu_{\hat \rho}=\operatorname{Tr} \left[\hat \rho^{2}\right]=\frac{1}{\sqrt{\operatorname{det}\left[{\boldsymbol{\mathrm V}}\right]}}.
	\end{equation}
	As a consequence of the last equation, we can write 
	\begin{equation}
		\operatorname{det}\left[{\boldsymbol{\mathrm V}}\right]=\left\{\begin{array}{l}
			+1 \Rightarrow \text { pure } \\
			>1 \Rightarrow \text { mixed }
		\end{array}\right..
	\end{equation}
	\section{Gaussian unitaries and the symplectic transformations} \label{Sec. 3.4}
	Quantum operations describe transformations between quantum states. For example, applying a unitary transformation $\hat U$ which transforms a state $\hat{\rho}$ according to the rule $\hat{\rho} \rightarrow \hat U \hat{\rho} \hat U^{\dagger}$, is a  quantum operation that corresponds to a change of basis or a symmetry transformation. In general, a quantum state undergoes a transformation called quantum operations. These operations are generalized by a linear maps, $\mathcal{E}: \hat{\rho} \rightarrow \mathcal{E}(\hat{\rho})$. It calls quantum channels when their mapping admits a preserving trace, i.e. $\operatorname{Tr}[\mathcal{E}(\hat{\rho})]=1$. Since Gaussian states are easy to characterize because their degree of freedom is limited only to the first and second moment, it is natural to ask whether the large class of transformations acting on these states is also easy to describe. In this context,  a quantum operation is Gaussian when it transforms Gaussian states onto another Gaussian state. In other words,  Gaussian channels (unitaries) are those channels that preserve the Gaussian character of a quantum state. All Gaussian unitaries are generated via $U=e^{-i \hat{H}}$, with Hamiltonians $\hat H$ being the second-order polynomials in the mode operators. Mathematically, if we define the vector of mode operators $\hat{\mathbf{a}}:=\left(\hat{a}_{1}, \ldots, \hat{a}_{N}\right)^{\top}$ and $\hat{\mathbf{a}}^{\dagger}:=\left(\hat{a}_{1}^{\dagger}, \ldots, \hat{a}_{N}^{\dagger}\right)^{\top}$, the Hamiltonian $\hat H$ must be in the following form:
	\begin{equation}
		\hat{H}=\hat{\mathbf{a}}^{\dagger} \boldsymbol{\alpha}+\hat{\mathbf{a}}^{\dagger} \mathbf{F} \hat{\mathbf{a}}+\hat{\mathbf{a}}^{\dagger} \mathbf{G} \hat{\mathbf{a}}^{\dagger T}+\text { h.c. }.
	\end{equation}
	where $\boldsymbol{\alpha} \in \mathbb{C}^{N}, \mathbf{F}$ and $\mathbf{G}$ are $N \times N$ complex matrices, and h.c. denoted the Hermitian conjugate. In the Heisenberg picture, this kind of unitary corresponds to a Bogoliubov transformation
	\begin{equation}\label{Eq. 3.30}
		\hat{\mathbf{a}} \rightarrow \hat U^{\dagger} \hat{\mathbf{a}} \hat U=\mathbf{A} \mathbf{a}+\mathbf{B} \hat{\mathbf{a}}^{\dagger}+\boldsymbol{\alpha}, 
	\end{equation}
	where $\mathbf{A}$ and $\mathbf{B}$ are $N \times N$ complex matrices satisfy, in order to preserve the commutation relations of Eq. (\ref{Eq. 3.11}), $\mathbf{A} \mathbf{B}^{T}=$ $\mathbf{B} \mathbf{A}^{T}$ and $\mathbf{A} \mathbf{A}^{\dagger}=\mathbf{B} \mathbf{B}^{\dagger}+\mathbb{1}_{N \times N}$.
	In order to respect the same analysis as the one followed previously, we will express this linear transformation at the level of the quadrature operators.
	\subsection{Symplectic transformations}
	Indeed, a Gaussian unitary, given in Eq. (\ref{Eq. 3.30}), is more simply described in terms of quadrature operators. It is easy to show this only take the conjugate transformation of Eq. (\ref{Eq. 3.30}) together with the definition of the quadrature operators given in Eq. (\ref{Eq. 3.12}), which leads to 
	\begin{equation}
		\hat{\mathbf{R}} \rightarrow \mathbf{S} \hat{\mathbf{R}}+\mathbf{R},
	\end{equation}
	where $\mathbf{R}$ is a real vector of $2N$ dimension and $\mathbf{S}$ is a real $2 N \times 2 N$ matrix. Once again, the commutation relations of (\ref{Eq. 3.13}) have to be preserved and this is respected if the matrix $\mathbf{S}$ is symplectic, that is if $\mathbf{S} \boldsymbol{\Omega} \mathbf{S}^{T}=\boldsymbol{\Omega}$. Clearly, the eigenvalues of $\boldsymbol{\hat R}$ must also follow the same transformation rule, i.e.
	\begin{equation}
		{\mathbf{R}} \rightarrow \mathbf{S} {\mathbf{R}}+\mathbf{R}.
	\end{equation}
	Thus, an arbitrary Gaussian unitary is equivalent to an affine symplectic map that acts on the phase space and depends on $\mathbf{S}$ and $\mathbf{R}$. Finally, in terms of the statistical moments $\mathbf{d}$ and ${\boldsymbol{\mathrm V}}$, the action of a Gaussian unitary is characterized by the following transformation rule
	\begin{equation}
		\mathbf{d} \rightarrow \mathbf{d}=\mathbf{S} \mathbf{d}+\mathbf{R}, \hspace{1cm} {\boldsymbol{\mathrm V}}\rightarrow {\boldsymbol{\mathrm V}}=\mathbf{S} {\boldsymbol{\mathrm V}} \mathbf{S}^{\dagger}.\label{Eq. 3.33}
	\end{equation}
	Thus, the action of a Gaussian unitary $\hat{U}\left(\boldsymbol{R}, \mathbf{S} \right)$ to a Gaussian state, of first and second moment $\mathbf{d}$ and $\boldsymbol{\mathrm V}$, is fully described by the transformations of Eq. (\ref{Eq. 3.33}).
	
	In summary, the complete description of any matrix density $\hat \rho$ acts in an infinite-dimensional Hilbert space can be provided by the characteristic function that acts in the finite phases space. In the case of a Gaussian state, the degree of freedom of this function is limited only to the first and second statistical moments of the quadrature field operators. Therefore, the infinite-dimensional Hilbert space problem is solved by an appropriate analysis in the phase space. In the following table, we summarize the correspondence of operations and tools between Hilbert and phase spaces.
	\begin{table}[H]
		\begin{tabular}{|R{4.4cm}||C{6cm}||C{6cm}|}
			\hline \centering{Properties} & Hilbert space $\mathcal{H}$ &  Phase space  $\mathcal{K}$  \\
			\hline
			\hline \centering{Dimension} & $\infty$ &  $2N$\\
			\hline
			\hline \centering{Structure} & $\otimes$ &  $\oplus $\\
			\hline
			\hline  \centering{Description}  & $\hat{\rho}$ & $\mathbf{d}$ and $\boldsymbol{\mathrm V}$\\
			\hline
			\hline \centering{Condition} & $\hat{\rho}\geq 0$ & 
			$\boldsymbol{\mathrm V}+i\boldsymbol{ \Omega}\geq 0$\\
			\hline 
			\hline \centering{Unitary transformation}  &  $\hat{\rho} \rightarrow \hat U \hat{\rho} \hat U^{\dagger}$ & $\mathbf{d} \mapsto \mathbf{S}\mathbf{d}, \boldsymbol{\mathrm V} \mapsto \mathbf{S}\boldsymbol{\mathrm V} \mathbf{S}^{\top}$\\
			\hline
		\end{tabular}
		\captionsetup{justification=centering}
		\caption{ A schematic table summarizes the comparable properties of Hilbert space and phase space pictures for $N$-mode Gaussian states.}
	\end{table}
	
	Due to their ready availability, the ubiquitous nature of Gaussian operations, and their ease of description, Gaussian states are obvious major candidates for quantum metrology in continuous-variable systems. This provides a compelling reason for investigating the {QFI} of Gaussian states, which is the subject of the next section (The results we will discuss in the following are recently published in our paper \cite{bakmou2020multiparameter})
	\section{Quantum metrology in continuous-variable systems}\label{Sec. 3.5}
	Quantum systems with continuous variables are quantum systems obeying the canonical commutation relations given in Eq. (\ref{Eq. 3.13}). As we have seen previously in Sec. (\ref{Sec 3.2}), such systems always require the adoption of infinite-dimensional Hilbert spaces, even when considering a finite number of degrees of freedom. This fact implies that the general study of quantum metrology with continuous variables systems urns out to be particularly difficult.
	
	A widely known way to make continuous variable systems more tractable is to consider the restriction to Gaussian states. Already well known in the older quantum optics literature, the investigation of Gaussian systems has more recently become a central theme in quantum information with continuous-variable, including quantum metrology.
	
	The main goal of this chapter is to provide a formulation of the central quantities in quantum metrology, namely the {SLD} and {RLD-QFIM}s\footnote{Note here that the multiparameter estimation model is more general than the single estimation model. For this reason and to avoid repetition, we need to focus only on the derivation of {QFIM}s, which can be reduced to {QFI}s if we have a single estimation. }, in terms of the statistical moments of the Gaussian state. The most efficient and appropriate way to achieve this goal is to use a phase-space analysis. To make our following analysis more comfortable, we make a change of variable at the level of the Weyl operators (\ref{Eq. 3.17}) by setting $\boldsymbol{\tilde R} = \boldsymbol{\Omega R}$ the Weyl operator becomes
	\begin{equation}\label{Eq. 3.34}
		\hat{D}(\boldsymbol{R})=e^{-i \boldsymbol{\tilde R}^{\top} \hat{\boldsymbol{R}}}. 
	\end{equation}
	In this new setting, the characteristic function of the Gaussian state (\ref{Eq. 3.20}) has taken the following form;
	\begin{equation} \label{Eq. 3.35}
		\chi_{\hat{\rho}}(\mathbf{R})=\exp \left[-\frac{1}{4} \tilde{\mathbf{R}}^{T} \boldsymbol{\mathrm V} \tilde{\mathbf{R}}+i \tilde{\mathbf{R}}^{T} \mathbf{d}\right].
	\end{equation}
	
	Now, let $\mathcal{S} = \left\{\hat\rho\left(\boldsymbol{\theta }\right)\right\}$ be a Gaussian quantum statistical model with the set of parameters $\boldsymbol{\theta}=\left(\theta_1, \theta_{2}, \cdots, \theta_{m} \right)$. Clearly, any such model is fully described by the first and second moments, i.e. $\mathcal{S} = \left\{\boldsymbol{d}_{\boldsymbol{\theta}}, \boldsymbol{\mathrm V}_{\boldsymbol{\theta }}\right\}$. Now that we have the full description of the Gaussian statistical model, we can proceed to express the {SLD} and {RLD-QFIM}s in the phase space formalism. We start by deriving the {SLD-QFIM}.
	\subsection{ SLD quantum Fisher information matrix for Gaussian states}
	Technically, to determine the {SLD-QFIM} given in Eq. (\ref{SLDQFIM}) for Gaussian states, we must first derive the expression of the associated {SLD} operators defined in Eq. (\ref{SLD}).  For a set of $N$-mode Gaussian states, we shall put forward the ansatz that the {SLD} operator must be at most quadratic in the canonical operators and write \footnote{We adopt here and also in what follows Einstein’s convention of summation over repeated indices.}
	\begin{equation}\label{Eq. 3.36}
		\hat{\mathcal{L}}_{\theta \mu}^{(S)}=\hat{\mathcal{L}}^{(S)^{(0)}}+\hat{\mathcal{L}}_{l}^{(S)^{(1)}} \hat{R}_{l}+\hat{\mathcal{L}}_{j k}^{(S)^{(2)}} \hat{R}_{j} \hat{R}_{k}, 
	\end{equation}
	where $\hat{\mathbf{R}}=\left(\hat{Q}_{1}, \hat{P}_{1}, \ldots, \hat{Q}_{n}, \hat{P}_{n}\right)^{T}$ is the vector of canonical operators, $\hat {\mathcal{L}}^{(S)^{(0)}} \in \mathbb{R}, {\boldsymbol{\hat {\mathcal{L}}}}^{(S)^{(1)}}$ is a vector in $\mathbb{R}^{2n}$ and ${\boldsymbol{\hat{\mathcal L}}}^{(S)^{(2)}}$ is a symmetric, real $2N \times 2N$ matrix (whose symmetry ensures the overall Hermiticity of \textbf{SLD}-operator). For a given set of the parameters $\theta_{\mu}$, we prove in Appendix. (\ref{AppA}) that the quantities $\hat{\mathcal{L}}_{\theta_{\mu}}^{(S)^{(0)}}$, $\boldsymbol{\hat{\mathcal{L}}}_{\theta_{\mu}}^{(S)^{(1)}}$ and $\boldsymbol{\hat{\mathcal{L}}}_{\theta_{\mu}}^{(S)^{(2)}}$ are, respectively, written as follows
	\begin{equation}
		\hat{\mathcal{L}}_{\theta_{\mu}}^{(S)^{(0)}}=-\frac{1}{2} \operatorname{Tr}\left[\boldsymbol{\mathrm V} \hat{\mathcal{L}}_{\theta_{\mu}}^{(S)^{(2)}}\right]-\mathbf{d}^{\top} \hat{\mathcal{L}}_{\theta_{\mu}}^{(S)^{(1)}}-\mathbf{d}^{\top} \hat{\mathcal{L}}_{\theta_{\mu}}^{(S)^{(2)}} \mathbf{d}.
	\end{equation}
	\begin{equation}
		\hat{\mathcal{L}}_{\theta_{\mu}}^{(S)^{(1)}}=2 {\boldsymbol{\mathrm V}}^{-1} \partial_{\theta_{\mu}} \mathbf{d}-2 \hat{\mathcal{L}}_{\theta_{\mu}}^{(S)^{(2)}} \mathbf{d}.
	\end{equation}
	\begin{equation}\label{Eq. 3.39}
		\operatorname{vec}\left[\hat{\mathcal{L}}_{\theta_{\mu}}^{(S)^{(2)}}\right]=\left(\boldsymbol{\mathrm V} \otimes \boldsymbol{\mathrm V}+\boldsymbol{ \Omega} \otimes \boldsymbol{\Omega}\right)^{+} \operatorname{vec}\left[\partial_{\theta_{\mu}} \boldsymbol{\mathrm V}\right]. 
	\end{equation}
	
	Now that we have obtained a formula for the Gaussian {SLD}-operators, $	\hat{\mathcal{L}}_{\theta_{\mu}}^{(S)}$, we can proceed to insert it into the expression of {SLD}-{QFIM} that is given in Eq. (\ref{QFIMSLD1}). Appendix. (\ref{AppB}) is devoted to the details of this task. In this Appendix, we have shown that the elements of {SLD}-{QFIM} take the following form:
	\begin{equation}\label{Eq. 3.40}
		\left[\mathcal{F}_Q^{(S)}\right]_{\theta_{\mu} \theta_{\nu}}=\frac{1}{2} \operatorname{vec}\left[\partial_{\theta_{\mu}} \boldsymbol{\mathrm V} \right]^{\dagger}\left(\boldsymbol{\mathrm V} \otimes \boldsymbol{\mathrm V}+\boldsymbol{\Omega} \otimes \boldsymbol{\Omega}\right)^{+} \operatorname{vec}\left[\partial_{\theta_{\nu}} \boldsymbol{\mathrm V} \right]+2 \partial_{\theta_{\mu}} \mathbf{d}^{\top} \boldsymbol{\mathrm V}^{-1} \partial_{\theta_{\nu}} \mathbf{d}.
	\end{equation}
	In the case where $\left(\boldsymbol{\mathrm V} \otimes \boldsymbol{\mathrm V}+\boldsymbol{\Omega} \otimes \boldsymbol{\Omega}\right)$ is invertible, the {SLD-QFIM} can be calculated as
	\begin{equation}\label{Eq. 3.41}
		\left[\mathcal{F}_Q^{(S)}\right]_{\theta_{\mu} \theta_{\nu}}=\frac{1}{2} \operatorname{vec}\left[\partial_{\theta_{\mu}} \boldsymbol{\mathrm V} \right]^{\dagger}\left(\boldsymbol{\mathrm V} \otimes \boldsymbol{\mathrm V}+\boldsymbol{\Omega} \otimes \boldsymbol{\Omega}\right)^{-1} \operatorname{vec}\left[\partial_{\theta_{\nu}} \boldsymbol{\mathrm V} \right]+2 \partial_{\theta_{\mu}} \mathbf{d}^{\top} \boldsymbol{\mathrm V}^{-1} \partial_{\theta_{\nu}} \mathbf{d}. 
	\end{equation}
	As a particular case, in single estimate models, Eqs. (\ref{Eq. 3.40}), (\ref{Eq. 3.41}) are reduced to the scalars quantities such that
	\begin{equation}\label{Eq. 3.42}
		\mathcal{F}_Q^{(S)}\left(\theta\right)=\frac{1}{2} \operatorname{vec}\left[\partial_{\theta} \boldsymbol{\mathrm V} \right]^{\dagger}\left(\boldsymbol{\mathrm V} \otimes \boldsymbol{\mathrm V}+\boldsymbol{\Omega} \otimes \boldsymbol{\Omega}\right)^{+} \operatorname{vec}\left[\partial_{\theta} \boldsymbol{\mathrm V} \right]+2 \partial_{\theta} \mathbf{d}^{\top} \boldsymbol{\mathrm V}^{-1} \partial_{\theta} \mathbf{d}.
	\end{equation}
	\begin{equation}\label{Eq. 3.43}
		\mathcal{F}_Q^{(S)}\left(\theta\right)=\frac{1}{2} \operatorname{vec}\left[\partial_{\theta} \boldsymbol{\mathrm V} \right]^{\dagger}\left(\boldsymbol{\mathrm V} \otimes \boldsymbol{\mathrm V}+\boldsymbol{\Omega} \otimes \boldsymbol{\Omega}\right)^{-1} \operatorname{vec}\left[\partial_{\theta} \boldsymbol{\mathrm V} \right]+2 \partial_{\theta} \mathbf{d}^{\top} \boldsymbol{\mathrm V}^{-1} \partial_{\theta} \mathbf{d}.
	\end{equation}
	These expressions ((\ref{Eq. 3.40}), (\ref{Eq. 3.41})) are the most general forms of the {SLD}-{QFIM} in the Gaussian state formalism. It allows computing the precision bounds via the {SLD}-{QCRB} for several estimations problems. They are also express the {SLD}-{QFIM} in a form amenable for numerical computation, overcoming the difficulties posed by the infinite-dimensional Hilbert space.
	\subsection{RLD quantum Fisher information matrix for Gaussian states}
	Similarly, the expression of the {RLD}-quantum Fisher information matrix requires the explicit formula of the {RLD}-operator, which is defined in Eq. (\ref{RLD}). In order to find the RLD-operator, we also write it in the quadratic form of canonical operators such as
	\begin{equation}\label{Eq. 3.44}
		\hat{\mathcal{L}}_{\theta \mu}^{(R)}=\hat{\mathcal{L}}^{(R)^{(0)}}+\hat{\mathcal{L}}_{l}^{(R)^{(1)}} \hat{R}_{l}+\hat{\mathcal{L}}_{j k}^{(S)^{(2)}} \hat{R}_{j} \hat{R}_{k}, 
	\end{equation}
	with, in this case, $\hat{\mathcal{L}}^{(R)^{(0)}}\in \mathbb{C}$, $\hat{\mathcal{L}}^{(R)^{(1)}}$ is a complex vector in $\mathbb{C}^{2n}$ and  $\hat{\mathcal{L}}^{(R)^{(2)}}$  is $2N\times 2N$ complex matrix, which is not necessarily symmetric since the {RLD} is not always Hermitian. Given an $n$-mode Gaussian state $\hat \rho \left(\boldsymbol{\theta}\right)$, depending on the set of parameters $\boldsymbol{\theta}$. The quantities, $\hat{\mathcal{L}}^{(R)^{(0)}}$, $\hat{\mathcal{L}}^{(R)^{(1)}}$ and $\hat{\mathcal{L}}^{(R)^{(2)}}$ for a parameter $\theta_{\mu}$ are derived respectively, such as
	\begin{equation}
		\mathcal{L}_{\theta_{\mu}}^{(R)^{(0)}}=-\frac{1}{2} \operatorname{Tr}\left[\mathfrak{M} \hat{\mathcal{L}}_{\theta_{\mu}}^{(R)^{(2)}}\right]-\mathbf{d}^{\top} \hat{\mathcal{L}}_{\theta_{\mu}}^{(R)^{(1)}}-\mathbf{d}^{\top} \hat{\mathcal{L}}_{\theta_{\mu}}^{(R)^{(2)}} \mathbf{d}.
	\end{equation}
	\begin{equation}
		\hat{\mathcal{L}}_{\theta_{\mu}}^{(R)^{(1)}}=2\mathfrak{M}^{+} \partial_{\theta_{\mu}} \mathbf{d}-2 \hat{\mathcal{L}}_{\theta_{\mu}}^{(R)^{(2)}} \mathbf{d}.
	\end{equation}
	\begin{equation}
		\operatorname{vec}\left[\hat{\mathcal{L}}_{\theta_{\mu}}^{(R)^{(2)}}\right]=\left(\mathfrak{M}^{\dagger} \otimes \mathfrak{M}\right)^{+} \operatorname{vec}\left[\partial_{\theta_{\mu}} \boldsymbol{\mathrm V}\right].
	\end{equation}
	where $\mathfrak{M}=\boldsymbol{\mathrm V}+i\boldsymbol{ \Omega}$. The details of the calculation of these quantities are mentioned in Appendix. (\ref{AppC}). Now that we have derived the {RLD} operator, we can proceed to insert it into the definition of {RLD}-{QFIM} given in Eq. (\ref{QFIMRLD1}). The details of this task are devoted to Appendix. (\ref{AppD}). In this appendix, we have shown that the elements of {RLD -QFIM} have taken the following form
	\begin{equation}\label{Eq. 3.48}
		\left[\mathcal{F}_Q^{(R)}\right]_{\theta_{\mu} \theta_{\nu}}=\frac{1}{2} \operatorname{vec}\left[\partial_{\theta_{\mu}} \boldsymbol{\mathrm V}\right]^{\dagger} \left(\mathfrak{M}\otimes\mathfrak{M}\right)^{+} \operatorname{vec}\left[\partial_{\theta_{\nu}} \boldsymbol{\mathrm V}\right]+2 \partial_{\theta_{\mu}} \mathbf{d}^{\top} \mathfrak{M}^{+} \partial_{\theta_{\nu}} \mathbf{d}. 
	\end{equation}
	In the case in which $\mathfrak{M}$ is invertible (non-singular) the {RLD}-{QFIM} can be expressed as
	\begin{equation}\label{Eq. 3.49}
		\left[\mathcal{F}_Q^{(R)}\right]_{\theta_{\mu} \theta_{\nu}}=\frac{1}{2} \operatorname{vec}\left[\partial_{\theta_{\mu}} \boldsymbol{\mathrm V}\right]^{\dagger} \left(\mathfrak{M}\otimes\mathfrak{M}\right)^{-1} \operatorname{vec}\left[\partial_{\theta_{\nu}} \boldsymbol{\mathrm V}\right]+2 \partial_{\theta_{\mu}} \mathbf{d}^{\top} \mathfrak{M}^{-1} \partial_{\theta_{\nu}} \mathbf{d}. 
	\end{equation}
	We noted that $\mathfrak{M}=\boldsymbol{\mathrm V}+i\boldsymbol{ \Omega}$, which exactly is the uncertainty principle given in Eq. (\ref{Eq. 3.25}). This principle is saturated for coherent states \cite{molmer1997optical, van2001quantum}, which, in turn, have a Gaussian formalism. In this case, the evaluation of {RLD-QFIM} is not appropriate by using (\ref{Eq. 3.49}), but it is evaluated by using the Moore-Penrose pseudo-inverse form (\ref{Eq. 3.48}). In fact, Eqs. ((\ref{Eq. 3.48}), (\ref{Eq. 3.49})) provides the compact expression of {RLD-QFIM} in the families of Gaussian states. It allows us to derive the {RLD-QCRB} of precision that has an important role specifically in the case of multiparameter estimation models since the {SLD-QCRB} is not always attainment.
	\subsection{Attainment the lower bound of QCRB for Gaussian states}
	Although, in the phase space analysis, the different {QCRB}s of the accuracy can be computed via the {SLD} and {RLD-QFIM}s developed above. However, the saturation of these bounds, specifically the {SLD-QCRB}, which ensured the attainment lower bound of precision, remains without any comment. For this reason, in this section, we will derive for Gaussian states the expression that determines the saturability of the { SLD-QCRB}. Indeed, the \textit{compatibility condition} given by Eq. (\ref{EQ. 2.104}) can also be written as follows\footnote{Recall that the imaginary part of a complex number $Z$ is $\mathfrak{Im}\left[Z\right]=1/2i (\bar{Z}-{Z}).$}: 
	\begin{equation}\label{Eq. 3.50}
		\left[U\right]_{\theta_{\mu} \theta_{\nu}}=-\frac{i}{2}\operatorname{Tr}\left[\hat \rho \left( \boldsymbol{\theta}\right)\left[\mathcal{\hat L}_{\theta_{\mu}}^{((S))}, \mathcal{\hat L}_{\theta_{\nu}}^{(S)}\right]\right]=\mathfrak{Im}\left(\operatorname{Tr}\left[\hat \rho \left( \boldsymbol{\theta}\right) \mathcal{\hat L}_{\theta_{\mu}}^{(S)}\mathcal{\hat L}_{\theta_{\nu}}^{(S)}\right]\right).
	\end{equation}
	For a set of n-mode Gaussian statistical models, $\mathcal{S} = \left\{\hat\rho\left(\boldsymbol{\theta }\right)\right\} \equiv \left\{\boldsymbol{d}_{\boldsymbol{\theta}}, \boldsymbol{\mathrm V}_{\boldsymbol{\theta }}\right\}$, following the same approach as for the derivation of {SLD} and {RLD-QFIM}s, we can derive the attainment condition given in Eq. (\ref{Eq. 3.50})  as following \cite{nichols2018multiparameter}
	\begin{equation}\label{Eq. 3.51}
		\left[U\right]_{\theta_{\mu} \theta_{\nu}}=2\operatorname{Tr}\left[ \boldsymbol{\mathrm V} \mathcal{\hat L}_{\theta_{\mu}}^{(S)}\boldsymbol{\Omega} \mathcal{\hat L}_{\theta_{\nu}}^{(S)}\right]+ 2 \partial_{\theta_{\mu}} \mathbf{d}^{T}  \boldsymbol{\mathrm V}^{-1} \boldsymbol{\Omega}  \boldsymbol{\mathrm V}^{-1} \partial_{\theta_{\nu}} \mathbf{d}. 
	\end{equation}
	To express this result in an elegant compact form, we had to introduce some properties of the tensor product, which are:
	\begin{equation}
		\left(A \otimes B\right)\left(C \otimes D\right)=A C \otimes B D,
	\end{equation}
	\begin{equation}
		\begin{aligned}
			\operatorname{Tr}\left[(A D)^{\dagger} B C\right] &=\operatorname{vec}[A D]^{\dagger} \operatorname{vec}[B C] \\
			&=\operatorname{vec}[A]^{\dagger}(D \otimes \mathbb{1})(\mathbb{1} \otimes B) \operatorname{vec}[C]\\
			&=\operatorname{vec}[A]^{\dagger}(D \otimes B) \operatorname{vec}[C],
		\end{aligned}
	\end{equation}
	where $A, B, C$ and $C$ are the real (complex) arbitrary matrix. The use of this property, together with the result of Eq. (\ref{Eq. 3.39}), leads to evolving the first term of Eq. (\ref{Eq. 3.51}), and one  gets 
	\begin{eqnarray}
		\begin{aligned}
			\operatorname{Tr}\left[ \boldsymbol{\mathrm V}\mathcal{\hat L}_{\theta_{\mu}}^{(S)}\boldsymbol{\Omega} \mathcal{\hat L}_{\theta_{\nu}}^{(S)}\right]&=\operatorname{vec}\left[\mathcal{\hat L}_{\theta_{\mu}}^{(S)}\right]\left(\boldsymbol{\mathrm V} \otimes \boldsymbol{\Omega}\right) \operatorname{vec }\left[\mathcal{\hat L}_{\theta_{\nu}}^{(S)}\right] \\
			&=\operatorname{vec}\left[\partial_{ \theta_{\mu}} \boldsymbol{\mathrm V}\right]^{\dagger} \boldsymbol{\Sigma }^{+}(\boldsymbol{\mathrm V} \otimes \boldsymbol{\Omega}) \boldsymbol{\Sigma}^{+} \operatorname{vec}\left[\partial_{\theta_{\nu}} \boldsymbol{\mathrm V}\right].
		\end{aligned}
	\end{eqnarray}
	Here we put $\boldsymbol{\Sigma}=\boldsymbol{\mathrm V} \otimes \boldsymbol{\mathrm V}+\boldsymbol{\Omega} \otimes \boldsymbol{\Omega}$. Therefore, the condition to attain the lower bound of {SLD-QCRB} is expressed in terms of $\boldsymbol{d}$, $\boldsymbol{{\mathrm V}}$, and their derivative with respect to the estimated parameters as
	\begin{equation}
		\left[U\right]_{\theta_{\mu} \theta_{\nu}}=2\operatorname{vec}\left[\partial_{ \theta_{\mu}} \boldsymbol{\mathrm V}\right]^{\dagger} \boldsymbol{\Sigma }^{+}(\boldsymbol{\mathrm V} \otimes \boldsymbol{\Omega}) \boldsymbol{\Sigma}^{+} \operatorname{vec}\left[\partial_{\theta_{\nu}} \boldsymbol{\mathrm V}\right]+ 2 \partial_{\theta_{\mu}} \mathbf{d}^{T}  \boldsymbol{\mathrm V}^{-1} \boldsymbol{\Omega}  \boldsymbol{\mathrm V}^{-1} \partial_{\theta_{\nu}} \mathbf{d}.
	\end{equation}
	In the case where $\boldsymbol{\Sigma}$ is invertible, $\boldsymbol{\Sigma}^{+}$ can be can be replaced by $\boldsymbol{\Sigma}^{-1}$ the last equation.
	
	In the following section, we will illustrate these results with some examples that should provide the reader with some intuition on how to apply the above findings in practical cases. In these examples, we will focus on the single-mode Gaussian states.
	\section{Quantum metrology with single-mode Gaussian states}\label{Sec. 3.6}
	Indeed, the formulas developed above can be applied in various directions and diverse areas of quantum physics. On the more practical side, the employment of these results in the study of Gaussian unitary operations is, in fact, a natural progression. In this section, we treat an protocol Gaussian unitary channel to illustrate the application of these results. In this example, we will try to treat the optical phase estimation problem.
	\subsection{Estimation of optical phase}
\begin{figure}[H]      
		\tikzset{every picture/.style={line width=0.75pt}} 
		\begin{tikzpicture}[x=0.93pt,y=0.9pt,yscale=-1,xscale=1]
			
			\draw  [color={rgb, 255:red, 74; green, 144; blue, 226 }  ,draw opacity=1 ][line width=2.25]  (34,76.5) .. controls (34,45.85) and (47.66,21) .. (64.5,21) .. controls (81.34,21) and (95,45.85) .. (95,76.5) .. controls (95,107.15) and (81.34,132) .. (64.5,132) .. controls (47.66,132) and (34,107.15) .. (34,76.5) -- cycle ;
			\draw  [color={rgb, 255:red, 74; green, 144; blue, 226 }  ,draw opacity=1 ][line width=2.25]  (95,58.63) -- (164.6,58.63) -- (164.6,40.75) -- (211,76.5) -- (164.6,112.25) -- (164.6,94.38) -- (95,94.38) -- cycle ;
			\draw  [color={rgb, 255:red, 74; green, 144; blue, 226 }  ,draw opacity=1 ][line width=2.25]  (211,22) -- (294,22) -- (294,128) -- (211,128) -- cycle ;
			\draw  [color={rgb, 255:red, 74; green, 144; blue, 226 }  ,draw opacity=1 ][line width=2.25]  (294,56.75) -- (354.6,56.75) -- (354.6,39) -- (395,74.5) -- (354.6,110) -- (354.6,92.25) -- (294,92.25) -- cycle ;
			\draw  [color={rgb, 255:red, 74; green, 144; blue, 226 }  ,draw opacity=1 ][line width=2.25]  (396,23) -- (431,23) .. controls (450.33,23) and (466,46.73) .. (466,76) .. controls (466,105.27) and (450.33,129) .. (431,129) -- (396,129) -- cycle ;
			\draw   (10,14) -- (548,14) -- (548,142) -- (10,142) -- cycle ;
			
			\draw (469,58) node [anchor=north west][inner sep=0.75pt]   [align=left] {\textbf{Estimation} \\\hspace{0.7cm}\textbf{of} $\varphi$};
			\draw (56,69) node [anchor=north west][inner sep=0.75pt]    {$\hat \rho_{inp}$};
			\draw (215,64) node [anchor=north west][inner sep=0.8pt]    {$\small{\hat U_{\varphi}=e^{-i\varphi\hat{a}^{\dagger}\hat{a}}}$};
		\end{tikzpicture}
		\centering
		\captionsetup{justification=centerlast, singlelinecheck=false}\captionof{figure}{Scheme of the Gaussian single-mode metrology. The state of the Gaussian probe undergoes a phase shift, which is the parameter that will be estimated.} \label{Fig. 5}
	\end{figure}
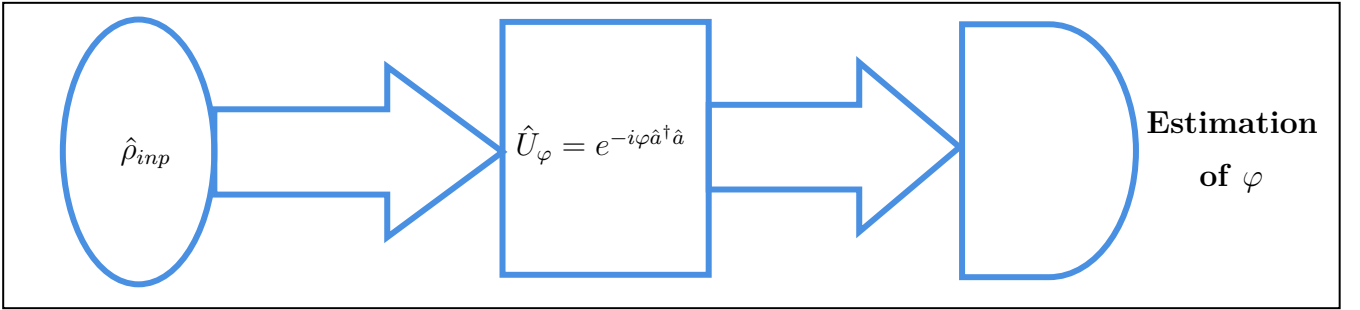

	In this example, we take the Gaussian state as a probe to consider the estimation of the single optical phase, which acts as a unitary phase shift operation as depicted in Fig. (\ref{Fig. 5}). This scheme aims to infer the actual value of the phase shift parameter by processing the different kinds of Gaussian probe states.
	\underline{\textbf{Estimation of $\varphi$ using the coherent states as a probe:}}
	An experimenter can generate a coherent state $\hat D\ket{0}$, in labs,  by applying a displacement operator to the vacuum state. Let $\ket{\psi}_{inp}=\hat D\ket{0}$ evolve under the Gaussian channel, which acts as a phase rotation operator, $\hat U_{\varphi}\hat\rho_{inp}\hat U_{\varphi}^{\dagger}$. It is characterized by an unknown phase parameter $\varphi$ that will be estimated. In the phase space,  this operator corresponds to a symplectic matrix 
	\begin{equation}
		\mathcal{R}_{\varphi}=\left(\begin{array}{cc}
			\cos \varphi & \sin \varphi \\
			-\sin \varphi & \cos \varphi
		\end{array}\right).
	\end{equation}
	To estimate the actual value of the $\varphi$ parameter that would need to be performed to calibrate the phase shifter, the {SLD-QFI}\footnote{Here the statistical model is a single estimation problem, so we only need to evaluate the {SLD-QFI} and not the {RLD-QFI}.} associated with that parameter must be determined. Thus, to calculate the {SLD-QFI}, we will use directly the expression of {SLD-QFI} given in (\ref{Eq. 3.43}). On the other hand, the initial coherent state is described by the first and second moments, which are
	\begin{equation}
		\boldsymbol{d}_{inp} = \left( {\begin{array}{*{20}{c}}
				q\\
				p
		\end{array}} \right), \hspace{1cm}\boldsymbol{{\mathrm V}}_{inp} = \mathbb{1}_{2\times2}.
	\end{equation}
	After the rotation phase, the output state is described by
	\begin{equation}
		\boldsymbol{d}_{out}=\mathcal{R}_{\varphi}\boldsymbol{d}_{inp} = \left( {\begin{array}{*{20}{c}}
				q\cos\varphi+p\sin\varphi\\
				-q\sin\varphi+p\cos\varphi
		\end{array}}\right), \hspace{1cm}\boldsymbol{{\mathrm V}}_{out} =\mathcal{R}_{\varphi}\mathcal{R}_{\varphi}^{\dagger} =\mathbb{1}_{2\times2}.
	\end{equation}
	Since the covariance matrix is constant, $\boldsymbol{{\mathrm V}}=cst$, then the first term in Eq. (\ref{Eq. 3.43}), depending on the derivative of the covariance, does not contribute to the {SLD-QFI}, and we can rewrite 
	\begin{equation}\label{Eq. 3.59}
		\mathcal{F}_Q^{(S)}\left(\varphi\right)=2 \partial_{\varphi} \mathbf{d}_{out}^{\top} \boldsymbol{\mathrm V}_{out}^{-1} \partial_{\varphi} \mathbf{d}_{out},
	\end{equation}
	which leads to 
	\begin{equation}\label{Eq. 3.60}
		\mathcal{F}_Q^{(S)}\left(\varphi\right)=2 \left(q^2+p^2\right)=2{\left| \alpha  \right|^2}=\mathtt{N},
	\end{equation}
	where $\mathtt{N}=\left\langle {{a^\dag }a} \right\rangle$ is the average number of excitation in the probe coherent state. Therefore, the {QCRB} of the optical phase estimation is
	\begin{equation}\label{Eq. 3.61}
		\mathtt{Var}_{\varphi}\left[\varphi^{est}\right]\ge\frac{1}{\mathtt{N}}.
	\end{equation}
	Clearly, the bound of precision in the estimation of the optical phase is independent of the value of the parameter $\varphi$. It is dependent on the average number of photons in the probe's coherent state. Hence, a stronger laser with a larger amplitude allows for better discrimination of $\varphi$.
	
	\underline{\textbf{Estimation of $\varphi$ using the squeezing states as a probe:}}
	Like the coherent states, the squeezing state is also produced in the laboratory by applying the squeezing operator to the vacuum state, i.e. $\ket{\psi}_{inp}=\hat{S}(r)\ket{0}$, where $\hat S(r)$ is the single-mode squeezing operator, which is defined as
	\begin{equation}
		\hat S\left(r\right)=\exp\left(\frac{r}{2}\left(\hat{a}^2-\hat{a}^{{\dagger}^2}\right)\right),
	\end{equation}
	with $r\in \mathbb{R}$ is called the squeezing parameter. In the Heisenberg picture, the annihilation operator is transformed by the linear unitary Bogoliubov transformation $\hat{a} \rightarrow(\cosh r) \hat{a}-(\sinh r) \hat{a}^{\dagger}$, and the quadrature operators $\boldsymbol{\hat{R}}$ by the symplectic map $\boldsymbol{\hat{R}}\rightarrow \mathbf{S} \boldsymbol{\hat{R}}$, where $\mathbf{S}$ is the symplectic matrix corresponding to $\hat S(r)$
	\begin{equation}
		\mathbf{S}\left(r\right)=\left(\begin{array}{cc}
			e^{-r}  & 0  \\
			0  & e^{r} 
		\end{array}\right).
	\end{equation}
	Hence, the squeezing state is characterized by the first and second moments, which are
	\begin{equation}
		\boldsymbol{d}_{inp} = 0 \hspace{2cm}\boldsymbol{{\mathrm V}}_{inp} = \mathbf{S}\left(r\right)\mathbf{S}\left(r\right)^{\dagger}=\mathbf{S}\left(2r\right).
	\end{equation}
	Now that we have prepared the squeezing state, we can proceed to its evolution under the rotation phase operator, which acts on it as 
	\begin{equation}
		\boldsymbol{d}_{out}=0 \hspace{1cm}\boldsymbol{{\mathrm V}}_{out} =\mathcal{R}_{\varphi}\mathbf{S}\left(2r\right)\mathcal{R}_{\varphi}^{\dagger}.
	\end{equation}
	Note that the first moment is zero, which means that the second term in Eq. (\ref{Eq. 3.43}) does not appear in the {SLD-QFI} in this case. Then we can write
	\begin{equation}
		\mathcal{F}_Q^{(S)}\left(\varphi\right)=\frac{1}{2} \operatorname{vec}\left[\partial_{\varphi} \boldsymbol{\mathrm V}_{out} \right]^{\dagger}\left(\boldsymbol{\mathrm V}_{out} \otimes \boldsymbol{\mathrm V}_{out}+\boldsymbol{\Omega} \otimes \boldsymbol{\Omega}\right)^{-1} \operatorname{vec}\left[\partial_{\varphi} \boldsymbol{\mathrm V}_{out} \right].
	\end{equation}
	Insert the derivation of $\boldsymbol{\mathrm V}_{out}$ with respect to $\varphi$ in the last equation, leading to
	\begin{equation}
		\mathcal{F}_Q^{(S)}\left(\varphi\right)=8\sinh^2 r\left(\sinh^2r+1\right)=8\mathtt{N}\left(\mathtt{N}+1\right).
	\end{equation}
	where, $\mathtt{N}=\left\langle {{a^\dag }a} \right\rangle=\sinh^2r$, in this case, denotes the mean number of photons in the probe squeezing state. Thus, the {SLD-QCRB} of the optical phase is given by
	\begin{equation}\label{Eq. .68}
		\mathtt{Var}_{\varphi}\left[\varphi^{est}\right]\ge\frac{1}{\mathtt{N}\left(\mathtt{N}+1\right)}\propto \frac{1}{\mathtt{N}^2}. 
	\end{equation}
	If we compare this bound with the one obtained above in Eq. (\ref{Eq. 3.61}), we find that the bound obtained by the squeezing state is better and smaller than the one obtained by employing the coherent state as a probe. This result means that the squeezing probe is most suitable for estimating, with maximal precision, the optical phase parameter.
	
	The bound given in Eq. (\ref{Eq. 3.61}) is proportional to $1/\mathtt{N}$, so using a more powerful laser with a large photons number decreases the uncertainty in our estimate of $\varphi$. This scaling $1/\mathtt{N}$ is known as the \textbf{ Standard Quantum Limit} ({SQL}). The origins of this scaling lie in the central limit theorem from a probability distribution. And are possible classically without the invocation of quantum mechanics, i.e. the best classical possible scaling is the standard quantum limit, and never any classical strategy can overcome this scaling. Therefore, quantum mechanics is the ultimate and most fundamental barrier to the precision of an estimation scheme. This unavoidable limit ({SQL}) is set by quantum vacuum fluctuations and can only be overcome by invoking quantum mechanical techniques alone. Quantum probes such as a squeezing state can go beyond the {SQL} and attain a scaling proportional to $1/\mathtt{N}^2$. This scaling is called the \textbf{Heisenberg Limit} ({HL}) and can only be reached by employing quantum mechanics resources or non-classical properties of quantum states. This improved scaling leads, in fact, to a more precise estimation and is at the root of the requirement of quantum metrology. 
	\section{conclusion}
	
	As a matter of fact, the {QCRB} is the fundamental tool employed to estimate unknown parameters in quantum statistical models. These bounds are determined from the corresponding {QFIM}s. In this chapter, for the multi-mode quantum Gaussian states, we have derived the expressions of the {RLD} and {SLD-QFIM}s by computing the {RLD} and {SLD}-operators corresponding. We have also expressed the attainment condition of the lower bound of {SLD-QCRB}. These results are discovered as functions of the displacement vector and covariance matrix of Gaussian states. We have illustrated the derived formalism by treating the optical phase estimation problem as an example. This protocol is applicable experimentally in diverse tasks in labs. Notice that all these results are discovered as functions of the displacement vector and the covariance matrix of Gaussian states. This remarkable advantage is an incentive to provide more practical applications in quantum metrology with Gaussian formalism. 
	
	As the example of optical phase estimation shows, the precision limit can be set in Gaussian unitary channels by the type of probe initially used. Some probes reach the limit beyond the {SQL}, such as the squeezing state that attains the {HL}. Nevertheless, this focuses on the single-parameter case encoded in the Gaussian unitary channels. Therefore, we can ask what happens in the case of single or multiparameter Gaussian noisy channels? There are resources most applicable to overcome the inevitable existence of environmental fluctuations or at least to limit their inevitable effect? In the next chapter, we address the motivation behind the usage of Gaussian quantum resources and their advantages in overcoming the {SQL} under realistic noise.
	
	\chapter{Ultimate precision limit under noisy Gaussian environment}\label{Ch. 4}
	
	\section{Introduction}
	Beyond their fundamental interest in quantum electrodynamics and its generalizations of the standard model, in the quantum optics and condensed matter theory, continuous variable systems are beginning to play a remarkable role in quantum information and communication theory \cite{weedbrook2012gaussian, braunstein2005quantum, adesso2014continuous}. In the context of quantum optics, this role has been demonstrated in surprising applications such as deterministic teleportation schemes and quantum key distribution protocols \cite{paterson2005experimental, grosshans2003quantum}. In all these practical instances, the information encoded in a given quantum state, so valuable for performing a specific task, is constantly threatened by the inevitable interaction with the environment. These interactions connect the system of interest with the environment and lead to the dispersion and loss of information. Moreover, the environmental fluctuations are out of experimental control, which means that information is irreversibly lost. The overall process, which corresponds to a non-unitary evolution of the system, is commonly called decoherence \cite{caldeira1983path, zurek2002decoherence}. It is thus of crucial importance to address how we can estimate unknown parameters under the effect of decoherence.
	
	As discussed previously, the major problem of quantum estimation theory is to find an ultimate measurement scheme that allows overcoming the {SQL} of precision. In fact,  it is not probable to overcome this precision limit using every quasi-classical estimation measurement. Although, in some specific quantum protocols without environmental noise, the ultimate sensitivity of a multiparameter quantum estimation can beat the {SQL} \cite{giovannetti2004quantum, nagata2007beating, okamoto2008beating, becerra2013experimental}. However, the presence of noise imposes constraints on the enhancement of precision. This fact is due to the inevitable existence of environmental fluctuations. Here, in this chapter, we address the motivation behind the usage of Gaussian quantum resources and their advantages in reaching the standard quantum limits under realistic noise. In this context, we aim to explore the ultimate limits of precision for the simultaneous estimation of a pair of parameters that characterize the displacement channel acting on Gaussian probes and subjected to open dynamics. We will investigate the role of quantum entanglement and the purity of the preparing Gaussian states to improve the simultaneous precision of measurements. We shall limit our analysis to an arbitrary two-mode Gaussian probe state that undergoes under a displacement operator. That last is acts only on one of the two modes and is subject to a Gaussian noise environment. We start in Sec. (\ref{Sec. 4.2}) by offering a brief review of the Gaussian non-unitary channels. At first, we focus our attention on the evolution of single-mode radiation. Then we extend our study to the case of evolving an n-mode state. Next, we present in Sec. (\ref{Sec. 4.3}) some relevant measurements that can be performed on continuous variable (CV) systems. These include both homodyne and heterodyne measurement schemes. We formulate in Sec. (\ref{Sec. 4.4}), the general framework to study the joint estimation of parameters encoded in the displacement operator acting on the probe Gaussian states and evolving in a Gaussian noise environment. And then, we investigate the estimation performance over time $t$ for the various state: two modes squeezed vacuum state, two-modes squeezed thermal state, two-modes coherent state, and two-modes coherent thermal state. In Sec. (\ref{Sec. 4.5}), we discuss the role of entanglement as one of the most critical resources of quantum information theory that lead to beating the SQL even in the existence of environmental fluctuation. Finally, we end this chapter with a conclusion. We note that the main results of this chapter are published in Ref. \cite{bakmou2022ultimate}.
	\section{Gaussian states in noisy channels} \label{Sec. 4.2}
	In the last chapter specifically, in section (\ref{Sec. 3.4}), we have dealt with the expansive class of physical operations or transformations that preserve the Gaussian characteristic of the initial states. Here in this section, we describe what could loosely be considered the continuous-time version of such a transformation. More precisely, we address the dissipative dynamics of a Gaussian state in the Gaussian environment or noisy channels that have a completely positive map and decreasing trace. Mathematically, the linear map $\mathcal{E}: \hat{\rho} \rightarrow \mathcal{E}(\hat{\rho})$ of noisy channels is characterized by  $\operatorname{Tr}[\mathcal{E}(\hat{\rho})]\le 1$. This dynamic can be reduced to a suitable transformation of the first and second moments that describe the initial Gaussian state. At first, we focus our attention on the evolution of a single-mode of radiation Gaussian. Then we extend our description to the evolution of n-mode Gaussian states, which will be treated as the evolution of global channels made of n non-interacting different channels.
	\subsection{Single-mode Gaussian states in noisy channels}
	Before starting this section, we would like to confirm that our principal purpose is to study quantum metrology in noisy Gaussian channels. Thus, we do not need to go deeply into the details of the open diffusion dynamics of quantum states under interaction with the environment ( an in-depth analysis may be found in Refs. \cite{ferraro2005gaussian, serafini2005quantifying}).
	
	In the interaction picture with the Markovian approximation, the dynamic of a single-mode of radiation, described by a quantum state $\hat \rho \left(t\right)$ through a noisy environment, is governed by the following master equation;
	\begin{equation}\label{EQ. 4.1}
		\partial_{t}{\hat \rho}\left(t\right)=\frac{\gamma}{2}\left\{(\mathtt{N}_e+1) \mathcal{\hat L}[\hat{a}]+\mathtt{N}_e \mathcal{\hat L}\left[\hat{a}^{\dagger}\right]-\mathtt{M}_e^{*} \mathcal{\hat D}[\hat{a}]-\mathtt{M}_e \mathcal{\hat D}\left[\hat{a}^{\dagger}\right]\right\} \hat \rho\left(t\right),
	\end{equation}
	where $\gamma$ is the overall damping rate, while $\mathtt{N}_e \in \mathbb{R}$ and $\mathtt{M}_e \in \mathbb{C}$ are respectively the effective photons number and the	squeezing parameter of the reservoir (bath). $\mathcal{L}[\hat{O}]  \hat \rho\left(t\right)=2 \hat{O}  \hat \rho\left(t\right) \hat{O}^{\dagger}-\hat{O}^{\dagger} \hat{O}  \hat \rho\left(t\right)- \hat \rho\left(t\right) \hat{O}^{\dagger} \hat{O}$ and  $\mathcal{D}[\hat{O}]  \hat \rho\left(t\right)=2 \hat{O}  \hat \rho\left(t\right) \hat{O}-\hat{O} \hat{O}  \hat \rho\left(t\right)- \hat \rho\left(t\right) \hat{O} \hat{O}$ are \textit{Lindblad super-operators}. The terms proportional to $\mathcal{L}[a]$ and to $\mathcal{L}\left[a^{\dagger}\right]$ describe losses and linear, phase insensitive, amplification processes, respectively. While, the terms proportional to $\mathcal{D}[a]$ and $\mathcal{D}\left[a^{\dagger}\right]$ describe phase dependent fluctuations. The positivity of the density matrix, $\hat \rho\left(t\right)$, imposes the constraint $|\mathtt{M}_e|^{2} \leq \mathtt{N}_e(\mathtt{N}_e+1)$. At thermal equilibrium, i.e. for
	$\mathtt{M}_e= 0$, $\mathtt{N}_e$ coincides with the average number of thermal photons in the bath. In order to derive the first and second moments of the Gaussian state $\hat \rho\left(t\right)$, we transform the master equation given in (\ref{EQ. 4.1}) into the Fokker-Planck equation for the Wigner function ${W_{\hat \rho\left(t\right) }}\left(\boldsymbol{R}  \right)$, which leads to  
	\begin{equation}\label{EQ. 4.2}
		\partial_{t} {W_{\hat \rho\left(t\right) }}\left(\boldsymbol{R}  \right)=\frac{\gamma}{2}\left(\partial_{\boldsymbol{R}}^{\top} R+\partial_{\boldsymbol{R}}^{\top} \boldsymbol{\mathrm{V}}_{\infty} \partial_{\boldsymbol{R}}\right) {W_{\hat \rho\left(t\right) }}\left(\boldsymbol{R}  \right), 
	\end{equation}
	where $\boldsymbol{R} \equiv(q, p)^{\top}, \partial_{\boldsymbol{R}} \equiv\left(\partial_{q}, \partial_{p}\right)^{\top}$, while $\boldsymbol{\mathrm{V}}_{\infty}$ is the diffusion covariance matrix that given by 
	\begin{equation}\label{EQ. 4.3}
		\boldsymbol{\mathrm{V}}_{\infty}=\left(\begin{array}{cc}
			\left(2\mathtt{N}_e+1\right)+\mathfrak{Re}[\mathtt{M}_e] & \mathfrak{Im}[\mathtt{M}_e] \\
			\mathfrak{Im}[\mathtt{M}_e] & \left(2\mathtt{N}_e+1\right)-\mathfrak{Re}[\mathtt{M}_e]
		\end{array}\right), 
	\end{equation}
	This diffusion matrix is determined only by the bath parameters $\mathtt{N}_e$ and $\mathtt{M}_e$.  
	
	Despite its interaction with the environment, the Gaussian initial state $\hat \rho\left(0\right)$ of the first and second moments $\boldsymbol{d}\left(0\right)$ and $\boldsymbol{{\mathrm V}}\left(0\right)$ is still Gaussian with the new first and second moments, which are 
	\begin{equation}\label{EQ. 4.4}
		{\boldsymbol{d}}\left(t\right)=e^{\frac{-\gamma }{2}t}{\boldsymbol{d}}\left(0\right) \hspace{1cm} \text { and } \hspace{1cm}	\boldsymbol{\mathrm{V}}\left(t\right)=e^{-\gamma t} \boldsymbol{\mathrm{V}}\left(0\right)+\left(1-e^{-\gamma t}\right) \boldsymbol{\mathrm{V}}_{\infty}, 
	\end{equation}
	We substitute that $\eta\left(t\right) = e^{\frac{-\gamma }{2}t}$, we can rewrite the last equation as
	\begin{equation}\label{EQ. 4.5}
		{\boldsymbol{d}}\left(t\right)=\eta\left(t\right) {\boldsymbol{d}}\left(0\right) \hspace{1cm} \text { and } \hspace{1cm}	\boldsymbol{\mathrm{V}}\left(t\right)=\eta\left(t\right) \boldsymbol{\mathrm{V}}\left(0\right)\eta\left(t\right)+ \boldsymbol{\mathrm{V}}_{\infty}\left(t\right), 
	\end{equation}
	where $\boldsymbol{\mathrm{V}}_{\infty}\left(t\right)=\left(1-e^{-\gamma t}\right) \boldsymbol{\mathrm{V}}_{\infty}$. In particular, focusing on second moments, Eq. (\ref{EQ. 4.5}) shows that the evolution imposed by the Master equation is a Gaussian map with $\boldsymbol{\mathrm{V}}_{\infty}\left(t\right)$ as the asymptotic covariance matrix. The covariance matrix  $\boldsymbol{\mathrm{V}}\left(t\right)$ is a real and symmetric matrix that must satisfy the uncertainty principle given in (\ref{Eq. 3.25}), as well as $\boldsymbol{\mathrm{V}}\left(t\right)$ and $\boldsymbol{\mathrm{V}}_{\infty}\left(t\right)$ are satisfied. 
	\subsection{Extension to $N$-mode Gaussian states in noisy channels}
	In this subsection, we extend the above description to the case when evolving an arbitrary $N$-mode Gaussian state under noisy channels. In this case, the dynamic of $\hat \rho\left(t\right)$ is governed by the following Master equation
	\begin{equation}\label{EQ. 4.6}
		\partial_{t}\hat \rho\left(t\right)=\sum_{i=1}^{N} \frac{\gamma_{i}}{2}\left\{\left(\mathtt{N}_{ei}+1\right) \mathcal{\hat L}\left[\hat a_{i}\right]+\mathtt{N}_{ei} \mathcal{\hat L}\left[\hat a_{i}^{\dagger}\right]-\mathtt{M}_{ei}^{*} \mathcal{\hat D}\left[\hat a_{i}\right]-\mathtt{M}_{ei} \mathcal{\hat D}\left[\hat a_{i}^{\dagger}\right]\right\} \hat \rho\left(t\right),
	\end{equation}
	As for the single-mode case, the Master equation (\ref{EQ. 4.6}) can be equivalently recast as a Fokker Planck equation for the Wigner function as follows;
	\begin{equation}\label{EQ. 4.7}
		\partial_{t}{W_{\hat \rho\left(t\right) }}\left(\boldsymbol{R}  \right)=\frac{1}{2}\left(\partial_{\boldsymbol{R}}^{\top} \mathbb{\Gamma} \boldsymbol{R}+\partial_{\boldsymbol{R}}^{\top} \mathbb{\Gamma} \boldsymbol{\mathrm{V}}_{\infty} \partial_{\boldsymbol{R}}\right) {W_{\hat \rho\left(t\right) }}\left(\boldsymbol{R}  \right),
	\end{equation}
	where $\mathbb{\Gamma}=\bigoplus_{i=1}^{N} \gamma_{i} \mathbb{1}$. Eq. (\ref{EQ. 4.7})  is formally identical to Eq. (\ref{EQ. 4.2}), but now\\ $\boldsymbol{R} =\left(q_{1}, p_{1}, \ldots, q_{N}, p_{N}\right)^{\top}$, $\partial_{\boldsymbol{R}}=\left(\partial_{q_{1}}, \partial_{p_{1}}, \ldots, \partial_{q_{N}}, \partial_{p_{N}}\right)^{\top}$ and the diffusion matrix is given by $\boldsymbol{\mathrm{V}}_{\infty}=\bigoplus_{i=1}^{N} \boldsymbol{\mathrm{V}}_{i, \infty}$ with $\boldsymbol{\mathrm{V}}_{i, \infty}$ is the asymptotic covariance matrix of the $i^{th}$ channel that given in Eq. (\ref{EQ. 4.3}). It is easy to see that the Eq. (\ref{EQ. 4.7}) is a generalization of Eq. (\ref{EQ. 4.2}). Therefore, the general solution of (\ref{EQ. 4.7}) is also the generalization of the solution of Eq. (\ref{EQ. 4.2}). Thus, also for the $n$-mode case, we have that Gaussian states remain Gaussian at any time. The first and seconds moments of these states  are given by 
	\begin{equation}\label{EQ. 4.8}
		{\boldsymbol{d}}\left(t\right)=\mathcal{G}\left(t\right) {\boldsymbol{d}}\left(0\right), \hspace{1cm} \text { and } \hspace{1cm}	\boldsymbol{\mathrm{V}}\left(t\right)=\mathcal{G}\left(t\right) \boldsymbol{\mathrm{V}}\left(0\right)\mathcal{G}\left(t\right)+ \boldsymbol{\mathrm{V}}_{\infty}\left(t\right),
	\end{equation}
	where $\mathcal{G}\left(t\right)=\bigoplus_{i=1}^{N}\eta_i\left(t\right)=\bigoplus_{i=1}^{N} e^{\frac{-\gamma_i }{2}t}$. In the last equation (\ref{EQ. 4.8}), the covariance matrix describes the evolution of an initially Gaussian state of $\boldsymbol{\mathrm{V}}\left(0\right)$ into a Gaussian environment of $\boldsymbol{\mathrm{V}}_{\infty}$.
	
	Since the full description of Gaussian states depends only on the first and second moments, then starting from ${\boldsymbol{d}}\left(t\right)$ and $\boldsymbol{\mathrm{V}}\left(t\right)$, one can effortlessly evaluate the evolution of all the quantities addressed in the previous chapter, specifically the {RLD} and {SLD-QFIM}s and the corresponding {QCRB}s. Therefore, we can perform the measurement and then estimate unknown parameters even in the presence of environmental fluctuations. But, what are the limits of ultimate sensitivity that may be achieved in these types of channels? And what kinds of measurements can be performed? In other words, are there specific types of measurements for these classes of quantum states?  Will be addressed all these questions in what follows. We start by reviewing the kind of Gaussian measurements.
	\section{Gaussian measurements} \label{Sec. 4.3}
	In the previous section, we reviewed the kinematics and evolution of the Gaussian state throughout noisy channels. In order to make this preface of Gaussian noisy channels as complete as possible, we will now look at general Gaussian measurements, such as homodyne and heterodyne detection, that are performed on Gaussian states. The role of these types of measurements is incredibly robust in quantum information processing because they are the basis of quantum teleportation. In addition, they have a prominent place in engineering and signal processing. Let us start our discussion of Gaussian measurements with homodyne detection.
	\subsection{Homodyne detection}
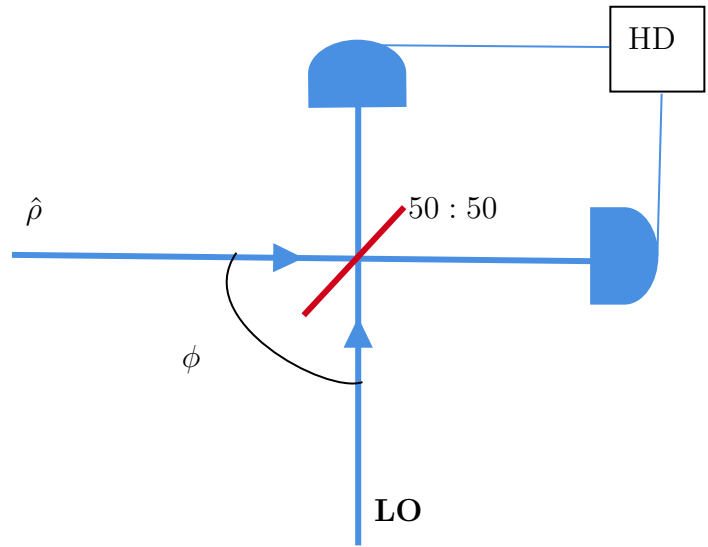
\begin{figure}[H]   
		\tikzset{every picture/.style={line width=0.75pt}} 
		\begin{tikzpicture}[x=0.8pt,y=0.8pt,yscale=-1,xscale=1]
			
			\draw [color={rgb, 255:red, 74; green, 144; blue, 226 }  ,draw opacity=1 ][line width=2.25]    (38,119) -- (311,122) ;
			\draw [shift={(174.5,120.5)}, rotate = 180.63] [fill={rgb, 255:red, 74; green, 144; blue, 226 }  ,fill opacity=1 ][line width=0.08]  [draw opacity=0] (14.29,-6.86) -- (0,0) -- (14.29,6.86) -- cycle    ;
			\draw [color={rgb, 255:red, 74; green, 144; blue, 226 }  ,draw opacity=1 ][line width=2.25]    (200,41) -- (200,256) ;
			\draw [shift={(200,148.5)}, rotate = 90] [fill={rgb, 255:red, 74; green, 144; blue, 226 }  ,fill opacity=1 ][line width=0.08]  [draw opacity=0] (14.29,-6.86) -- (0,0) -- (14.29,6.86) -- cycle    ;
			\draw [color={rgb, 255:red, 208; green, 2; blue, 27 }  ,draw opacity=1 ][line width=2.25]    (174.5,147.5) -- (221.5,96.5) ;
			\draw  [color={rgb, 255:red, 74; green, 144; blue, 226 }  ,draw opacity=1 ][fill={rgb, 255:red, 74; green, 144; blue, 226 }  ,fill opacity=1 ] (309,97) -- (324.5,97) .. controls (333.06,97) and (340,107.07) .. (340,119.5) .. controls (340,131.93) and (333.06,142) .. (324.5,142) -- (309,142) -- cycle ;
			\draw  [color={rgb, 255:red, 74; green, 144; blue, 226 }  ,draw opacity=1 ][fill={rgb, 255:red, 74; green, 144; blue, 226 }  ,fill opacity=1 ] (177.15,49.22) -- (177,33.72) .. controls (176.92,25.16) and (186.92,18.12) .. (199.35,18) .. controls (211.77,17.88) and (221.92,24.72) .. (222,33.28) -- (222.15,48.78) -- cycle ;
			\draw [color={rgb, 255:red, 74; green, 144; blue, 226 }  ,draw opacity=1 ]   (191.35,20) -- (317.35,21) ;
			\draw   (318,2) -- (362,2) -- (362,42) -- (318,42) -- cycle ;
			\draw [color={rgb, 255:red, 74; green, 144; blue, 226 }  ,draw opacity=1 ]   (342,43) -- (340,119.5) ;
			\draw [color={rgb, 255:red, 0; green, 0; blue, 0 }  ,draw opacity=1 ]   (202,179) .. controls (185,185) and (121,150) .. (143,118) ;
			
			\draw (116,160) node [anchor=north west][inner sep=0.75pt]    {$\phi$};
			\draw (325,10) node [anchor=north west][inner sep=0.75pt]   [align=left] {HD};
			\draw (43,90) node [anchor=north west][inner sep=0.75pt]    {$\hat \rho$};
			\draw (206,233) node [anchor=north west][inner sep=0.75pt]   [align=left] {\textbf{LO}};
			\draw (222,91) node [anchor=north west][inner sep=0.75pt]    {$50:50$};
		\end{tikzpicture}
		\centering
		\captionsetup{justification=centerlast, singlelinecheck=false}\captionof{figure}{Schematic diagram illustre the principe of the homodyne detector.} \label{Fig. 6}   
	\end{figure}

	Homodyne detection schemes are developed to provide single-mode quadrature measurement by mixing the signal under investigation on a $50:50$ beam splitter with a coherent state (commonly called local oscillator (\textbf{LO})) of the same frequency. After the beam splitter, the two outputs are collected on detectors and then perform the measurement. Fig .(\ref{Fig. 6}) is devoted to illustrating this process.
	
	Let us now assume to perform a Gaussian measurement on one of the output modes, which is a measurement described by a {POVM} with a Gaussian characteristic function. Without loss of generality, we can consider that the measurement involves mode $(1)$, and we can then write the corresponding characteristic function such that:
	\begin{equation}
		\chi_{{m}}\left(\boldsymbol{\xi}_{1}\right)=\pi^{-1} \exp \left\{-\frac{1}{2} \boldsymbol{\xi}_{1}^{T} \mathrm{V}_{{m}} \boldsymbol{\xi}_{1}-i \boldsymbol{\xi}_{1}^{T} \boldsymbol{d}_{m}\right\},
	\end{equation}
	where $\boldsymbol{d}_{m}$ and $\mathrm{V}_{m}$ are, respectively, the first and second moments of mode (1) after the measurement, or, more precisely, the outcome of the measurement. For simplicity, we write the vector $\boldsymbol{\xi}$ and the covariance matrix $\mathrm{V}$ of the global system in the following block form:
	\begin{equation}
		\boldsymbol{\xi}  = {\left( {{\xi _1},{\xi _2}, \ldots ,{\xi _N}} \right)^{\top}} = {\left( {{\xi _1},\widetilde \xi } \right)^{\top}},\quad {\text{ and }}\quad \boldsymbol{{\mathrm{V}}} = \left( {\begin{array}{*{20}{c}}
				{\boldsymbol{A}}&\boldsymbol{C}\\
				{{\boldsymbol{C}^{\top}}}&\boldsymbol{B}
		\end{array}} \right),
	\end{equation}
	where $\boldsymbol{A} \in \mathbb{R}^{2} \times \mathbb{R}^{2}$ and $\boldsymbol{B} \in \mathbb{R}^{2(N-1)} \times \mathbb{R}^{2(N-1)}$ are symmetric, and $\boldsymbol{C} \in \mathbb{R}^{2} \times \mathbb{R}^{2(N-1)}$, making the mode undergoing the measurement obvious. After the measurement, the conditional characteristic function of the system with the outcome $\boldsymbol{R}_m$ is 
	\begin{equation}
		{\chi _{\hat \rho }}( \boldsymbol{\widetilde \xi} ) = \frac{1}{{p(\boldsymbol{R})}}\int_{{\mathbb{X}^2}} {\frac{{{d^2}{\boldsymbol{\xi} _1}}}{{2\pi }}} \chi \left( {{\boldsymbol{\xi} _1},\boldsymbol{\widetilde \xi} } \right){\chi _{\rm{m}}}\left( { - {\boldsymbol{\xi} _1}} \right),
	\end{equation}
	where $p\left(\boldsymbol{R}_m\right)$ is the normalized probability density function associated with the "general"-dyne\footnote{ "general"-dyne is a terminology used to refer to both homodyne and heterodyne detection.} outcome, and it's given by
	\begin{equation}
			p(\boldsymbol{R}_m) =\frac{{\left\langle {{\psi _G}} \right|\hat \rho \left| {{\psi _G}} \right\rangle }}{{{{\left( {2\pi } \right)}^{2N}}}}=\int_{\mathbb{R}^{2 n}} \frac{d^{2} \boldsymbol{\xi}_{1} d^{2(N-1)} \tilde{\boldsymbol{\xi}}}{(2 \pi)^{N}} \chi\left(\boldsymbol{\xi}_{1}, \tilde{\boldsymbol{\xi}}\right) \chi_{m}\left(-\boldsymbol{\xi}_{1}\right)(2 \pi)^{(N-1)} \delta(-\tilde{\boldsymbol{\xi}}),
	\end{equation}
which can be evaluate such as \footnote{It is easily deduced from the use of the trace rule of two Gaussian states: $$\operatorname{Tr}\left(\hat \rho_{1} \varrho_{2}\right)=\frac{2^{N}}{\sqrt{\operatorname{Det}\left(\boldsymbol{{\mathrm{V}}}_{1}+\boldsymbol{{\mathrm{V}}}_{2}\right)}} \mathrm{e}^{-\left(\mathbf{d}_{1}-\mathbf{d}_{2}\right)^{\top}\left(\boldsymbol{{\mathrm{V}}}_{1}+\boldsymbol{{\mathrm{V}}}_{2}\right)^{-1}\left(\mathbf{d}_{1}-\mathbf{d}_{2}\right)}$$} 
	\begin{equation}\label{EQ. 4.12}
	p(\boldsymbol{R}_m)=\frac{{\exp \left\{ { - \frac{1}{2}{{\left( {{\boldsymbol{R}_m} - \boldsymbol{d}} \right)}^{\top}}{{\left( {{\boldsymbol{A}} + {{\boldsymbol{{\mathrm{V}}}}_{\rm{m}}}} \right)}^{ - 1}}\left( {{\boldsymbol{R}_m} - \boldsymbol{d}} \right)} \right\}}}{{\pi \sqrt {\det \left[ {{\bf{A}} + {{\boldsymbol{{\mathrm{V}} }}_{\rm{m}}}} \right]} }},
	\end{equation}
	where $\delta(-\tilde{\xi})=\prod_{k=2}^{N} \delta^{(2)}\left(-\Lambda_{k}\right)$ is the product of Kronecker deltas in $\mathbb{R}^{2}$. Here, we noted that:
	\begin{equation}
		\chi\left(\boldsymbol{\xi}_{1}, \tilde{\boldsymbol{\xi}}\right) \chi_{{m}}\left(-\boldsymbol{\xi}_{1}\right)=\pi^{-1} \exp \left\{-\frac{1}{2}\left(\boldsymbol{\xi}_{1}, \tilde{\boldsymbol{\xi}}\right)^{\top} \boldsymbol{{\mathrm{V}}}\left(\boldsymbol{\xi}_{1}, \tilde{\boldsymbol{\xi}}\right)+i \boldsymbol{\xi}_{1}^{\top} \boldsymbol{R}\right\},
	\end{equation}
	with
	\begin{equation}
		\boldsymbol{{\mathrm{V}}}=\left(\begin{array}{cc}
			\boldsymbol{A}+\boldsymbol{{\mathrm{V}}}_{m} & \boldsymbol{C} \\
			\boldsymbol{C}^{\top} & \boldsymbol{B}
		\end{array}\right).
	\end{equation}
	Analogously, if we carry out the measurement on the mode (N), the global state is still Gaussian, but with the covariance matrix  
	\begin{equation}
		\boldsymbol{{\mathrm{V}}}=\left(\begin{array}{cc}
			\boldsymbol{A} & \boldsymbol{C} \\
			\boldsymbol{C}^{\top} & \boldsymbol{B}+\boldsymbol{{\mathrm{V}}}_{m}
		\end{array}\right).
	\end{equation}
	where $\boldsymbol{A} \in \mathbb{R}^{2(N-1)} \times \mathbb{R}^{2(N-1)}$ and $\boldsymbol{B} \in \mathbb{R}^{2} \times \mathbb{R}^{2}$ are symmetric, and $\boldsymbol{C} \in \mathbb{R}^{2(N-1)} \times \mathbb{R}^{2}$, and the probability of the outcome $\boldsymbol{R}_m$ is given, in this case, by
	\begin{equation}
		p(\boldsymbol{R}_m)=\frac{{\exp \left\{ { - \frac{1}{2}{{\left( {{\boldsymbol{R}_m} - \boldsymbol{d}} \right)}^{\top}}{{\left( {{\boldsymbol{B}} + {{\boldsymbol{{\mathrm{V}}}}_{\rm{m}}}} \right)}^{ - 1}}\left( {{\boldsymbol{R}_m} - \boldsymbol{d}} \right)} \right\}}}{{\pi \sqrt {\det \left[ {{\boldsymbol{B}} + {{\boldsymbol{{\mathrm{V}} }}_{\rm{m}}}} \right]} }}.
	\end{equation}
	
	Now, we restrict our description to the case of single-mode projective Gaussian measurements. As the most general single-mode Gaussian state is a displaced squeezed vacuum state, then the corresponding covariance matrix $\boldsymbol{{\mathrm{V}}}_{m}$ can be written as  
	\begin{equation}
		\boldsymbol{{\mathrm{V}}}_{m}\left( {r,\phi } \right) =\mathcal{\hat R}(\phi )\hat S\left( r \right)\hat S{\left( r \right)^\dag }\mathcal{\hat R}{(\phi )^\dag },
	\end{equation}
	with the corresponding symplectic transformation  
	\begin{equation}\label{EQ. 4.18}
		\boldsymbol{{\mathrm{V}}}_{m}\left( {s,\phi } \right) = \left( {\begin{array}{*{20}{c}}
				{\cos \phi }&{\sin \phi }\\
				{ - \sin \phi }&{\cos \phi }
		\end{array}} \right)\left( {\begin{array}{*{20}{c}}
				s&0\\
				0&{1/s}
		\end{array}} \right){\left( {\begin{array}{*{20}{c}}
					{\cos \phi }&{\sin \phi }\\
					{ - \sin \phi }&{\cos \phi }
			\end{array}} \right)^ \top },
	\end{equation}
	where $s=e^{-2r}$ is the squeezing parameter of light. In the principle of homodyne detection in the single mode, the homodyne measurement of
	$\hat Q$ would correspond to the limit $\boldsymbol{{\mathrm{V}}}_{m}^{hom}\left( {s,\phi } \right)=\mathop {\lim }\limits_{s \to 0}\boldsymbol{{\mathrm{V}}}_{m}\left( {s,\phi } \right)$.  While, the measurement of $\hat P$ would correspond to $s\to \infty$, and the argument generalizes to an arbitrary phase space direction by rotating this covariance matrix through $\mathcal{R}_{\phi}$. In this limit, the matrix quantity $\left(\boldsymbol{{\mathrm{V}}}_{m}+\boldsymbol{A}\right)^{-1}$ that appears in the expression of the normalized probability density function of "general"-dyne tends to $\operatorname{daig}\left(A_{11}^{-1},0\right)$. Hence, only one of the two real-valued readings, $q$ in this example, since we have $\boldsymbol{R}_m=\left(q,p\right)^{\top}$ that ordinarily labels the general-dyne outcomes of single mode. Thus,  the homodyne measurements give a single real outcome per mode $\boldsymbol{R}_m=\left(q,0\right)^{\top}$ with $\boldsymbol{d}=\left(\left\langle {\hat Q} \right\rangle,0\right)$.  And then, the probability density function of Eq. (\ref{EQ. 4.12}) reduces, in the homodyne detection, to
	\begin{equation}\label{EQ. 4.19}
		p\left( q \right) = \frac{{{{\rm{exp}}\left\{{ - \frac{{{{\left( {q - \left\langle {\hat Q} \right\rangle } \right)}^2}}}{{{2A_{11}}}}}\right\}}}}{{\pi \sqrt {{A_{11}}} }}.
	\end{equation}
	Similarly, in the case of measurement $\hat P$, the homodyne probability density function of the outcome $p$ is  
	\begin{equation}\label{EQ. 4.20}
		p\left( p \right) = \frac{{{{\rm{exp}}\left\{{ - \frac{{{{\left( {p - \left\langle {\hat P} \right\rangle } \right)}^2}}}{{{2A_{11}}}}}\right\}}}}{{\pi \sqrt {{A_{11}}} }}.
	\end{equation}
	\subsection{Heterodyne detection}
	In fact, the optical heterodyne detection schema is carried out through a scheme analogous to homodyne detection.  But, in this case, the mode that will be measured is mixed with a laser field at a different frequency, whence the adjective “heterodyne”. In the previous subsection, we have described in detail the homodyne detection
	scheme that, in the limit of a strong local oscillator, allows one to approximate the measurement of any quadrature operator on a single mode $\hat Q$ or $\hat P$. Let us now consider a  heterodyne detection scheme where a single-mode input state $\hat \rho$, to
	be measured, is prepared in the vacuum state, i.e. $\hat \rho=\ket{0}\bra{0}$. in this case, the outcome measurement given in Eq. (\ref{EQ. 4.18}) has been reduces into
	\begin{equation}
		\boldsymbol{{\mathrm{V}}}_{m}^{(het)}=\mathop {\lim }\limits_{s \to 1}\boldsymbol{{\mathrm{V}}}_{m}\left( {s,\phi } \right)=
		\left( {\begin{array}{*{20}{c}}
				{\cos \phi }&{\sin \phi }\\
				{ - \sin \phi }&{\cos \phi }
		\end{array}} \right)\left( {\begin{array}{*{20}{c}}
				1&0\\
				0&1
		\end{array}} \right){\left( {\begin{array}{*{20}{c}}
					{\cos \phi }&{\sin \phi }\\
					{ - \sin \phi }&{\cos \phi }
			\end{array}} \right)^ \top }.
	\end{equation}
	Thus, the two real-valued readings, $q$, and $p$. And then, the heterodyne and general-dyne yield two real values per mode. While the homodyne measurements give a single real outcome per mode $q$ or $p$. Therefore, the normalized probability density function with the heterodyne measurements is 
	\begin{equation}
		p({{\bf{R}}_m})=\frac{{\exp \left\{ { - \frac{1}{2}{{\left( {{{\bf{R}}_m} - {\bf{d}}} \right)}^ \top }{{\left( {{\bf{A}} + {{\rm{V}}_{\rm{m}}}\left( {1,\phi } \right)} \right)}^{ - 1}}\left( {{{\bf{R}}_m} - {\bf{d}}} \right)} \right\}}}{{\pi \sqrt {\det \left[ {{\bf{A}} + {{\rm{V}}_{\rm{m}}}\left( {1,\phi } \right)} \right]} }}.
	\end{equation}
	
	In the case of Gaussian noise channels, the measurement matrix of Eq. (\ref{EQ. 4.18}) is computed via the action of a dual noise map on the projective measurement of the covariance matrix \cite{mari2014quantum, giovannetti2014ultimate, genoni2017cramer}
	\begin{equation}
		\boldsymbol{{\mathrm{V}}}_{m}^{(\text{ineff})}\left(t,s,\phi\right)=\boldsymbol{X}^{*}\boldsymbol{{\mathrm{V}}}_{m}\left(s,\phi\right)\boldsymbol{X}^{{*}^{\dagger}}+\boldsymbol{Y}^{*},
	\end{equation}
	where $\boldsymbol{X}^{*}=e^{\frac{\gamma t}{2}}\mathbb{1}_{2\times 2}$ and $\boldsymbol{Y}^{*}=\left(e^{\gamma t}-1\right)\mathbb{1}_{2\times 2}$ are two real $2\times 2$-matrices fulfilling
	\begin{equation}
		\boldsymbol{Y}^{*} + i\boldsymbol{\Omega}  - i\boldsymbol{X}^{{*}^{\dagger}}\boldsymbol{\Omega} \boldsymbol{X}^{*} \ge 0.
	\end{equation}
	In order to extend this Gaussian measurement, either unitary or non-unitary, into the case of two-mode or even multi-mode, the overall measurement covariance matrix $\boldsymbol{{\mathrm{V}}}_{m}$  is evaluated by taking the direct sum of the single modes measurement developed above, i.e. 
	\begin{equation}
		\boldsymbol{{\mathrm{V}}}_{m}\left( {s,\phi } \right) = \mathop  \oplus \limits_{i = 1}^N {\boldsymbol{{\mathrm{V}}}_{m \hspace{0.1cm} \left( i \right)}}\left( {s,\phi } \right).
	\end{equation}
	\section{Estimation of joint quadrature parameters under a noisy Gaussian channel}\label{Sec. 4.4}
	In the previous chapter, we have clarified how to use the scalar {SLD} and the corresponding {QCRB} in the estimation f a single parameter. While how to use the {SLD} and {RLD-QFIM}s and the corresponding {QCRB}s will be now apparent in the case of multiparameter estimation. Here, we take a further step forward in this path by providing ultimate quantum-enhanced strategies to estimate the two conjugated parameters characterizing a phase-space displacement under a noisy Gaussian environment. The operation of displacing a state in the phase space is represented by the Weyl displacement operator
	\begin{equation}
		\hat D\left(\theta_{1}, \theta_{2}\right)=\exp \left(i \theta_{2} \hat{Q}-i \theta_{1} \hat{P}\right)
	\end{equation}
	A question that arises is, from an initial preparation state p that undergoes an unknown displacement and evolves in a Gaussian environment, how accurately can we jointly estimate the two conjugate parameters $\theta_{1}$ and $\theta_{2}$ of the displacement operator? One possibility is to use coherent states as the initial probe state, followed by heterodyne detection as the measurement strategy at the output of the displacement transformation. On the other hand, one may ask whether entanglement, in the form of \textbf{Einstein-Podolsky-Rosen }(\textbf{EPR}) correlations \cite{reid2009colloquium}, could lead to a better estimation precision for this issue, as was suggested in Refs. \cite{genoni2013optimal, bradshaw2018ultimate} for the unitary channels.
	
\begin{figure}[H]
		\tikzset{every picture/.style={line width=0.75pt}} 
		\begin{tikzpicture}[x=0.92pt,y=0.8pt,yscale=-1,xscale=1]
			\draw  [fill={rgb, 255:red, 155; green, 155; blue, 155 }  ,fill opacity=0.75 ] (49,170) .. controls (49,120.29) and (57.73,80) .. (68.5,80) .. controls (79.27,80) and (88,120.29) .. (88,170) .. controls (88,219.71) and (79.27,260) .. (68.5,260) .. controls (57.73,260) and (49,219.71) .. (49,170) -- cycle ;
			\draw    (81,99) -- (80,100) -- (125,99) ;
			\draw  [fill={rgb, 255:red, 184; green, 233; blue, 134 }  ,fill opacity=0.76 ] (123,76) -- (196,76) -- (196,130) -- (123,130) -- cycle ;
			\draw    (198,101) -- (234,101) -- (245,101) ;
			\draw  [fill={rgb, 255:red, 128; green, 128; blue, 128 }  ,fill opacity=0.7 ] (232,165) .. controls (232,109.77) and (258.86,65) .. (292,65) .. controls (325.14,65) and (352,109.77) .. (352,165) .. controls (352,220.23) and (325.14,265) .. (292,265) .. controls (258.86,265) and (232,220.23) .. (232,165) -- cycle ;
			\draw    (406,100) -- (436,100) ;
			\draw    (406,239) -- (431,238) ;
			\draw [color={rgb, 255:red, 74; green, 144; blue, 226 }  ,draw opacity=0.98 ][line width=3]    (463,168) -- (497,168) ;
			\draw    (436,100) -- (478,166) ;
			\draw    (433,238) -- (480,168) ;
			\draw    (480,166) -- (518,102) ;
			\draw    (518,102) -- (544,102) ;
			\draw    (518,242) -- (540.5,242) ;
			\draw  [fill={rgb, 255:red, 155; green, 155; blue, 155 }  ,fill opacity=0.75 ] (540,79) -- (562.5,79) .. controls (574.93,79) and (585,119.74) .. (585,170) .. controls (585,220.26) and (574.93,261) .. (562.5,261) -- (540,261) -- cycle ;
			\draw  [color={rgb, 255:red, 0; green, 0; blue, 0 }  ,draw opacity=1 ][fill={rgb, 255:red, 155; green, 155; blue, 155 }  ,fill opacity=0.75 ] (372,169.5) .. controls (372,119.52) and (381.63,79) .. (393.5,79) .. controls (405.37,79) and (415,119.52) .. (415,169.5) .. controls (415,219.48) and (405.37,260) .. (393.5,260) .. controls (381.63,260) and (372,219.48) .. (372,169.5) -- cycle ;
			\draw    (333,239) -- (380,239) ;
			\draw    (339,100) -- (379,99) ;
			\draw    (80,242) -- (253,242) ;
			\draw    (479,168) -- (520,242) ;
			
			\draw (130,90) node [anchor=north west][inner sep=0.85pt] {$\hat D\left(\theta_1,\theta_2\right)$};
			\draw (435,161) node [anchor=north west][inner sep=0.75pt]    {$\text{BS}$};
			\draw (552,240) node [anchor=north west][inner sep=0.75pt]  [rotate=-270.1] [align=left] {Homodyne detection};
			\draw (248,136) node [anchor=north west][inner sep=0.75pt]   [align=left] {\textbf{ \ Gaussian}\\\textbf{ \ \ \ thermal }\\\textbf{envirnoment}};
			\draw (57.51,179.79) node [anchor=north west][inner sep=0.75pt]  [rotate=-271.09]  {$\hat\rho_{inp}$};
			\draw (380.53,179.79) node [anchor=north west][inner sep=0.75pt]  [rotate=-271.03]  {$\hat \rho_{out}$};
		\end{tikzpicture}
		\centering
		\captionsetup{justification=centerlast, singlelinecheck=false}\captionof{figure}{Schematic to illustrate the adaptive protocol for estimating the displacement parameters under the noises Gaussian environment. Initially, we prepare the probe state $\hat \rho_{inp}$, using the various Gaussian operations.  Next, the probe state is submitting to an unknown displacement $\hat D\left(\theta_{1}, \theta_{2}\right)$. After this displacement, the two-mode is evolving under a Gaussian thermal environment. The two modes of output state $\hat \rho_{out}$ are mixing with a beam splitter (BS), and then the homodyne detection measurement is performed.}\label{Fig. 7}
	\end{figure}
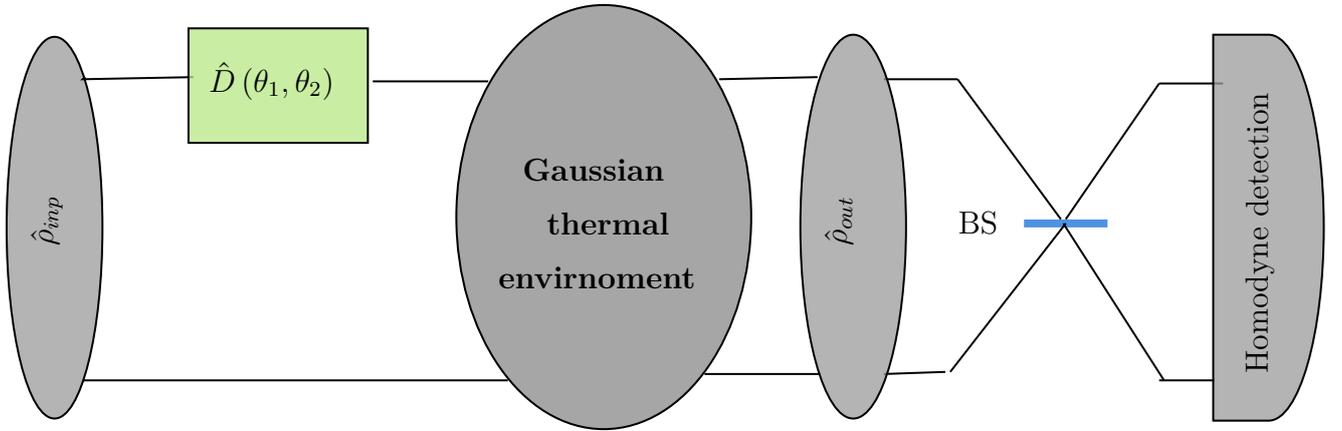

	This section is devoted to formalizing a framework to study and analyze a measurement scheme in which entanglement phenomena play a paramount role in improving the estimation precision of joint quadratic parameters encoded in the displacement operator under a noisy Gaussian environment. For this purpose, we will prepare the probe state as two-mode Gaussian states. One of the two modes is displaced by the acting displacement operator ${\hat D}\left(\theta_{1}, \theta_{2}\right)$ that contains the two unknown parameters. After this displacement, the two modes are subject to the effects of a Gaussian environment. The latter can be considered an unavoidable part of the detection procedure. The state of the system, after the interaction with the environment, is given by $\hat \rho_{out}$. The two modes of the output state are mixed with a beam splitter, and the Gaussian homodyne measurement is carried out. Fig. (\ref{Fig. 7}) is devoted to illustrating this metrological proposed schema. 
	For the homodyne measurement, one can be exploited the probability distribution function given in Eqs. ((\ref{EQ. 4.19}), (\ref{EQ. 4.20}))  to evaluate the {CFIM} given in Eq. (\ref{CFIM}) and then the corresponding {CCRB}. We will refer to this bound in the following as the homodyne detection bound ({HDB}). Since the {HDB} is assessed by the classical estimation approach, i.e. classical Fisher information. Then, undoubtedly, there is a gap between the precision of the estimate obtained by the homodyne measurement and that achievable by evaluating the {SLD} and {RLD}-{QCRB}s. This is not surprising since, in general, we know that the {SLD} and {RLD}-{QCRB}s dos not always saturated and then are not within tight bounds. Logically, this raises two fundamental questions: The first one is, from the input state displaced by the action of $\hat D\left(\theta_{1}, \theta_{2}\right)$ and evolving under a noise environment, how can we estimate simultaneously, with precision, the pair of parameters $\theta_{1}$ and $\theta_{2}$? In other words, can we derive a tight bound of precision even under the existence of environmental fluctuations? We addressed these questions for a general two-mode Gaussian probe, i.e. a two-mode squeezed displaced thermal state of the following density matrix
	\begin{equation}\label{EQ. 4.27}
		{\hat \rho _{inp}} = {\hat S_2}\left(\zeta   \right)\hat D\left( \alpha  \right)\left( {{\hat \rho _{th}} \otimes {\hat \rho _{th}}} \right)\hat D{\left( \alpha  \right)^\dag }{\hat S_2}{\left( \zeta  \right)^\dag }, 
	\end{equation}
	where ${\hat S_2}\left( \zeta  \right) = \exp \left( {\zeta {{\hat a_1}^\dag }{{\hat a_2}^\dag } - {\zeta ^ * }\hat a_1\hat a_2} \right)$ is the two-mode squeezing operator with the squeezed parameter $r$ and the rotation angle $\varphi$ ($\zeta=r e^{i\varphi}$), and $\hat D\left( \alpha  \right) = \exp \left( {{\alpha _1}\hat a_1^\dag  - \alpha _1^*{{\hat a}_1} + {\alpha _2}\hat a_2^\dag  - \alpha _2^*{{\hat a}_2}} \right)$ is the two-mode displacement operator, with $ \alpha_{k=(1,2)}=q_k+ip_k$ is the parameter of coherent light, and $\hat \rho_{th}$ denote the thermal states 
	\begin{equation}\label{EQ. 4.28}
		{\hat \rho _{th}} = \sum\limits_N {\frac{{{{\mathtt{\bar N}_{th}}^N}}}{{{{\left( {\mathtt{\bar N}_{th} + 1} \right)}^{N+1}}}}\left| N \right\rangle } \left\langle N \right|, 
	\end{equation}
	where ${\mathtt{\bar N}}_{th} = \left\langle {\hat a^\dag {\hat a}} \right\rangle= \operatorname{Tr}\left[\hat \rho_{th}\hat a^\dag {\hat a} \right]$  is the mean number of photons in the bosonic mode, which is expressed in terms of the temperature effect as ${\mathtt{\bar N}_{th}} = {\left( {{e^{\frac{1}{k_{B}T}} }- 1} \right)^{ - 1}}$ with $k_B$ is the Boltzmann constant. In the limit of zero temperature, one recovers the two-modes pure vacuum state $\left| 0 0 \right\rangle \left\langle 0 0 \right|$.
	
	From the general probe state of Eq. (\ref{EQ. 4.27}), we can derive two particular types of Gaussian probes: The first consists of pure Gaussian states, which, in turn,  are decomposed into two-mode squeezed vacuum state and two-mode displacement vacuum state. The second one is the mixed probes Gaussian states, which are the two-mode squeezed thermal state and two-mode displacement thermal states. All these classes of states are typical states that can be implemented and used for processing quantum information in \textbf{CV} systems. For example, coherent states are very relevant in \textbf{CV} implementations of quantum information protocols. They provide a very suitable description of the states produced by a laser: they are states with well-defined amplitude and phase and with minimal fluctuations in both quadratures, while the squeezed states present non-classic features such as the potential ability to generate quantum entanglement, which is a robust resource for performing the diverse protocols in different disciplines of quantum information theory. In this context, one may ask how entanglement may influence the estimation of accuracy? In other terms, what is the relationship between the entangled states and the limit of precision? There is another question related to the role of purity in this problem: Which of the input states leads to a better improvement, the pure entangled state or the mixed entangled state?
	To answer these questions, we will try to treat each state individually. In what follows, we assume that the two modes are identical, and the squeezing parameter of baths takes the value zero ($\mathtt{M}_e = 0$). This assumption means that the photon number of the Gaussian environment corresponds to the photon number of thermal states, i.e. $\mathtt{\bar N}_e = \mathtt{\bar N}_{th}$.
	\subsection{Pure Gaussian probe states}
	Let us now discuss, in this subsection, the estimation of the conjugate parameters $\theta_{1}$ and $\theta_{2}$, considering the two types of pure Gaussian probe states. We start with the two-modes squeezed vacuum state ({TMSV})
	\subsubsection{Two-modes squeezed vacuum state}
	When we pump a nonlinear crystal with a bright laser, some of the pump with frequency $2\omega$ are split into pairs of photons with frequency $\omega$. Whenever the matching conditions for a degenerate optical parametric amplifier (OPA) are satisfied, the outgoing
	mode is ideally composed of a superposition of even number states $\ket{2N}$. This process is allowed to generate the single-mode squeezing operator. In the opposite case, when a non-linear crystal is pumped in a non-degenerate OPA regime, pairs of photons of different frequencies are generated, and then the two different modes, which are called the signal and the idler.  The process of the latter regime leads to the generation of the two-mode compression operator $\hat S_2\left(\zeta\right)$. The corresponding symplectic transformation in the phase space is 
	\begin{equation}
		{S_2}\left( r \right) = \left( {\begin{array}{*{20}{l}}
				{\cosh r \mathbb{1}_{2 \times 2}}&{\sinh r \mathcal{R}_{\varphi}}\\
				{\sinh r \mathcal{R}_{\varphi}}&{\cosh r \mathbb{1}_{2\times 2}} 
		\end{array}} \right), \quad \text{with}\quad \mathcal{R}_\varphi  = \left( {\begin{array}{*{20}{l}}
				{\cos \varphi }&{\sin \varphi }\\
				{\sin \varphi }&{ - \cos \varphi }
		\end{array}} \right).
	\end{equation}
	The action of $\hat S_2\left(\zeta\right)$ into couple of vacuum leads immediately to generate the two mode squeezed vacuum state, also known as the {EPR} state. It is derived from the general density matrix of Eq. (\ref{EQ. 4.27}) by setting $\alpha=0$ and $\mathtt{\bar N}_{th}=0$, then it's written as 
	\begin{equation}
		\hat{\rho}_{i n p}=\hat{S}_{2}(\zeta)\ket{00}\bra{00} \hat{S}_{2}(\zeta)^{\dagger},
	\end{equation}
	where $\hat{S}_{2}(\xi)\ket{00}$ is expressed in Fock space by
	\begin{equation}
		\hat{S}_{2}(\zeta)\ket{00}=\frac{1}{\cosh r} \sum_{N=0}\left(-e^{i \varphi} \tanh r\right)^{N}\ket{N,N}.
	\end{equation}
	The the mean number of photons in this state is ${\mathtt{\bar N}} = \left\langle {\hat a^\dag {\hat a}} \right\rangle=\sinh^2 r$. It has a Gaussian Wigner function with the zero mean and covariance matrix
	\begin{equation}\label{EQ. 4.32}
		{{\boldsymbol{{\mathrm{V}}}}_{{\rm{EPR}}}}\left( r \right) = \left( {\begin{array}{*{20}{c}}
				{\cosh \left( {2r} \right)\mathbb{1}_{2\times2}}&{\sinh \left( {2r} \right)\mathcal{R}_{\varphi}}\\
				{\sinh \left( {2r} \right)\mathcal{R}_{\varphi}}&{\cosh \left( {2r} \right)\mathbb{1}_{2\times2}}
		\end{array}} \right).
	\end{equation}
	The diagonalization of (\ref{EQ. 4.32}) easily yields 
	\begin{equation}
		\mathtt{Var}\left(\hat Q_{-}\right)=\mathtt{Var}\left(\hat P_{+}\right)=e^{-2r}.
	\end{equation}
	where $\hat Q_{-}=\left(\hat Q_{1}-\hat Q_{2}\right)/{\sqrt{2}}$ and $\hat P_{+}=\left(\hat P_{1}+\hat P_{2}\right)/{\sqrt{2}}$. For $r=0$, the {EPR} state corresponds to a two-mode vacuum state, and the previous variances are equal to $1$, corresponding to the quantum shot-noise or {SQL}. For every two-mode squeezing with $r>0$, we have $\mathtt{Var}\left(\hat Q_{-}\right)=\mathtt{Var}\left(\hat P_{+}\right)<1$, which means that the correlations between the quadratures of the two systems beat the {SQL}. These correlations are known as {EPR} correlations, and they imply the presence of bipartite entanglement. In the limit of $r\to \infty$, we have an ideal {EPR} state with maximum entangled and perfect correlations.
	
\begin{figure}[H]
		\centering \includegraphics[scale=1.6]{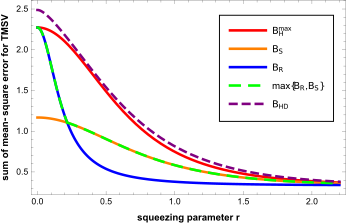}
		\captionsetup{justification=centerlast, singlelinecheck=false}\captionof{figure}{The plot of the average sum of variances for the two-modes squeezed vacuum probe state. The different bounds are plotted as the functions of the initial squeezing parameter of the {TMSV} state, with fixing the other parameters such as $\mathtt{\bar N}_{e}=0.5, \gamma=1, t=0.2$, the homodyne angle of BS is fixed at $\phi=\pi/2$.}\label{Fig. 8}
	\end{figure}

	Now that we have a complete description of the {TMSV} state, we can proceed to evolve it in our schema adapted above in Fig. (\ref{Fig. 7}). After the action of the displacement operator $\hat{D}\left(\theta_{1}, \theta_{2}\right)$ and the evolution under a noise environment, the output state of the system is characterized by the first and second moments such as
	\begin{equation}\label{EQ. 4.34}
		\mathbf{d}_{\text {out }}(t)=\eta\left(t\right)\left(\theta_{1}, \theta_{2}, 0,0\right)^{\top} \quad \boldsymbol{\mathrm{V}}_{out}\left(t\right)=\eta\left(t\right) \boldsymbol{\mathrm{V}}\left(0\right)\eta\left(t\right)+ \boldsymbol{\mathrm{V}}_{\infty}\left(t\right),
	\end{equation}
	where $\boldsymbol{\mathrm{V}}\left(0\right)=\boldsymbol{\mathrm{V}}_{\rm{EPR}}$ and $\boldsymbol{\mathrm{V}}_{\infty}\left(t\right)=\left(1-e^{-\gamma t}\right)\left(2\mathtt{N}_{e}+1\right) \mathbb{1}_{4\times 4}$, since the noise environment is thermal, and previously we assumed that $\mathtt{M}_{e}=0$. To estimate the unknown parameters $\theta_{1}$ and $\theta_{2}$, we proceed to evaluate the {SLD} and {RLD}-{QFIM}s and the corresponding {QCRB}s. For this purpose, we shall be using the results developed in the last chapter. In view of Eq. (\ref{Eq. 3.40})  and Eq. (\ref{Eq. 3.48}) as well as Eq. (\ref{EQ. 4.34}), we can conclude that the first terms of Eq. (\ref{Eq. 3.40}) and Eq. (\ref{Eq. 3.48}) do not contribute to the expression of {SLD} and {RLD}-{QFIM}s. Therefore, we can write the {SLD} and {RLD-QFIM}s, respectively, as 
	\begin{equation}\label{EQ. 4.35}
		\left[	\mathcal{F}_Q^{(S)}\left(\boldsymbol{\theta}\right)\right]_{jk}=2 \partial_{\theta_j} \mathbf{d}_{\text {out }}(t)^{\top} \boldsymbol{\mathrm{V}}_{out}\left(t\right)^{-1} \partial_{\theta_k} \mathbf{d}_{\text {out }} \quad \text{with} \quad\boldsymbol{\theta}=\left(\theta_{1},\theta_{2}\right)^{\top}.
	\end{equation}
	\begin{equation}\label{EQ. 4.36}
		\left[\mathcal{F}_Q^{(R)}\left(\boldsymbol{\theta}\right)\right]_{jk}=2 \partial_{\theta_j} \mathbf{d}_{\text {out }}(t)^{\top} \mathfrak{M}\left(t\right)_{out}^{-1} \partial_{\theta_k} \mathbf{d}_{\text {out }} \quad \text{with} \quad\boldsymbol{\theta}=\left(\theta_{1},\theta_{2}\right)^{\top}.
	\end{equation}
	where $\mathfrak{M}_{out}\left(t\right)=\boldsymbol{\mathrm{V}}_{out}\left(t\right)+i\boldsymbol{\Omega}$. Using the equations ((\ref{BS}), (\ref{BR}), we can easily evaluate the two different {QCRB}s associated, respectively, with (\ref{EQ. 4.35}) and (\ref{EQ. 4.36}) 
	\begin{equation}\label{EQ. 4.37}
		{B_S}\left( t \right) = \left( {{{\rm{e}}^{t\gamma }} - 1} \right)\left( {1 + 2 \mathtt{\bar N}_{e}} \right) + \cosh \left( {2r} \right) - \frac{{\sinh {{\left( {2r} \right)}^2}}}{{\left( {{{\rm{e}}^{t\gamma }} - 1} \right)\left( {1 + 2 \mathtt{\bar N}_{e}} \right) + \cosh \left( {2r} \right)}}, 
	\end{equation}
	\begin{equation}\label{EQ. 4.38}
		{B_R}\left( t \right) = 2{{\rm{e}}^{t\gamma }}\left( {1 + \mathtt{\bar N}_{e}} \right) + \cosh \left( {2r} \right) - \left( {2 \mathtt{\bar N}_{e} + 1} \right) - \frac{{\sinh {{\left( {2r} \right)}^2}}}{{2\left( {{{\rm{e}}^{t\gamma }} - 1} \right) \mathtt{\bar N}_{e} + \cosh \left( {2r} \right) - 1}}.
	\end{equation}
	In order to check where are these bounds of precision are attainable or not, it is necessary to evaluate the quantumness parameter $\mathcal{R}_{Q}$, given in Eq. (\ref{Eq. 2.134}), that quantifies the degree of incompatibility between the parameters $\theta_{1}$ and $\theta_{2}$. Thus,  by using Eq. (\ref{Eq. 2.134}), we can derive $\mathcal{R}_{Q}$ such as 
	\begin{equation}\label{EQ. 4.39}
		\mathcal{R}_{Q} = \frac{{{{\rm{e}}^{t\gamma }}}}{{{{\rm{e}}^{t\gamma }} - 2\mathtt{\bar N}_{e}\left( {1 - {{\rm{e}}^{t\gamma }}} \right) + 2\sinh^2r}}=\frac{{{{\rm{e}}^{t\gamma }}}}{{{{\rm{e}}^{t\gamma }} - 2\mathtt{N}_{e}\left( {1 - {{\rm{e}}^{t\gamma }}} \right) + 2\mathtt{\bar N}}}. 
	\end{equation}
	Therefore, $\mathcal{R}\ne 0$, which means that the model considered is incompatible, and then the limits of (\ref{EQ. 4.37}) and (\ref{EQ. 4.38}) are not attainable in general, except in the case where $\mathtt{\bar N}\to\infty$. In this limit, we have $\mathcal{R} \to 0$. Thus,  the multiparameter model is compatible. In this case, the indeterminacy that arises from the quantum nature of the system disappears. This limit is known as the asymptotic limit, and the multiparameter model is called the asymptotically classical quantum statistical model. Outside this limit, the performance of any measurement is very poor and does not support the ultimate estimation precision, even the homodyne detection. To predict the degree of performance possible in the estimation of $\theta_{1}$ and $\theta_{2}$, we proceed to compute the upper bound of {HCRB}, which is tightly bound. Then, by exploiting the results of  Eq. (\ref{Eq. tight}) and Eqs. (\ref{EQ. 4.37}), (\ref{EQ. 4.39}), one gets
	\begin{equation}
		B_H^{\max }\left(t\right) = C\left(t\right)\left( {\left( {{{\rm{e}}^{t\gamma }} - 1} \right)\left( {1 + 2{\mathtt{\bar N}_{e}}} \right) + \cosh \left( {2r} \right) - \frac{{\sinh {{\left( {2r} \right)}^2}}}{{\left( {{{\rm{e}}^{t\gamma }} - 1} \right)\left( {1 + 2{\mathtt{\bar N}_{e}}} \right) + \cosh \left( {2r} \right)}}} \right)
	\end{equation}
	where $C\left(t\right)=\left( {1 + \frac{{{{\rm{e}}^{t\gamma }}}}{{\left( {{{\rm{e}}^{t\gamma }} - 1} \right)\left( {1 + 2{\mathtt{\bar N}_{e}}} \right) + \cosh \left( {2r} \right)}}} \right)$. Fig. (\ref{Fig. 8}) is devoted to illustrating these results. We plot the different precision bounds that we have already evaluated.
	Fig. (\ref{Fig. 8}) shows the behavior of {SLD} and{ RLD-QCRB}s of Eqs. ((\ref{EQ. 4.37}), (\ref{EQ. 4.38})) and the behavior of {HDB} that evaluated by classical Fisher information. From this Fig, we notice that the sum {MSE} of the homodyne measurement does not reach the {SLD} and {RLD} bounds for most values of $r$. In fact, this means that either the measurement is not optimal or the {SLD} and {RLD} bounds are not tight. This is not surprising since the $\mathcal{R}_{Q}\ne 0$, then the model is incompatible, and we can not estimate the two-parameter $\theta_{1}$ and $\theta_{2}$ simultaneously with precision. For this, we evaluated the upper bound of {HCRB}, which is tight, and then helped us predict the limit of ultimate precision. By inspecting the upper bound of {HCRB}, we notice that it is a decreasing functions of the initial squeezing parameter $r$ (and thus of the entanglement of the probe state). For most values of $r$, we found that the upper bound of {HCRB} almost coincides with that of {HDB}, which means that the upper bound of {HCRB} achieves the ultimate accuracy that was not possible with the {SLD} and {RLD} bounds. As meaningful results, in the limit of large values of $r$, we see that all bounds coincide, and the sum of {MSE} reaches the minimum values and then reaches the ultimate accuracy of $\theta_{1}$ and $\theta_{2}$. In this limit, the initial state has entangled.
	\subsubsection{Two-modes displacement vacuum state}
	We proceed now to consider the case in which the initial state is prepared as a two-mode coherent state, also known as a two-mode displacement vacuum state ({TMDV}). In this case, the general input state of Eq. (\ref{EQ. 4.27}) is reduced to 
	\begin{equation}
		\hat \rho _{inp} = \hat D\left( \alpha  \right)\left| {00} \right\rangle \left\langle {00} \right|\hat D{\left( \alpha  \right)^\dag}, 
	\end{equation}
	where $\hat D\left( \alpha  \right)\left| {00} \right\rangle$ is the displacement vacuum state, that is expressed in Fock space by 
	\begin{equation}
		\left| {{\alpha _1},{\alpha _2}} \right\rangle  = {e^{ - \frac{{{{\left| {{\alpha _1}} \right|}^2}}}{2}}}{e^{ - \frac{{{{\left| {{\alpha _2}} \right|}^2}}}{2}}}\sum\limits_{{N} = 0} {\frac{{{\alpha _1}^{{N}}{\alpha _2}^{{N}}}}{{ {{N}!} }}} \left| {{N},{N}} \right\rangle,\label{EQ. 4.42}
	\end{equation}
	where $\alpha_{k=(1,2)}$ is the parameter of coherent light in the mode $k$. The mean number of photons in this state is $\mathtt{\bar N}=\left\langle {\hat a_1^\dag {{\hat a}_1}} \right\rangle  = \left\langle {\hat a_2^\dag {{\hat a}_2}} \right\rangle  = {\left| {{\alpha _1}} \right|^2} = {\left| {{\alpha _2}} \right|^2}= {\left| {{\alpha}} \right|^2}$. It has a Gaussian characteristic function with the following first and second moments 
	\begin{equation}
		{\mathbf{d}_{inp}} = {\left( {{q_1},{p_1},{q_2},{p_2}} \right)^{\top}}, \hspace{0.7cm}  \hspace{1cm} {\boldsymbol{{\mathrm{V}}}_{\text{inp}}} = {\mathbb{1}_{4 \times 4}}.
	\end{equation}
	After the action of the displacement operator and making the input state of Eq. (\ref{EQ. 4.42}) evolve under the interaction with the Gaussian environment, its output state is characterized by 
	\begin{equation}
		{\boldsymbol{d}}_{out}\left( t \right) = \eta\left(t\right){\left( {{q_1} + {\theta _1},{p_1} + {\theta _2},{q_2},{p_2}} \right)^{\top}};\quad \boldsymbol{\mathrm{V}}_{out}\left(t\right)=\eta\left(t\right) \boldsymbol{\mathrm{V}}\left(0\right)\eta\left(t\right)+ \boldsymbol{\mathrm{V}}_{\infty}\left(t\right),
	\end{equation}
	where $\boldsymbol{\mathrm{V}}\left(0\right)=\boldsymbol{\mathrm{V}}_{\text{inp}}$. In order to estimate the two parameters $\theta_{1}$ and $\theta_{2}$, in this case, we will first evaluate the {SLD} and {RLD}-{QFIM}s and the corresponding {QCRB}s. Similarly, in this case, the first terms of Eq. (\ref{Eq. 3.40}) and Eq. (\ref{Eq. 3.48}) do not contribute to the expression of {SLD} and {RLD}-{QFIM}. Thus, {SLD} and {RLD-QFIM}s have the same expression of Eq. (\ref{EQ. 4.35}) and Eq. (\ref{EQ. 4.36}), respectively. The precision {SLD} and {RLD-QCRB}s, in this case, are evaluated as
	\begin{equation}\label{EQ. 4.45}
		{B_S}\left( t \right) = {{\rm{e}}^{t\gamma }}\left( {1 + 2 \mathtt{ \bar N}_e} \right) - 2  \mathtt{\bar N}_e, 
	\end{equation}
	\begin{equation}
		{B_R}\left( t \right) = {{\rm{e}}^{t\gamma }}\left( {1 + 2 \mathtt{\bar N}_e} \right) + {{\rm{e}}^{t\gamma }} - 2\mathtt{\bar N}_e. \label{EQ. 4.46}
	\end{equation}
	The order of incompatibility between $\theta_{1}$ and $\theta_{2}$ is quantified by the quantumness parameter $\mathcal{R}_Q$, which evaluate in this case as
	\begin{equation}\label{EQ. 4.47}
		\mathcal{R} = \frac{{{{\rm{e}}^{t\gamma }}}}{{{{\rm{e}}^{t\gamma }} - 2 \mathtt{\bar N}_e\left( {1 - {{\rm{e}}^{t\gamma }}} \right)}}, 
	\end{equation}
	which does not vanish, then the model is incompatible and can never estimate both parameters $\theta_{1}$ and $\theta_{2}$ simultaneously with accuracy using the {SLD} and {RLD} bounds. This fact requires looking into the evaluation of the upper bound of {HCRB}, which is derived directly using Eqs. (\ref{EQ. 4.47}), (\ref{EQ. 4.45})
	\begin{equation}\label{EQ. 4.48}
		B_H^{\max }\left(t\right) = {{\rm{e}}^{t\gamma }}\left( {1 + 2{\mathtt{\bar N}_e}} \right) +{\rm{e}}^{t\gamma}- 2{\mathtt{\bar N}_e}, 
	\end{equation}
	which exactly corresponds to the {RLD-QCRB} of Eq. (\ref{EQ. 4.46}) this means that $B_H= B_R$ and therefore the model is D-invariant. Inspecting these precision bounds, we note that they do not all depend on the average energy of the two-mode probe coherent state. More precisely, by increasing or decreasing the mean photon number of the coherent state, one does not obtain any enhancement in the estimation precision. Thus, the two-mode coherent probe state is not suitable for estimating the displacement parameter under the interaction with the environment.
	
	\subsection{Mixed Gaussian probe states}
	That we have treat the estimation of $\theta_{1}$ and $\theta_{2}$ by using the Gaussian pure states as probes, next let us consider the more general case, in where the input states are mixed. We will focus on two of the most interesting cases: the first concerns a two-modes squeezed thermal state, and the second case corresponds to a two-modes coherent thermal state.
	\subsubsection{Two-modes squeezed thermal state}
	In this case, the general input state of Eq. (\ref{EQ. 4.27})  has reduced to 
	\begin{equation}\label{EQ. 4.49}
		{\hat \rho _{inp}} = {\hat S_2}\left( \zeta  \right)\left( {{\hat \rho _{\text{th}}} \otimes {\hat \rho _{\text{th}}}} \right){\hat S_2}{\left( \zeta  \right)^\dag }, 
	\end{equation}
	where $\hat \rho_{\text{th}}$ is the thermal state that given in Eq. (\ref{EQ. 4.28}). The mean number of photons in this state is $\left\langle {\hat a_k^\dag {{\hat a}_k}} \right\rangle = {\sinh ^2}r + \mathtt{ \bar N}_{\text{th}}$. This input state has a Gaussian Wigner function with zero first moment and the covariance matrix
	\begin{equation}\label{EQ. 4.50}
		\boldsymbol{\mathrm{V}} _{\text{inp}} = (2\mathtt{ \bar N}_{\text{th}} + 1)\left( {\begin{array}{*{20}{c}}
				{\cosh \left( {2r} \right){\mathbb{1}_{2 \times 2}}}&{\sinh \left( {2r} \right){ \mathcal{R}_\varphi }}\\
				{\sinh \left( {2r} \right){ \mathcal{R}_\varphi }}&{\cosh {{\left( {2r} \right)}\mathbb{1}_{2 \times 2}}}
		\end{array}} \right).
	\end{equation}
	The corresponding output state is still described by a Gaussian Wigner function but with the following first ad second moments
	\begin{equation}
		{\mathbf{d}_{out}}\left( t \right)  = \eta\left(t\right){\left( {{\theta _1},{\theta _2},0,0} \right)^{\top}}, \quad \boldsymbol{\mathrm{V}}_{out}\left(t\right)=\eta\left(t\right) \boldsymbol{\mathrm{V}}\left(0\right)\eta\left(t\right)+ \boldsymbol{\mathrm{V}}_{\infty}\left(t\right),
	\end{equation}
	where $\boldsymbol{\mathrm{V}}\left(0\right)$, here, is the covariance matrix given in Eq. (\ref{EQ. 4.50}). Since the covariance matrix also in this case is not depend on the estimate parameters, then the {SLD} and {RLD-QFIM}s are the same expressed above in Eq. (\ref{EQ. 4.35}) and Eq. (\ref{EQ. 4.36}). The corresponding {SLD} and {RLD}-{QCRB}s are derived, respectively, as
	\begin{equation}\label{EQ. 4.52}
		{B_S}\left( t \right) = \frac{{\left( {1 + 2\mathtt{ \bar N}_{\text{th}}} \right)\left( {2 - 2{{\rm{e}}^{t\gamma }} + {{\rm{e}}^{2t\gamma }} + 2\left( {{{\rm{e}}^{t\gamma }} - 1} \right)\cosh \left( {2r} \right)} \right)}}{{\cosh \left( {2r} \right) + {{\rm{e}}^{t\gamma }} - 1}}, 
	\end{equation} 
	\begin{equation}\label{EQ. 4.53}
		{B_R}\left( t \right) = \frac{{2\mathtt{ \bar N}_{\text{th}}\left( {1 + \mathtt{ \bar N}_{\text{th}}} \right){{\rm{e}}^{2t\gamma }} + 2\left( {{{\rm{e}}^{t\gamma }} - 1} \right){{\left( {1 + 2\mathtt{ \bar N}_{\text{th}}} \right)}^2}\sinh {{\left( r \right)}^2}}}{{\mathtt{ \bar N}_{\text{th}}{{\rm{e}}^{t\gamma }} + \left( {1 + 2\mathtt{ \bar N}_{\text{th}}} \right)\sinh {{\left( r \right)}^2}}}.
	\end{equation}
	Now that we have compute the SLD and RLD bounds, we proceed to evaluate the quantumness parameter $\mathcal{R}_{Q}$ in order to quantify the degree of incompatibility between the estimates parameters $\theta_{1}$ and $\theta_{2}$
	\begin{equation} \label{Eq. 67}
		\mathcal{R}_{Q} = \frac{{{{\rm{e}}^{t\gamma }}}}{{2\mathtt{ \bar N} + 2\mathtt{ \bar N}_{e}\left( {{{{\mathop{\sinh}\nolimits} }^2}r - 2 + {{\rm{e}}^{t\gamma }}} \right) + {{\rm{e}}^{t\gamma }}}}.
	\end{equation}
	In the limit of $\mathtt{ \bar N} \to \infty $, we have $\mathcal{R} \to 0$. In this limit, the model is asymptotically classical and then we have attainable the optimal estimation precision of $\theta_{1}$ and $\theta_{2}$. But, in general, the parameter $\mathcal{R}_{Q}$ is not vanish and the estimation model is incompatible. In this case, it should be to looking in evaluate the upper bound of HCRB 
	\begin{equation}
		B_H^{\max }\left( t \right) = \frac{{\left( {2 - 2{{\rm{e}}^{t\gamma }} + {{\rm{e}}^{2t\gamma }} + 2\left( {{{\rm{e}}^{t\gamma }} - 1} \right)\cosh \left( {2r} \right)} \right)\left( {2{{\rm{e}}^{t\gamma }}\left( {1 + {\mathtt{ \bar N}_{\text{th}}}} \right) + \left( {2 + 4{\mathtt{ \bar N}_{\text{th}}}} \right)\sinh {{\left( {2r} \right)}^2}} \right)}}{{{{\left( {{{\rm{e}}^{t\gamma }} + \cosh \left( {2r} \right) - 1} \right)}^2}}}.
	\end{equation}
	In order to illustrate our result, in this case, we plot in the following the different evaluate precision bounds together with the HDB
	\begin{figure}[H]
	\centering	\includegraphics[scale=1.6]{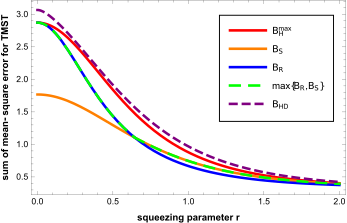}
		\captionsetup{justification=centerlast, singlelinecheck=false}\captionof{figure}{The plot of the average sum of variances for the two-modes squeezed thermal probe state. The different bounds are plotted as the functions of the initial squeezing parameter of the {TMST} state, with fixing the other parameters such as $\mathtt{\bar N}_{e}=\mathtt{\bar N}_{th}=0.5, \gamma=1, t=0.2$, the homodyne angle of BS is fixed at $\phi=\pi/2$.} \label{Fig. 9}
	\end{figure}
	As shown in Fig. (\ref{Fig. 9}), the sum of {MSE} achieved by {HDB} does not coincide with {SLD} and {RLD}-{QCRB}s for most values of $r$. This confirmed that both {SLD} and {RLD}-{QCRB}s are not tight. This fact is not surprising since $\mathcal{R}_{Q} \ne$. However, the upper bound on {HCRB} almost coincides with {HDB} for most values of $r$, except for the limit when $r$ tends to zero. This implies that the upper bound of {HCRB} is tightly bound and that the limit of $r\to 0$ is not profitable for estimating the displacement parameters $\theta_{1}$ and $\theta_{2}$. In this limit, the mixed probe state is separable, which means that the separability criterion of the probe state is not appropriate for the estimation of $\theta_{1}$ and $\theta_{2}$. In the opposite limit, $r$ becomes large, all precision bounds coincide, and the sum of {MSE} for estimating $\theta_{1}$ and $\theta_{2}$ reaches the minimum value. This implies that the precision has achieved the ultimate enhancement, and this limit is associated with a mixed probe entangled state. These results are similar to those obtained in the case of using the pure {TMSV} state as a probe, the predicted difference between them will be in the degree of accuracy achieved. We will illustrate this difference in section (\ref{Sec. 4.5}). This similarity is relevant to the common entangled feature between these kinds of quantum states.
	\subsubsection{Two-modes displacement thermal state}
	Let us now consider the second case of a mixed probe state, i.e. a mixed two-mode coherent thermal state. It is also called a two-mode displaced thermal state ({TMDT}), which is derived from the general state of Eq. (\ref{EQ. 4.27}), by setting  $\zeta=0$. Thus, we gets
	\begin{equation}\label{EQ. 4.56}
		{\hat \rho _{inp}} = \hat D\left( \alpha  \right)\left( {{\hat\rho _{\text{th}}} \otimes {\hat \rho _{\text{th}}}} \right)\hat D{\left( \alpha \right)^\dag }. 
	\end{equation}
	The mean number of photons in this state is $\left\langle {\hat a_k^\dag {{\hat a}_k}} \right\rangle  = {\left| \alpha  \right|^2} + \mathtt{\bar{N}}_{\text{th}}$ (for $k=1, 2$). It has an Wigner function with the flowing first and second moments 
	\begin{equation}\label{EQ. 4.57}
		\mathbf{d}_{inp}\left( t \right)  =\eta\left(t\right){\left( {q_1 +, p_1, q_2, p_2 }\right)^{\top}}, \quad 	 {\boldsymbol{{\mathrm{V}}}_{\text{inp}}} =\left(2 \mathtt{ \bar N}_{\text{th}}+1\right) {\mathbb{1}_{4 \times 4}}.
	\end{equation}
	The corresponding output state is also described by a Gaussian Wigner function with the following moments
	\begin{equation}
		\mathbf{d}_{out}\left( t \right)  =\eta\left(t\right){\left( {q1 + {\theta _1}, q2+ \theta_{2}, q_2, p_2 }\right)^{\top}}, \quad \boldsymbol{\mathrm{V}}_{out}\left(t\right)=\eta\left(t\right) \boldsymbol{\mathrm{V}}\left(0\right)\eta\left(t\right)+ \boldsymbol{\mathrm{V}}_{\infty}\left(t\right),
	\end{equation}
	where, in this case, $\boldsymbol{\mathrm{V}}\left(0\right)$ is the seconds moment given in Eq. (\ref{EQ. 4.57}). Followed a similar approach, we can evaluate the {SLD} and {RLD}-{QCRB}s such as
	\begin{equation}\label{EQ. 4.59}
		{B_S}\left( t \right) = {{\rm{e}}^{t\gamma }}\left( {1 + 2 \mathtt{\bar{N}}_{\text{th}}} \right), 
	\end{equation}
	\begin{equation}\label{EQ. 4.60}
		{B_R}\left( t \right) = {{\rm{e}}^{t\gamma }}\left( {1 + 2 \mathtt{\bar{N}}_{\text{th}}} \right) + {{\rm{e}}^{t\gamma }}. 
	\end{equation}
	Now that we have evaluated the {SLD} and {RLD}-{QCRB}s bounds, we will determine the order of incompatibility between the estimation parameters. In this case, the quantumness parameter is derived as
	\begin{equation}\label{EQ. 4.61}
		\mathcal{R} = \frac{1}{{1 + 2\mathtt{ \bar N}_{\text{th}}}}.
	\end{equation}
	Thus, the parameter $\mathcal{R}_{Q}$ does not vanish, and then the estimation model is generally incompatible. For this, we are going to evaluate the upper bound of {HCRB}. By using the results of Eq. (\ref{EQ. 4.61}) and Eq. (\ref{EQ. 4.59}), one gets
	\begin{equation}\label{EQ. 4.62}
		{B_{H}^{max}}\left( t \right) = {{\rm{e}}^{t\gamma }}\left( {1 + 2 \mathtt{ \bar N}_{\text{th}}} \right) + {{\rm{e}}^{t\gamma }}, 
	\end{equation}
	which coincides with the {RLD-QCRB} that was derived in Eq. (\ref{EQ. 4.60}). This means that $B_H^{max}=B_R=B_H$ and therefore the model is D-invariant quantum statistical model. From Eqs. (\ref{EQ. 4.59}), (\ref{EQ. 4.60}) and ((\ref{EQ. 4.62})), it is clear that all these precision bounds do not depend entirely on the average energy of the probe state. This fact means that preparing this kind of state as a probe does not yield any improvement in the estimation of displacement parameters. These results are the same ones obtained in the case where we apply the pure {TMDV} state as the probe. The weakness of these families of quantum states can be interpreted by their inability to be entangled.
	\section{Role of entanglement}\label{Sec. 4.5}
	
	Entanglement is one of the most important phenomena that has attracted considerable attention in quantum mechanics, being a central feature in most quantum information protocols. Given this importance, one may ask how to differentiate between entangled and non-entangled quantum systems? In other words, how do we define the quantum entangled state? In order to answer this question, we start by considering two bosonic systems $A$ with $N$-modes and $B$ with $M$-modes having Hilbert spaces $\mathcal{H}_{A}$ and $\mathcal{H}_{B}$, respectively. Then, the global bipartite system $A+B$ of $N+M$-modes has acted in Hilbert space $\mathcal{H}=\mathcal{H}_{A}+\mathcal{H}_{B}$. By definition, a quantum state $\hat \rho \in \mathcal{H}$ is called a separable state if it can be written as a convex combination of product states. Mathematically, it can be written as
	\begin{equation}\label{EQ. 4.63}
		\hat{\rho}=\sum_{i} p_{i} \hat{\rho}_{i}^{A} \otimes \hat{\rho}_{i}^{B}, \quad \hat{\rho}_{i}^{A(B)} \in \mathcal{H}_{A(B)}, 
	\end{equation}
	where $p_{i} \geq 0$ and $\sum_{i} p_{i}=1$. Note that the index can also be continuous. In this case, we replace the previous sum with an integral, and the probabilities are replaced by the probability density function. The physical interpretation of Eq. (\ref{EQ. 4.63}) means that the separable state is prepared via local operations and classical communications (\textbf{LOCC}s). Thus, we can define the entangled state as every quantum state that is not separable. More precisely, in an entangled state, the correlations between $A$ and $B$ are so strong that they cannot be created by any strategy based on {LOCC}s. Now that we have the complete definition of the entangled quantum state, the emerging question is: Is the state entangled? If the answer is yes, then how much entanglement does it have? In other words, how do we measure and quantify the amount of entanglement contained in such a state? In fact, there are many quantifiers measures of entanglement neither in discrete nor in continuous variables systems. Among these quantifiers, we have cited entanglement entropy \cite{bennett1996concentrating},  entanglement of formation \cite{wolf2004gaussian}, logarithmic negativity \cite{plenio2005logarithmic}, etc. Of course, each of these quantifiers measures has advantages and disadvantages. Fortunately, we do not need to delve into the details of entanglement quantifying, as our principal focus is to discuss their role in improving estimation accuracy.
	
	In the remainder of this section, we will focus on discussing the role of the entanglement also the purity of the probe state to achieve a precision bound that beat the {SQL}. In our measurement scheme, which is devoted to estimating the displacement parameters under the noise environment, we have used the entangled Gaussian states, {TMSV} and {TMST}, as probe states. To verify that the precision bound achieved by these classes of states can beat {SQL}, we first proceed to define a relationship between the sum of MSE and the inseparable criteria. For this purpose, we take a generic two-mode Gaussian state with quadratic operators $\hat Q_{i}$ and $\hat P_{i}$. Let $a$ be an arbitrary real non-zero, and we define $\hat u$ and $\hat v$ as $\hat u = \left| a \right|{\hat Q_1} + \frac{1}{a}{\hat Q_2},\quad \hat v = \left| a \right|{\hat P_1} - \frac{1}{a}{\hat P_2}$, in such case,  Duan et al. [44] proved that 
	\begin{equation}
		\mathtt{Var}\left( u \right) + \mathtt{Var}\left( v \right) < {a^2} + \frac{1}{{{a^2}}}.
	\end{equation}
	This is a sufficient condition for the inseparability feature of the probe state. The probe coherent is the one that allows saturating this inequality, and the sum of {MSE} in such case corresponds to the {SQL}. We will now project this crucial result into our measurement scheme, so we will check whether the sum of {MSE} for {TMSV} and {TMTS} can be written as
	\begin{equation}
		\mathtt{Var}\left( \theta_{1} \right) + \mathtt{Var}\left( \theta_{2} \right) < B_{\text{SQL}}.
	\end{equation}
	
	By inspecting the upper and bottom bound of {HCRB} in Fig. (\ref{Fig. 10}), we can clarify the results obtained from the various probes states.
	\begin{figure}[H]
		\centering
		\begin{subfigure}{.49\textwidth}
			\centering
			\includegraphics[width=7.7cm]{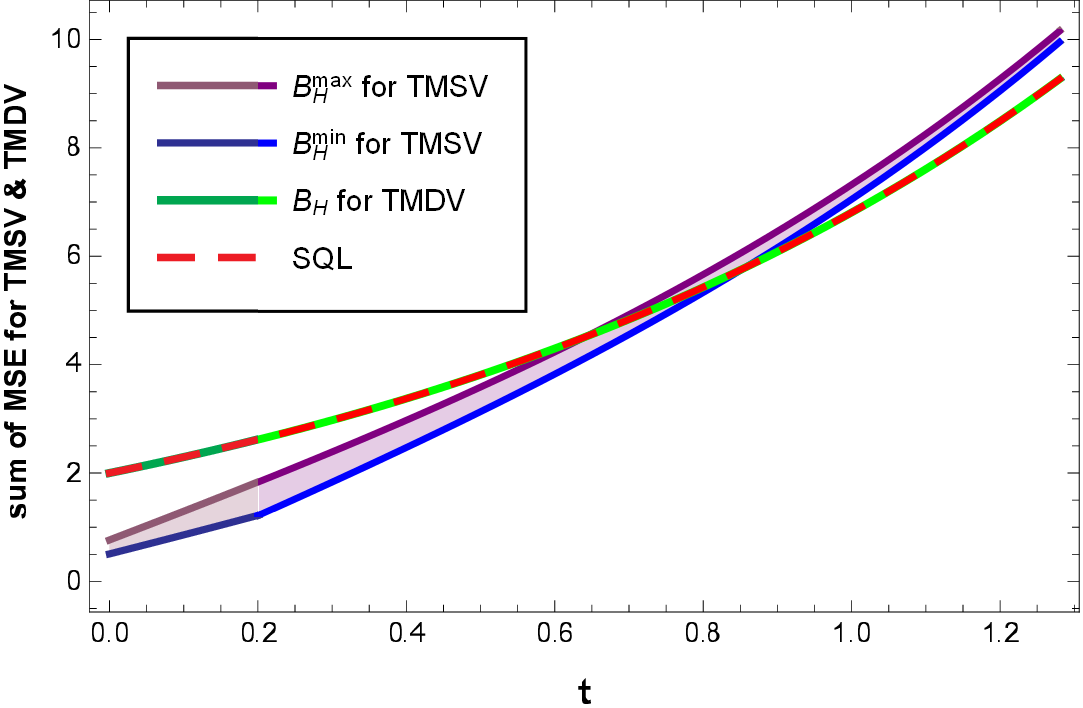}
			\caption{}\label{Fig. 5a}
		\end{subfigure}
		\begin{subfigure}{.49\textwidth}
			\centering
			\includegraphics[width=7.8cm]{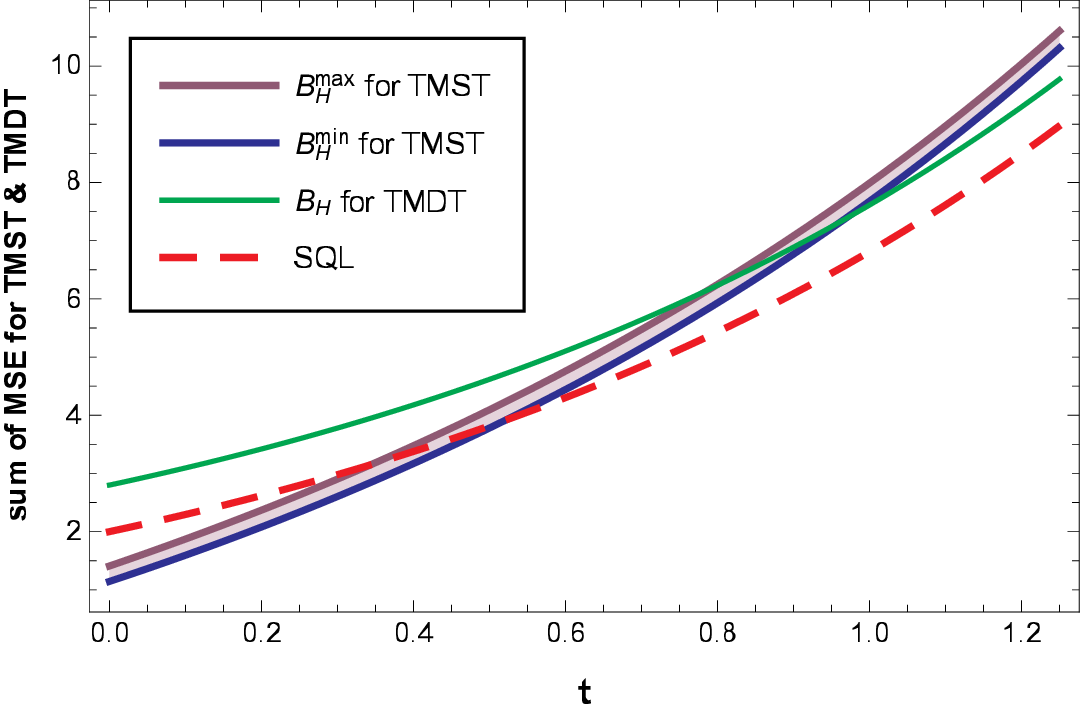}
			\caption{}\label{Fig. 5b}
		\end{subfigure}
		\addtocounter{figure}{-1}
		\captionsetup{justification=raggedright, singlelinecheck=false, labelfont=sc} \captionof{figure}{comparative study of the sum of MSE for the various probe states. Fig. (\ref{Fig. 5a}) represents the sum of MSE for pure states as functions of time $t$. While Fig. (\ref{Fig. 5b}) represents the sum of MSE for different mixed probe states as functions of time $t$. These results are plotted by setting the other parameters such as $\mathtt{ \bar N}_{e}=0.5$, $r=0.4$ and $\gamma=1$.}\label{Fig. 10}
	\end{figure}
	From Fig. (\ref{Fig. 5a}), we observe that, for {TMSV}, the upper and bottom bound of {HCRB} reached a minimum that goes beyond the standard quantum limit ({SQL}), which is evaluated by applying the vacuum state or a coherent state as a probe state. This fact is the best result that one can reach in estimation precision. While, for the {TMDV}, the {HCRB} coincides with the {SQL} without exceeding it. This means that the {TMDV} did not provide any addition to improving the estimation precision of displacement parameters. In the case of mixed states, Fig. (\ref{Fig. 5b}) shows that the upper and lower {HCRB} limit of the mixed {TMST} state exceeds the {SQL}. On the other hand, the {HCRB} of the mixed {TMDT} varies above the {SQL}. Both results are similar to those obtained in the pure state but with better precision for the latter. As a result, we finally conclude that the {TMDT} state and the {TMDV} state are not suitable for improving the estimation precision of displacement parameters. One can be explained these results by the independence of the precision bounds on the average energy of the probe states. Conversely, the {TMST} state and the {TMSV} state are the best archetypes for improving the estimation precision of displacement parameters under a Gaussian thermal environment. That is essentially due to the entanglement encompassed in such classes of states.
	
	\section{Conclusion}
	
	Quantum Gaussian states are one of the building blocks of a quantum system with a continuous spectrum. They have been used in various applications due to their simplicity of production and manipulation in the laboratory. Moreover, it is easy to deal with their behavior in the face of losses induced by environmental effects. In this chapter, we presented and discussed a measurement scheme that precisely estimates the two real parameters characterizing the displacement operator under the interaction with the environment. In this measurement scheme, we exploited Gaussian states as probes, and we used homodyne detection as the performed Gaussian measurement. We have investigated the limits of the ultimate possible accuracy by evaluating the different {SLD} and {RLD-QCRB}s as well as the upper bound of {HCRB} and the {HDB}. As expected, we found that the accuracy estimate is reduced under the effect of the environment. This is the case when the pure {TMDV} and mixed {TMDT} are employed as probe states.
	Alternatively,  in the case of using the pure {TMST} and a mixed {TMST} as probe states, we get the required improvement in estimation accuracy beyond the {SQL} even in the presence of environmental fluctuations. Furthermore, we have found that the {TMSV} state has been providing more precision than {TMST} with the same squeezing parameter and the same average number of thermal photons. Finally, we have emphasized that the obtained results have relevance to the role of entangled probes states in the beat of the {SQL}. This shows that the entanglement is a robust resource that allows achieving a precision limit beyond the {SQL}. Thus, it  permits us to overcome the constraints imposed by loss evolution in a Gaussian thermal environment.
	\chapter{General conclusion and open questions}
As a matter of fact, quantum mechanics has penetrated almost every nook and cranny of modern science in the last decades, not only as a fundamental theory but also as modern technology. It has harnessed its resources to accomplish many achievements, mainly driven by the perspective of quantum computers \cite{ladd2010quantum} and quantum communications \cite{gisin2007quantum}. Quantum metrology bears substantial evidence to support these claims since it is the most brand-new technology promising that can enter into practice shortly. Indeed, quantum metrology provides a natural extension of classical statistics and aims to drive performance improvement using quantum mechanics terms. In this thesis, we have provided the fundamental technique used in quantum metrology.
	
	Believing that it is not easy to understand quantum metrology without a deep understanding of its classical counterparts, we have dedicated the first chapter \ref{Ch. 1} of this thesis to discuss the basic concepts of classical metrology. We have presented this chapter like the textbook with many application examples, which we hope will help readers follow and clearly understand the estimation problems. It is a powerful approach that provides a method to visualize the estimation process and helps to comprehend how to evaluate different concepts. More precisely, this chapter presented the fundamental concepts of classical estimation theory, such as the probability theory of a random variable, classical statistical model, and estimator notion, MSE. All these concepts construct the classical estimation problem. We have introduced and proved CR which is the key to setting the lower bound on MSE or variance. Also, we have presented the CFI and CFIM in both the single-parameter and multi-parameter cases, as they have a substantial role in evaluating the CRB. We have also discussed the MLE as an appropriate principle for finding the efficient estimator reaching the CRLB. Given the importance of improving accuracy in estimation problems,  it is necessary to understand what resources are needed to improve the precision of measurements. Quantum mechanics and their operations are invaluable resources in this direction.

	In the second chapter \ref{Ch. 2}, we have reviewed the methods and approaches followed to express all central quantities of the classical estimation theory in quantum mechanics terms. We have introduced the quantum statistical model defined using the density operator instead of the PDF used in the classical ones. Due to the non-commutativity nature of quantum mechanics, many families of QFIs have been derived. For the single parameter, we reviewed the SLD and RLD quantum Fisher information and the corresponding QCRBs, and we have proved that the RLD-QCRB is not practical in this case because it is not tight. In the multiparameter case, we have derived the SLD and RLD-QFIMs and the corresponding QCRBs. Due to their importance in evaluating different QCRBs, we have discussed explicit SLD and RLD-QFIM forms for finite-dimensional quantum systems. These forms allow us to compute QCRBs and then extract the estimated values of the unknown parameters. We have provided a comparison of these bounds and discussed their attainability. Specifically, we pointed out that the commonly used SLDs and RLD-QCRBs for several parameters are often not simultaneously saturable, which adds to the difficulty of extracting optimal measurements of unknown parameters. The Holevo Cramér-Rao bound then proves to be the most informative alternative bound, although it is defined with an untraceable form for arbitrary quantum states. This untraceable form of HCRB is still an open question that provides an area of current research interest. We have also discussed the different classifications of quantum statistical models, which are; classical, \textit{quasi-classical}, asymptotically classical, and D-invariant models. Despite all these successes, this chapter has left without discussing the quantum statistical modes described by the density operator living in the Hilbert space of infinite-dimensional systems. This issue is the subject that we have addressed in chapter \ref{Ch. 3}.
	
	As a matter of fact, many practical physical applications have quantum statistical models described by the state of light. This state is often a probe field of continuous and infinity spectrum acting in the Hilbert space of infinity-dimensional systems. The most strong archetypes of this probe are the Gaussian states. In chapter \ref{Ch. 3} of this thesis, we have integrated quantum metrology with the formalism of  Gaussian states. More precisely, we have reviewed in first the fundamental formalism of continuous-variable systems, specifically,  Gaussian states and their operations. Then, in the next, we have proved the expressions of the SLD and RLD- QFIMs by explicitly calculating the SLD and RLD-operators corresponding to the multi-mode quantum Gaussian states. Also, we have derived the attainable condition of QCRBs associated with the SLD operators. We have illustrated the derived formalism by treating the optical phase estimation problem as an example. This example was used to clarify the SQL and HL. These are the limits of precision accessible by quantum metrology that are not possible with the resources of classical statistics. The results of this chapter have incredible advantages to apply to all Gaussian protocols since they are expressed in terms of the first and second moments of the Gaussian state. Therefore, we have exploited them to treat, in Chapter \ref{Ch. 4}, a quantum metrology protocol evolving under a noisy Gaussian environment.

	The common problem between quantum metrology and the process of open quantum systems is to develop a measurement scheme that leads to going beyond SQL with less influence from the unavoidable interaction with the environment. By exploiting Gaussian states and their operations, we have proposed and analyzed, in Chapter \ref{Ch. 4}, an adaptive scheme of a protocol that aims at estimating the parameters characterizing the displacement operator under the effect of environmental noise. We have first offered a brief review of the Gaussian non-unitary channels in the single and multi-mode cases. Then, we have reviewed the relevant Gaussian measurements, which are homodyne and heterodyne detection. Using homodyne detection with SLD, RLD-QCRBs, and HCRB, we have investigated the ultimate limit of precision measurement in estimating the displacement parameters under a noisy Gaussian environment. As expected, we found that the accuracy of measurement is reduced due to the effect of the environment. This fact occurs when evolving the coherent states as probe states. Alternatively, if we choose a squeezed state as the probe state, we get the required enhancement in the estimation precision beyond the SQL even when the existence of environmental noise. Finally, we have explained the results obtained by the role of quantum entanglement, which shows again that it is a crucial resource to reach the SQL and then allows to overcome the constraints imposed by the inevitable interaction with the environment.
	
	Before closing this thesis, we list many elusive issues that deserve further study in the future. The first of them concerns the quantum multiparameter case. Although, many schemes of multiparameter estimation have shown advantages over the single-parameter ones. However, the evaluation of HCRB  remains an elaborate obstacle in these metrological schemes. The second open question arises from the results of Chapter \ref{Ch. 3}, in which we have focused on expressing the SLDs, RLD-QFIMs, and then the corresponding QCRBs for multi-mode Gaussian states. Therefore, it is interesting to see how to extend these results to non-Gaussian states. In Chapter \ref{Ch. 4}, we discussed the role of entanglement as a resource needed to improve accuracy in the Gaussian noise environment. Thus, it is interesting to see where accuracy improvement is related to quantum correlations beyond entanglement \cite{slaoui2019comparative, lahlou2022quantifying, slaoui2018universal, slaoui2020influence, slaoui2018dynamics}. Also, along these lines, can QFIM be exploited to quantify the amount of entanglement? Throughout this thesis, we have only discussed SLD and RLD-QFIM. We expect that other QFIM families, such as MAX-QFI \cite{hayashi2002two}, and anti-symmetric logarithmic derivative-QFI (ALD-QFI) \cite{roy2019fundamental}, have also given another QCRB and a new classification of the quantum statistical model. In addition to quantum metrology, QFIM also connects to other aspects of quantum physics. For example, the quantum phase transition \cite{wang2014quantum, gu2010fidelity}, the Fubini study metric, and a Kahler metric in the Hilbert complex projective space \cite{sidhu2020geometric, liu2019quantum}. All these connections encourage us to think about contributing and developing other manuscripts in these areas.
	
	Finally, we hope that the analysis and approaches followed in this thesis have proven helpful to the reader in understanding quantum metrology and its applications.
	\chapter{Supplementary}\label{Ch. 5}

	\section{Appendix A: SLD-operator}\label{AppA}
	To determine the expressions of the \textbf{SLD} and \textbf{RLD}-operators and the corresponding \textbf{QFIM}s, we have to use many properties of the characteristic function of Gaussian states. Indeed, this function can be rewritten as follows;
	\begin{equation}\label{Eq. 4.1}
		\chi_{\hat{\rho}}(\mathbf{R})=\operatorname{Tr}[\hat{D} \hat{\rho}] \stackrel{(\ref{Eq. 3.34})}{=\joinrel=\joinrel=\joinrel=} \operatorname{Tr}\left[e^{i \tilde{\mathbf{q}}^{\top} \hat{\mathbf{Q}}} e^{i \tilde{\mathbf{p}}^{\top} \hat{\mathbf{P}}} e^{\frac{i}{2} \tilde{\mathbf{q}}^{\top} \tilde{\mathbf{p}}} \hat{\rho}\right]=\operatorname{Tr}\left[e^{i\tilde{\mathbf{p}}^{\top} \hat{\mathbf{P}}} e^{i \tilde{\mathbf{q}}^{\top} \hat{\mathbf{Q}}} e^{-\frac{i}{2} \tilde{\mathbf{q}}^{\top} \tilde{\mathbf{p}}} \hat{\rho}\right] 
	\end{equation}
	where we defined the $n$-dimensional vectors $\hat{\mathbf{Q}}(\hat{\mathbf{P}})$ and $\tilde{\mathbf{q}}(\tilde{\mathbf{p}})$ as the vectors of odd (even) entries of the parent vectors $\hat{\mathbf{R}}$ and $\tilde{\mathbf{R}}$, such that $\hat{\mathbf{Q}} \oplus \hat{\mathbf{P}}=\hat{\mathbf{R}}$ and $\tilde{\mathbf{q}} \oplus \tilde{\mathbf{p}}=\tilde{\mathbf{R}}$. This decomposition follows from the Baker-Campbell-Hausdorff formula. Now, differentiating the last two equalities of (\ref{Eq. 4.1}) with respect to $\tilde{R}_{k}(\tilde{q}_{k},\tilde{p}_{k})$ yields
	\begin{equation}
		\partial_{\tilde{q}_{k}} \chi_{\hat{\rho}}=i \operatorname{Tr}\left[\mathrm{e}^{i \tilde{\mathbf{R}}^{\top} \hat{\mathbf{R}}}\hspace{0.1cm} \hat{\rho}\hspace{0.1cm}
		\hat{Q}_{k}\right]+\frac{i}{2} \tilde{p}_{k} \chi_{\hat{\rho}}=i \operatorname{Tr}\left[\mathrm{e}^{i \tilde{\mathbf{R}}^{\top} \hat{\mathbf{R}}}\hspace{0.1cm} \hat{Q}_{k} \hspace{0.1cm}\hat{\rho}\right]-\frac{i}{2} \tilde{p}_{k} \chi_{\hat{\rho}}
	\end{equation}
	\begin{equation}
		\partial_{\tilde{p}_{k}} \chi_{\hat{\rho}}=i \operatorname{Tr}\left[\mathrm{e}^{i \tilde{\mathbf{R}}^{\top} \hat{\mathbf{R}}}\hspace{0.1cm} \hat{\rho}\hspace{0.1cm} \hat{P}_{k}\right]-\frac{i}{2} \tilde{q}_{k} \chi_{\hat{\rho}}=i \operatorname{Tr}\left[\mathrm{e}^{i \tilde{\mathbf{R}}^{\top} \hat{\mathbf{R}}}\hspace{0.1cm} \hat{P}_{k}\hspace{0.1cm} \hat{\rho}\right]+\frac{i}{2} \tilde{q}_{k} \chi_{\hat{\rho}}
	\end{equation}
	hence
	\begin{equation}\label{Eq. 4.4}
		\partial_{\tilde{q}_{k}} \chi_{\hat{\rho}}=\frac{i}{2}\operatorname{Tr}\left[\mathrm{e}^{i \tilde{\mathbf{R}}^{\top} \hat{\mathbf{R}}}\left(\hat{Q}_{k} \hspace{0.1cm}\hat{\rho}+\hat{\rho}\hspace{0.1cm} \hat{Q}_{k}\right)\right],\hspace{0.5cm} \partial_{\tilde{p}_{k}} \chi_{\hat{\rho}}=\frac{i}{2}\operatorname{Tr}\left[\mathrm{e}^{i \tilde{\mathbf{R}}^{\top} \hat{\mathbf{R}}}\left(\hat{P}_{k} \hspace{0.1cm}\hat{\rho}+\hat{\rho}\hspace{0.1cm} \hat{P}_{k}\right)\right]
	\end{equation}
	\begin{equation}\label{Eq. 4.5}
		\tilde{p}_{k} \chi_{\hat{\rho}}= \operatorname{Tr}\left[\mathrm{e}^{i \tilde{\mathbf{R}}^{\top} \hat{\mathbf{R}}}\left(\hat{Q}_{k} \hspace{0.1cm}\hat{\rho}-\hat{\rho}\hspace{0.1cm} \hat{Q}_{k}\right)\right], \hspace{0.7cm} \tilde{q}_{k}\chi_{\hat{\rho}}=-\operatorname{Tr}\left[\mathrm{e}^{i \tilde{\mathbf{R}}^{\top} \hat{\mathbf{R}}}\left(\hat{P}_{k} \hspace{0.1cm}\hat{\rho}-\hat{\rho}\hspace{0.1cm} \hat{P}_{k}\right)\right].
	\end{equation}
	Going back to the variables $\boldsymbol{\tilde{R}}$, the general correspondence of the last equalities ((\ref{Eq. 4.4}), (\ref{Eq. 4.5})) are
	\begin{equation}
		\partial_{\tilde{R}_{k}} \chi_{\hat{\rho}}=\frac{i}{2}\operatorname{Tr}\left[\mathrm{e}^{i \tilde{\mathbf{R}}^{\top} \hat{\mathbf{R}}}\left(\hat{R}_{k} \hspace{0.1cm}\hat{\rho}+\hat{\rho}\hspace{0.1cm} \hat{R}_{k}\right)\right]\hspace{0.7cm} 	\tilde{R}_{k} \chi_{\hat{\rho}}=\Omega_{k'k} \operatorname{Tr}\left[\mathrm{e}^{i \tilde{\mathbf{R}}^{\top} \hat{\mathbf{R}}}\left(\hat{R}_{k'} \hspace{0.1cm}\hat{\rho}-\hat{\rho}\hspace{0.1cm} \hat{R}_{k'}\right)\right]
	\end{equation}
	which leads to 
	\begin{equation}\label{Eq. 4.7}
		\operatorname{Tr}\left[\hat{D}\hspace{0.1cm} \hat{\rho} \hspace{0.1cm}\hat{R}_{k}\right]=\left(-i \partial_{\tilde{R}_{k}}-\frac{1}{2} \Omega_{k k^{\prime}} \tilde{R}_{k^{\prime}}\right) \chi_{\hat{\rho}}  \hspace{0.7cm} 	\operatorname{Tr}\left[\hat{D}\left(\hat{\rho}\hspace{0.1cm} \hat{R}_{k}+\hat{R}_{k}\hspace{0.1cm} \hat{\rho}\right)\right]=-2 i \partial_{\tilde{R}_{k}} \chi_{\hat{\rho}}
	\end{equation}
	Following the same approach for the second derivative, we obtain
	\begin{equation}
		\begin{aligned}
			&\operatorname{Tr}\left[\hat{D}\hspace{0.1cm} \hat{\rho} \hspace{0.1cm}\hat{R}_{j}\hspace{0.1cm} \hat{R}_{k}\right]=\left(-i \partial_{\tilde{R}_{k}}-\frac{1}{2} \Omega_{k k^{\prime}} \tilde{R}_{k^{\prime}}\right)\left(-i \partial_{\tilde{R}_{j}}-\frac{1}{2} \Omega_{j j^{\prime}} \tilde{R}_{j^{\prime}}\right) \chi_{\hat{\rho}}\quad\\
			&\operatorname{Tr}\left[\hat{D}\hspace{0.1cm} \left(\hat{\rho}\hspace{0.1cm} \hat{R}_{j}\hspace{0.1cm} \hat{R}_{k}+\hat{R}_{k}\hspace{0.1cm} \hat{R}_{j}\hspace{0.1cm} \hat{\rho}\right)\right]=\frac{1}{2}\left(\Omega_{j j^{\prime}} \tilde{R}_{j^{\prime}} \Omega_{k k^{\prime}} \tilde{R}_{k^{\prime}}-4 \partial_{\tilde{R}_{k}} \partial_{\tilde{R}_{j}}\right) \chi_{\hat{\rho}}
		\end{aligned}\label{Eq. 4.8}
	\end{equation}
	The next step is the derivative of the Gaussian characteristic function given in Eq. (\ref{Eq. 3.35}) with respect to the estimation parameter $\theta_{\mu=1,2,...,m}$\footnote{Recall here that $\boldsymbol{d}$ and $\boldsymbol{\mathrm V}$ are the functions depending on $\boldsymbol{\theta}$.}, and with respect to $\tilde{R}$
	\begin{equation}\label{Eq. 4.9}
		\begin{aligned}
			\partial_{\theta_{\mu}}	\chi_{\hat \rho} &=\left(i \tilde{R}_{p} \partial_{\theta_{\mu}}	d_{p}-\frac{1}{4} \partial_{\theta_{\mu}}{\mathrm V}_{l m}\tilde{R}_{l} \tilde{R}_{m}\right) \chi_{\hat \rho},\\
			\partial_{\tilde{R}_{j}} \chi_{\hat \rho} &=\left(i d_{j}-\frac{1}{2} {\mathrm V}_{j j^{\prime}} \tilde{R}_{j^{\prime}}\right) \chi_{\hat \rho} \\
			\partial_{\tilde{R}_{k}} \partial_{\tilde{R}_{j}} \chi_{\hat \rho} &=\left[\left(i d_{k}-\frac{1}{2} {\mathrm V}_{k k^{\prime}} \tilde{R}_{k^{\prime}}\right)\left(i d_{j}-\frac{1}{2} {\mathrm V}_{j j^{\prime}} \tilde{R}_{j^{\prime}}\right)-\frac{1}{2} {\mathrm V}_{j k}\right] \chi_{\hat \rho},
		\end{aligned}
	\end{equation}
	Most of the above equalities play a central role in computing the explicit expression of the \textbf{SLD} operator. We will now determine the expression of the \textbf{SLD} operator. In each path, we will mention the form used 
	\begin{equation}
		\begin{aligned}
			\partial_{\theta_{\mu}} \chi_{\hat{\rho}} &=\operatorname{Tr}\left[\hat{D}\hspace{0.1cm} \partial_{\theta_{\mu}} \hat{\rho}\right] \\
			& \stackrel{(\ref{SLD})}{=\joinrel=\joinrel=\joinrel=} \operatorname{Tr}\left[\hat{D}\hspace{0.1cm} \hat{\rho} \hspace{0.1cm}\hat{\mathcal{L}}_{\theta_{\mu}}^{(S)}\right] \\
			& \stackrel{\left(\ref{Eq. 3.36}\right)}{=\joinrel=\joinrel=\joinrel=} \mathcal{ L}^{(S)^{(0)}} \operatorname{Tr}[\hat{D}\hspace{0.1cm} \hat{\rho}]+\mathcal{ L}_{l}^{(S)^{(1)}} \operatorname{Tr}\left[\hat{D}\hspace{0.1cm} \hat{\rho}\hspace{0.1cm} \hat{R}_{l}\right]+\mathcal{ L}_{jk}^{(S)^{(2)}} \operatorname{Tr}\left[\hat{D} \hspace{0.1cm}\hat{\rho}\hspace{0.1cm} \hat{R}_{j}\hspace{0.1cm} \hat{R}_{k}\right] \\
			& \stackrel{\left((\ref{Eq. 4.7}), (\ref{Eq. 4.8})\right)}{=\joinrel=\joinrel=\joinrel=\joinrel=\joinrel=\joinrel=} \mathcal{L}^{(S)^{(0)}} \chi_{\hat{\rho}}+\mathcal{ L}_{l}^{(S)^{(1)}}\left(-i \partial_{\tilde{R}_{l}}-\frac{1}{2} \Omega_{ll^{\prime}} \tilde{R}_{l^{\prime}}\right) \chi_{\hat{\rho}}\\
			&\hspace{3.6cm} +\mathcal{ L}_{j k}^{(S)^{(2)}}\left(-i \partial_{\tilde{R}_{k}}-\frac{1}{2} \Omega_{k k^{\prime}} \bar{R}_{k^{\prime}}\right)\left(-i \partial_{\tilde{R}_{j}}-\frac{1}{2} \Omega_{j j^{\prime}} \tilde{R}_{j^{\prime}}\right) \chi_{\hat{\rho}}
		\end{aligned}
	\end{equation}
	Now we will insert the result of (\ref{Eq. 4.9})  into last equality, we get 
	\begin{eqnarray}
		\begin{aligned}
			\left(i \tilde{R}_{p} \partial_{\theta_{\mu}} d_{p}-\frac{1}{4} \partial_{\theta_{\mu}} {\mathrm V}_{lm} \tilde{R}_{l} \tilde{R}_{m}\right) \chi_{\hat{\rho}}&=\mathcal{ L}^{(S)^{(0)}} \chi_{\hat{\rho}}+\mathcal{ L}_{l}^{(S)^{(1)}}\left(\frac{i}{2} {\mathrm V}_{ll^{\prime}} \tilde{R}_{l^{\prime}}+d_{l}-\frac{1}{2} \Omega_{ll'} \tilde{R}_{l'}\right) \chi_{\hat{\rho}}\\
			&+\mathcal{ L}_{jk}^{(S)^{(2)}}\left(\left(\frac{1}{2} {\mathrm V}_{j j^{\prime}} \tilde{R}_{j^{\prime}}-i d_{j}\right)\left(-\frac{1}{2} {\mathrm V}_{k k^{\prime}} \tilde{R}_{k^{\prime}}+i d_{k}\right)\right) \chi_{\hat{\rho}}\\
			&+\mathcal{ L}_{jk}^{(S)^{(2)}}\left(\frac{1}{2} {\mathrm V}_{j k}+\frac{i}{2} \Omega_{j k}-\frac{i}{4} \Omega_{jj^{\prime}} {\mathrm V}_{k k^{\prime}} \tilde{R}_{j^{\prime}} \tilde{R}_{k^{\prime}}\right)\chi_{\hat{\rho}}\\
			&-\mathcal{ L}_{j k}^{(S)^{(2)}}\left(\frac{i}{4} \Omega_{k k^{\prime}} \sigma_{j j^{\prime}} \tilde{R}_{k^{\prime}} \tilde{R}_{j^{\prime}}+\frac{1}{2} \Omega_{j j^{\prime}} \tilde{R}_{j^{\prime}} d_{k}\right) \chi_{\hat{\rho}}\\
			&+\mathcal{ L}_{jk}^{(S)^{(2)}}\left(\frac{1}{4} \Omega_{j j^{\prime}} \Omega_{k k^{\prime}} \tilde{R}_{j^{\prime}} \tilde{R}_{k^{\prime}}-\frac{1}{2} \Omega_{k k^{\prime}} \tilde{R}_{k^{\prime}} d_{j}\right)\chi_{\hat{\rho}}
		\end{aligned}
	\end{eqnarray}\label{Eq. 4.11}
	We have divided by $\chi_{\hat{\rho}}$, which is allowed since it is never zero. This must be true for all $\boldsymbol{\tilde{R}}$. Thus, we can equalize the different orders of the last equation independently. It is convenient to return to a geometric representation, without indices, of the matrices involved. The identification of second-order terms of Eq. (\ref{Eq. 4.11}) leads to: 
	\begin{equation}
		\partial_{\theta_{\mu}}\boldsymbol{\mathrm V}=\boldsymbol{\mathrm V}\mathcal{\hat L}^{(S)^{(2)}} \boldsymbol{\mathrm V}-\boldsymbol{\Omega} \mathcal{\hat L}^{(S)^{(2)}} \boldsymbol{\Omega}
	\end{equation}
	To determine the expression of $\mathcal{\hat L}^{(S)^{(2)}}$, one employs the property
	\begin{equation}\label{Eq. 4.13}
		\operatorname{vec}[A B C]=\left(C^{\dagger} \otimes A\right) \operatorname{vec}[B]
	\end{equation}
	where $A, B$ and $C$ are the arbitrary matrices. Thus, we get
	\begin{equation}\label{Eq. 4.14}
		\operatorname{vec}\left[\mathcal{ L}_{\theta_{\mu}}^{S^{(2)}}\right]=\left(\boldsymbol{\mathrm V} \otimes \boldsymbol{\mathrm V}+\boldsymbol{\Omega} \otimes \boldsymbol{\Omega}\right)^{+} \operatorname{vec}\left[\partial_{\theta_{\mu}} \boldsymbol{\mathrm V}\right].
	\end{equation}
	The identification of first-order terms of Eq. (\ref{Eq. 4.11}) leads to:
	\begin{equation}
		2 \partial_{\theta_{\mu}} d_{p}=\mathcal{ L}_{l}^{S^{(2)}} {\mathrm V}_{ll'}+\mathcal{ L}_{jk}^{S^{(2)}}\left(d_{j} {\mathrm V}_{k k^{\prime}}+d_{k} {\mathrm V}_{j j^{\prime}}\right),
	\end{equation}
	and the corresponding matrix form is 
	\begin{equation}
		\mathcal{\hat L}_{\theta_{\mu}}^{(S)^{(1)}}=2 \boldsymbol{\mathrm V}^{-1} \partial_{\theta_{\mu}} \mathbf{d}-2 \mathcal{\hat L}_{\theta_{\mu}}^{(S)^{(2)}} \mathbf{d}.
	\end{equation}
	The identification of zero-order terms of Eq. (\ref{Eq. 4.11}) leads to :
	\begin{equation}
		2 \mathcal{ L}^{(S)^{(0)}}+2 \mathcal{ L}_{l}^{(S)^{(1)}} d_{l}+\mathcal{ L}_{j k}^{(S)^{(2)}} {\mathrm V}_{jk}+2 \mathcal{ L}_{j k}^{(S){(2)}} d_{j} d_{k}=0^,
	\end{equation}
	which takes the following matrix form
	\begin{equation}
		\mathcal{\hat L}_{\theta_{\mu}}^{(S)^{(0)}}=-\frac{1}{2} \operatorname{Tr}\left[\mathcal{\hat L}_{\theta_{\mu}}^{(S)^{(2)}} \boldsymbol{\mathrm V}\right]-\mathbf{d}^{\top}\mathcal{\hat L}_{\theta_{\mu}}^{(S)^{(1)}}-\mathbf{d}^{\top} \mathcal{\hat L}_{\theta_{\mu}}^{(S)^{(2)}} \mathbf{d}.
	\end{equation}
	Now that we have obtained the different order of the \textbf{SLD} operator, we can proceed to insert it into Eq. (\ref{QFIMSLD1}) to find the expression of \textbf{SLD-QFIM}. This task is the subject of the next appendix.
	\section{Appendix B: SLD-quantum Fisher information matrix}\label{AppB}
	Before proceeding to the insertion of the \textbf{SLD} operator developed above into the definition of \textbf{SLD-QFIM} (\ref{QFIMSLD1}), it is worth recalling that the characteristic function of Gaussian states has verified the following property
	\begin{equation}\label{Eq. 4.19}
		\operatorname{Tr}[\hat{\rho}]={\left. {{\mathop{\rm Tr}\nolimits} [\hat D\hat \rho ]} \right|_{\boldsymbol{\tilde{R}} = 0}}=\left.\chi_{\hat{\rho}}\right|_{\tilde{\mathbf{R}}=0}=1,
	\end{equation}
	which is the fundamental key in the derivation of the elements of \textbf{SLD-QFIM}. Now we are going to insert the \textbf{SLD} operator into the \textbf{SLD-QFIM} defined in (\ref{QFIMSLD1}), then we have  
	\begin{eqnarray}
		\begin{aligned}
			\left[\mathcal{F}_Q^{(S)}\right]_{\theta_{\mu} \theta_{\nu}}&= \operatorname{Tr}\left[\partial_{\theta_{\mu}} \hat{\rho}\hspace{0.1cm} \mathcal{\hat{L}}_{\theta_{\nu}}^{(S)}\right]\\
			&\stackrel{(\ref{Eq. 3.36})}{=\joinrel=\joinrel=\joinrel=} \operatorname{Tr}\left[\partial_{\theta_{\mu}} \hat{\rho}\left(\mathcal{{L}}^{(S)^{(0)}}+\mathcal{{L}}_{l}^{(S)^{(1)}} \hat{R}_{l}+\mathcal{{L}}_{j k}^{(S)^{(2)}} \hat{R}_{j} \hat{R}_{k}\right)\right]\\
			&=\mathcal{{L}}^{(S)^{(0)}} \operatorname{Tr}\left[\partial_{\theta_{\mu}} \hat{\rho}\right]+\mathcal{{L}}_{l}^{(S)^{(1)}} \operatorname{Tr}\left[\partial_{\theta_{\mu}} \hat{\rho}\hspace{0.1cm} \hat{R}_{l}\right]+\mathcal{{L}}_{j k}^{(S)^{(2)}} \operatorname{Tr}\left[\partial_{\theta_{\mu}} \hat{\rho}\hspace{0.1cm} \hat{R}_{j} \hat{R}_{k}\right]\\
			&\left.\stackrel{(\ref{Eq. 4.19})}{=\joinrel=\joinrel=\joinrel=}  \mathcal{{L}}^{S^{(0)}} \partial_{\theta_{\mu}} \chi_{\hat{\rho}}\right|_{\boldsymbol{\tilde{R}}=0}+\left.\mathcal{{L}}_{l}^{(S)^{(1)}} \operatorname{Tr}\left[\hat{D} \hspace{0.1cm}\hat{\rho}\hspace{0.1cm} \hat{R}_{l}\right] \partial_{\theta_{\mu}} \chi_{\hat{\rho}}\right|_{\boldsymbol{\tilde{R}}=0}\\
			&\hspace{4cm}+\left.\mathcal{{L}}_{j k}^{(S)^{(2)}} \operatorname{Tr}\left[\hat{D}\hspace{0.1cm} \hat{\rho}\hspace{0.1cm} \hat{R}_{j} \hat{R}_{k}\right] \partial_{\theta_{\mu}} \chi_{\hat{\rho}}\right|_{\boldsymbol{\tilde{R}}=0}\\
			&\stackrel{((\ref{Eq. 4.7}), (\ref{Eq. 4.8}))}{=\joinrel=\joinrel=\joinrel=\joinrel=\joinrel=\joinrel=\joinrel=}\left.\mathcal{{L}}_{l}^{(S)^{(1)}}\left(-i \partial_{\tilde{R}_{l}}-\frac{1}{2} \Omega_{l l^{\prime}} \tilde{R}_{l^{\prime}}\right) \partial_{\theta_{\mu}} \chi_{\hat{\rho}}\right|_{\boldsymbol{\tilde{R}}=0}\\
			&\hspace{3cm}+\left.\mathcal{{L}}_{j k}^{(S)^{(2)}}\left(-i \partial_{\tilde{R} j}-\frac{1}{2} \Omega_{j j^{\prime}} \tilde{R}_{j^{\prime}}\right)\left(-i \partial_{\tilde{R}_{k}}-\frac{1}{2} \Omega_{k k^{\prime}} \tilde{R}_{k^{\prime}}\right) \partial_{\theta_{\mu}} \chi_{\hat{\rho}}\right|_{\boldsymbol{\tilde{R}}=0}\\
			&\stackrel{(\ref{Eq. 4.9})}{=\joinrel=\joinrel=\joinrel=} \left.\mathcal{{L}}_{l}^{(S)^{(1)}}\left(-i \partial_{\tilde{R}_{l}}-\frac{1}{2} \Omega_{ll'}\tilde{R}_{l'}\right)\left(i \tilde{R}_{p} \partial_{\theta_{\mu}} d_{p}-\frac{1}{4} \partial_{\theta_{\mu}} {\mathrm V}_{p m} \tilde{R}_{p} \tilde{R}_{m}\right) \chi_{\hat{\rho}}\right|_{\boldsymbol{\tilde{R}}=0}+\\
			&\hspace{-0.3cm}\left.\mathcal{{L}}_{j k}^{(S)^{(2)}}\left(-i \partial_{\tilde{R}_{j}}-\frac{1}{2} \Omega_{j j^{\prime}} \tilde{R}_{j^{\prime}}\right)\left(-i \partial_{\tilde{R}_{k}}-\frac{1}{2} \Omega_{k k^{\prime}} \tilde{R}_{k^{\prime}}\right)\left(i \tilde{R}_{p} \partial_{\theta_{\mu}} d_{p}-\frac{1}{4} \partial_{\theta_{\mu}} {\mathrm V}_{p m} \tilde{R}_{p} \tilde{R}_{m}\right) \chi_{\hat{\rho}}\right|_{\boldsymbol{\tilde{R}}=0} .
		\end{aligned}
	\end{eqnarray}
	Evaluating the last equation when $\tilde{\mathbf{R}}=0$, one gets
	\begin{equation}
		\left[\mathcal{F}_Q^{(S)}\right]_{\theta_{\mu} \theta_{\nu}}=\mathcal{{L}}_{l}^{(S)^{(1)}} \partial_{\theta_{\mu}} d_{l}+\frac{1}{2} \mathcal{{L}}_{j k}^{(S)^{(2)}} \partial_{\theta_{\mu}} {\mathrm V}_{j k}+2 \mathcal{{L}}_{j k}^{(S)^{(2)}} \partial_{\theta_{\mu}} d_{j} d_{k},
	\end{equation}
	which rewrites in the matrix representation as
	\begin{equation}
		\left[\mathcal{F}_Q^{(S)}\right]_{\theta_{\mu} \theta_{\nu}}=\partial_{\theta_{\mu}} \mathbf{d}^{\top} \mathcal{\hat{L}}_{\theta_{\nu}}^{(S)^{(1)}}+\frac{1}{2} \operatorname{Tr}\left[\partial_{\theta_{\mu}} \boldsymbol{{\mathrm V}}\hspace{0.1cm} \mathcal{\hat{L}}_{\theta_{\nu}}^{(S)^{(2)}}\right]+2 \partial_{\theta_{\mu}} \mathbf{d}^{T} \mathcal{\hat{L}}_{\theta_{\nu}}^{(S)^{(2)}} \mathbf{d}.
	\end{equation}
	Replacing $\mathcal{\hat{L}}_{\theta_{\nu}}^{(S)^{(2)}}$ by its expression,  which is derived in Eq. (\ref{Eq. 4.14}),  together with using the property of trace\footnote{$\rm Tr\left(A^{\dagger} B\right)=\operatorname{vec}[A]^{\dagger} \operatorname{vec}[B]$}, leads to 
	\begin{equation}
		\left[\mathcal{F}_Q^{(S)}\right]_{\theta_{\mu} \theta_{\nu}}=\frac{1}{2} \operatorname{vec}\left[\partial_{\theta_{\mu}} \boldsymbol{{\mathrm V}}\right]^{\dagger} \left(\boldsymbol{\mathrm V} \otimes \boldsymbol{\mathrm V}+\boldsymbol{\Omega} \otimes \boldsymbol{\Omega}\right)^{+} \operatorname{vec}\left[\partial_{\theta_{\nu}} \boldsymbol{{\mathrm V}}\right]+2 \partial_{\theta_{\mu}} \mathbf{d}^{\top} \boldsymbol{{\mathrm V}}^{-1} \partial_{\theta_{\nu}} \mathbf{d}
	\end{equation}
	if $\left(\boldsymbol{\mathrm V} \otimes \boldsymbol{\mathrm V}+\boldsymbol{\Omega} \otimes \boldsymbol{\Omega}\right)$ is invertible, then we have
	\begin{equation}
		\left[\mathcal{F}_Q^{(S)}\right]_{\theta_{\mu} \theta_{\nu}}=\frac{1}{2} \operatorname{vec}\left[\partial_{\theta_{\mu}} \boldsymbol{\mathrm V} \right]^{\dagger}\left(\boldsymbol{\mathrm V} \otimes \boldsymbol{\mathrm V}+\boldsymbol{\Omega} \otimes \boldsymbol{\Omega}\right)^{-1} \operatorname{vec}\left[\partial_{\theta_{\nu}} \boldsymbol{\mathrm V} \right]+2 \partial_{\theta_{\mu}} \mathbf{d}^{\top} \boldsymbol{\mathrm V}^{-1} \partial_{\theta_{\nu}} \mathbf{d}. 
	\end{equation}
	\section{Appendix C: RLD-operator}\label{AppC}
	Analogously, to find the explicit expression of the \textbf{RLD}-quantum Fisher information matrix, one must first determine the analytic expression of the corresponding \textbf{RLD}-operator. To express the \textbf{RLD}-operator, we consider it to be quadratic in the canonical operators, as illustrated in Eq. (\ref{Eq. 3.44}). And the central goal of this appendix is the finding the different components of $\hat{\mathcal{L}}^{(R)}_{\theta_{\mu}}$, which are $\hat{\mathcal{L}}_{\theta_{\mu}}^{(R)^{(0)}}$, $\hat{\mathcal{L}}_{\theta_{\mu}}^{(R)^{(1)}}$ and $\hat{\mathcal{L}}_{\theta_{\mu}}^{(R)^{(2)}}$. We will follow the same guidelines as for \textbf{SLD}-operator, and we will start with
	\begin{eqnarray}
		\begin{aligned}
			\partial_{\theta_{\mu}} \chi_{\hat{\rho}} &=\operatorname{Tr}\left[\hat{D}\hspace{0.1cm} \partial_{\theta_{\mu}} \hat{\rho}\right] \\
			& \stackrel{(\ref{RLD})}{=\joinrel=\joinrel=\joinrel=} \operatorname{Tr}\left[\hat{D}\hspace{0.1cm} \hat{\rho}\hspace{0.1cm} \hat{\mathcal{L}}_{\theta_{\mu}}^{(R)}\right] \\
			& \stackrel{(\ref{Eq. 3.44})}{=\joinrel=\joinrel=\joinrel=} \mathcal{L}^{(R)^{(0)}} \operatorname{Tr}[\hat{D}\hspace{0.1cm} \hat{\rho}]+\mathcal{L}_{l}^{(R)^{(1)}} \operatorname{Tr}\left[\hat{D}\hspace{0.1cm} \hat{\rho}\hspace{0.1cm} \hat{R}_{l}\right]+\mathcal{L}_{j k}^{(R)^{(2)}} \operatorname{Tr}\left[\hat{D}\hspace{0.1cm} \hat{\rho} \hspace{0.1cm}\hat{R}_{j}\hspace{0.1cm} \hat{R}_{k}\right] \\
			& \stackrel{((\ref{Eq. 4.7}), (\ref{Eq. 4.8}))}{=\joinrel=\joinrel=\joinrel=\joinrel=\joinrel= \joinrel=} \mathcal{L}^{(R)^{(0)}} \chi_{\hat{\rho}}+\mathcal{L}_{l}^{(R)^{(1)}}\left(-i \partial_{\tilde{R}_{l}}-\frac{1}{2} \Omega_{l l^{\prime}} \tilde{R}_{l^{\prime}}\right) \chi_{\hat{\rho}}\\
			&\hspace{3cm}+\mathcal{L}_{j k}^{(R)^{(2)}}\left(-i \partial_{\tilde{R}_{k}}-\frac{1}{2} \Omega_{k k^{\prime}} \tilde{R}_{k^{\prime}}\right)\left(-i \partial_{\tilde{R}_{j}}-\frac{1}{2} \Omega_{j j^{\prime}} \tilde{R}_{j^{\prime}}\right) \chi_{\hat{\rho}} .
		\end{aligned}
	\end{eqnarray}
	Substituting the result of (\ref{Eq. 4.9}) into the last equation, we obtain
	\begin{eqnarray}\label{Eq. 4.26}
		\begin{aligned}
			\left(i \tilde{R}_{p} \partial_{\theta_{\mu}} d_{p}-\frac{1}{4} \partial_{\theta_{\mu}} {\mathrm V}_{m p} \tilde{R}_{m} \tilde{R}_{p}\right) \chi_{\hat{\rho}}=& \mathcal{L}^{(R)^{(0)}} \chi_{\hat{\rho}}+\mathcal{L}_{l}^{(R)^{(1)}}\left(\frac{i}{2} {\mathrm V}_{l l^{\prime}} \tilde{R}_{l^{\prime}}+d_{l}-\frac{1}{2} \Omega_{ll'} \tilde{R}_{l'}\right) \chi_{\hat{\rho}}\\
			&\hspace{-0.4cm} +\mathcal{L}_{j k}^{(R)^{(2)}}\left(\left(\frac{1}{2} {\mathrm V}_{j j^{\prime}} \tilde{R}_{j^{\prime}}-i d_{j}\right)\left(-\frac{1}{2} {\mathrm V}_{k k^{\prime}} \tilde{R}_{k^{\prime}}+i d_{k}\right)\right) \chi_{\hat{\rho}}\\
			&\hspace{-1.3cm}-\mathcal{L}_{j k}^{(R)^{(2)}}\left(\frac{i}{4} \Omega_{j j^{\prime}} {\mathrm V}_{k k^{\prime}} \tilde{R}_{j^{\prime}} \tilde{R}_{k^{\prime}}+\frac{i}{4} \Omega_{k k^{\prime}} {\mathrm V}_{j j^{\prime}} \tilde{R}_{k^{\prime}} \tilde{R}_{j^{\prime}}+\frac{1}{2} \Omega_{j j^{\prime}} \tilde{R}_{j^{\prime}} d_{k}\right) \chi_{\hat{\rho}}\\
			&\hspace{-1.3cm} +\mathcal{L}_{j k}^{(R)^{(2)}}\left(\frac{1}{4} \Omega_{j j^{\prime}} \Omega_{k k^{\prime}} \tilde{R}_{j^{\prime}} \tilde{R}_{k^{\prime}}+\frac{i}{2} \Omega_{j k}+\frac{1}{2} {\mathrm V}_{j k}-\frac{1}{2} \Omega_{k k^{\prime}} \tilde{R}_{k^{\prime}} d_{j}\right)\chi_{\hat{\rho}}
		\end{aligned}
	\end{eqnarray}
	We divided by $\chi_{\hat{\rho}}$, which is allowed since it is never zero. Then we can equate the different orders of the last equation independently, which is convenient to switch back to a geometric representation of the matrices involved without indexes. We start with the second-order we obtain
	\begin{equation}
		-\partial_{\theta_{\mu}} {\mathrm V}_{m p} =\mathcal{L}_{j k}^{(R)^{(2)}}\left(\Omega_{j j^{\prime}} \Omega_{k k^{\prime}}-{\mathrm V}_{j j^{\prime}} {\mathrm V}_{k k^{\prime}}+ {\mathrm V}_{k k^{\prime}} \Omega_{j j^{\prime}}\Omega_{k k^{\prime}} {\mathrm V}_{j j^{\prime}} \tilde{R}_{j^{\prime}}\right)
	\end{equation}
	Using a matrix representation (without index), one finds
	\begin{eqnarray}
		\begin{aligned}
			\partial_{\theta_{\mu}} \boldsymbol{\mathrm V}&=\boldsymbol{\mathrm V} \mathcal{\hat L}^{(R)^{(2)}} \boldsymbol{\mathrm V}-\boldsymbol{\Omega} \mathcal{\hat L}^{(R)^{(2)}} \boldsymbol{\Omega}+i \boldsymbol{\mathrm V}\mathcal{\hat L}^{(R)^{(2)}} \boldsymbol{\Omega}+i \boldsymbol{\Omega} \mathcal{\hat L}^{(R)^{(2)}} \boldsymbol{\mathrm V}\\
			&=\left(\boldsymbol{\mathrm V}+i\boldsymbol{\Omega}\right)\mathcal{\hat L}^{(R)^{(2)}}\left(\boldsymbol{\mathrm V}+i\boldsymbol{\Omega}\right)\\
			&=\mathfrak{M}\hspace{0.1cm}\mathcal{\hat L}^{(R)^{(2)}}\mathfrak{M}
		\end{aligned}
	\end{eqnarray}
	where $\mathfrak{M}=\boldsymbol{\mathrm V}+i\boldsymbol{\Omega}$. Using (\ref{Eq.  4.13}), we can solve that last equation, and we have
	\begin{equation}\label{Eq. 4.29}
		\operatorname{vec}\left[\mathcal{\hat L}^{(R)^{(2)}}\right]=\left(\mathfrak{M}^{\dagger} \otimes \mathfrak{M}\right)^{+} \operatorname{vec}\left[\partial_{\theta_{\mu}} \boldsymbol{\mathrm V}\right]. 
	\end{equation}
	The identification of first-order terms of (\ref{Eq. 4.26}) leads to:
	\begin{equation}
		i  \partial_{\theta_{\mu}} d_{p}=\mathcal{L}_{l}^{(R)^{(1)}}\left(\frac{i}{2} {\mathrm V}_{l l'}-\frac{1}{2} \Omega_{ll'}\right)+\mathcal{L}_{j k}^{(R)^{(2)}}\left(\frac{i}{2} {\mathrm V}_{j j^{\prime}} d_{k}+\frac{i}{2} {\mathrm V}_{k k^{\prime}}  d_{j}-\frac{1}{2} \Omega_{j j^{\prime}}  d_{k}-\frac{1}{2} \Omega_{k k^{\prime}} d_{j}\right),
	\end{equation}
	which is rewritten in matrix form as
	\begin{eqnarray}
		\begin{aligned}
			\partial_{\theta_{\mu}} \mathbf{d}&=\frac{1}{2}\left(\boldsymbol{\mathrm V}+i\boldsymbol{\Omega}\right) \mathcal{\hat L}^{(R)^{(1)}}+\left(\boldsymbol{\mathrm V}+i\boldsymbol{\Omega}\right) \mathcal{\hat L}^{(R)^{(2)}} \mathbf{d}\\
			&=\frac{1}{2} \mathfrak{M} \mathcal{\hat L}^{R^{(1)}}+\mathfrak{M} \mathcal{\hat L}^{(R)^{(2)}} \mathbf{d}
		\end{aligned}
	\end{eqnarray}
	Solve this equation leads to
	\begin{equation}
		\mathcal{\hat L}^{(R)^{(1)}}=2 \mathfrak{M}^{+} \partial_{\theta_{\mu}} \mathbf{d}-2 \mathcal{\hat L}^{(R)^{(2)}} \mathbf{d}
	\end{equation}
	Finally, the identification of zero-order terms of (\ref{Eq. 4.26}) leads to:
	\begin{equation}
		0=\mathcal{L}^{(R)^{(0)}}+\mathcal{L}_{l}^{R^{(1)}} d_{l}+\mathcal{L}_{j k}^{(R)^{(2)}}\left(d_{j} d_{k}+\frac{1}{2} {\mathrm V}_{j k}+\frac{i}{2} \Omega_{j k}\right),
	\end{equation}
	which takes the following matrix form
	\begin{equation}
		0=\mathcal{\hat L}^{(R)^{(0)}}+\mathbf{d}^{\top}\mathcal{\hat L}^{(R)^{(1)}} +\mathbf{d}^{\top} \mathcal{\hat L}^{(R)^{(2)}} \mathbf{d}+\frac{1}{2} \operatorname{Tr}\left[\mathfrak{M} \mathcal{L}^{(R)(2)}\right]
	\end{equation}
	Thus, the expression of $\mathcal{\hat L}^{R^{(0)}}$ is 
	\begin{equation}
		\mathcal{\hat L}^{(R)^{(0)}}=-\frac{1}{2} \operatorname{Tr}\left[\mathfrak{M} \mathcal{\hat L}^{(R)^{(2)}}\right]-\mathbf{d}^{\top}\mathcal{\hat L}^{(R)^{(1)}} -\mathbf{d}^{\top} \mathcal{\hat L}^{(R)^{(2)}}\mathbf{d}.
	\end{equation}
	Now that we have obtained the different components of the \textbf{RLD}-operator, we can insert it into (\ref{RLD FI}) in order to derive the elements of the \textbf{RLD}-quantum Fisher information matrix. This task will be performed in the next appendix.
	\section{Appendix D: RLD quantum Fisher information matrix}\label{AppD}
	This appendix is devoted to the derivation of\textbf{ RLD-QFIM}. To realize this purpose, we proceed to insert the expression of \textbf{RLD}-operator that developed above into the definition of \textbf{RLD-QFIM}, then we get
	\begin{eqnarray}
		\begin{aligned}
			\left[\mathcal{F}_Q^{(R)}\right]_{\theta_{\mu} \theta_{\nu}}&={{\rm{Tr}}\left[ {\hat \rho \left( \boldsymbol{\theta}  \right)\hat {\cal L}_{\theta_{\mu}}^{(R)}{{\left( {\hat {\cal L}_{\theta_{\nu}}^{(R)}} \right)}^\dag }} \right]}= \operatorname{Tr}\left[\partial_{\theta_{\mu}} \hat{\rho}\hspace{0.1cm} \left(\mathcal{\hat{L}}_{\theta_{\nu}}^{(R)}\right)^{\dagger}\right]\\
			&\stackrel{(\ref{Eq. 3.44})}{=\joinrel=\joinrel=\joinrel=} \mathcal{\bar L}^{(R)^{(0)}} \operatorname{Tr}\left[\partial_{\theta_{\mu}} \hat{\rho}\right]+\mathcal{\bar L}_{l}^{(R)^{(1)}}  \operatorname{Tr}\left[\partial_{\theta_{\mu}} \hat{\rho}\hspace{0.1cm} \hat{R}_{l}\right]+\mathcal{\bar L}_{jk}^{(R)^{(2)}}  \operatorname{Tr}\left[\partial_{\theta_{\mu}} \hat{\rho}\hspace{0.1cm} \hat{R}_{j} \hat{R}_{k}\right]\\
			&\stackrel{(\ref{Eq. 4.19})}{=\joinrel=\joinrel=\joinrel=} \left.\mathcal{\bar L}^{(R)^{(0)}} \partial_{\theta_{\mu}} \operatorname{Tr}[\hat{D} \hat{\rho}]\right|_{\tilde{\boldsymbol{R}}=0}+\left.\mathcal{\bar L}_{l}^{(R)^{(1)}} \partial_{\theta_{\mu}} \operatorname{Tr}\left[\hat{D}\hspace{0.1cm} \hat{\rho}\hspace{0.1cm} \hat{R}_{l}\right]\right|_{\tilde{\boldsymbol{R}}=0}\\
			&\hspace{4cm}+\left.\mathcal{\bar L}_{jk}^{(R)^{(2)}} \partial_{\theta_{\mu}} \operatorname{Tr}\left[\hat{D}\hspace{0.1cm} \hat{\rho}\hspace{0.1cm} \hat{R}_{k} \hat{R}_{j}\right]\right|_{\tilde{\boldsymbol{R}}=0}\\
			&\left.\stackrel{((\ref{Eq. 4.7}), (\ref{Eq. 4.8}))}{=\joinrel=\joinrel=\joinrel=\joinrel=\joinrel=\joinrel=\joinrel=} \mathcal{\bar L}^{(R)^{(0)}} \partial_{\theta_{\mu}} \chi_{\hat{\rho}}\right|_{\tilde{\boldsymbol{R}}=0}+\left.\mathcal{\bar L}_{l}^{(R)^{(1)}}\left(-i \partial_{\tilde{R}_{l}}-\frac{1}{2} \Omega_{l l^{\prime}} \tilde{R}_{l^{\prime}}\right) \partial_{\theta_{\mu}} \chi_{\hat{\rho}}\right|_{\tilde{\boldsymbol{R}}=0}\\
			&\hspace{2.5cm}+\left.\mathcal{\bar L}_{jk}^{(R)^{(2)}}\left(-i \partial_{\tilde{R}_{j}}-\frac{1}{2} \Omega_{jj^{\prime}} \tilde{R}_{j^{\prime}}\right)\left(-i \partial_{\tilde{R}_{k}}-\frac{1}{2} \Omega_{k k^{\prime}} \tilde{R}_{k^{\prime}}\right) \partial_{\theta_{\mu}} \chi_{\hat{\rho}}\right|_{\tilde{\boldsymbol{R}}=0}.
		\end{aligned}
	\end{eqnarray}
	Replacing $\partial_{\theta_{\mu}} \chi_{\hat{\rho}}$ with the corresponding expression given in (\ref{Eq. 4.9}), one gets 
	\begin{eqnarray}
		\begin{aligned}
			\left[\mathcal{F}_Q^{(R)}\right]_{\theta_{\mu} \theta_{\nu}}&=\left.\mathcal{\bar L}^{(R)^{(0)}}\left(i \tilde{R}_{p} \partial_{\theta_{\mu}} d_{p}-\frac{1}{4} \partial_{\theta_{\mu}} {\mathrm V}_{p m} \tilde{r}_{p} \tilde{R}_{m}\right) \chi_{\hat{\rho}}\right|_{\tilde{\boldsymbol{R}}=0}\\
			&+\left.\mathcal{\bar L}_{l}^{(R)^{(1)}}\left(-i \partial_{\tilde{R}_{l}}-\frac{1}{2} \Omega_{l l^{\prime}} \tilde{R}_{l^{\prime}}\right)\left(i \tilde{R}_{p} \partial_{\theta_{\mu}} d_{p}-\frac{1}{4} \partial_{\theta_{\mu}} {\mathrm V}_{p m} \tilde{R}_{p} \tilde{R}_{m}\right) \chi_{\hat{\rho}}\right|_{\tilde{\boldsymbol{R}}=0}\\
			&\hspace{-1cm}+\left.\mathcal{\bar L}_{jk}^{(R)^{(2)}}\left(-i \partial_{\tilde{R}_{j}}-\frac{1}{2} \Omega_{jj^{\prime}} \tilde{R}_{j^{\prime}}\right)\left(-i \partial_{\tilde{R}_{k}}-\frac{1}{2} \Omega_{k k^{\prime}} \tilde{R}_{k^{\prime}}\right)\left(i \tilde{R}_{p} \partial_{\theta_{\mu}} d_{p}-\frac{1}{4} \partial_{\theta_{\mu}} {\mathrm V}_{p m} \tilde{R}_{p} \tilde{R}_{m}\right) \chi_{\hat{\rho}}\right|_{\tilde{\boldsymbol{R}}=0}
		\end{aligned}
	\end{eqnarray}
	Using the expression of $\partial_{\tilde{R}_j} \chi_{\hat{\rho}}$ given by (\ref{Eq. 4.9}) together with evaluating the result when $\boldsymbol{\tilde{R}}$, leads to
	\begin{equation}
		\left[\mathcal{F}_Q^{(R)}\right]_{\theta_{\mu} \theta_{\nu}}=\mathcal{\bar L}_{jk}^{(R)^{(1)}} \partial_{\theta_{\mu}} d_{l}+\mathcal{\bar L}_{jk}^{(R)^{(2)}}\left(\frac{1}{2} \partial_{\theta_{\mu}} {\mathrm V}_{k j}+2 \partial_{\theta_{\mu}} d_{k} d_{j}\right)
	\end{equation}
	Using the matrix representation without indices, one finds
	\begin{equation}
		\left[\mathcal{F}_Q^{(R)}\right]_{\theta_{\mu} \theta_{\nu}}=\frac{1}{2} \operatorname{Tr}\left[\partial_{\theta_{\mu}} \boldsymbol{{\mathrm V}} \left(\mathcal{L}_{\theta_{\nu}}^{(R)^{(2)}}\right)^{\dagger}\right]+\left(\mathcal{L}_{\theta_{\nu}}^{(R)^{(1)}}\right)^{\dagger} \partial_{\theta_{\mu}} \mathbf{d}+2 \partial_{\theta_{\mu}} \mathbf{d}^{\top} \left(\mathcal{L}_{\theta_{\nu}}^{(R)^{(2)}}\right)^{\dagger} \mathbf{d}
	\end{equation}
	Finally, using the results of equation (\ref{Eq. 4.29}) and the trace property mentioned in the footnote above, we obtain
	\begin{equation}\label{Eq. 4.40}
		\left[\mathcal{F}_Q^{(R)}\right]_{\theta_{\mu} \theta_{\nu}}=\frac{1}{2} \operatorname{vec}\left[\partial_{\theta_{\mu}} \boldsymbol{{\mathrm V}}\right]^{\dagger}\left(\mathfrak{M}^{\dagger} \otimes \mathfrak{M}\right)^{+} \operatorname{vec}\left[\partial_{\theta_{\nu}} \boldsymbol{{\mathrm V}}\right]+2 \partial_{\theta_{\mu}} \mathbf{d}^{\top} \mathfrak{M}^{+} \partial_{\theta_{\nu}} \mathbf{d}.
	\end{equation} 
	if $\mathfrak{M}$ is invertible, then Eq. (\ref{Eq. 4.40}) reduces to
	\begin{equation}\label{Eq. 4.41}
		\left[\mathcal{F}_Q^{(R)}\right]_{\theta_{\mu} \theta_{\nu}}=\frac{1}{2} \operatorname{vec}\left[\partial_{\theta_{\mu}} \boldsymbol{{\mathrm V}}\right]^{\dagger}\left(\mathfrak{M}^{\dagger} \otimes \mathfrak{M}\right)^{-1} \operatorname{vec}\left[\partial_{\theta_{\nu}} \boldsymbol{{\mathrm V}}\right]+2 \partial_{\theta_{\mu}} \mathbf{d}^{\top} \mathfrak{M}^{-1} \partial_{\theta_{\nu}} \mathbf{d}.
	\end{equation}

\newgeometry{left=1.5cm,right=1.5cm,lines=60,top=0in}
\setstretch{0.9}
\AddToShipoutPicture*{
	\unitlength=0.5cm
	\put(2,5){
		\parbox[b][\paperheight]{\paperwidth}{%
			\vfill
			\centering
			{\transparent{0.5}\includegraphics[width=1.4\textwidth]{logo.jpg}}
		}
	}
}
\vspace{-2cm}
\begin{tikzpicture}[overlay,remember picture]
	\draw [line width=2pt,rounded corners=7pt]
	($ (current page.north west) + (.2cm,-.2cm) $)
	rectangle
	($ (current page.south east) + (-.2cm,.2cm) $);
	\draw [line width=1pt,rounded corners=7pt]
	($ (current page.north west) + (.3cm,-.3cm) $)
	rectangle
	($ (current page.south east) + (-.3cm,.3cm) $);
\end{tikzpicture}
	\vspace{-.5cm}
\begin{figure}[H]
	\begin{center}
		\includegraphics[width=0.5\linewidth]{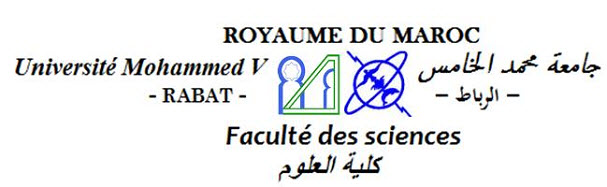}
	\end{center}
\end{figure}
\vspace{-1.5cm}
\begin{center}
	\begin{minipage}{17cm}
		\begin{center}
			{\textcolor{blue}{\fontfamily{pnc}{\selectfont
						{ \textit{CENTRE D’ETUDES DOCTORALES - SCIENCES ET TECHNOLOGIES}}
						\vskip .2cm
						\noindent\hrule height 2pt\vskip 0.2ex\nobreak
						{\textcolor{green}{
								\noindent\hrule height 2pt \vskip 0.2ex}}
			}}}
		\end{center}
	\end{minipage}
\end{center}
\vspace{-.8cm}
\begin{center}
	{\Large \textbf{Abstract}}
\end{center} 
\vspace{-.3cm}
\lettrine[lines=2]Q{uantum} estimation theory is a reformulation of random statistical theory with the modern language of quantum mechanics. Since the mathematical language of quantum mechanics is operator theory, then the probability distribution functions of conventional statistics are replaced by the density operator appearing in its quantum counterpart. Thus, the density operator plays a role similar to that of probability distribution functions in classical probability theory and statistics. However, the use of the probability distribution functions in classical theories is founded on premises that seem intuitively clear enough. Whereas in quantum theory, the situation with operators is different due to its non-commutativity nature. By exploiting this difference, quantum estimation theory aims to attain ultra-measurement precision that would otherwise be impossible with classical resources. In this thesis, we reviewed all the fundamental principles of classical estimation theory. Next, we extend our analysis to quantum estimation theory. Due to the non-commutativity of quantum mechanics, we prove the different families of QFIs and the corresponding QCRBs. We compared these bounds and discussed their accessibility in the single-parameter and multiparameter estimation cases. We also introduce HCRB as the most informative alternative bound suitable for multiparameter estimation protocols. Since the quantum state of light is the most accessible in practice, we studied the quantum estimation theory with the formalism of these types of quantum states. We formulate, with complete generality, the quantum estimation theory for Gaussian states in terms of their first and second moments.  Furthermore, we address the motivation behind using Gaussian quantum resources and their advantages in reaching the standard quantum limits under realistic noise. In this context, we propose and analyze a measurement scheme that aims to exploit quantum Gaussian entangled states to estimate the displacement parameters under a noisy Gaussian environment.\\
\underline{\bf Keywords:} Classical estimation theory, Quantum estimation theory, Gaussian state, Gaussian noise channels, Standard quantum limit,  Entanglement\\
\vspace{-1cm}
\begin{center}
	{\textcolor{black}{\fontfamily{pnc}{\selectfont
				\noindent\hrule height 1.1pt\vskip 0.2ex\nobreak
	}}}
\end{center}
\vspace{-0.5cm}
\begin{center}
	{\Large \textbf{Résumé}}
\end{center} 
\vspace{-.3cm}
\lettrine[lines=2]L{a théorie} de l'estimation quantique est une reformulation de la théorie statistique aléatoire avec le langage moderne de la mécanique quantique. Puisque le langage mathématique de la mécanique quantique est basé sur la théorie des opérateurs, la fonction de densité de probabilité des statistiques conventionnelles est remplacée par l'opérateur de densité apparaissant dans sa contrepartie quantique. Ainsi, l'opérateur de densité joue un rôle similaire à celui de la fonction de densité de probabilité dans la théorie classique des probabilités et des statistiques. Cependant, l'utilisation des fonctions de distribution de probabilité dans les théories classiques est fondée sur des prémisses qui semblent intuitivement assez claires. Alors qu'en théorie quantique, la situation des opérateurs est différente en raison de leur nature non-commutative. En exploitant cette différence, la théorie de l'estimation quantique vise à atteindre une ultra-précision de mesure qui serait autrement impossible avec les ressources classiques. Dans cette thèse, nous avons passé en revue tous les principes fondamentaux de la théorie de l'estimation classique. Ensuite, nous étendons notre analyse à la théorie de l'estimation quantique. En raison de la non-commutativité de la mécanique quantique, nous prouvons les différentes familles de QFIs et les QCRBs correspondants. Nous avons comparé ces bornes et discuté de leur accessibilité dans les cas d'estimation à un et plusieurs paramètres. Nous présentons également le HCRB comme la limite alternative la plus informative adaptée aux protocoles d'estimation multiparamètres. L'état quantique de la lumière étant le plus accessible en pratique, nous avons étudié la théorie de l'estimation quantique avec le formalisme de ces types d'états quantiques. Nous formulons, avec une généralité complète, la théorie de l'estimation quantique pour les états gaussiens en termes de leurs premiers et seconds moments.  En outre, nous abordons la motivation derrière l'utilisation des ressources quantiques gaussiennes et leurs avantages pour atteindre les limites quantiques standard sous un bruit réaliste. Dans ce contexte, nous proposons et analysons un schéma de mesure qui vise à exploiter les états quantiques gaussiens intriqués pour estimer les paramètres de déplacement dans un environnement gaussien bruyant.\\
\underline{\bf Mots clés:} Théorie de l'estimation classique, Théorie de l'estimation quantique, Etat Gaussien, Canaux gaussiens bruyants, Limite quantique standard, Intrication.\\
	\vspace{.7cm}
\begin{center}
	\begin{minipage}{17cm}
		\begin{center}
			{\textcolor{black}{\fontfamily{pnc}{\selectfont
				{ Année Universitaire : {2021/2022}}
				\vspace{0.2cm}
				\noindent\hrule height 1.79pt\vskip 0.2ex\nobreak
				}}}
		\end{center}
	\end{minipage}

	{ \XBox} Faculté des Sciences, avenue Ibn Battouta, BP. 1014 RP, Rabat –Maroc\\
	\phone \hspace*{0.2cm}00212(0) 37 77 18 76, \hspace*{0.1cm} {\large \bell}Fax:\hspace*{0.1cm} 00212(0) 37 77 42 61 ; http://www.fsr.um5.ac.ma
\end{center}

\end{document}